\documentclass[12pt]{article} % 12-point font

\usepackage[margin=1in]{geometry} % set page to 1-inch margins
\usepackage{amsmath} % for math
\usepackage{amssymb} % like \Rightarrow
\usepackage{setspace, lscape}
\usepackage{caption, subcaption}
\usepackage{pdfsync}
\usepackage{graphicx}  % for figures
\usepackage{float} % Put figure exactly where I want [H]

\def\elbo{\text{ELBO}}
\def\miss{\text{missing}}
\def\sigmoid{\text{sigmoid}}
\newcommand{\dquote}[1]{\lq\lq #1\rq\rq}

\usepackage{natbib}

\usepackage{bm}
\usepackage{bbm}
\newcommand{\ds}{\displaystyle}
\newcommand{\p}[1]{\left(#1\right)}
\newcommand{\bk}[1]{\left[#1\right]}
\newcommand{\bc}[1]{ \left\{#1\right\} }
\newcommand{\abs}[1]{ \left|#1\right| }
\newcommand{\G}{ \text{Gamma} }

\newcommand{\Uniform}{ \text{Unif} }
\newcommand{\E}{ \text{E} }
\newcommand{\iid}{\overset{iid}{\sim}}
\newcommand{\ind}{\overset{ind}{\sim}}

\def\bet{\bm{\eta}}
\allowdisplaybreaks

\def\logit{\text{logit}}

\def\N{\text{N}}
\def\G{\text{Ga}}
\def\IG{\text{IG}}
\def\Dir{\text{Dir}}
\def\Ber{\text{Ber}}
\def\Be{\text{Be}}
\def\lin{\lambda_{i,n}}
\def\btheta{\bm{\theta}}
\def\y{\bm{y}}
\newcommand\m{\bm{m}}
\def\mus{\mu^\star}

\def\rest{\text{rest}}
\def\Z{\bm{Z}}

\newcommand{\true}{{\mbox{\tiny TR}}}
\newcommand{\bZ}{\mbox{\boldmath $Z$}}

\newcommand{\bw}{\mbox{\boldmath $w$}}

\usepackage[dvipsnames,usenames]{color}

\usepackage{hyperref}
\hypersetup{
  pdfborder={0 0 0} % removes red box from links
}
\usepackage{pdfsync}
\begin{document}

\title{
  A Bayesian Feature Allocation Model for Identification of Cell Subpopulations
 Using Cytometry Data
}

\author{
  Arthur Lui\thanks{Address for correspondence:
                    Department of Statistics,
                    Baskin School of Engineering,
                    University of California Santa Cruz,
                    1156 High Street,
                    Santa Cruz, CA 95064 USA.
                    E-mail: alui2@ucsc.edu.},
  Juhee Lee\\
    {\small Dep.\ of Statistics,
    Univ.\ of California at Santa Cruz, Santa Cruz, CA} \and
  Peter F.\ Thall\\
    {\small Dep.\ of Biostatistics,
            M.D. Anderson Cancer Center, Houston, TX} \and
  May Daher, Katy Rezvani, Rafet Barar\\
    {\small Dep.\ of Stem Cell Transplantation and Cellular Therapy,} \\
    {\small M.D. Anderson Cancer Center, Houston, TX}
}

\date{\today}
\maketitle

% TODO: Last
\begin{abstract}
  \noindent
  % \hh haven't read. please make changes as needed \ech
  A Bayesian feature
  allocation model (FAM) is presented for identifying cell subpopulations based
  on multiple samples of cell surface or intracellular marker expression level data obtained
  by cytometry by time of flight (CyTOF).
  Cell subpopulations are characterized by differences in expression patterns of makers, and individual cells are clustered into the subpopulations based on the {\it patterns} of their observed expression levels.
  % A finite Indian buffet process is used to model cell phenotypes as
  % latent features.
  A finite Indian buffet process is used to model subpopulations as latent
  features,  and a model-based method based on these latent feature subpopulations is used  to construct cell clusters within each sample. %Each
  %cell in each sample belongs to one subpopulation, but each surface marker can
  %belong to more than one subpopulation.
  Non-ignorable missing
  data due to technical artifacts in mass cytometry instruments are accounted
  for by defining a static missingship mechanism. In contrast to
  conventional cell clustering methods  based on observed marker expression
  levels that are applied separately to different samples,
  % by using latent variables to identify and place probability distributions
  % on phenotypes,
 the FAM based  method can be applied simultaneously to multiple samples,
 and can identify important cell subpopulations likely to be missed by conventional clustering.
  The proposed FAM based method is applied to jointly analyze
  three datasets, generated by CyTOF, to study natural killer (NK) cells. %, a lymphocyte subset that plays a critical role in   cancer immune surveillance and provides a first-line of defense against viruses and transformed tumor cells.
  Because the subpopulations identified by the FAM may define novel NK cell subsets, this statistical analysis may provide  useful  information about  the biology of NK cells and their potential role in cancer immunotherapy which may lead, in turn, to development of improved cellular therapies. Simulation studies of the proposed method's behavior under two cases of known subpopulations also are presented, followed by analysis of the CyTOF NK cell surface marker data.


\end{abstract}

%\newpage

%%% BEGINNING_OF_SECTIONS -- DO NOT REMOVE!!! %%%
\section{Introduction}\label{sec:intro}
Mass cytometry data have been used for high-throughput characterization of cell subpopulations based
on unique combinations of surface or intracellular markers that may be expressed by each cell.
Cytometry by time-of-flight (CyTOF) is new technology that can rapidly
quantify a large number of biological, phenotypic, or functional markers on single cells through use of metal-tagged antibodies.  For example, CyTOF can
identify up to 40 cell surface or intracellular markers in less time and at a higher
resolution than previously available methods, such as fluorescence cytometry
\citep{cheung2011screening}.
%
%When a given marker is not expressed on a cell event, it is given a value of zero for that given channel. This is because no metal ion signal is associated with that particular marker. To bring down the height of the resulting zero peak, the default setting in the software distributes the zeroes randomly between 0 and -1.
%
Because CyTOF can reveal cellular diversity and heterogeneity that
could not be seen previously, it has the potential to rapidly advance the
study of cellular phenotype and function in immunology.

Despite the potential of CyTOF, analysis of the data that it generates is
computationally expensive and challenging, and
statistical tools for making inferences about cell subpopulations identified
by CyTOF are quite limited. Manual \lq\lq gating\rq\rq\ is a traditional
method in which homogeneous cell clusters are sequentially identified and
refined based on a given set of surface markers. Manual gating has several
severe shortcomings, however, including its inherent subjectivity due to the
fact that it requires manual analysis, and being unscalable for high
dimensional data with large numbers of markers. While manual gating is used
very commonly in practice, a variety of computational methods that
automatically identify cell clusters have been proposed to analyze
high-dimensional cytometry data. Many existing automated methods use
dimension reduction techniques and/or clustering methods, such as density-based
or model-based clustering. For example, FlowSOM in \cite{van2015flowsom} uses an
unsupervised neural-network-based method, called a self-organizing map (SOM),
for clustering and dimension reduction. A low-dimensional representation of
the marker vectors is obtained by using unsupervised neural networks for easy
visualization in a graph called a map. FlowSOM is fast and can be used either
as a starting point for manual gating, or as a visualization tool after
gating has been performed. Other common approaches are density-based
clustering methods, including DBSCAN \citep{ester1996density} and ClusterX
\citep{chen2016cytofkit}, and model-based clustering methods, including
flowClust \citep{lo2009flowclust} and BayesFlow
\citep{johnsson2016bayesflow}, among many others. More sophisticated
clustering methods based on Bayesian nonparametric models also have been
proposed, for example by \cite{soriano2019mixture}).
\cite{weber2016comparison} performed a study to compare several clustering methods for high-dimensional cytometry data. They
analyzed six publicly available cytometry datasets and compared identified
cell subpopulations to cell population identities known from expert manual
gating. They found that, in many scenarios, FlowSOM had significantly shorter
runtimes, and in many studies where manual gating was performed FlowSOM
produced the best clusterings in terms of a metric that characterizes how
well a clustering algorithm performs, compared to cell clustering by
manual gating.

% Cell markers are a very useful way to identify a specific cell population.
% However, they will often be expressed on more than one cell type. The
% immunophenotypic profiles of leukemia cell populations should preferably be
% based on a detailed comparison of the phenotypes of individual cells for all
% markers together.

\begin{figure}[t!]
%\begin{figure}[th!]
  \begin{center}
  \includegraphics[width=0.7\columnwidth]{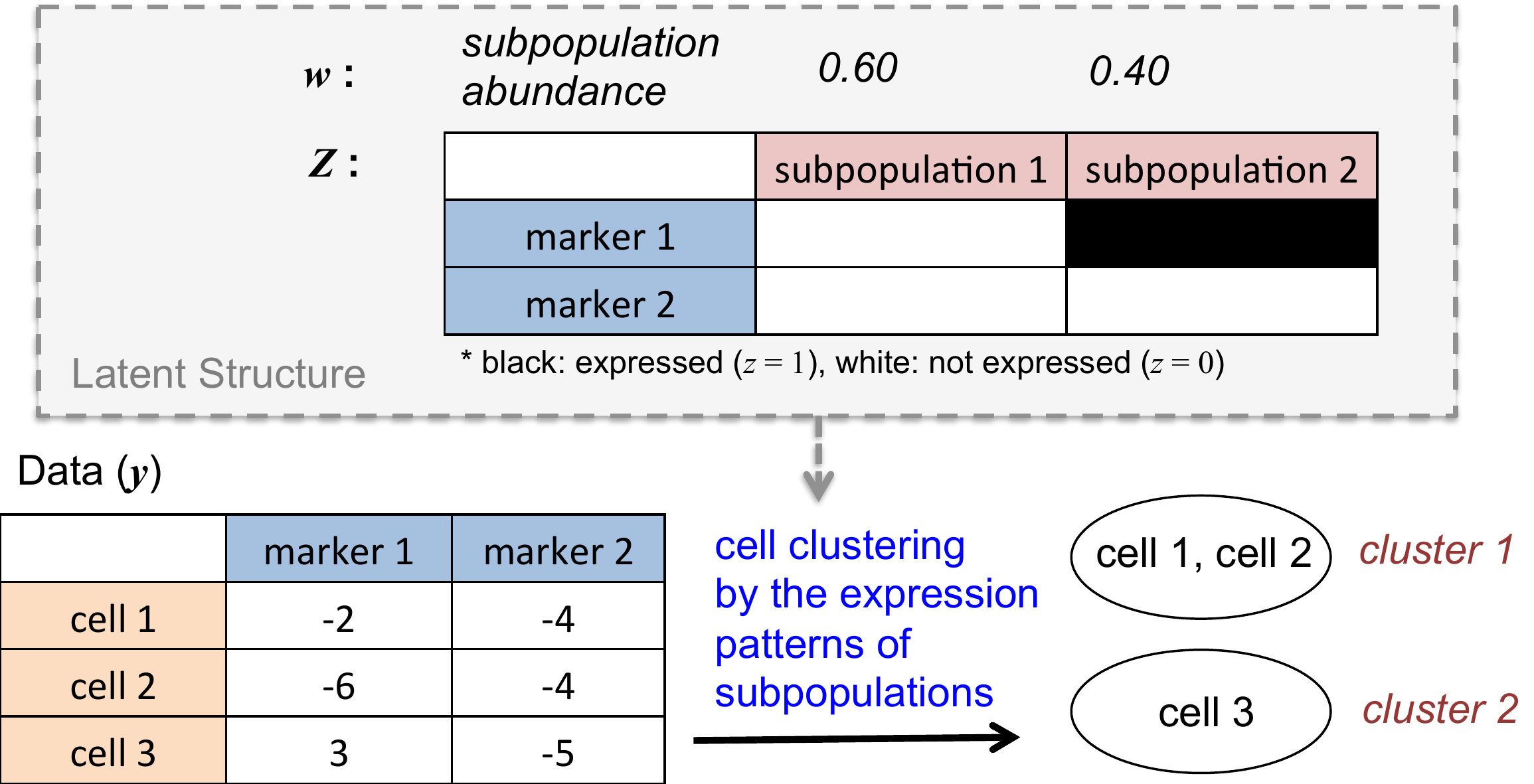}
  \end{center}
  \vspace{-0.1in}
  \caption{\small A stylized overview of the proposed feature allocation
  model (FAM). $\bZ$ is a binary matrix whose columns define latent
  subpopluations, and $\bw$ is a vector of abundances of the cell subpopluations.
  Two subpopluations are constructed in $\bZ$ based on their marker expression
  patterns. Cells are clustered to the subpopluations based on the patterns of their observed expression levels. }
\label{fig:overview}
\end{figure}

While conventional clustering methods identify subgroups of cells with
similar marker expression values, they often fail to provide direct inference
on the identification and characterization of cell subpopulations. With
clustering methods, cells are clustered together if their expression levels
are similar, and it is assumed implicitly that underlying cell subpopulations can be identified and constructed
from clusters estimated directly from the marker expression levels. The
usefulness of such conventional clustering approaches also is limited by the
fact that observed numerical marker expression values may differ substantially due to
variability between samples or between markers. Fig~\ref{fig:overview}
illustrates a toy example. Suppose that the respective log expression levels
of markers 1 and 2 are -2 and -4 on a given cell, and that the respective log
expression levels of the markers on a second cell are -6 and -4. A negative
(positive) log expression level implies that it is unlikely (likely) that a
surface marker is expressed. Although their expression patterns are similar
and have the same subpopulation, a conventional clustering method is unlikely to
include these two cells in the same cluster because their marker 1 expression
levels are very different. Furthermore, expression levels can differ
significantly between samples, often due to technical variation in the
cytometry measurement process, and cell clusters based on actual expression values may not serve as a useful surrogate for cell subpopulations. As a result, most existing clustering methods are used to
analyze different samples separately.

%%%%%%%%%%%%%%%%%%%%%%%%
%%%%%%%%%%%%%%%%%%%%%%%%%

In this paper, we propose a Bayesian feature allocation model (FAM) to
identify and place probabilities on cell subpopluations  based on multiple
cytometry samples of cell surface marker expression values. Our proposed FAM characterizes cell subpopluations as latent features defined in
terms of their expression patterns, and cluster individual cells to one of the identified subpopulations. We will refer to each latent feature
as a \lq\lq subpopulation." Markers often are
expressed in more than one cell subpopulation, and different subpopluations can be
characterized by distinctive patterns of marker expressions. To represent subpopluation configurations, we introduce a
random binary matrix $\bZ$ whose rows and columns correspond to markers and subpopluations,
respectively. We let 0 and 1 represent the expression and non-expression of a
marker in a subpopluation, respectively. Using the toy example in Fig~\ref{fig:overview},
in contrast to clustering methods, the FAM constructs latent
subpopluations based on marker expression patterns as in $\bZ$ (top of the
figure).  It assigns cells 1 and 2 to subpopulation 1, for which neither marker is expressed, and it
assigns cell 3 to a subpopulation where marker 1 is expressed and marker 2 is not expressed.
(bottom right).
We assume a finite Indian buffet process
(IBP), as a prior distribution for $\bZ$.
The IBP is a popular model for latent binary features, and may be obtained by taking the infinite limit of a
Beta-Bernoulli process \citep{ghahramani2006infinite}. Applications of the
IBP as FAMs for a range of biological applications are given by
\cite{hai2011inferring, chen2013phylogenetic, xu2013nonparametric,
sengupta2014bayclone, xu2015mad, lee2015bayesian, lee2016bayesian,
ni2018bayesian}. \cite{griffiths2011indian} reviews some earlier applications
of the IBP.
%%%
%%%
Furthermore, we introduce a vector of subpopulation abundances $\bw,$ and allow the cell
samples to have different values of $\bw$. This approach provides a framework for joint
analysis of multiple samples, and includes structures to account for large
sample-to-sample variation and abnormalities, such as missing values due to
technical artifacts in the cytometry data, while quantifying uncertainty in
posterior inferences.

This work is motivated by a dataset comprised  of three CyTOF samples of surface
marker expression levels in umbilical cord blood (UCB)--derived
natural killer (NK) cells.

 NK cells play a critical role in cancer immune
surveillance, and are the first line of defense against viruses and
transformed tumor cells. NK cells have the intrinsic ability to infiltrate
cancer tissues. Recently,  NK cells have been used therapeutically to treat a variety of diseases \citep{wu2003natural, lanier2008up}.
In particular, NK cells have emerged as a potentially powerful
treatment modality for advanced cancers refractory to conventional therapies \citep{rezvani2015application,
suck2016natural, shah2017phase, miller2005successful, lupo2019natural}.
 Because cell-surface protein expression levels are used as
markers to examine the behavior of NK cells, accurate identification of
diverse NK-cell subpopulations along with their composition is crucial to the
process of obtaining more complete characterizations of their biological
processes and functions. The goal of our statistical analysis is to
identify and characterize NK cell subpopulations and functions across heterogeneous
collections of these cells. This may provide critical information for guiding selective {\it ex vivo}
expansion of UCB-derived NK cells for treating specific cancer types.

The remainder of the paper is organized as follows. We present the proposed
statistical model in \S~\ref{sec:prob-model}, simulation studies in
\S~\ref{sec:sim-study}, and an analysis of the NK cell mass cytometry data in
\S~\ref{sec:cb-analysis}. We close with concluding remarks in
\S~\ref{sec:conclusions}.

\section{Probability Model}\label{sec:prob-model}
\subsection{Sampling Model}
Index cell samples by $i = 1,2,...,I$.  Suppose that $N_i$ cells, indexed by
$n=1, \ldots, N_i$, are obtained from the $i^{th}$ sample, and the expression
levels of $J$ markers on each cell within each sample are measured.
Let $\tilde{y}_{i,n,j} \in \mathbb{R}^+$ denote the raw measurement of the
expression level of marker $j$ on cell $n$ in sample $i$. While raw measurement
values reflect actual expression or non-expression of markers on cells,  they
also vary between cells and between samples for several reasons, including biological heterogeneity in the range of expression among different populations, as well as experimental artifacts or batch effects, such as instrument fluctuations or signal crosstalk among channels designed for
different markers.
While, compared to conventional flow cytometry and the use of fluorescent antibodies, the use of pure metal isotopes minimizes spectral overlap among measurement channels in CyTOF, crosstalk still may be observed due to the presence of isotopic impurity, oxide formation, and properties related to the mass cytometer.
Raw measurements are normalized using cutoff values computed by a flow
(rather than mass) cytometry algorithm called flowDensity
\citep{malek2014flowdensity}, which aims to gate predefined cell populations
of interest, in settings where the gating strategy is known.
This frees practitioners from the need to manually gate analysis results, but it
relies substantially on user-provided information to produce good results.
Consequently, cutoffs obtained from such algorithms are crude, but are useful as
a starting point for our analysis.
Let $c_{i,j}$ denote the cutoff obtained for marker $j$ in sample $i$.  A
marker of a cell is likely to be expressed if its observed expression level
$\tilde{y}_{i,n,j} > c_{i,j}$, while a value  $\tilde{y}_{i,n,j} < c_{i,j}$
may imply that marker $j$ is not expressed on cell $n$ in sample $i$.  To
reduce skewness of the marker distributions, we will consider the log
transformed values
% $\tilde{y}_{i,n,j}$ by $c_{i,j}$,
$
y_{i,n,j}=\log\p{\tilde{y}_{i,n,j}/c_{i,j}} \in \mathbb{R}.
$
This transformation makes 0 the reference point for dichotomizing
marker expression and non-expression.  To account for the fact that some
$y_{i,n,j}$ may be missing due to experimental artifacts, we define the binary
indicator $m_{i,n,j} = 1$ if $y_{i,n,j}$ is observed, and $m_{i,n,j} = 0$ if
missing.  Denote the probability that $y_{i,n,j}$ is observed by
Pr$(m_{i,n,j}=1)=1-\rho_{i,n,j}(y_{i,n,j})$.  Below, we will define the latent
subpopulation membership indicator, $\lambda_{i,n},$ of cell $n$ in sample $i.$ For each cell in the
$i^{th}$ sample, we assume conditional independence of the cell's $J$ marker
values given its  latent subpopulation,  formally $y_{i,n,1},\cdots,  y_{i,n,J} \mid
\lambda_{i,n}$  are independent, and we assume the following joint model for
$y_{i,n,j}$ and $m_{i,n,j}$,
\begin{eqnarray}
y_{i,n,j} \mid \mu_{i,n,j}, s_{i,n}^2, \lambda_{i,n} \ind
  \N(\mu_{i,n,j}, s^2_{i,n}), \mbox{ and }
m_{i,n,j} \mid \rho_{i,n,j}(y_{i,n,j}), \lambda_{i,n} \ind
  \Ber(1-\rho_{i,n,j}(y_{i,n,j})).
\label{eq:joint-like}
\end{eqnarray}
Below, we will relate the mean expression $\mu_{i,n,j}$ to the configuration of
cell subpopulation $\lambda_{i,n}$.   To reflect expert biological knowledge of the
investigators, a model for $\rho_{i,n,j}$ as a function of $y_{i,n,j}$ will be
given in the following section.

\subsection{Priors}\label{priors}
\paragraph*{Priors for latent cell subpopulation}\

\noindent We assume that each sample has a heterogeneous cell population, and
denote the number of different latent subpopulations by $K$. The cell subpopulations
are defined by columns of a $J \times K$ (marker, subpopulation) stochastic binary
matrix $\Z$. The element $z_{j, k} \in \{0, 1\}$ of $\Z$ determines marker
expression by subpopulation,  with $z_{j,k}=0$ if marker $j$ is not expressed  and
$z_{j,k}=1$ if it is expressed for subpopulation $k$. We construct a {\it feature
allocation prior} for $\Z$ as follows: For $j=1, \ldots J$ and $k=1, \ldots,
K,$
\begin{eqnarray}
   z_{j,k} \mid v_k  \ind \Ber\p{v_k}  ~~ \mbox{ and }~~
  v_k \mid \alpha \iid \Be(\alpha/K, 1). \label{eq:FAM}
\end{eqnarray}
As $K \rightarrow \infty$, the limiting distribution of $\Z$ in \eqref{eq:FAM}
is the IBP \citep{ghahramani2006infinite} with parameter $\alpha$, after
removing all columns that contain only zeros.  We assume hyperprior  $\alpha
\sim \text{Gamma}(a_\alpha, b_\alpha)$ with mean $a_\alpha/b_\alpha$.
The IBP, which is one of the most popular FAMs, thus defines a distribution
over binary matrices having an unbounded number of columns (features).  In the
present context, this Bayesian model provides a very useful statistical tool
for identifying marker expression patterns to define
latent cell subpopulations.  %This provides a basis for making inferences about
%underlying cell population structures.

%Realizations from an IBP are sparse with an unbouned number of columns
%(features). This is desireable in our application as we anticipate
%a relatively small number of unique cell types ($\ll N_i$) to
%be present in each sample.

We assume that each of the $K$ cell subpopulations is possible in each sample, but
allow their cellular fractions to differ between samples. In addition, we
include a special, $(K+1)^{st}$ cell type, called a \lq\lq noisy cell,\rq\rq\
to address the problem that some cells do not belong to any of the $K$ cell
subpopulations.  In sample $i$, let $0 < \epsilon_i < 1$ denote the proportion of
noisy cells and $(1-\epsilon_i)w_{ik}$ the proportion of subpopulation $k$, where
$\bw_i$ =$(w_{i,1},\ldots, w_{i,K})$ with $\sum_{k=1}^K w_{i,k}=1$ and
$w_{i,k}>0,$ is a probability distribution on $\{1,\cdots,K\}.$   We assume
priors $\epsilon_i \iid \Be(a_\epsilon, b_\epsilon)$ with fixed hyperparameters
$a_\epsilon$ and $b_\epsilon$, and $\bw_i \mid K \iid \Dir_K(d/K)$ with fixed
hyperparameter $d$. For cell $n=1,\ldots, N_i$ in sample $i=1, \ldots, I,$ we
introduce stochastic {\it latent subpopulation indicators} (equivalently, cell cluster memberships) $\lambda_{i,n} \in \{0,
1, \ldots, K\}$.  We set  $\lin=0$ if cell $n$ in sample $i$ does not belong to
any of the cell subpopulations  in $\Z$, and set $\lambda_{i,n}=k>0$  if cell $n$ in
sample $i$ belongs to subpopulation $k$.  For the latent subpopulation indicators, we
assume $\Pr(\lin=0 \mid \epsilon_i) = \epsilon_i$ to account for noisy cells,
and $\Pr(\lin=k \mid \lin \neq 0, \bw_i) = w_{ik}$.  Within each sample
$i=1,\cdots,I,$ assigning cells to subpopulations using $\{\lin,\ i=1,\cdots,N_i\}$
induces cell clusters.  Thus, in contrast with
clustering methods that infer only cell clusters in the $i^{th}$ sample based
on $\{y_{i,n,j}\},$ our proposed  method produces direct inferences on both characterization of cell
subpopulations and cell clusters, simultaneously for all samples.   This is highly
desirable because a primary aim is to identify and make inferences about cell
subpopulations.

Since the number of columns containing non-zero entries under the IBP is
random,  the dimensions of $\Z$ and $\bw_i$ may vary during posterior
computation. Because this dimension change may cause a high computational
cost, especially for big datasets such as those obtained by CyTOF,  we use a
finite version of the IBP by fixing $K$.  To accommodate the fact that the
number of latent subpopulations is not known {\it a
priori}, we consider a set of different values for $K,$ from which we select
one value of $K$ using Bayesian model selection criteria. We will discuss this
selection process in detail below.

\paragraph*{Priors for mean expression level}\

\noindent
The mean expression level $\mu_{i,n,j}$ of marker $j$ on cell $n$ in sample $i$
in \eqref{eq:joint-like} is determined by characterizing the cell's
latent subpopulation. Recall that a cell $n$ either belongs to a subpopulation $\lin=k >0$ in column
$k$ of $\bZ,$ or the noisy cell subpopulation $\lin=0$.
%%%
For cells with a noisy cell subpopulation, we fix $\mu_{i,n,j}=0$ for all $j$ and
$s^2_{i,n}=s^2_\epsilon$, where $s^2_\epsilon$ is fixed at a large value.
%%%
For a cell with $\lin \in \{1,\cdots,K\}$, if the marker is not expressed in
cell subpopulation $\lin$ (i.e., $z_{j, \lambda_{i,n}}=0$), we let its mean
expression level take a negative value, $\mu_{i,n,j} < 0$.
%{\bf This accommodates cases where $y_{i,n,j}< c_{i,j},$ since each cutoff
%$c_{i,j}$ is determined by a data preprocessing algorithm that may include
%estimation errors}.
In particular, for $(i,n,j)$ with $z_{j, \lambda_{i,n}}=0$, we introduce a set
of means for expression levels of markers not expressed, $\mus_{0,\ell}=
\sum_{r=1}^\ell \delta_{0,r}$, where $\delta_{0, \ell} \iid \text{TN}^-(\psi_0,
\tau_0^2)$, $\ell=1, \ldots, L_0$ with fixed $L_0$. Here $\text{TN}^-(\psi,
\tau^2)$ denotes the normal distribution with mean $\psi$ and variance $\tau^2$
truncated above at zero.  This construction induces the ordering $0 >
\mus_{0,1} > \ldots > \mus_{0,L_0}$.  We then let $\mu_{i,n,j}=\mus_{0, \ell}$
with probability $\eta^0_{i,j,\ell}$.  Note that even for a marker not expressed,
positive $y_{i,n,j}$ can be observed due to measurement error or
estimation error in the cutoff $c_{i,j}$, and the model accounts for such cases
through $s^2_{i,n}$.
%%%
Similarly, we assume that the mean expression level of marker $j$ takes a
positive value ($\mu_{i,n,j} > 0$) if the marker is expressed ($z_{j,
\lambda_{i,n}}=1$).  For cases with $z_{j, \lambda_{i,n}}=1$, we construct
another set of $\delta$, $\delta_{1, \ell} \iid \text{TN}^+(\psi_1, \tau_1^2)$,
$\ell=1, \ldots, L_1$ with fixed $L_1$, where $\text{TN}^+(\psi, \tau^2)$
denotes the normal distribution truncated below at zero with mean $\psi$ and
variance $\tau^2.$ We let $\mus_{1,\ell}= \sum_{r=1}^\ell \delta_{1,r}$, so $0
< \mus_{1,1} < \ldots < \mus_{1,L_1}$.  We then let $\mu_{i,n,j}=\mus_{1,\ell}$
with probability $\eta^1_{i,j,\ell}$. We also let $s_{i,n}^2=\sigma_i^2$ for
$\lin >0$ and assume $\sigma^2_i \ind \IG(a_\sigma, b_\sigma)$.  This leads to
a mixture of normals for $y_{i,n,j}$ whose location parameters are determined
by the cell's (latent) subpopulation,
\begin{eqnarray}
y_{i,n,j} \mid z_{j,\lambda_{i,n}}=z, \bm\mus_z, \bm\eta^z_{i,j}, \sigma^2_i
~\ind~
F_{i,j}^z=\sum_{\ell=1}^{L_z} \eta^z_{i,j,\ell} \N(\mus_{z,\ell}, \sigma_i^2),
~ z \in \{0, 1\}, ~ k > 0.
\label{eq:mixture-like}
\end{eqnarray}
Finally, we let $\bm\eta^z_{i,j}
\iid\Dir_{L_z}(a_{\eta^z}/L_z)$, $z=0, 1$, $i=1, \ldots, I$ and $j=1, \ldots, J$.

The mixture model in \eqref{eq:mixture-like} encompasses a wide class of
distributions, such as multi-modal or skewed distributions. It captures
virtually any departure from a conventional distribution, such as a parametric
exponential family model, that may appear to give a good fit to the
log-transformed expression values.  A key property of  (\ref{eq:mixture-like})
is that it allows cells with very different numerical expression values to have
the same subpopulation if their  marker expression/non-expression pattern is the
same.  This provides a basis for obtaining a succinct representation of cell
subpopulations.
%%%
Because all $(i, n, j)$ share the locations $\mus_{z}$ in
\eqref{eq:mixture-like}, the model borrows strength across both samples and
markers, while $\bm \eta^z_{i,j}=(\eta^z_{i,j,1}, \ldots, \eta^z_{i,j,L^z})$ allows
the distribution of $y_{i,n,j}$ to vary across both sample $i$ and marker $j$.
The construction of $\mus_{z, \ell}$ through $\delta_{z, \ell}$ also ensures
ordering in $\mus_{z, \ell}$ and circumvents potential identifiability and
label-switching issues that may be present in conventional Bayesian
mixture models
% , where the mixture components are not identifiable without imposing further
% restrictions on the model parameters.
\citep{celeux2000computational, stephens2000dealing, jasra2005markov,
fruhwirth2006finite}.

\paragraph*{Model for data missingship mechanism}\

\noindent
We next build a model for the data missingship mechanism.  To do this, we
incorporate information provided by a subject area expert, that a marker
expression level is recorded as \lq\lq missing\rq\rq\ in a cell when the marker
has a very weak signal, strongly implying that the marker is not expressed on
that cell.
There is an extensive literature for analyzing data with observations missing
not at random, including methods for Bayesian data imputation and frequentist
multiple imputation (\cite{rubin1974characterizing, rubin1976inference,
allison2001missing, schafer2002missing, franks2016non}).

The dataset does not contain information for inferring the missingness mechanism, and any assumptions for the distribution of unobserved data are not testable. Consequently, it cannot be anticipated that
the imputed value of missing $y_{i,n,j},$ under any assumed missingness mechanism, is close to its potentially observed numerical values,  except for the key fact that the potential value is very likely negative. We thus focus on estimating the probability that a missing value is no expression of a marker, since the task of recovering $\bZ$, $\bm w$ and $\bm \lambda$, which are the primary interest,  is not affected.
We model missingship conditional on $y_{i,n,j}$ by
assuming a logit regression model for the probability $\rho_{i,n,j}$ that
$y_{i,n,j}$ is missing,
\begin{eqnarray}
\logit(\rho_{i,n,j}) =
\beta_{0i} + \beta_{1i} y_{i,n,j} + \beta_{2i} y_{i,n,j}^2. \label{eq:link}
\end{eqnarray}
This quadratic function of $y_{i,n,j}$ is assumed in the real-valued domain of
$\logit(\rho_{i,n,j})$ to allow values of the $\bm{\beta}_i$ = $(\beta_{0i},\beta_{1i},\beta_{2i})$ in
the $i^{th}$ sample for which $y_{i,n,j}<0$
yields a larger probability $\rho_{i,n,j}$ of being missing.  To specify values of $\bm{\beta}_i$ in
\eqref{eq:link} for each sample $i=1,\cdots,I$, we take an empirical approach, using the
minimum, first quartile, and median of negative $y_{i,n,j}$ values, set their
$\rho_{i,n,j}$ values to .05, .80 and .50, respectively, and solve for
$\bm{\beta}_i$.  Under this specification of $\bm{\beta}_i$,  imputed values of $y_{i,n,j}$
take a negative value with large probability and their distributions are very
similar to those of observed $y_{i,n,j}<0$ in sample $i.$
We performed sensitivity analyses to
the specification of values of the $\bm{\beta}_i$'s in this way, to examine robustness of the estimation of $\bZ$, $\bm w$ and $\bm \lambda$, the parameters of primary interest.  Additionally,  in our
simulation studies, missing values were generated under a mechanism different
from that in \eqref{eq:link} to further examine robustness.
\S~\ref{sec:sim-study} and \S~\ref{sec:cb-analysis} provide details of the
sensitivity analyses.

\def\cpo{\text{CPO}}
\def\lpml{\text{LPML}}
\def\data{\text{data}}

\paragraph*{Selection of $K$}\label{par:sel-K}\

\noindent
Instead of estimating $K$, we cast the problem of selecting a value for $K$ as
a model comparison problem. This approach reduces computational burden,
especially for large datasets, but identifying a value of $K$ that optimizes
model fit while penalizing for high model complexity is still challenging.  We
choose $K$ using two model selection criteria, the deviance criterion information
(DIC, \cite{dic}) and log pseudo marginal likelihood (LPML,
\cite{geisser1979predictive, gelfand1994bayesian}).
% TODO:
% \hh also include some early papers that actually develop LPML.  \ech
The DIC and LPML are commonly used to quantify goodness-of-fit for model
comparison in the Bayesian paradigm. The DIC measures posterior prediction
error based on deviance penalized by model complexity, with smaller values
corresponding to a better fit.  The LPML is a metric based on cross-validated
posterior predictive probability, and is defined as the sum of the logarithms
of conditional predictive ordinates (CPOs), with larger LPML corresponding to a
better fit to the data. Details of the computation of DIC and LPML are given in
Supp.\ \S \ref{sec:lpml-dic}.  In addition, we count the number of subpopulations having negligible weights, $\sum_{i,k}
\mbox{I}(w_{i,k} < 1\%)$, for each value of $K$ and plot the LPML against the
number of such subpopulations. A model with larger $K$ may produce cell subpopulations
with very small $w_{i,k}$ that only make subtle contributions to model fit in
terms of LPML.   We thus search for a value of $K$, where the change
rate of the increase in LPML drops.  \cite{miller2018robust} used
a similar calibration method to tune a model hyperparameter that determines how
much coarsening is required to obtain a model that maximizes model fit while
maintaining low model complexity.

\subsection{Posterior Computation}\label{sec:sampling-via-mcmc}
Let $\btheta=\bc{\bZ, \bw, \bm \delta_0, \bm \delta_1, \bm \sigma^2, \bm
\eta^0, \bm \eta^1, \bm \lambda, \bm v, \bm \epsilon, \alpha}$ denote all
model parameters. Let $\y$ and $\m$ denote the vectors of $y_{i,n,j}$ and $m_{i,n,j}$ values,
respectively, for all $(i,n,j)$. The joint posterior distribution is
\begin{eqnarray}
p(\btheta \mid \y, \m, K)
&\propto &
p(\btheta \mid K) \prod_{i,n,j} p(m_{i,n,j} \mid y_{i,n,j}, \btheta, K)
p(y_{i,n,j} \mid \btheta, K)
\nonumber\\
&\propto &
p(\btheta \mid K)
\prod_{i,n} \bk{
  \prod_j
  \rho_{i,n,j}^{1 - m_{i,n,j}} \times
  \sum_{\ell=1}^{L_{z_{j, \lambda_{i,n}}}} \eta^{z_{j, \lambda_{i,n}}}_{i,j,\ell}
  \phi(y_{i,n,j} \mid \mu^\star_{{z_{j, \lambda_{i,n}}}, \ell}, \sigma^2_i)
  %\frac{1}{\sqrt{2\pi\sigma^2_{i}}}
  %\exp\bc{-\frac{(y_{i,n,j}-\mu^\star_{{z_{j, \lambda_{i,n}}}, \ell})^2}{2\sigma^2_{i}}}
  }^{1(\lin>0)} \nonumber \\
&&
\qquad \times \bk{ \prod_j \rho_{i,n,j}^{1 - m_{i,n,j}} \times
\phi(y_{i,n,j} \mid 0, s^2_\epsilon)
%\frac{1}{\sqrt{2\pi s^2_\epsilon}}
%\exp\bc{-\frac{y_{i,n,j}^2}{2 s^2_{\epsilon}}}
}^{1(\lin=0)},
\label{eq:joint-post}
\end{eqnarray}
where $\phi(y \mid \mu, \sigma^2)$ is the density of the normal distribution with mean $\mu$ and variance $\sigma^2$ evaluated at $y$.
Since $\rho_{i,n,j}$ is a constant for a given $y$ with fixed
$\beta$'s, the terms $p(m_{i,n,j}=1)=(1-\rho_{i,n,j})^{m_{i,n,j}}$ for observed
$y_{i,n,j}$ do not appear in \eqref{eq:joint-post}.  Posterior simulation can
be done via standard Markov chain Monte Carlo (MCMC) methods with Gibbs and
Metropolis steps to draw samples from the posterior distribution. Each
parameter is updated by sampling from its full conditional distribution.
Details of the posterior simulation are described in Supp.\ \S
\ref{sec:post-comp}.

Summarizing the joint posterior distribution $p(\btheta \mid \y, \m, K)$ is
challenging, especially for $\Z$, which may be susceptible to label-switching
problems common in mixture models. The distributions of $\bw_i$ and $\bm \lambda_i$
depend on $\Z$. To summarize the posterior distribution of $(\Z,\bw_i,
\bm \lambda_i)$ with point estimates, we extend the sequentially-allocated
latent structure optimization (SALSO) method in \cite{salso} and incorporate
$\bw_i$. To summarize random feature allocation matrices, we first construct
$\bm A_i=\{A_{i,(j,j')}(\bZ)\}$, the $J \times J$ pairwise allocation matrix
corresponding to a binary matrix $\bZ$, where
\begin{eqnarray}
A_{i,(j, j')}(\bZ) =
\sum_{k=1}^K w_{i,k} \times 1(z_{j,k}=1) \times 1(z_{j',k}=1),
~~~\text{for } 1\leq j, j^\prime \leq J, \label{eq:compute-A}
\end{eqnarray}
is the number of active features that markers $j$ and $j'$ have in common,
weighted by $w_{i,k}$. The from of \eqref{eq:compute-A} encourages the selection of $\bZ$
based on subpopulations that are prevalent in samples.  We then use constrained
optimization to find a point estimate $\hat{\bZ}_i$ for sample $i$ that
minimizes the sum of the element-wise squared distances,
\begin{eqnarray*}
\text{argmin}_Z\sum_{j=1}^J\sum_{j'=1}^J(A(\bZ)_{i, (j,j')} - \bar{A}_{i, (j,j')})^2
\label{eq:salso}
\end{eqnarray*}
where $\bar A_{i, (j, j^\prime)}$ is the pairwise allocation matrix averaged by
the posterior distribution of $\bZ$ and $\bw_i$. We use posterior Monte Carlo
samples to obtain posterior point estimates $\hat{\Z}_i$ as follows.  Suppose
that we obtain $B$ posterior samples simulated from the posterior distribution
of $\btheta$. For the $b^{th}$ posterior sample of $\Z$ and $\bw_i$, we compute
a $J \times J$ adjacency matrix, $\bm A_i^{(b)} =\{A^{(b)}_{i,(j,j')}\}$, $b=1,
\ldots, B$ and then the mean adjacency matrix $\bar A_i = \sum_{b=1}^B
A_i^{(b)} / B$. We determine a posterior point estimate of $\bZ$ for sample
$i$ by minimizing the mean squared deviation,
$
\hat{\bm Z}_i = \text{argmin}_{\bm Z} \sum_{j,j'} (A_{i,j,j'}^{(b)} - \bar
A_{i,j,j'})^2,
$
where $\hat\Z_i \in \bc{\Z^{(1)} \dots \Z^{(B)}}$.
For $\hat{\Z}_i = \Z^{(b)}$, we report the posterior point estimates
$\hat{\bw}_i=\bw_i^{(b)}$ and $\hat{\lambda}_{i,n}=\lambda_{i,n}^{(b)}$.

In addition, we implemented variational inference (VI), which approximates the
posterior distribution of $\btheta$ through optimization
\citep{wainwright2008graphical, blei2017variational, zhang2018advances}. Because VI
tends to be faster than MCMC, it is a popular emerging alternative to MCMC,
especially when models are complex and/or a dataset is large.  In particular, we
used  automatic differentiation variational inference (ADVI) \citep{advi},
which makes use of automatic differentiation to simplify the process of
implementing variational inference for differentiable models. ADVI requires no
model-specific hand derivations, and is relatively simple to implement when
an automatic differentiation library such as PyTorch
\citep{paszke2017automatic}, Tensorflow \citep{tensorflow2015-whitepaper},
and Flux \citep{flux} is available. Details of the VI
implementation using ADVI are included in Supp.\ \S~\ref{sec:vi}.

\section{Simulation Studies}\label{sec:sim-study}

%%% sim-study-proj1 %%%
\begin{table}[t!]
    \begin{subtable}{.5\linewidth}
      \centering
        \begin{tabular}{c}
        \includegraphics[width=.95\linewidth]{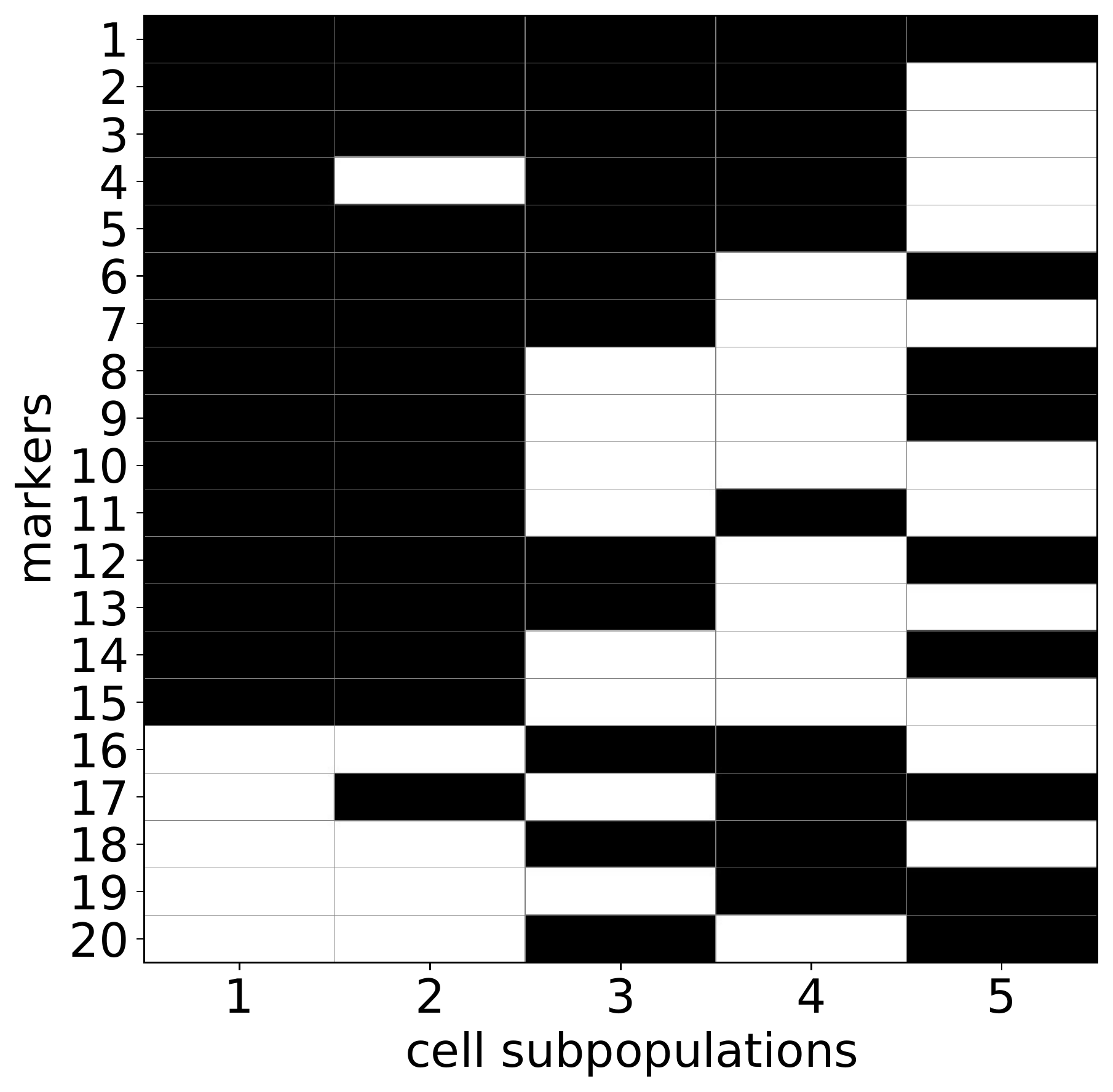}
        \end{tabular}
        \caption{ $\Z^\true$}
    \end{subtable}%
    \begin{subtable}{.5\linewidth}
      \centering
	    \begin{tabular}{|c|rrrrr|}
	      \hline
          & \multicolumn{5}{c|}{Cell Subpopulations}\\
          & $k=1$ & $k=2$ & $k=3$ & $k=4$ & $k=5$ \\
        \hline
          sample 1 & 0.068& 0.163& 0.351& 0.297& 0.118\\
          sample 2 & 0.194& 0.282& 0.066& 0.257& 0.199\\
          sample 3 & 0.112& 0.141& 0.224& 0.119& 0.402\\
        \hline
      \end{tabular}
      \caption{$\bw^\true$}
    \end{subtable}
    \caption{Design of Simulation 1. $\Z^\true$ and $\bw^\true$ are illustrated
    in (a) and (b), respectively. $K^\true=5$, $J=20$, and $I=3$ are assumed.
    In (a), black  represents $z^\true_{j,k}=1$ (marker expression) and white
    represents $z^\true_{j,k}=0$ (marker non-expression).}
\label{tab:sim1-tr}
\end{table}

In this section, we summarize simulations to  assess the performance of the
proposed FAM based method for identifying features and clustering cells within
each sample,   and compare it to an alternative model and method.  We simulated
data for three samples, each with 20 markers, consisting of 4000,
500, and 1000 cells, respectively. Thus, $I=3$, $J=20$, and $N_i=4000$, 500,
and 1000 for $i=1$, 2, and 3. We let the true number of latent
features (subpopulations) $K^\true=5$ and specified a $J \times K^\true$ (binary)
feature-allocation matrix $\bZ^\true$ and $K^\true$-dimensional vectors
$\bw^\true_i$ as follows:
%%% simualte Z
We first simulated $\bZ^\true$ by setting $z^\true_{j,k}=1$ with probability
0.6. If any column or row in $\bZ^\true$ was a column or row consisting of all 0's, the
entire matrix was re-sampled.
% Simulating W
We then simulated $\bw^\true_i$ from a Dirichlet distribution with parameters
being random permutations of $(1, \ldots, K^\true)$ for each $i$. This makes
it likely that the elements of $\bw^\true_{i}$ will contain both large and
small values. $\bZ^\true$ and $\bw^\true_i$ are shown in
Table~\ref{tab:sim1-tr}. We set abundances of the noisy cell types to be
$\epsilon_i^\true = 0.05$ for all $i$.
%% simulate y
We specified the mixture models for the expression levels by setting
$\bm\mu^{\star, \true}_{0} = (-1, -2.3, -3.5)$ and $\bm\mu^{\star, \true}_{1}
= (1, 2, 3)$ with $L^{0, \true}=L^{1, \true}=3$, and simulating mixture
weights $\bet_{i,j}^{z,\true}$ from a Dirichlet distribution with parameters
being a random permutation of $(1,\ldots, L^{z,\true})$, for $z \in \bc{0,
1}$ and for each $(i, j)$. The values of $\sigma^{2, \true}_i$ were set to
0.2, 0.1, and 0.3 for samples 1, 2, and 3, respectively. We then simulated
latent subpopulation indicators $\lambda_{i,n}^\true$ with probabilities
$\Pr(\lin^\true=0)=\epsilon_i^\true$ and $\Pr(\lin^\true=k \mid \lin^\true
\neq 0)=w_{ik}^\true$. We generated $y_{i,n,j} \iid \N(0, 9)$ for all $(i, n,
j)$ with $\lambda^\true_{i,n}=0$. Otherwise, we generated $y_{i,n,j}$ from a
mixture of normals, $\sum_{\ell=1}^{L^{z,\true}} \eta^{z, \true}_{i,j}\times
\N(\mu^{\star,\true}_{z\ell}, \sigma^{2, \true}_{i})$ given
$z^\true_{j\lin^\true}=z$ for each $(i, n, j)$.
%%% missing y
To simulate the missingship indicators, $m_{i,n,j}$, we first generated the
proportions $p_{i,j}$ of missing values for each $(i, j)$ from a $\Uniform\p{0,
0.7\sum_k w^\true_{i,k}(1-z^\true_{j,k})}$ and sampled $p_{i,j}\times N_i$
cells without replacement with probability proportional to $\{1 + \exp\p{-9.2
-2.3 y_{i,n,j}}\}^{-1}.$ Under the true missingness mechanism, a marker having
a lower expression level has a higher chance of being recorded as missing. Note
that the true mechanism is different from that assumed in \eqref{eq:link}.
Heatmaps of the simulated $\y$ are shown in Fig~\ref{fig:sim1-post}(b), (d)
and (f). The $y_{i,n,j}$'s are sorted within a sample according to their
posterior subpopulation indicator estimates $\hat{\lambda}_{i,n}$ (explained
later). The red, blue, and black colors represent high expression levels, low
expression levels, and missing values, respectively.

We fit the model separately for each $K = 2,3, \ldots, 10$.
For all $K$, we fixed $L^0=L^1=5$ and $s_\epsilon^2=10$.
We specified the remaining fixed hyper-parameters as follows:
$a_\alpha=b_\alpha=0.1$ for $\alpha$; $\psi_z=1$ and
$\tau^2_z=1$ for $\delta_{z, \ell}$; $a_\sigma=3$ and $b_\sigma=2$ for
$\sigma^2_i$; $a_{\eta^z}=1$ for $\bet_{i,j}$; $d=1$ for $\bw_i$;
$a_\epsilon=1$ and $b_\epsilon=99$ for $\epsilon_i$.  We used the empirical
approach in \S~\ref{sec:prob-model} to obtain values of $\bm\beta$ for the
missingship mechanism. For each $i$, we initialized the missing values at
$-\beta_{2i} / (2\beta_{1i})$, which correspond to the largest missing
probabilities {\it a priori}. To initialize $\lambda_{i,n}$, $\bw_i$, $\bZ$,
$\alpha$ and $\bet^z_{i,j}$, we applied density-based clustering via finite
Gaussian mixture models using the MClust package \citep{mclust}, and used the
resulting clustering of $y_{i,n,j}$.
% posterior inference
We then drew samples of $\btheta$ values and imputed missing values of
$y_{i,n,j}$ using MCMC simulation based on 16,000 iterations, discarding the
first 10,000 iterations as burn-in for each model, and then keeping every other
draw as thinning.  We diagnosed convergence and mixing of the described
posterior MCMC simulations using trace plots. We found no evidence of any
practical convergence problems. For a model with $K=5$, it took 38 minutes per
1000 iterations on an interactive Linux server with four Intel Xeon E5-4650
processors and 512 GB of random access memory.

%%%%%%%%%%%%%%%%%%%%%%%%
\begin{figure}
  \begin{center}
    \begin{tabular}{ccc}
    \includegraphics[width=.32\columnwidth]{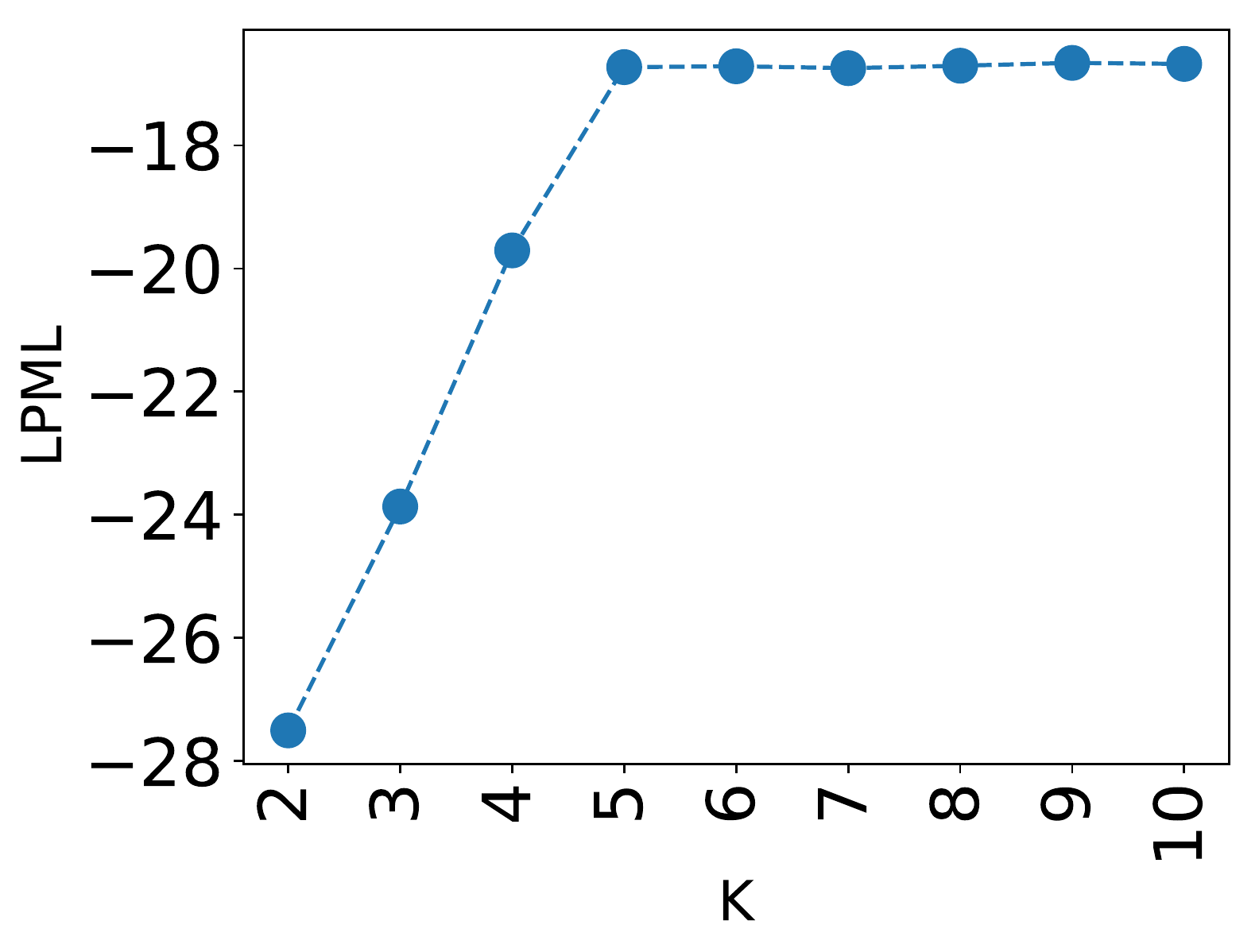} &
    \includegraphics[width=.32\columnwidth]{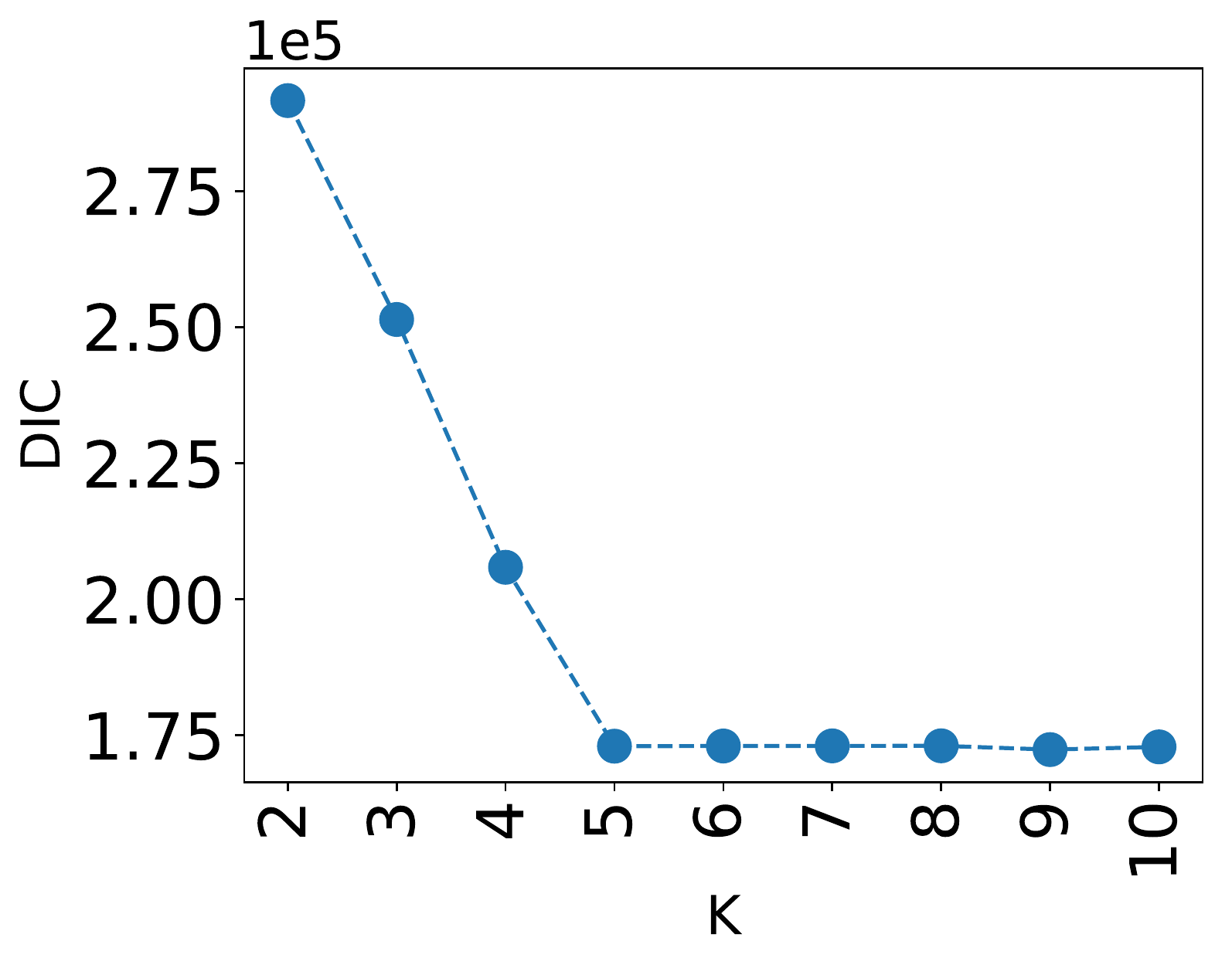} &
    \includegraphics[width=.32\columnwidth]{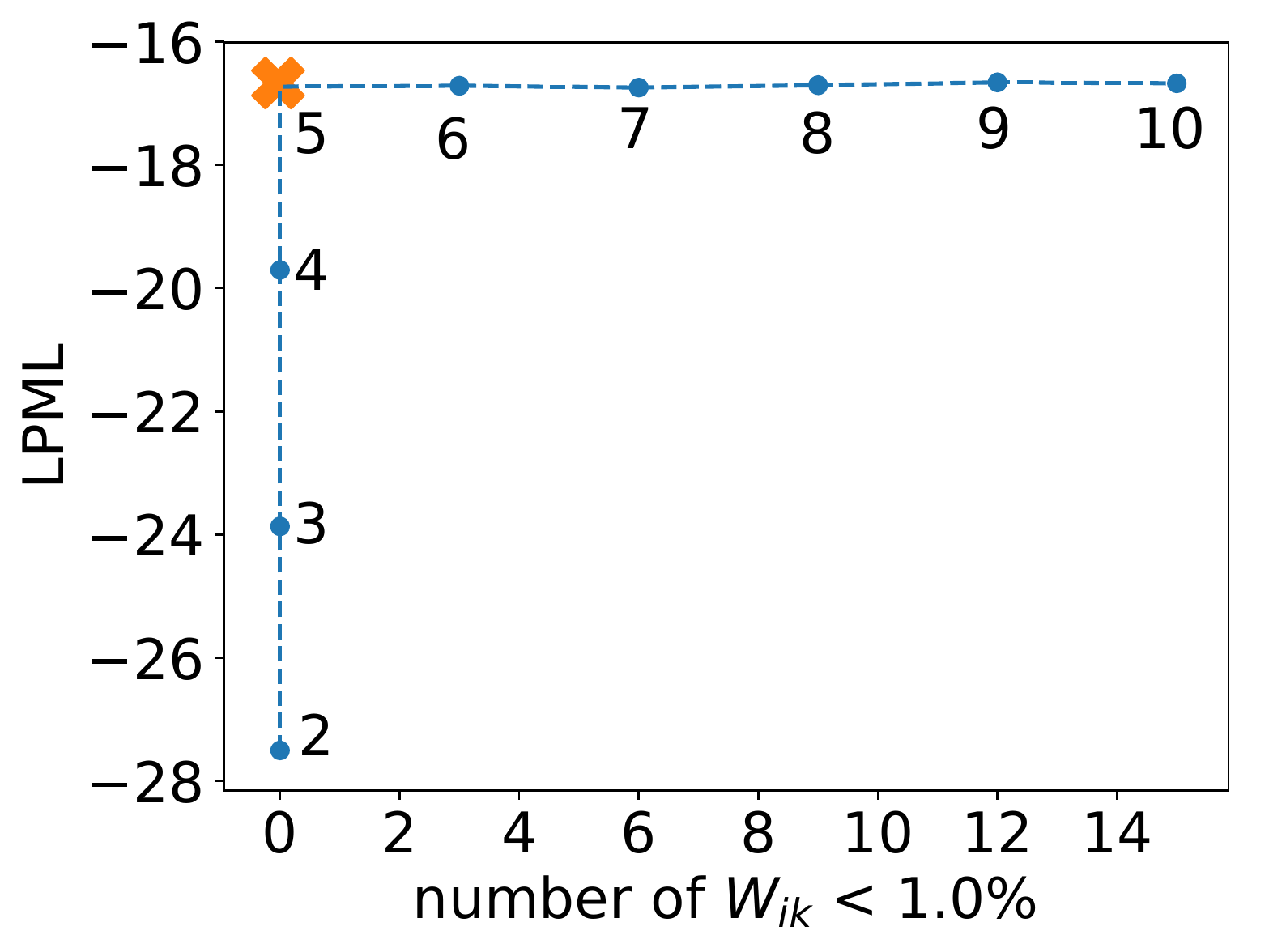} \\
    {(a) LPML} &
    {(b) DIC} &  {(c) Calibration of $K$} \\
    \end{tabular}
  \end{center}
    \vspace{-0.1in}
  \caption{ Results of Simulation 1. Plots of (a) LPML = log pseudo marginal
  likelihood, (b) DIC = deviance information criterion , and (c) calibration
  metric, for $K=2, \dots, 10$.}
  \label{fig:metrics-sim1}
\end{figure}
%%%%%%%%%%%%%%%%%%%%%%%%%%%%%%%%%%%

For each value of $K,$ we computed the LPML and DIC, and obtained point estimates
$\hat{\bZ}_i$, $\hat{\bw}_i$ and $\hat{\bm\lambda}_i$ using the method
described in \S~\ref{sec:sampling-via-mcmc}. Figures \ref{fig:metrics-sim1}(a) and (b) respectively show plots of LPML and DIC as functions of $K$.
Fig \ref{fig:metrics-sim1}(c) plots LPML against the number of subpopulations
with $\hat{w}_{i,k} < 1\%$. The increase in LPML is very minimal, while negligible
subpopulations are added for values of $K > 5$. The plots clearly indicate that
$\hat{K}=5$ yields a parsimonious model with good fit.
Fig~\ref{fig:sim1-post} illustrates $\hat{\bZ}_i$, $\hat{\bw}_i$ and
$\hat{\lambda}_{i,n}$ for $\hat{K}=5$.
Panels (a), (c) and (e) show $\hat{\bZ}_i$ and $\hat{\bw}_i$ for samples 1, 2,
and 3, respectively. The subpopulations with $\hat w_{ik} > 1\%$ are included in
the plots of $\hat{\bZ}_i$. The estimates  $\hat{\bZ}_i$ and $\hat{\bw}_i$ are close to
their truth values in Table~\ref{tab:sim1-tr} for all samples, implying that the
true cell population structure is well recovered.  The heatmaps of $y$
rearranged by cell clustering membership estimates $\hat{\lambda}_{i,n}$ are shown in
panels (b), (d), and (f) of Fig 3, where the colors, red, blue, and black
represent high, low, and missing expression levels, respectively. The
horizontal yellow lines separate cells by $\hat{\lambda}_{i,n}$. The figures
show that the cell clustering based on the estimated subpopulations captures the
true clustering of $y$ quite well.

%%%%%%%%%%%%%%%%%%%%%%%%
\begin{figure}[h!]
%\begin{figure}[th!]
  \begin{center}
  \begin{tabular}{cc}
  \includegraphics[width=0.45\columnwidth]{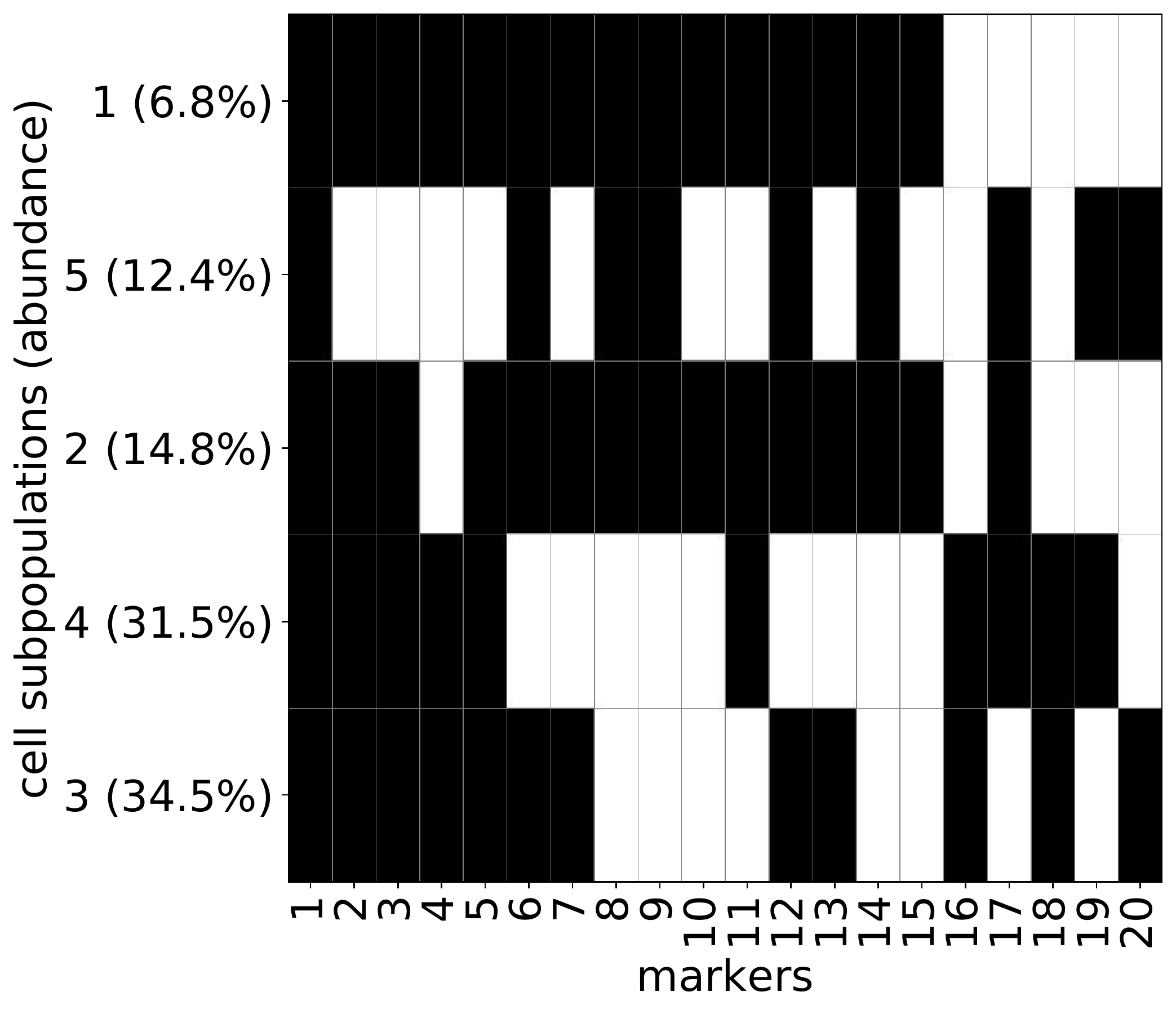} &
  \includegraphics[width=0.45\columnwidth]{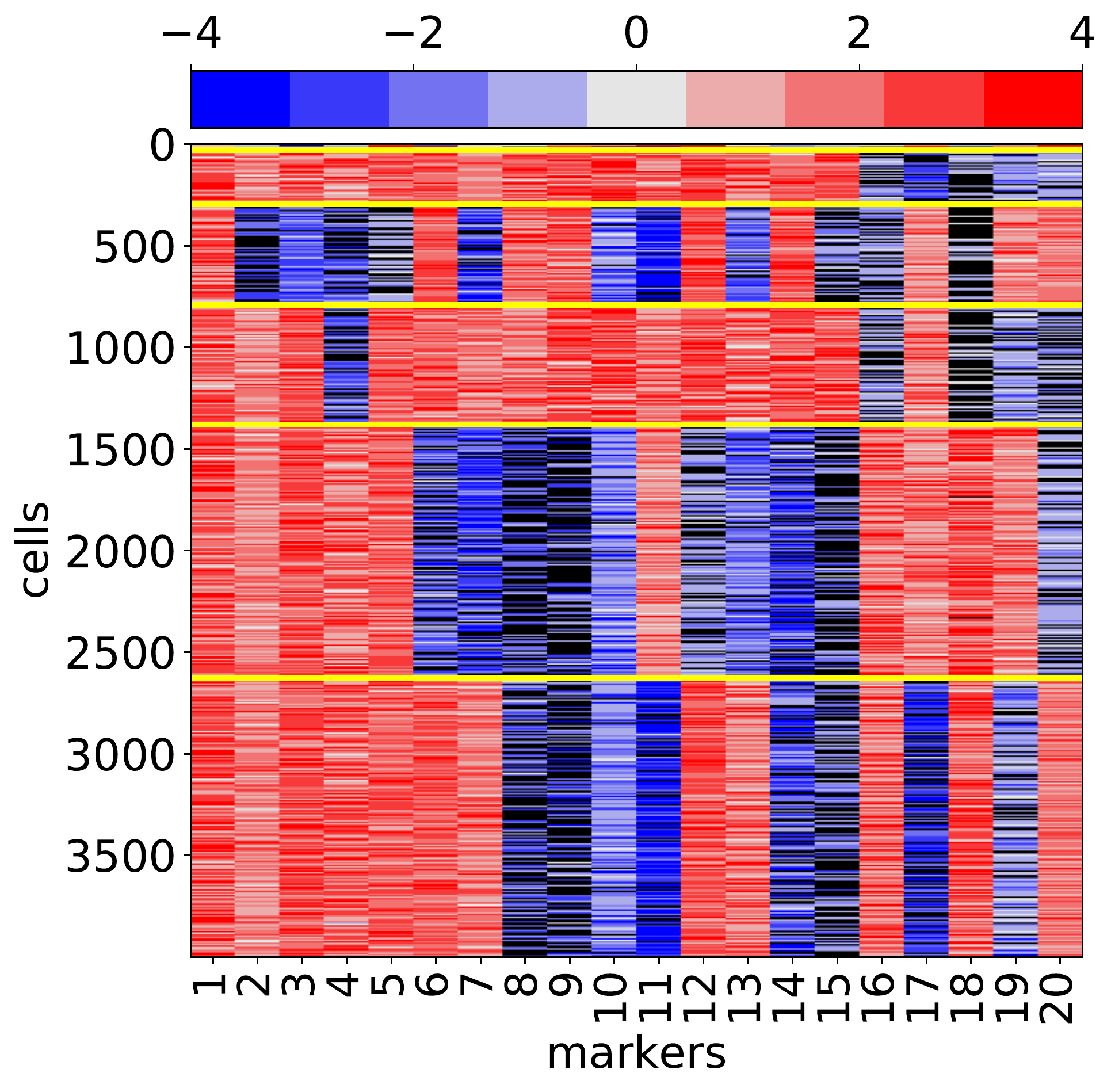} \\
  {(a) $\hat{\Z}^\prime_1$ \& $\hat{\bw}_1$} & {(b) heatmap of $y_{1nj}$} \\
  \includegraphics[width=0.45\columnwidth]{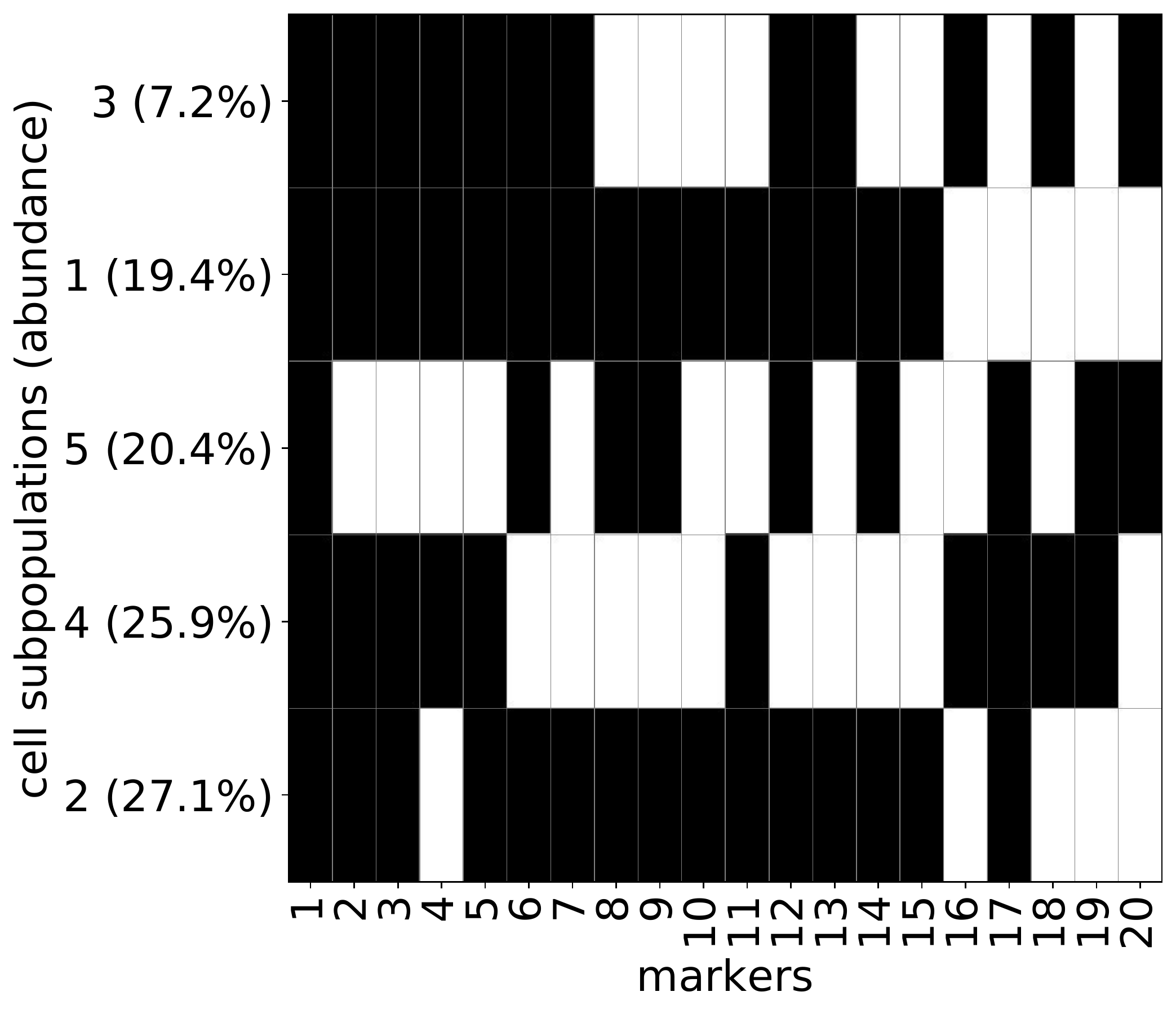} &
  \includegraphics[width=0.45\columnwidth]{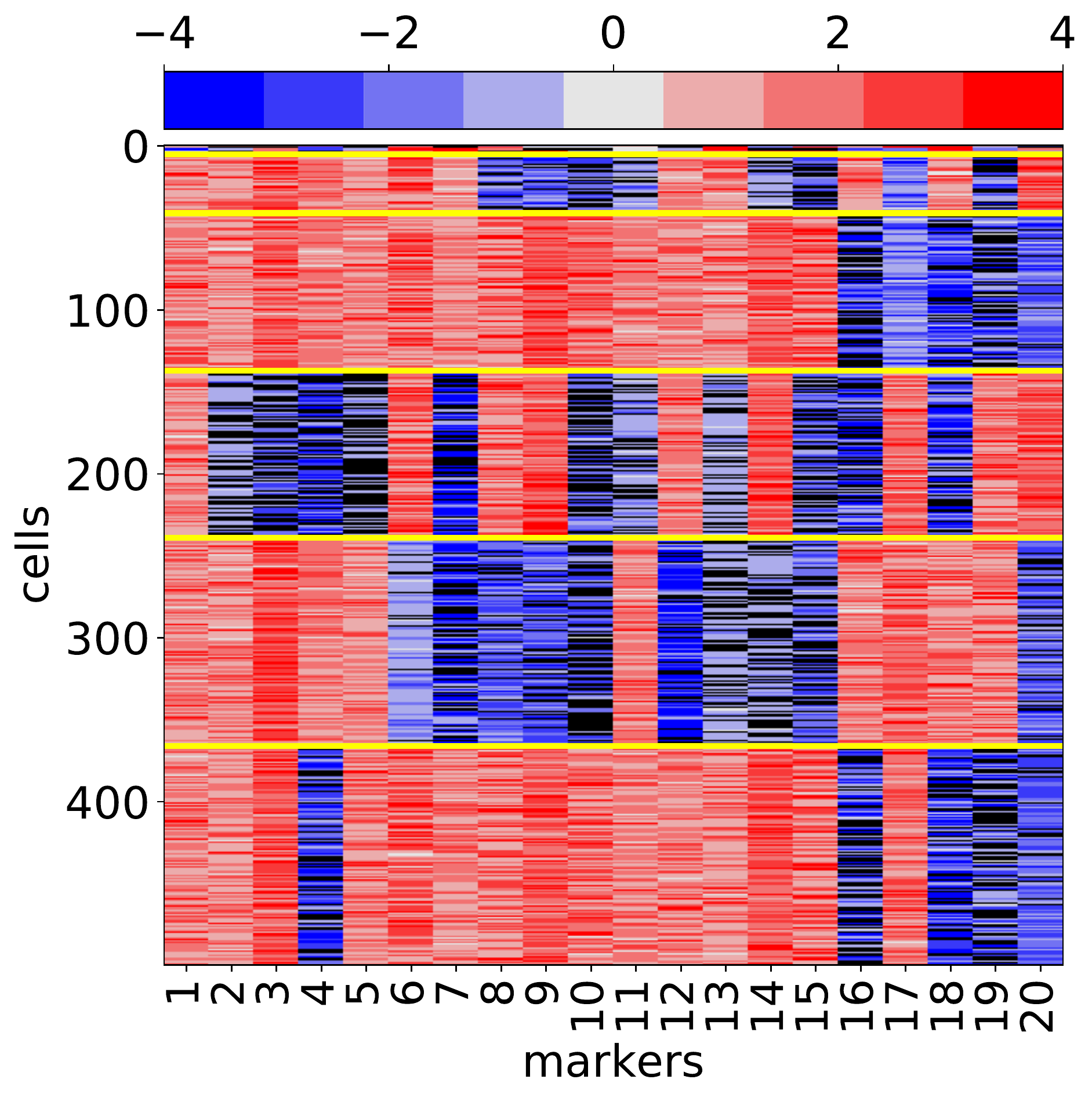} \\
  {(c) $\hat{\Z}^\prime_2$ \& $\hat{\bw}_2$} &  {(d) heatmap of $y_{2nj}$}\\
  \end{tabular}
  \end{center}
  \vspace{-0.1in}
  \caption{Results of Simulation 1. In (a) and (c), the transpose
  $\hat{\Z}^\prime_i$ of $\hat \bZ_i$ and $\hat{\bw}_i$ are shown for samples
  1 and 2, respectively, with markers that are expressed denoted by black and
  not expressed by white. Only subpopulations with $\hat{w}_{ik} > 1\%$ are
  included. Heatmaps of $\bm y_i$ are shown for sample 1 in (b) and sample 2
  in (d). Cells are ordered by posterior point estimates of their subpopulation indicators,
  $\hat{\lambda}_{i,n}$. Cells are given in rows and markers are given in
  columns. High and low expression levels are represented by red and blue,
  respectively, and black represents missing values. Yellow horizontal lines
  separate cells into five subpopulations.}
\label{fig:sim1-post}
\end{figure}
%%%%%%%%%%%%%%%%%%%%%%%%%

%%%%%%%%%%%%%%%%%%%%%%%%
\begin{figure}[h!]
%\begin{figure}[th!]
  \begin{center}
  \begin{tabular}{cc}
  \includegraphics[width=0.45\columnwidth]{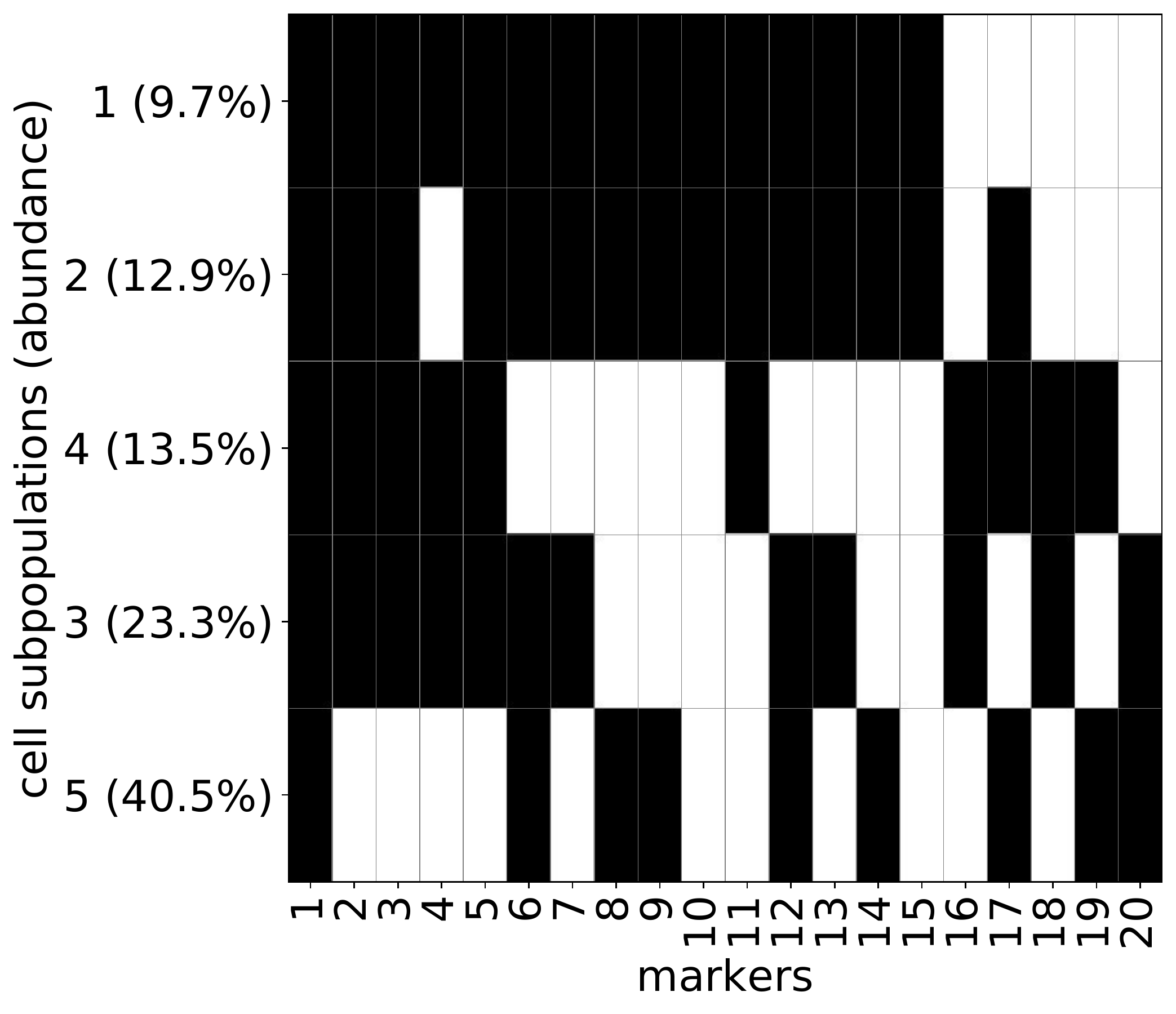} &
  \includegraphics[width=0.45\columnwidth]{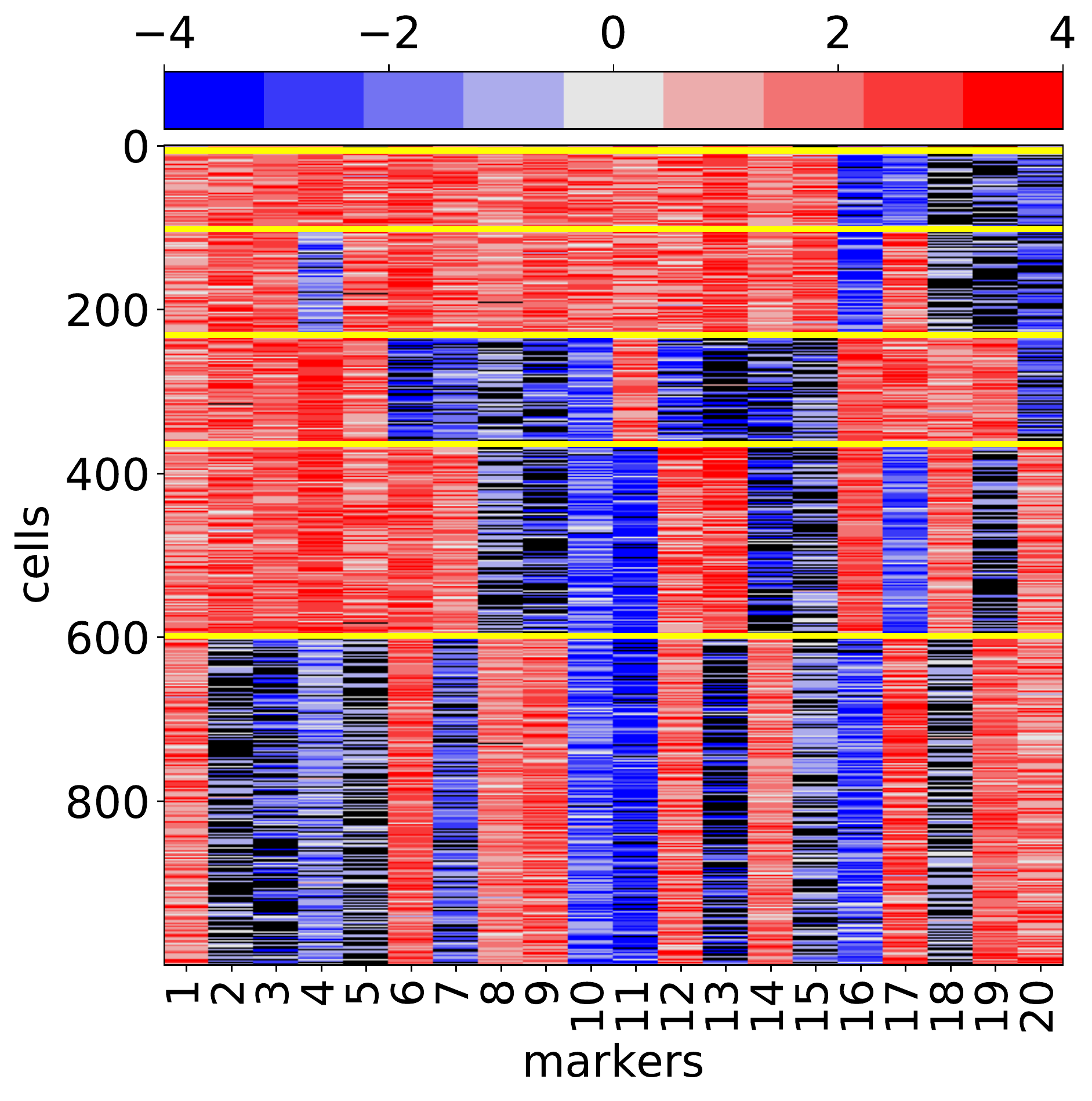} \\
  {(e) $\hat{\Z}^\prime_3$ \& $\hat{\bw}_3$} & {(f) heatmap of $y_{3nj}$}\\
  \end{tabular}
  \end{center}
  \vspace{-0.1in}
  \caption*{Fig~\ref{fig:sim1-post}: Results of Simulation 1 (continued). In
  (e), the transpose $\hat{\Z}^\prime_i$ of $\hat \bZ_i$ and $\hat{\bw}_i$
  are shown for sample 3, with markers that are expressed denoted by black and
  not expressed by white. Only subpopulations with $\hat{w}_{ik} > 1\%$ are
  included. Heatmaps of $\bm y_i$ for sample 3 is shown in (f). Cells are
  ordered by posterior point estimates of their subpopulation indicators,
  $\hat{\lambda}_{i,n}$. Cells are given in rows and markers are given in
  columns. High and low expression levels are represented by red and blue,
  respectively, and black represents missing values. Yellow horizontal lines
  separate cells into five subpopulations.}
\end{figure}
%%%%%%%%%%%%%%%%%%%%%%%%%

%We used the adjusted rand index (ARI) \citep{ari} to assess the accuracy of
%cluster assignments produced by the FAM. ARI is a measure of agreement
%between two sets of clusters, that takes values between -1 and 1. A larger
%ARI implies greater agreement between two sets of clusterings, and ARI = 1
%means that the two clusterings agree perfectly. In the case of random
%clusterings, ARI = 0. ARI can be negative in cases where the
%agreement between clusters is less than what is expected from random
%clusterings. The ARIs computed for clusterings obtained from our model are
%above 0.99 for all samples, indicating that the model recovers the true cell
%clusters determined by the subpopulations extremely well.
%%%

%%% ADVI
We also fit the model to the simulated data using ADVI, with a mini-batch size
of 2000,  $K=30$, and 20000 iterations. The time required to fit the model
was approximately 4 hours, which is substantially less than that of the
analogous MCMC method.  Supp.\ Fig \ref{fig:sim-vb-1} shows the
posterior estimates of $\bZ$, $\bw$ and $\lambda_{i,n}$ for ADVI. Inference for model
parameters using ADVI are similar to those using MCMC. The simulation truth for
the model parameters $\btheta$ are well recovered as in the MCMC
implementation.

% sensitivity to miss mech
We assessed sensitivity of the model to the specification of the data
missinship mechanism by fitting the FAM using different specifications of
$\bm\beta$ with $K=\hat{K}$, and comparing the inferences. The two different
specifications of $\bm\beta$ are given in Supp.\
Table~\ref{tab:missmechsen-sim}.  The estimates of $\bm\btheta$ do not
change significantly across different specifications of $\bm\beta$.   Point
estimates of $\bZ$, $\bw_i$, and $\lambda_{i,n}$ are shown in Supp.\
Figures \ref{fig:Z-w-sim1-missmechsen-1}
and~\ref{fig:Z-w-sim1-missmechsen-2}. The estimates $\hat \bZ$ remain the
same for all specifications of $\bm\beta$, and the $\hat \bw_i$ values also are very
similar. Supp.\ Table \ref{tab:missmechsen-sim} shows that LPML
and DIC are slightly better for the data missingship mechanisms that encourage
imputing smaller missing values $y_{i,n,j}$. This results in $\mu^\star_{0,
L_0}$ (the smallest of the mixture component locations for non-expressed
markers) being smaller than that obtained under the other specifications, accidentally more closely resembling the simulation truth. Details
of the sensitivity analysis are in Supp.\ \S \ref{sec:sim}.

%%%%%%%%%%%%%%%%%%%%%%%%
\begin{figure}[t!]
\begin{center}
  \begin{tabular}{ccc}
  \includegraphics[width=0.3\columnwidth]{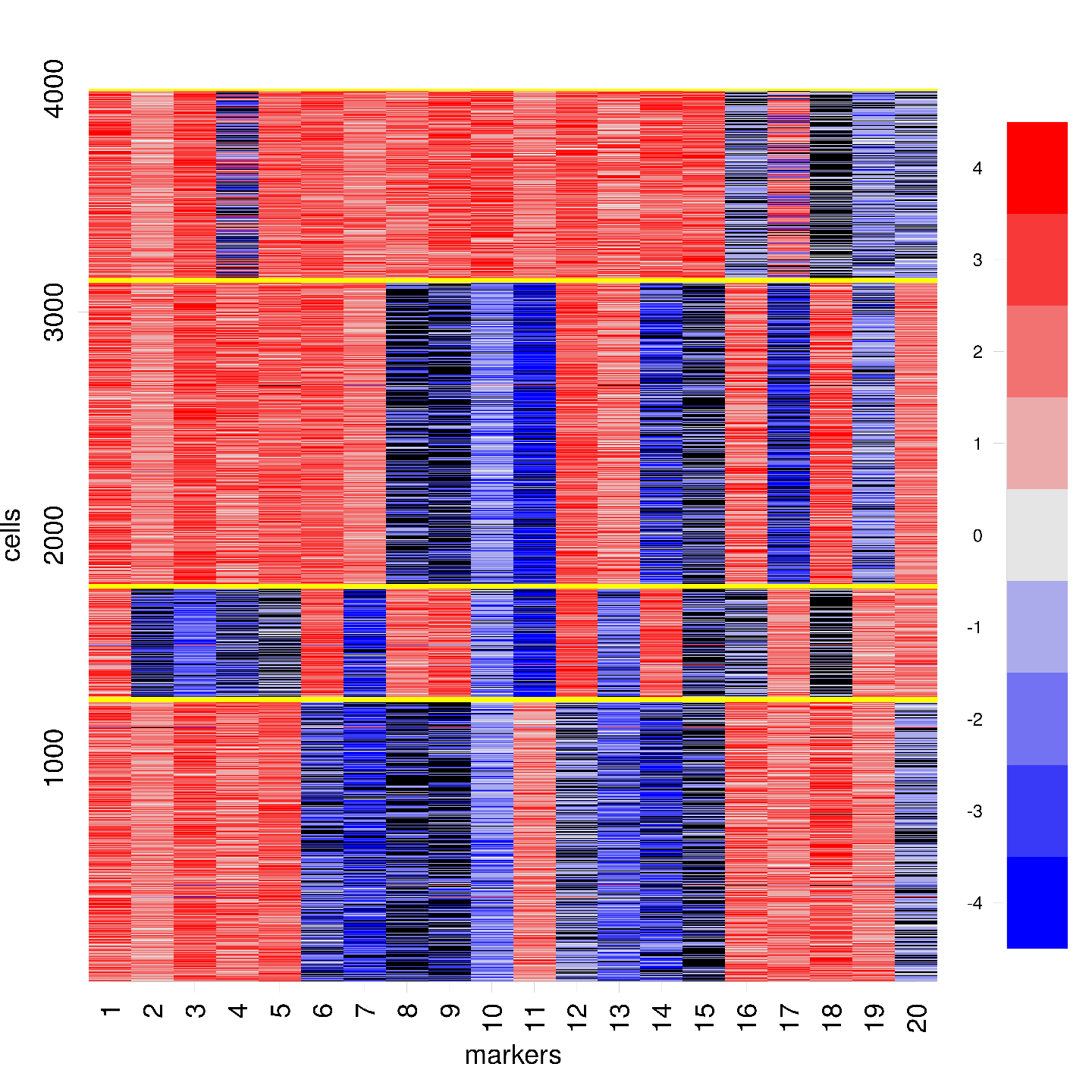}&
  \includegraphics[width=0.3\columnwidth]{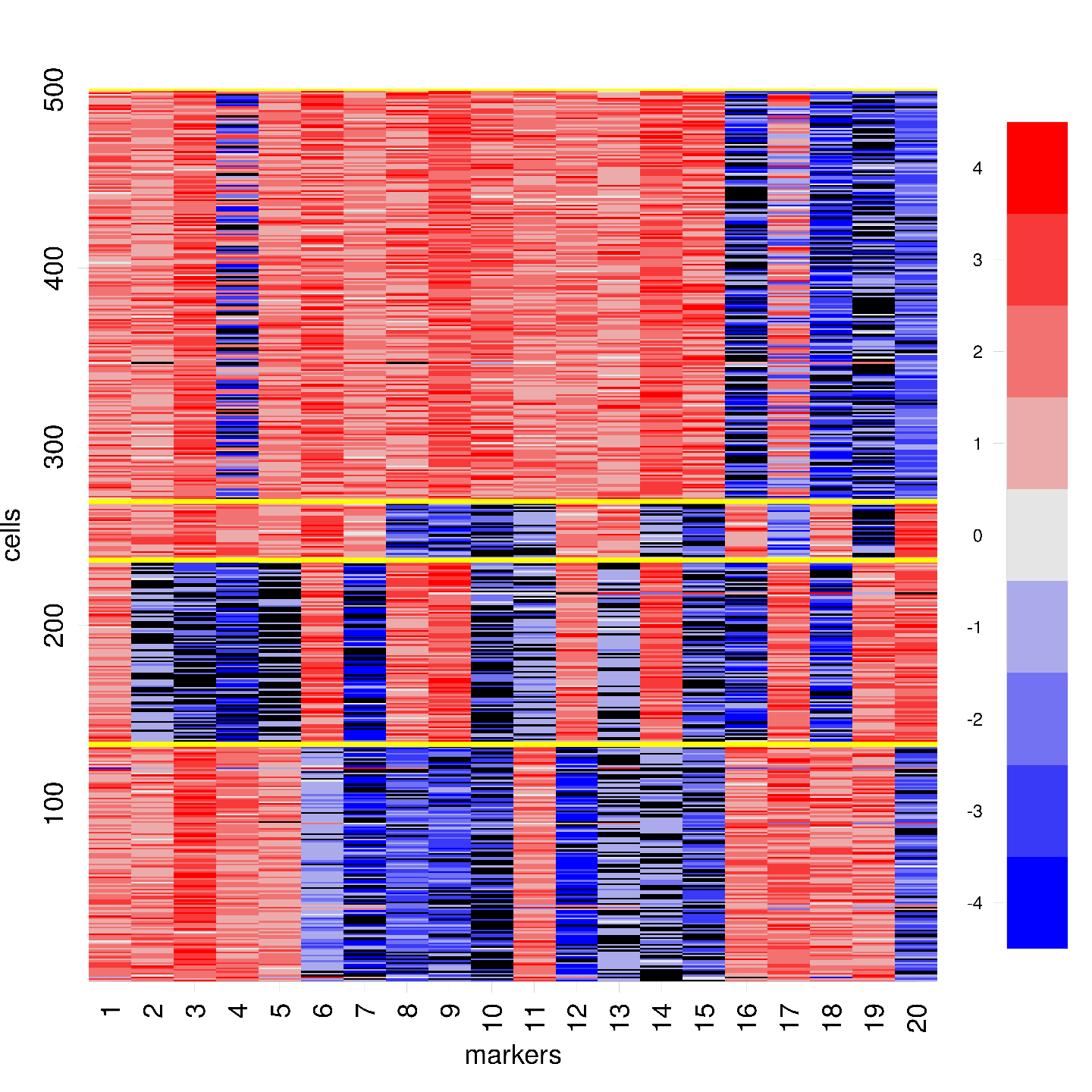}&
  \includegraphics[width=0.3\columnwidth]{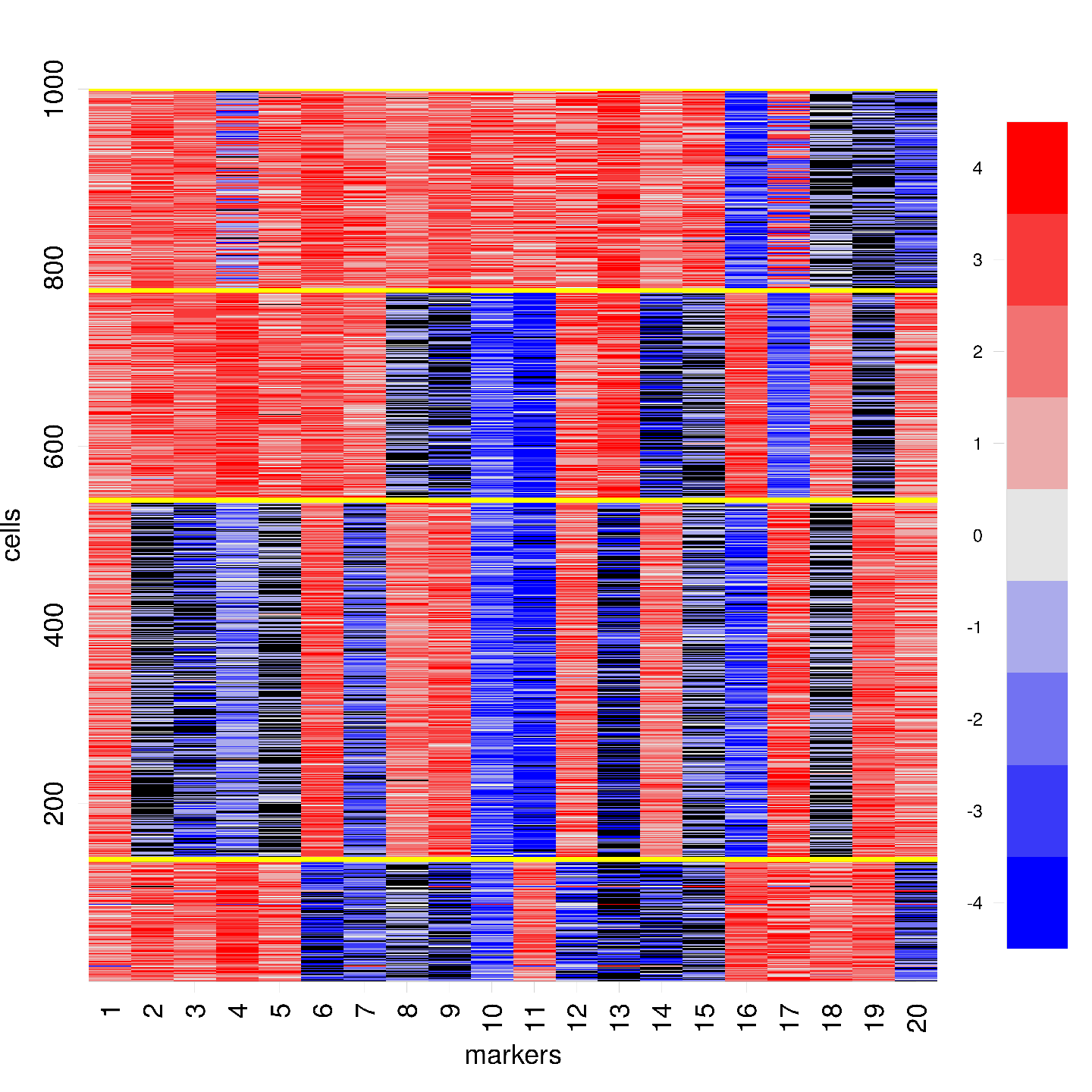}\\
  {\small (a) Sample 1} & {\small(b) Sample 2} & {\small (c) Sample 3} \\
  \end{tabular}
  \vspace{-0.1in}
  \caption{Results of Simulation 1 (continued). Heatmaps of $\bm y_{i}$ for
  clusters estimated by FlowSOM, with cells ordered by the cluster labels
  $\lambda_{i,n}.$  Cells are in rows and markers are in columns. High, low,
  and missing expression levels are in red, blue, and black, respectively.
  Yellow horizontal lines separate the identified cell clusters.}
  \label{fig:sim1-FlowSOM}
\end{center}
\end{figure}
%%%%%%%%%%%%%%%%%%%%%%%%%

% FlowSOM comparison
We compared our model via simulation to FlowSOM in \citep{van2015flowsom},
which is implemented in the R package FlowSOM \citep{rflowsom}. FlowSOM fits
a model with a varying number of clusters and selects a value of $K$ that
minimizes the within-cluster variance while also minimizing the number of
clusters via an \lq\lq elbow\rq\rq criterion, an {\it ad hoc} graphical
method that chooses $K$ such that $K+1$ does not substantially increase
percentage of variation explained. FlowSOM does not impute missing values, so
we used all $y$ assuming that there is no missing $y$. In practice,
missing values could be pre-imputed, or multiple imputation could be employed.
Note that FlowSOM does not account for variability between samples.
We combined the samples for analysis to avoid a further {\it ad-hoc} process
of finding common clusters among the samples. If desired, one might do
separate analyses for each of the samples.
FlowSOM was considerably faster
than our model, with a computation time of 11 seconds on the simulated
dataset.  FlowSOM identified four cell
clusters, as summarized in Fig~\ref{fig:sim1-FlowSOM}, where the cells are
rearranged by their cluster membership estimates in each sample. The fourth
cluster (shown near the top of the heatmaps) is a mix of the cells having the
true subpopulations 1 and 2 that differ only by markers 4 and 17, and its performance of cell clustering deteriorates.
%ARI is used to compare the clustering estimates
%produced by FlowSOM to the true clustering of cells. The ARIs for the cell
%clusterings obtained by FlowSOM were 0.945, 0.738, and 0.935 for samples 1,
%2, and 3, respectively. The ARIs for clusterings produced by the FAM are
%clearly greater than those produced by FlowSOM, implying a substantial
%improvement in cell clusterings under the FAM. The ARI in sample 2 is
%especially low for FlowSOM, as the two cell subpopulations grouped together by
%FlowSOM have a large cellular abundance in that sample.
More importantly, FlowSOM does not
model latent cell subpopulations, and no inference on cell subpopulations is
produced. For this simulation scenario,   the FAM easily recovers
the truth, but a clustering-based method such as FlowSOM may perform poorly
in inferring the cell population structure.

% SIMULATION 2 (LARGE)
We examined the performance of our model through an additional simulation
study, Simulation 2. In this simulation, we kept most of the set-up used in
Simulation 1, but assumed a more complex subpopluation structure with  much
larger numbers of cells, by assuming $K^\true=10$ and $N=(40000, 5000, 10000)$.
$\bZ^\true$ and $\bw^\true_i$ are illustrated in Supp.\
Fig~\ref{tab:sim2-tr}. We considered ten models with $K = 2, 4, \cdots,
20$. For the fixed hyperparameters, we let $L^0=L^1=5$, and the remaining
specifications for hyperparameters were the same as those in Simulation 1. The
model comparison metrics strongly suggest $\hat{K}=10$, for which the
posterior point estimates of the underlying structure including $\bZ$, $\bw$
and $\lin$  recover the simulation truth quite well, as shown in Supp.\
Fig~\ref{fig:sim2-post}. In contrast, in this case FlowSOM groups together cells in two
subpopulations that have similar configurations, similarly to Simulation 1, and
estimates nine cell clusters. The FAM provides direct inference on cell
subpopulations, and the cell clustering by subpopulations is better than that
under FlowSOM. Details of Simulation 2 including a sensitivity analysis for
the data missingship mechanism and fast computation using ADVI, are given in
Supp.\ \S~\ref{sec:sim-2}.

\section{Analysis  of Cord Blood Derived NK Cell Data}\label{sec:cb-analysis}

We next report an analysis of the CyTOF dataset of surface marker expression
levels on UCB-derived NK cells.
Identifying and characterizing NK cell subpopulations in terms of marker
expression may serve as a critical step to identifying NK cell
subpopulations to develop disease-specific therapies in a variety of severe
hematologic malignancies.
Our NK cell dataset consists of three samples collected from different cord blood donors, containing 41,474, 10,454, and
5,177 cells, respectively. 32 cell surface proteins were evaluated.
%
% used as markers to apply the FAM-based method for estimating phenotypes and
% clusters of NK cells.
%
We removed markers having positive values in more than 90\% of the cells in all
samples, or with missing or negative values in over 90\% of the cells in all
samples. We also removed all cells with an expression level $< -6$ for any
marker.
After this preprocessing, $J=20$ markers remained and the numbers of cells in
the samples were $N_i=$ 38,636, 9,555, and 4,827. Supp.\ Table~\ref{tab:marker-codes} lists the markers included in
the analysis.
Figures
\ref{fig:cb-post}(b), (d) and (e) show heatmaps of $\bm y$ after rearranging
the cells by posterior estimates $\hat{\lambda}_{in}$ of the cell clusterings for each sample.
%% t-SNE
We also visualize the data using a data visualization technique ``t-SNE (t-Distributed Stochastic Neighbor Embedding)'' in Supp.\ Fig~\ref{fig:CB-tsne}.
t-SNE is a popular method for visualization of high-dimensional data  in a two- or three-dimensional map through stochastic neighbor embedding \citep{maaten2008visualizing, van2014accelerating}. It also is used for detecting clusters in data.  We used Barnes-Hut SNE implemented in
the Python library sklearn to obtain two dimensional t-SNE embeddings separately for each sample.
%%%
We fit our FAM over a grid for $K$ from 3 to 33 in increments
of 3, with $L_0=5$ and $L_1=3$. We set priors and the data missingship
mechanism as outlined in \S~\ref{sec:sim-study}. Random parameters $\btheta$
also were initialized in a similar manner.
6000 samples from the posterior distribution of the model parameters were
obtained after a burn-in of 10000 iterations. The posterior samples were thinned by
selecting every other sample to yield a total of 3000 samples.

\begin{figure}[t]
  \begin{center}
    \begin{tabular}{ccc}
        \includegraphics[width=.31\columnwidth]{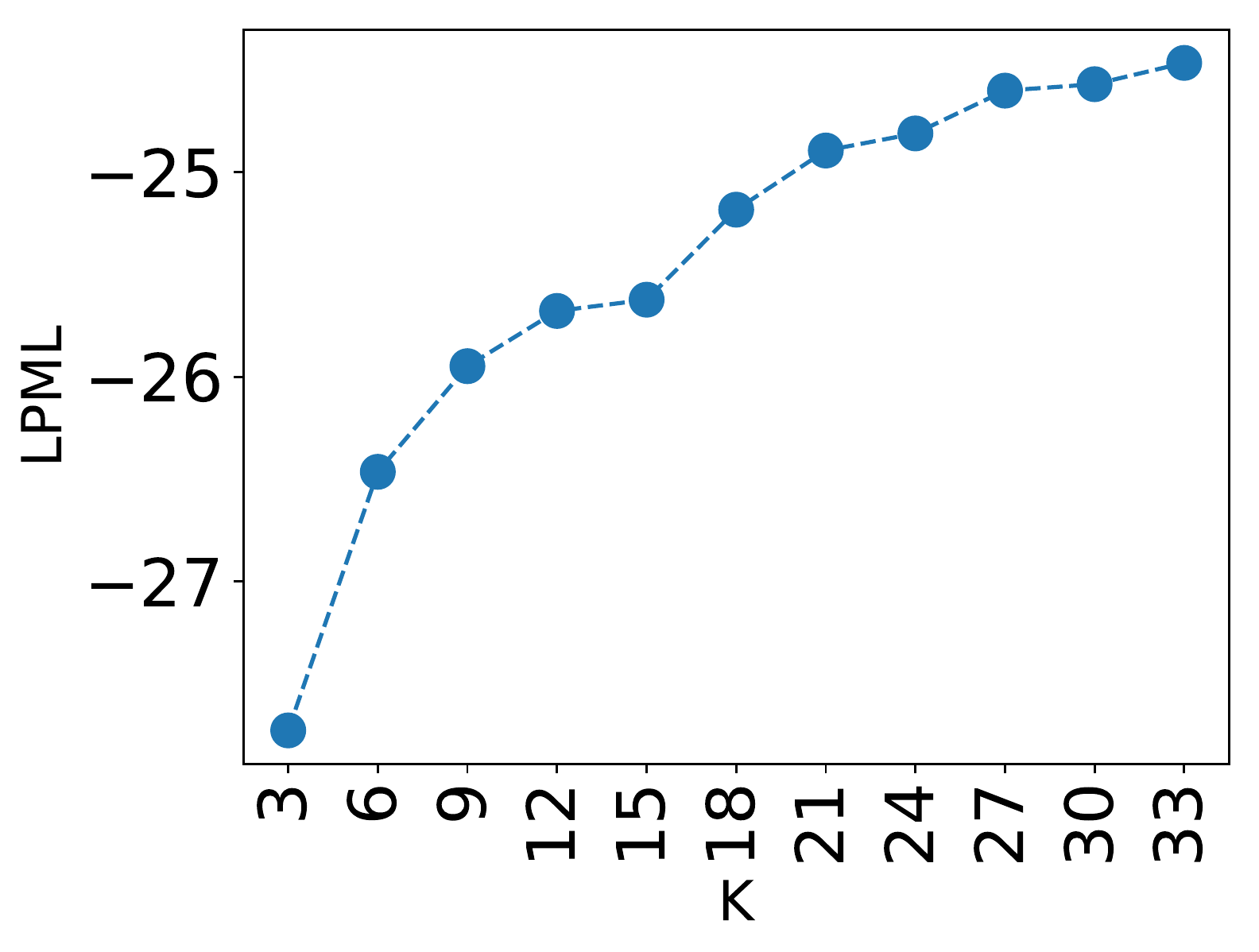} &
        \includegraphics[width=.31\columnwidth]{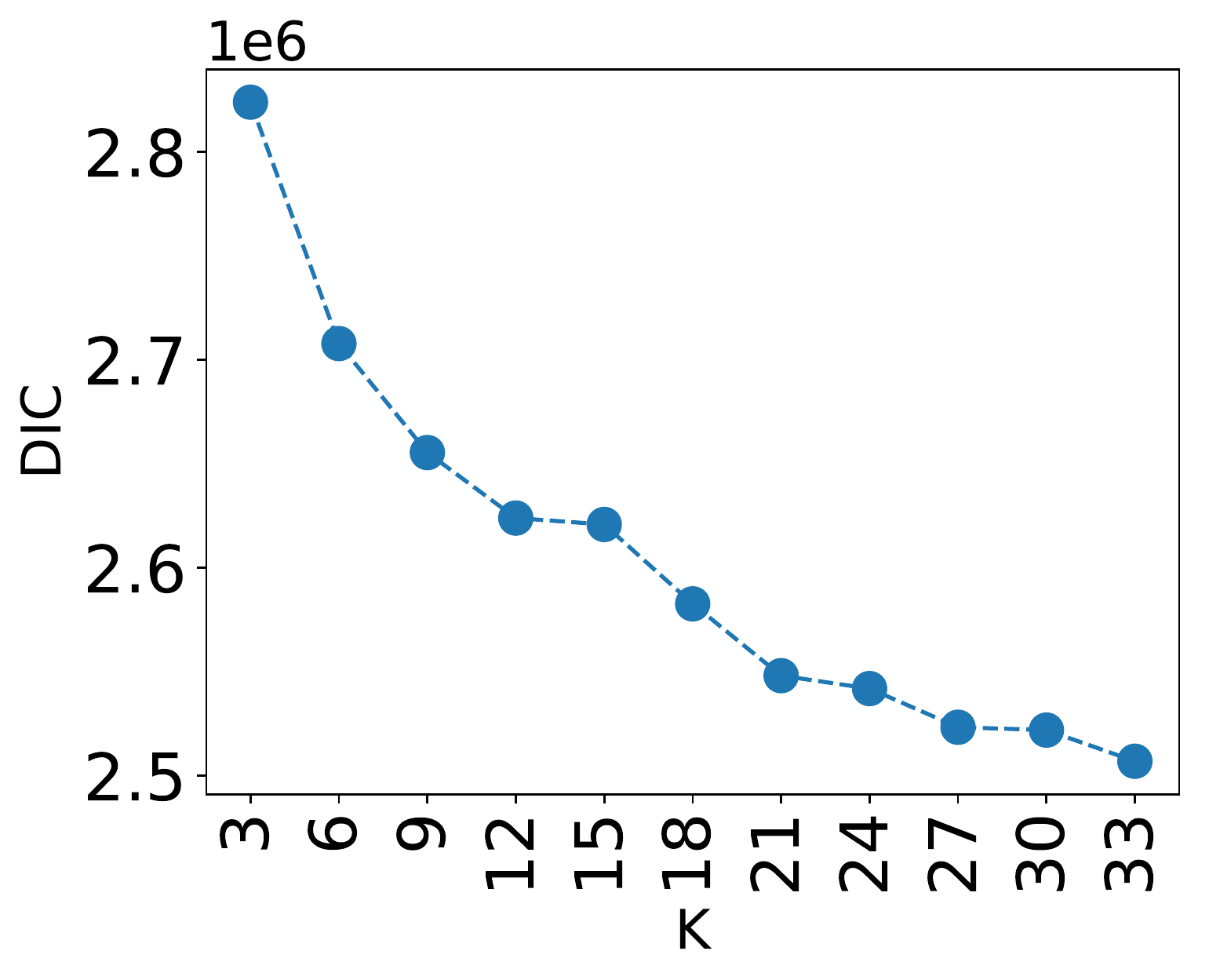} &
        \includegraphics[width=.31\columnwidth]{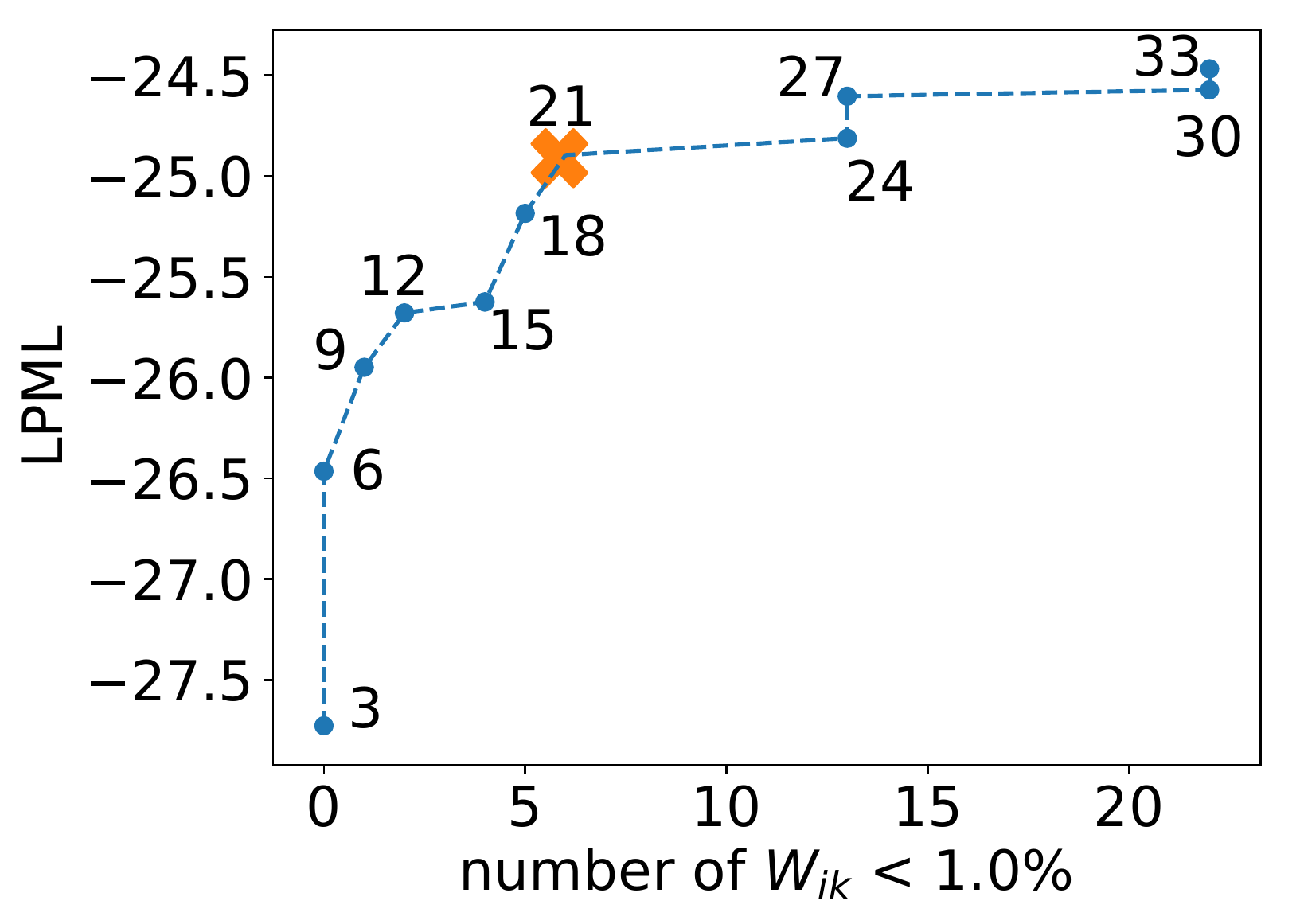} \\
        {(a) LPML} & {(b) DIC}  & {(c)} \\
    \end{tabular}
  \end{center}
  \vspace{-0.1in}
  \caption{\small Analysis of UCB-derived NK cell data. Plots of (a) LPML,
  (b) DIC, and (c) calibration metric, for $K=3, 6, \ldots, 33$.}
\label{fig:cb-select-K}
\end{figure}

\begin{figure}[t!]
  \begin{center}
  \begin{tabular}{cc}
  \includegraphics[width=0.45\columnwidth]{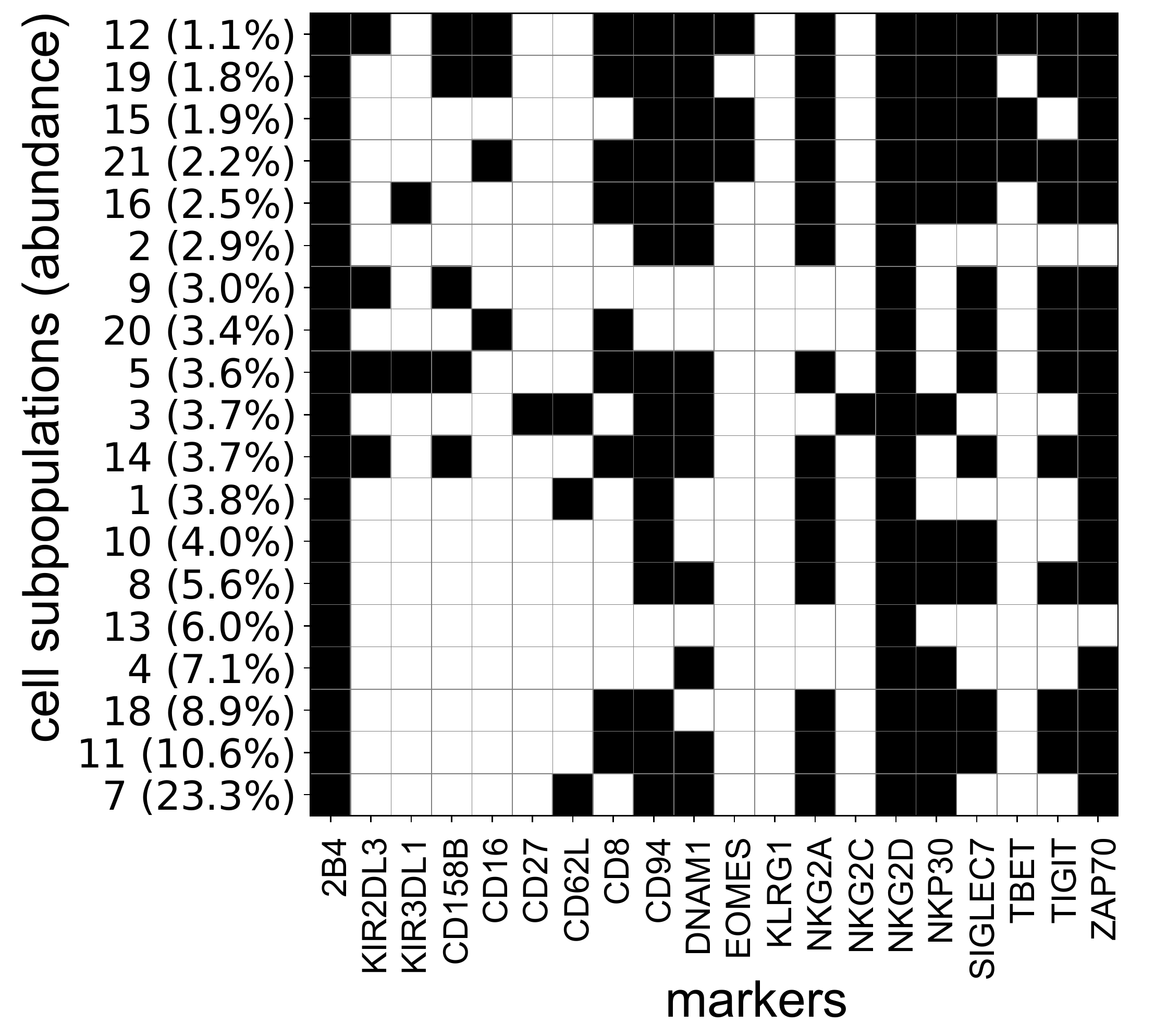}&
  \includegraphics[width=0.45\columnwidth]{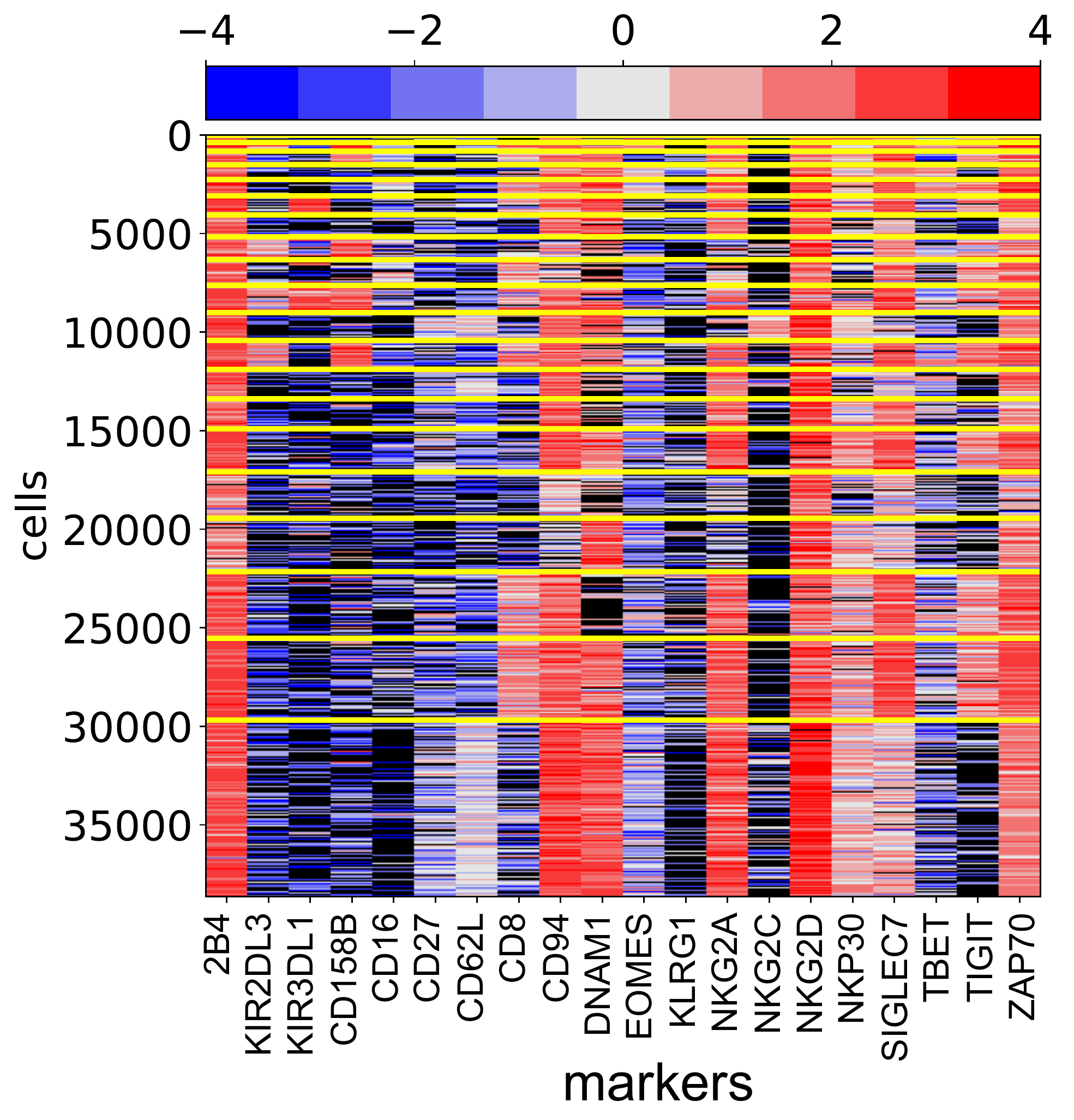}\\
   (a) $\hat{\Z}^\prime_1$ and $\hat{\bw}_1$ & (b) Clustering of  $y_{1nj}$\\
  \includegraphics[width=0.45\columnwidth]{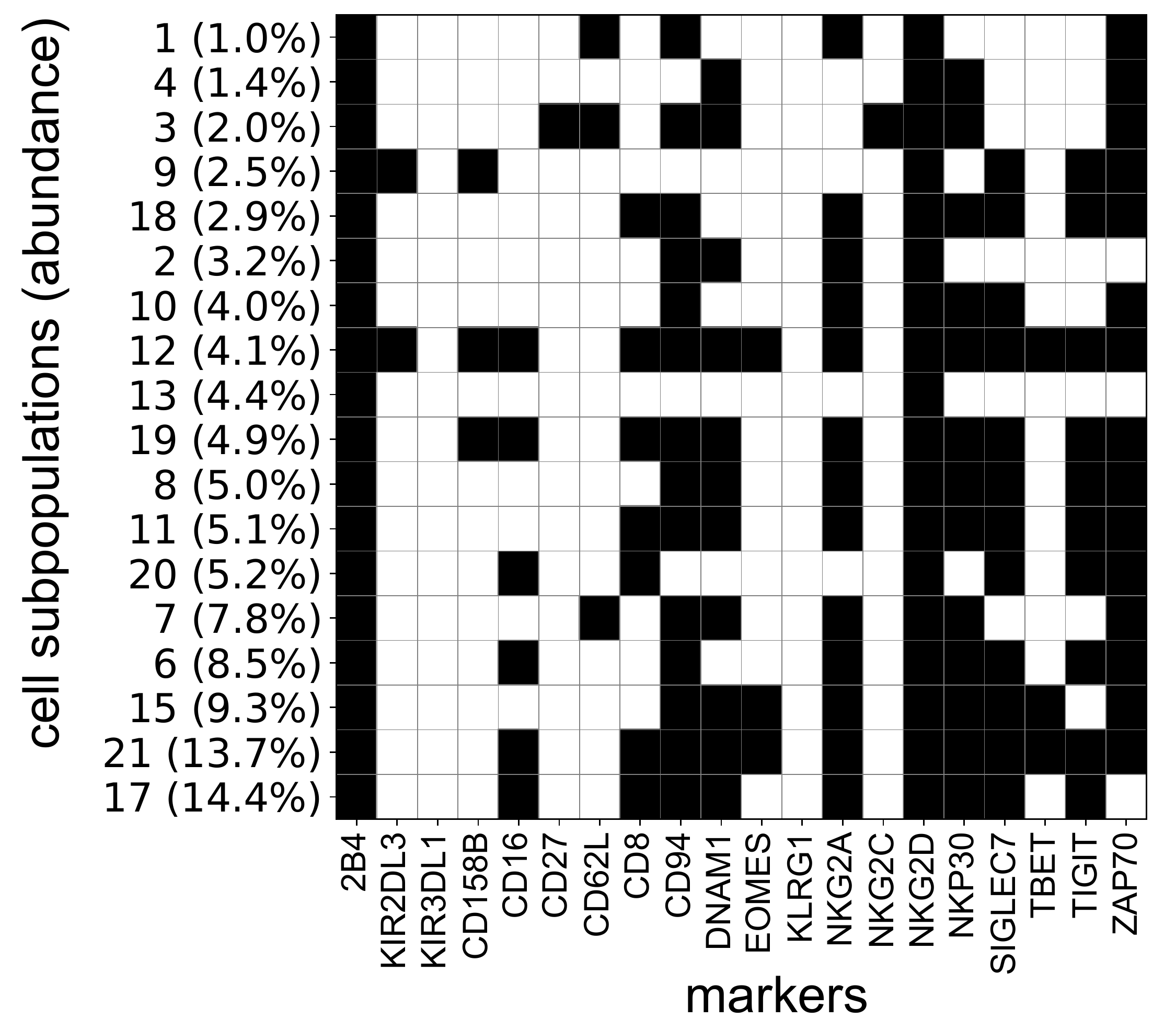}&
  \includegraphics[width=0.45\columnwidth]{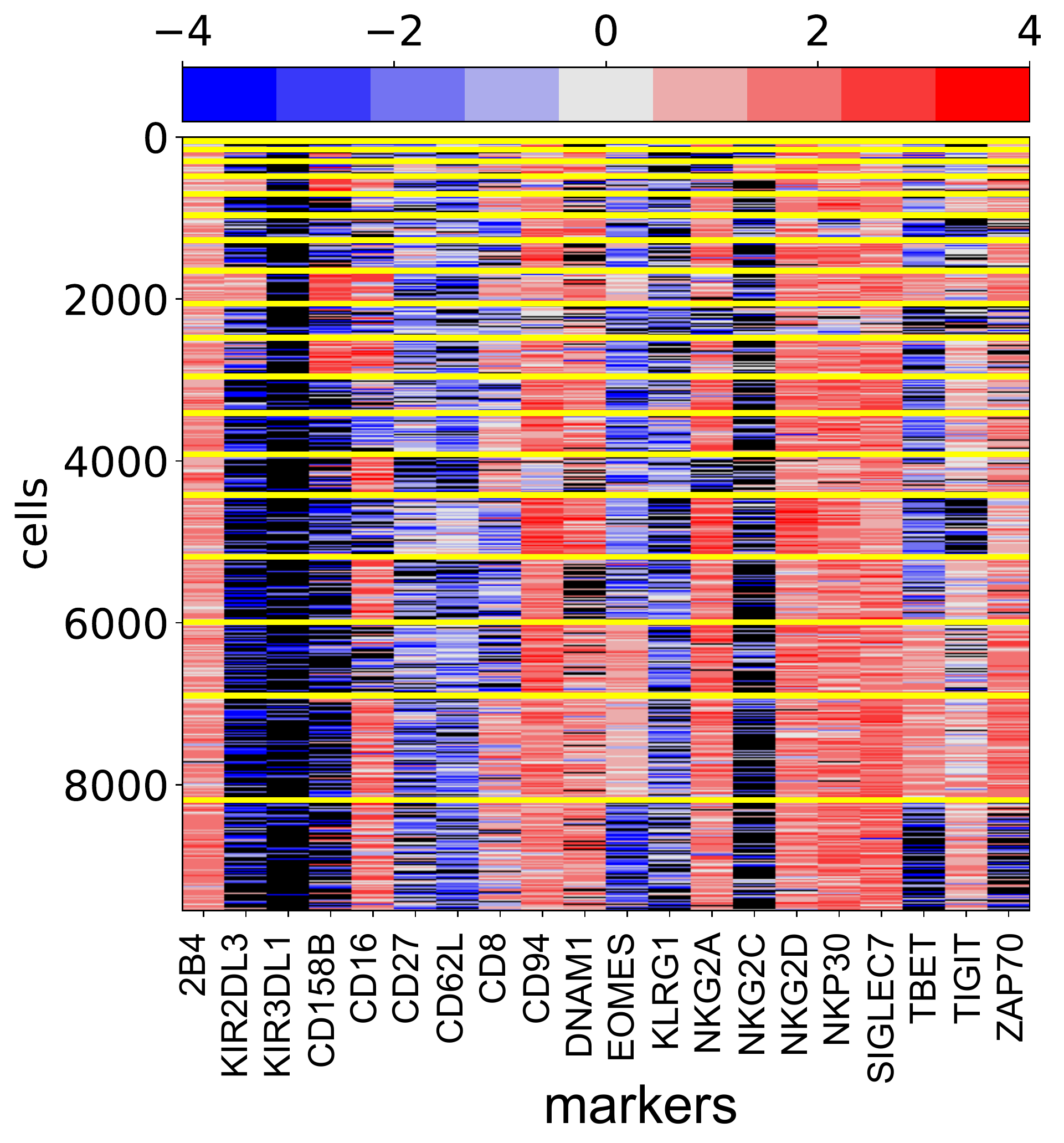}\\
   (c) $\hat{\Z}^\prime_2$ and $\hat{\bw}_2$ & (d) Clustering of $y_{2nj}$\\
  \end{tabular}
  \end{center}
  \vspace{-0.1in}
  \caption{Analysis of the UCB-derived NK cell data. $\hat{\Z}_i^{\prime}$ and
  $\hat{\bw}_i$ of samples $i=1$ and 2 are illustrated in panels (a) and (c),
  respectively, with markers that are expressed dented by black and not
  expressed by white.  Only subpopulations with $\hat{w}_{ik} > 1\%$ are included.
  Heatmaps of expression level $\bm y_i$ are shown in panels (b) and (d) for
  samples 1 and 2, respectively, with cells in rows and markers columns. Each
  column thus contains the expression levels of one marker for all cells in a
  sample. High, low, and missing expression levels are red, blue, and black,
  respectively. Cells are ordered by the posterior estimates of their
  clustering memberships, $\hat{\lambda}_{i,n}$. Yellow horizontal lines
  separate cells by different subpopulations.}
\label{fig:cb-post}
\end{figure}

Figures \ref{fig:cb-select-K} (a) and (b) display  LPML and DIC as functions of
 $K$.  The LPML changes sharply for small values of $K$, and
tapers at $K=21$, indicating that $\hat{K}=21$.  A similar pattern is seen
for DIC. As depicted in Fig \ref{fig:cb-select-K} (c), our additional
calibration method also indicates that the models with $K > 21$ include more
cell subpopulations comprising less than one percent of a sample (i.e.
$\sum_{i,k}\hat{w}_{i,k} < 1\%$ is larger), but improve fit only minimally.

%
%
%%%%%%%%%%%%%%%%%%%%%%%%
%%%%%%%%%%%%%%%%%%%%%%%%%
%
%%%%%%%%%%%%%%%%%%%%%%%%
\begin{figure}[t!]
  \begin{center}
  \begin{tabular}{cc}
  \includegraphics[width=0.45\columnwidth]{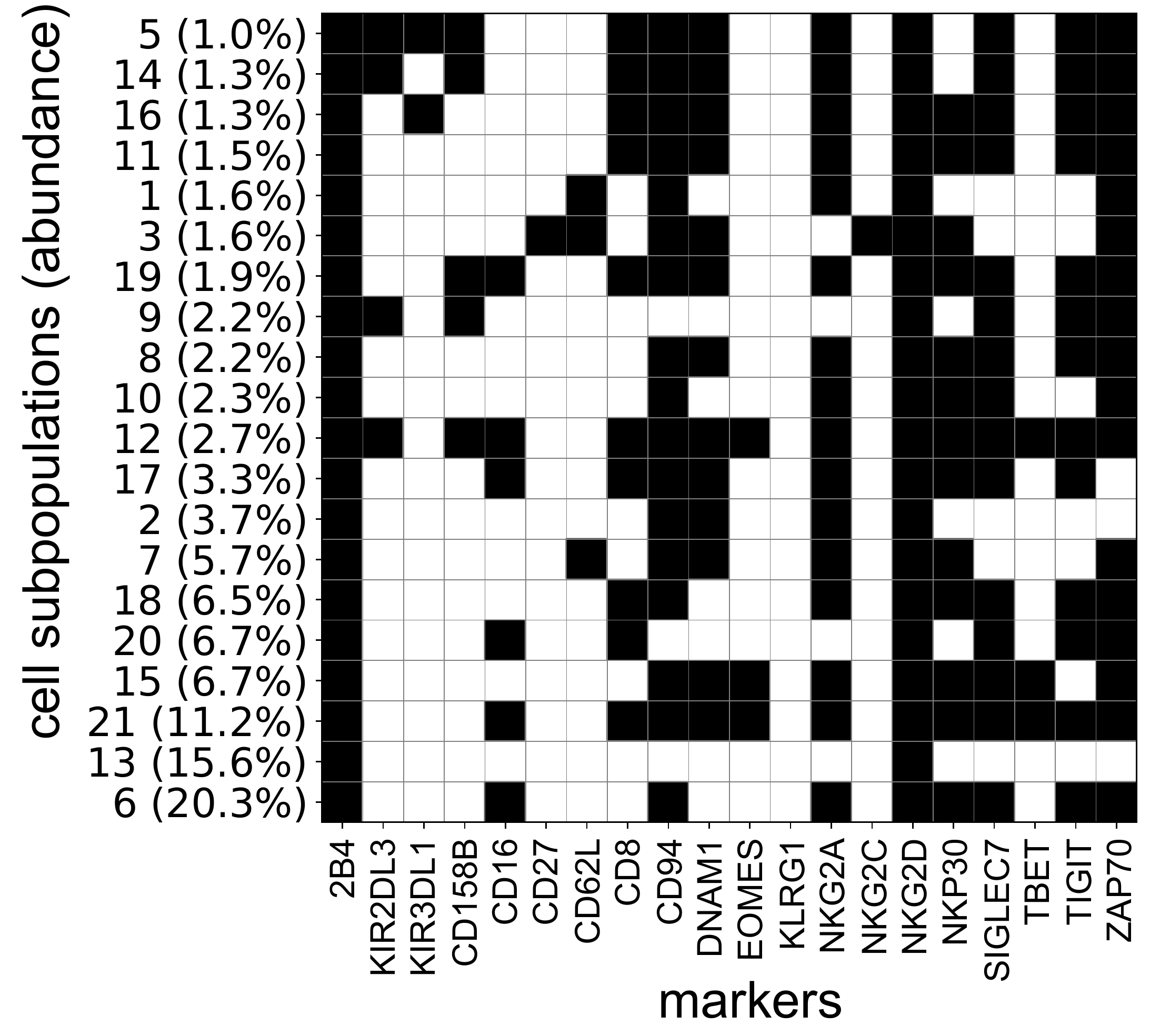}&
  \includegraphics[width=0.45\columnwidth]{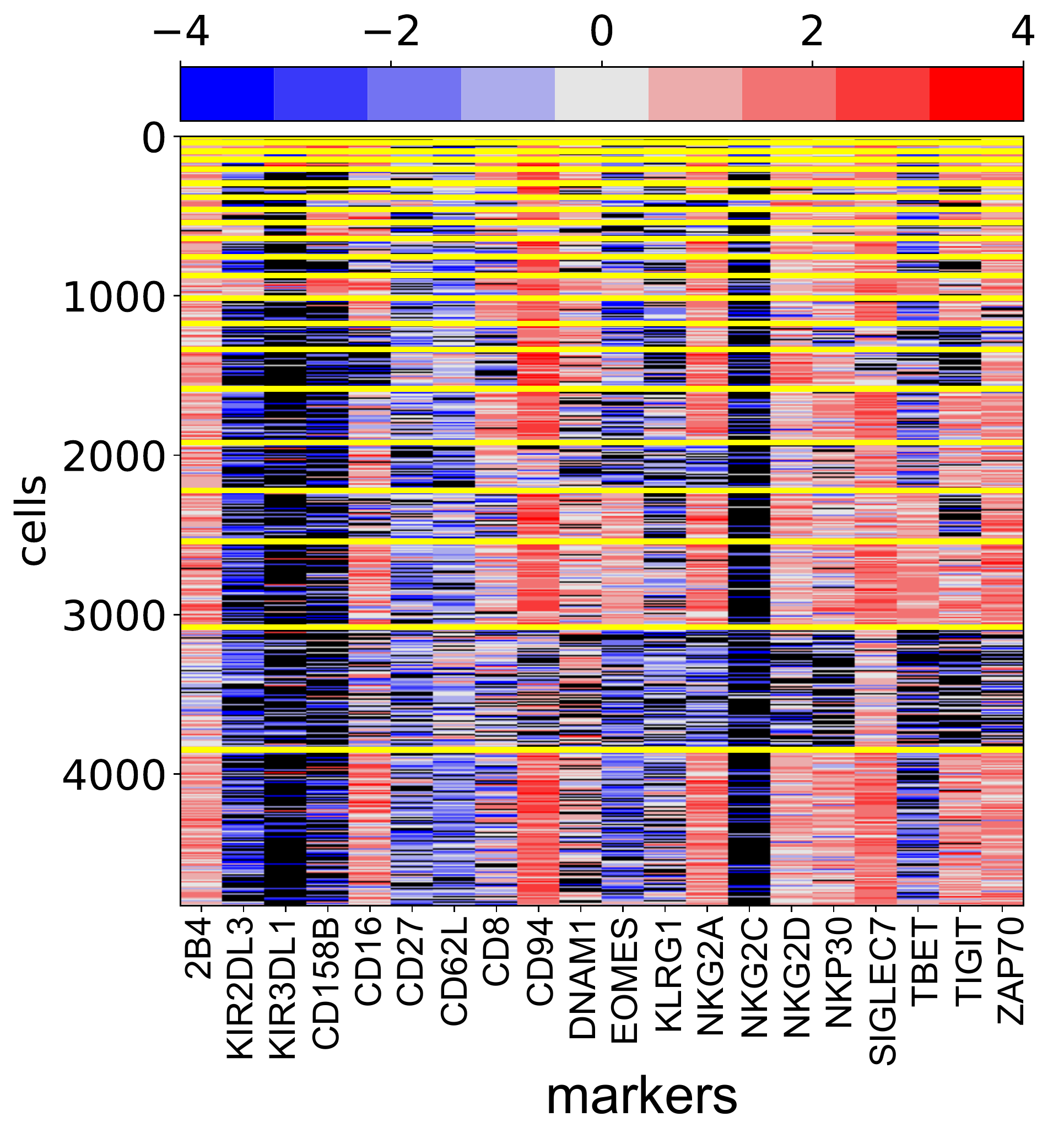}\\
   (e) $\hat{\Z}^\prime_3$ and $\hat{\bw}_3$ & (f) Clustering of $y_{3nj}$\\
  \end{tabular}
  \end{center}
  \vspace{-0.1in}
  \caption*{Fig~\ref{fig:cb-post} Analysis of the UCB-derived NK cell data
  (continued) $\hat{\Z}_i^{\prime}$ and $\hat{\bw}_i$ of sample 3 are
  illustrated in panel (e), with markers that are expressed dented by black and
  not expressed by white. Only subpopulations with $\hat{w}_{ik} > 1\%$ are
  included. Heatmaps of $\bm y_i$ are shown in panel (f) for sample 3. Cells
  are in rows and markers in columns. Each column contains the expression
  levels of a marker for all cells in the sample. High, low, and missing
  expression levels are red, blue, and black, respectively. Cells are ordered
  by the posterior estimates of their clustering memberships,
  $\hat{\lambda}_{i,n}$. Yellow horizontal lines separate cells by different
  subpopulations.}
\end{figure}
%%%%%%%%%%%%%%%%%%%%%%%%%
%

Fig \ref{fig:cb-post} summarizes posterior inference on the latent cell
population structure with $\hat{K}=21$. The cells are grouped by their estimated cell
subpopulation indicators $\hat{\lambda}_{i,n}$. The
figure shows the estimated cell subpopulations $\hat\bZ_i$ (in the left column) and clustered
marker expression levels $\bm y_i$ (in the right column) for the samples. Cells having
subpopulations with larger $\hat{w}_{i,k}$ are shown at the bottom of the
heatmaps. The subpopulations with the two largest $\hat{w}_{i,k}$ are different in the
samples. %Markers CD16, DNAM-1, and TIGIT have different expression patterns in the top
%two dominant subpopulations of the samples.
%
%For example, the first
%row shows the summarized inferences for the sample 1. The bottom row in
%Figure~\ref{fig:cb-post}~(a) is the most prevelant cell phenotype in sample 1, and
%comprises 23.3\% of the cells in the sample. The marker label (7) is a unique
%identifier simply used to match cell phenotypes common in different samples.
%In this case, phenotype 7 also appears in sample 2
%(Figure~\ref{fig:cb-post}~(c)) in row 5. The black grid cells indicate that a
%marker is expressed. That is, in phenotype 7, markers 1, 7, 9, 10, 13, 15,
%16, and 20 are expressed, while the other markers are not expressed. The
%corresponding cell clusters at the bottom of Figure~\ref{fig:cb-post}~(b) confirm
%that markers which are expressed (for cells in the cluster) have predominantly
%high expression levels (red), while markers that are not expressed have
%predominantly low expression levels (blue).
%
%NK cells are sorted into a dichotomy of expressions of few markers, such as markers CD56 and CD56. However, such classification of NK cells is limited to CD56/CD16 subsets as they can also be specified based on the expression of a rich variety of receptors including natural cytotoxicity receptors such as NKp30, NKp44 and NKp46, activating receptor CD94 as well as inhibitory receptors belonging to NKG2 groups which can recognize non-classical MHC, and killer cell immunoglobulin-like receptors (KIR) that can recognize classical MHC.
%
%
%
The resulting inference indicates that the composition of the NK cell
population varies across the samples, pointing to variations in the phenotype
of NK cells among different cord blood donors. We observe similarities in the
phenotypes of NK cells from samples 2 and 3, however, while sample 1 displays a
different phenotype and a distinct distribution of cell subsets. NK cells from
all three samples express 2B4, CD94, DNAM-1, NKG2A, NKG2D, Siglec-7, NKp30 and
Zap70 in the majority of their identified subpopulations.
These markers dictate NK cell functional status. While their  interactions are very complicated, taken together they provide a basis for determining whether NK cells have a normal function, and whether they are mature or not.

Despite great variability between cord blood 1 and the other two cord bloods, all three had a significant subset of cells with an immature phenotype. Cord blood 1 Cluster 7, cord blood 2 Cluster 17 and cord blood 3 Cluster 6 comprise the largest population of immature cells defined as EOMES (-), TBET (-), and KIR (-).  Markers, KIR2DL3 and KIR3DL1, belong to killer-cell immunoglobulin-like receptors (KIRs). These immature clusters of NK cells still retain expression of 2B4, NKG2A, NKG2D, CD94 and NKp30. In particular, NKp30 is natural cytotoxicity receptor, while KIR is not.
  This implies that, despite great
variability between sample 1 and the other two samples, all three have a
significant subset of cells with an immature phenotype.
Markers EOMES, TBET, Zap70 and KIR are not expressed in the largest subpopulation of
each sample, indicating that those are subsets of immature cells.  
An immature phenotype of NK cells usually is associated with low diversity and low effector function in the absence of exogenous cytokines, \citep{li2019, savaria2017},   while a mature NK cell phenotype has been linked to superior cytotoxicity and better clinical outcomes in cancer patients \citep{ilander2017,carlsten2019}. These immature clusters of NK cells still retain expression of 2B4, CD94, NKG2A, NKG2D, and NKp30.   

In addition, we identify three subpopulations (12, 15 and 21)
 that are conserved among the three samples (although at lower
percentages in sample 1).  In those subpopulations, EOMES and TBET are
expressed, indicating that those are a more mature phenotype. The subset
with expression of EOMES and TBET could be further divided into three
subpopulations based on the expressions of markers CD8, CD16, TIGIT, and KIR. Subpopulations 12 and 21 are very similar, sharing positivity for CD16,
CD8 and TIGIT and are differentiated by KIR expression, which are are negative in subpopulation 21 while
being positive in subpopulation 12. Subpopulation 15, however, is negative for
CD16, CD8, TIGIT and KIR, making EOMES and TBET its only differentiation
markers. These novel subsets of cord blood NK cells have not been described in
the literature previously, and may need to be further validated. We also identified
cluster 3 as an important conserved cluster among all 3 samples, which is
positive for NKG2C, CD62L and CD27 which could point towards a memory subset in
cord blood NK cells which has not been well described previously.  Taken
together, these data indicate that FAM allows not only the definition of
biologically recognized subsets of NK cells but also may be applied for the
discovery of novel NK cell subpopulations.

% of 2DL3 and 3DL1, which are
%killer-cell immunoglobulin-like receptors (KIRs) and are known to regulate the
%killing function.  Their expressions
% Subpopulation 7 of sample 1, subpopulation 17 of Sample 2 and Subpopulation 6 of sample 3 comprise the largest population of immature cells defined as TBET (-), EOMES (-), ZAP70 (-) and KIR (-).
%These immature clusters of NK cells still retain expression of 2B4, NKG2A, NKG2D, CD94 and NKp30.

%\begin{table}[t]
%  \centering
%  \begin{tabular}{|c|rrr|}
%    \hline
%    Data Missingship Mechanism & Sample 1 & Sample 2 & Sample 3 \\
%    \hline
%    I & 0.446 & 0.356 & 0.364 \\
%    II &  0.387 & 0.317 & 0.268 \\
%    \hline
%  \end{tabular}
%  \caption{ARI between clusterings resulting from the default (MM-0) and each of the
%  two other data missingship mechanisms (MM-I and MM-II) for each sample.}
%  \label{tab:ari-cb-missmech}
%\end{table}

Model sensitivity to the specification of the data missingship mechanism in
the NK cell data analysis was assessed by fitting the FAM under two additional
specifications of $\bm\beta$, which we call data missingship mechanisms (MM) I
and II. We will refer to the previous (default) missingship mechanism as MM-0.
Supp.\ Tables \ref{tab:missmechsen-cb} and
\ref{tab:missmechsen-cb-beta} list the different data
missingship mechanism specifications and the corresponding $\bm\beta$ values,  respectively.
Under the different specifications of $\bm\beta$, the estimates
$\hat{\bZ_i}$ and $\hat{\bw}_i$ are similar, as shown in Supp.\ Figures
\ref{fig:Z-w-CB-missmechsen-1} and \ref{fig:Z-w-CB-missmechsen-2}.
The subpopulations estimated under the different missingness mechanisms are the same or differ by fewer than three markers.
The subpopulations estimated under MM-I and MM-II are exactly the same or differ by no more than three markers, compared to those under MM-0.
%
%We also compared cell clusterings under MM-0 to those under MM-I and MM-II using ARI and
%summarized the results in Table \ref{tab:ari-cb-missmech}.
%
%Since MM-I resembles MM-0 more closely, the resulting cell clusterings between
%the two MMs are also more similar. The ARIs indicate that the cell
%clusterings resulted from different missingness mechanisms differ moderately.
%
%The difference in the clusterings may be due to the fact that the clusterings
%are compared based on point estimates rather than their posterior
%distributions.
%%
% \hh The numbers look a bit too small to me (meaning our model's clustering
% is sensitive to the missingness mechanism specification). I wonder why the
% clustering changes this much even though they have some common phenotypes. Do
% you have any insights? Does the difference come mainly from small phenotypes?
% It would be good that we can say the clustering is very similar. \ech
% RESPONSE: The difference in the ARI is mainly due to randomness in the
% $\lambda_{i,n}$. The ARI between random samples from the default missinship
% mechanism is already about 30\%. Consequently, the ARI between random
% samples for {\it different} mechanisms will be even greater.
%
%As a comparator,
We also fit the model to the UCB-derived NK cell data using ADVI with a
mini-batch size of 500 and $K=30$ for 20000 iterations. The runtime was 74
minutes on the previously described machine.
Supp.\ Fig \ref{fig:cb-vb-Z} summarizes the posterior
distribution of $\bZ$ and the posterior mode of cell clusterings
$\hat\lambda_{i,n}$. The cell subpopulations inferred by ADVI are similar to
those obtained by MCMC, but the cell clustering estimates are quite different.
Notably, subpopulations with large $\hat{w}_{ik}$ can be found in the
estimates obtained by both methods, e.g, the subpopulations with the two largest abundances
in sample 1. For subpopulations with small $\hat{w}_{ik}$, we do not find clear
matches. The cluster sizes obtained by ADVI are larger than those obtained from
MCMC and cells in the clusters are less homogeneous. It thus appears that ADVI should not be used
in this type of setting, and that its shorter runtime compared to MCMC is a false economy.

% due to larger estimates of $\sigma^2_i$ by ADVI.
%
%\hh are estimates of $\sigma^2$ larger?? \ech
%\bch Smaller, actually. MCMC: (.755, .787, 949).
%ADVI: (.676, .793, .730) \ech
%
%Comparing the cell clustering estimates by ADVI to those by MCMC, ARI are
%0.290, 0.144 and 0.224 for samples 1-3, respectively.

%, indicating that the clusterings by the estimation methods are not very similar.

%%%%%%%%%%%%%%%%%%%%%%%%
\begin{figure}[t]
  \begin{center}
  \begin{tabular}{cc}
  % sorted by abundance for each sample
  \includegraphics[width=0.48\columnwidth]{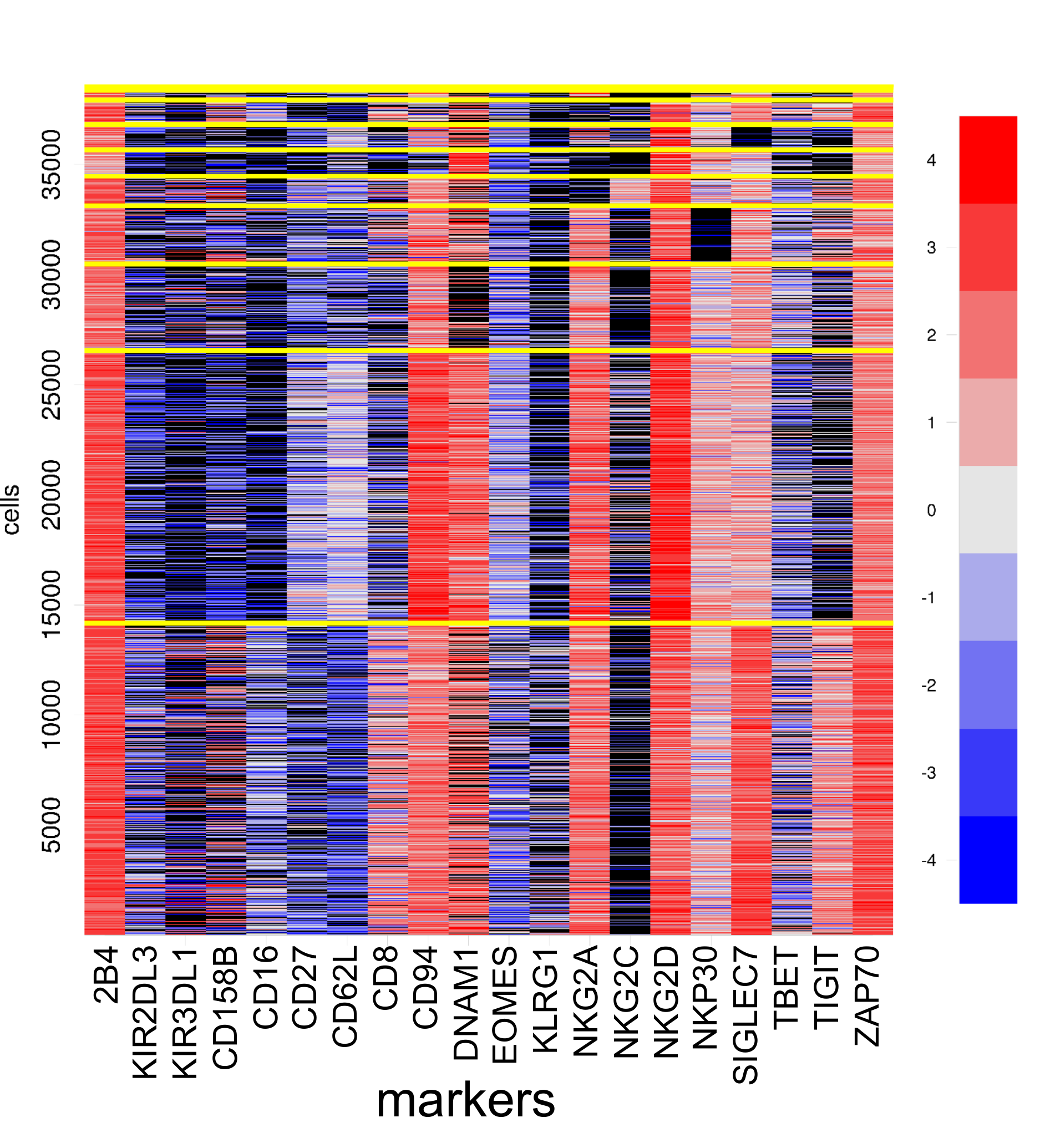}&
  \includegraphics[width=0.48\columnwidth]{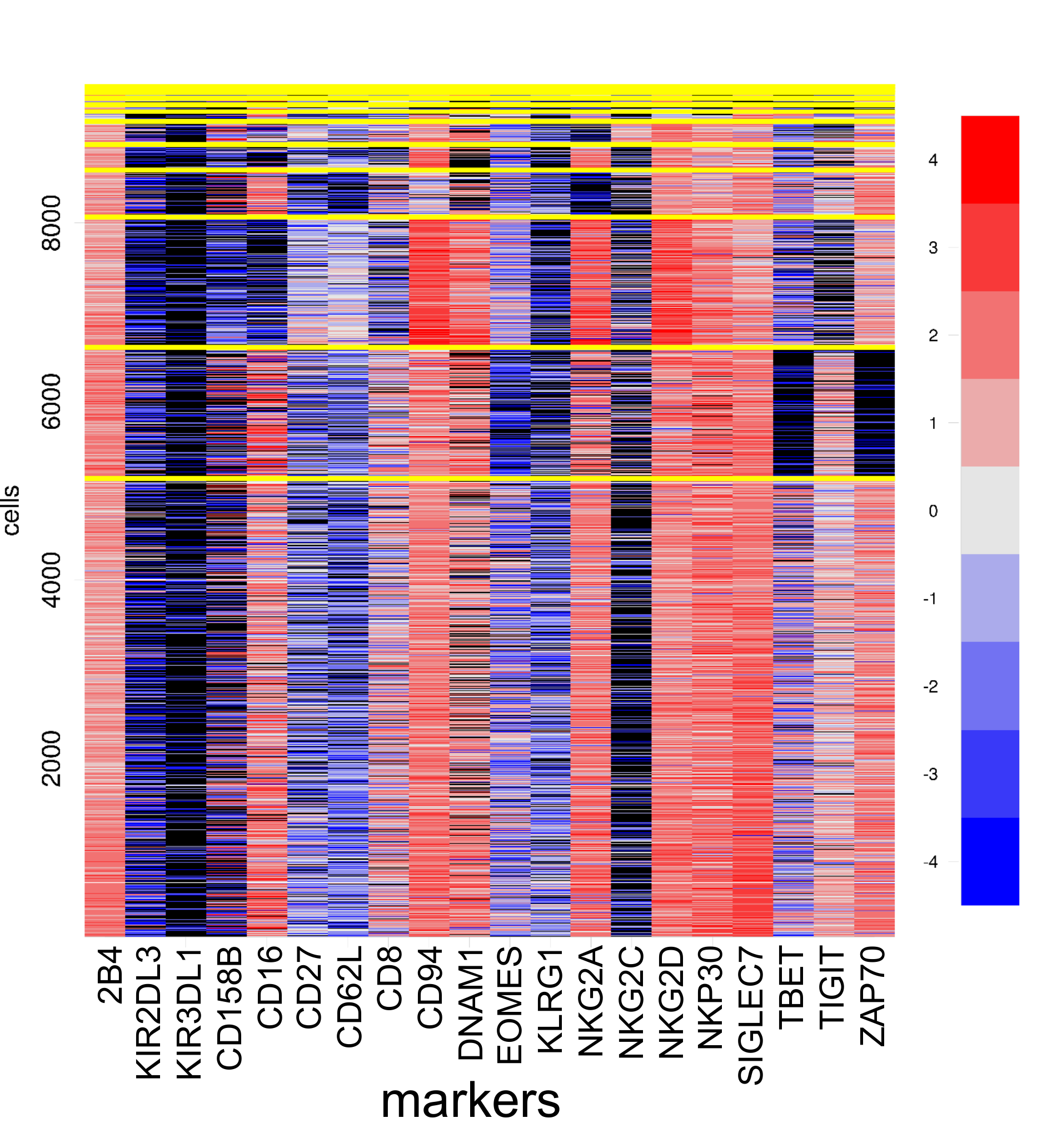}\\
  (a) Clustering of $y_{1nj}$ & (b) Clustering of $y_{2nj}$\\
  \includegraphics[width=0.48\columnwidth]{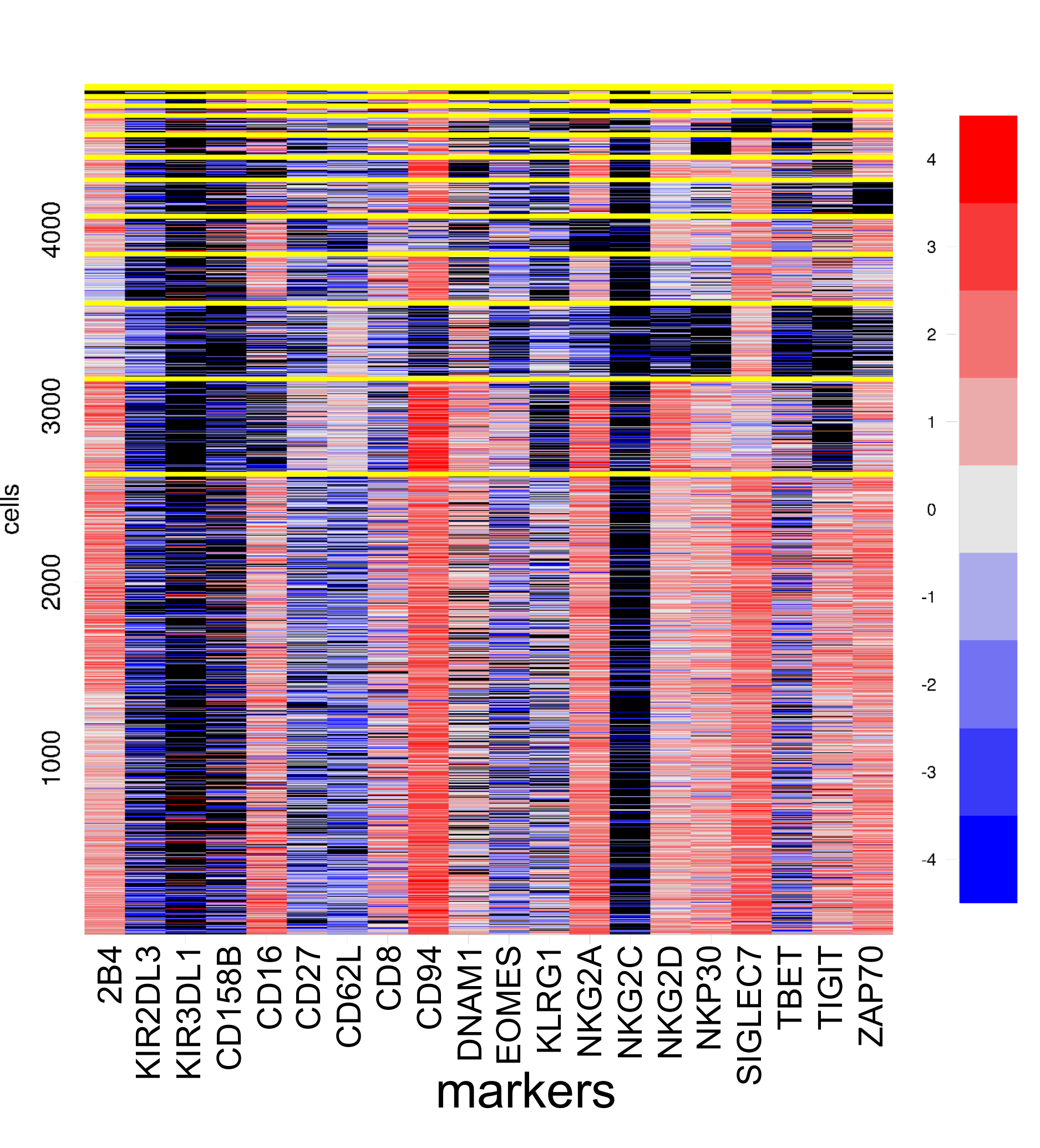} &
  \includegraphics[width=0.40\columnwidth]{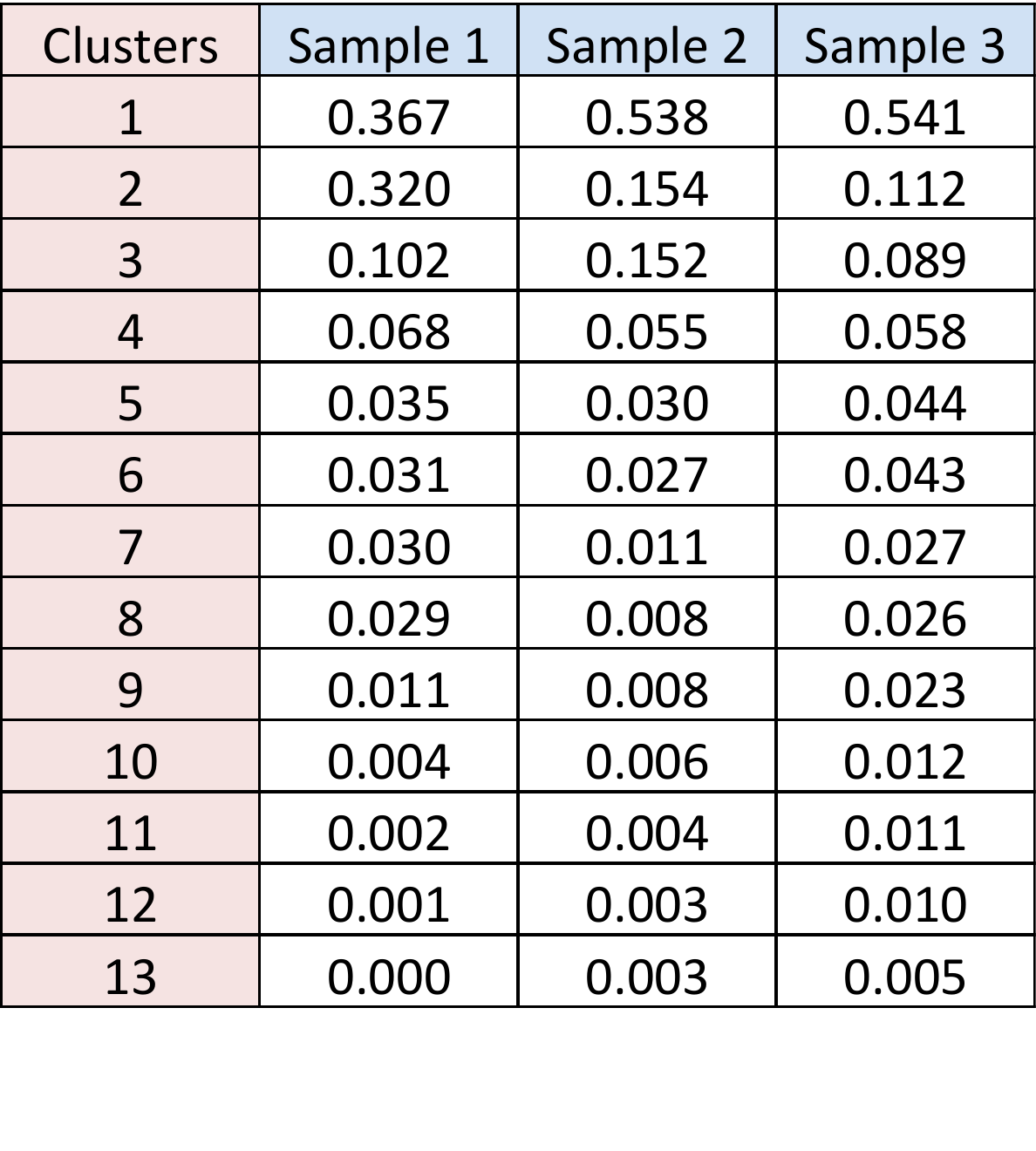} \\
  (c) Clustering of $y_{3nj}$ & (d) Proportions\\
  \end{tabular}
  \end{center}
  \vspace{-0.1in}
  \caption{[CB Data: Comparison to FlowSOM] Heatmaps of cells in (a)-(c) for
  samples 1-3, respectively. Cells are arranged by
  the cluster membership estimates by FlowSOM. The clusters are separated by yellow
  horizontal lines, with the most abundant clusters in each sample closer to
  the bottom. High, low, and missing expression levels are red, blue, and
  black, respectively.  The proportions of the cells in the estimated clusters
  are shown in (d).}
\label{fig:cb-flowsom}
\end{figure}
%%%%%%%%%%%%%%%%%%%%%%%%%

For comparison, we also applied FlowSOM to the UCB data. We fixed the missing
values of $y_{i,n,j}$ at the minimum of the negative observed values of $y$
for each $(i, j)$ prior to analysis.
FlowSOM identified 13 cell clusters in the samples.
Heatmaps of $y_{i,n,j}$ rearranged by cell clustering estimates by FlowSOM are
given in Fig \ref{fig:cb-flowsom} (a)-(c).  Heterogeneity between cells
within clusters estimated under FlowSOM is noticeably greater than that under
the proposed FAM shown in Fig~\ref{fig:cb-post}.  For example, marker 10
shows a mix of red, blue, and black colors for cluster 1, the largest cluster.
The proportions of cells assigned to the clusters are summarized in Fig
\ref{fig:cb-flowsom}(d).  The clusters are much larger than those under the
FAM. Particularly, cluster 1 under FlowSOM contains 36.7\%, 53.8\% and 54.1\%
of the cells in samples 1-3, respectively. Lastly, FlowSOM does
not produce an explicit inference on the characterization of subpopulations.

%
% We also compared the clustering by FlowSOM to those by MCMC using ARI.
% Computed ARI values are 0.201, 0.106 and 0.131 for the samples, respectively.
%

\section{Discussion}\label{sec:conclusions}
We have proposed a Bayesian FAM to identify and estimate cell subpopulations
using CyTOF data.  Our FAM identifies latent subpopulations, defined as functions
of the marker expression levels, and fits the data in multiple samples
simultaneously. The model accounts formally for missing values and
between-sample variability. The fitted FAM assigns each cell in each sample
to exactly one subpopulation, but each surface marker can belong to more than one
subpopulation. The method also yields cell clusters within each sample that are
defined in terms of the inferred subpopulations. We constructed a data missingship
mechanism based on expert knowledge, and we examined the robustness of the
model to the specification of the missingship mechanism through simulation.
This showed that inferences were not sensitive to changes in the
specification of the missingship mechanism. Compared to established
clustering methods, including FlowSOM, the proposed FAM is more effective at
discovering latent subpopulations when the underlying cell subpopulations are similar.

Our proposed  FAM can be extended to accommodate similar but more complex data structures,
in particular  including covariates. For example, samples with similar covariates may also
have similar cell subpopulation structures. The model can incorporate such
information by incorporating appropriate regression submodels,
to enhance inferences and study how the structures may change with
covariates.  One also may introduce the concept of \lq\lq repulsiveness\rq\rq\
to latent features and obtain a parsimonious representation of the latent
subpopulations by discouraging the creation of redundant subpopulations.  Repulsive
models, which are more likely to produce features that differ from each other substantially,
have been developed mostly in the context of mixture models (e.g., see
\cite{petralia2012repulsive, quinlan2018density, xie2019bayesian}).
\cite{xu2016bayesian} used the detrimental point process (DPP) for a repulsive
FAM that uses the determinant of a matrix as a repulsiveness metric. A model
that explicitly penalizes the inclusion of similar features also can be
developed to replace the IBP in our model.
%%% END_OF_SECTIONS -- DO NOT REMOVE!!! %%%

\vskip .2in

\noindent {\bf Acknowledgments}\hfil\break

This work was supported by NIH 1 R01 CA211044-01, 5 P01CA148600-03, and P50CA100632-16 (Katy Rezvani), a grant (CA016672) to the M.D. Anderson Cancer Center from the NIH (Katy Rezvani) and NSF grant DMS-1662427 (Juhee Lee).

% Title Settings
\newpage
\noindent {\Huge Supplementary Materials} \\

\section{Data and Code}
\noindent
Data used for this project is available at
\url{https://github.com/luiarthur/cytof-data}. \\

\noindent
This project was implemented in the Julia programming language. Code for this
project is available at \url{https://github.com/luiarthur/CytofResearch}. \\

%%% BEGINNING_OF_SECTIONS -- DO NOT REMOVE!!! %%%
\section{Posterior Computation}\label{sec:post-comp}

\subsection{MCMC Simulation}
Recall that $\btheta=\bc{\bZ, \bw, \bm \delta_0, \bm \delta_1, \bm \sigma^2,
\bm \eta^0, \bm \eta^1, \bm \lambda, \bm v, 
\bm \epsilon, \alpha}$ denotes all random parameters. We let expression
levels $\y$ and binary indicators $\m$ denote $y_{i,n,j}$ and $m_{i,n,j}$,
respectively, for all $(i,n,j)$. To facilitate the posterior sampling of
$\delta_{z,\ell}$, we introduce auxiliary indicators for normal mixture 
components $\gamma_{i,n,j} \in \{1, \ldots, L_{z_{j, \lin}}\}$ for each
$y_{i,n,j}$ when $\lin\neq 0$. That is, $p(\gamma_{i,n,j} = \ell \mid z_{j,\lin}=z,
\eta^z_{i,j,\ell}, \lin\neq 0) = \eta^z_{i,j,\ell}$, where $\ell \in \bc{1,
\ldots,L_{z_{j, \lin}}}$, and let $\mu_{i,n,j}=\mu^\star_{z_{j, \lin}, \gamma_{i,n,j}}$. We extend the vector of random parameters,
$\widetilde{\btheta}=(\btheta, \{\gamma_{i,n,j}\})$ by including $\gamma_{i,n,j}$ for more convenient
posterior simulation. Similar to the joint posterior distribution of
$\btheta$ in \eqref{eq:joint-post} of the main text, the joint posterior
probability model of $\widetilde{\btheta}$ under our Bayesian FAM model is
\begin{eqnarray}
p(\widetilde{\btheta} \mid \y, \m, K) 
&\propto&
p(\widetilde{\btheta}\mid K)
\prod_{i,n} \bk{
  \prod_j
  \rho_{i,n,j}^{1-m_{i,n,j}} \times 
  \frac{1}{\sqrt{2\pi\sigma^2_{i}}}
  \exp\bc{-\frac{(y_{i,n,j}-\mu_{i,n,j})^2}{2\sigma^2_{i}}}}^{1(\lin\neq 0)}
  \nonumber \\
&&
\times \bk{\prod_j \rho_{i,n,j}^{1 - m_{i,n,j}} \times
\frac{1}{\sqrt{2\pi s^2_\epsilon}}
\exp\bc{-\frac{y_{i,n,j}^2}{2 s^2_{\epsilon}}}}^{1(\lin=0)}.
\label{eq:joint-post-supp}
\end{eqnarray}
Posterior samples of $\widetilde{\btheta}$ are obtained by iteratively
drawing samples from each of the full conditionals using the most recent
estimate of the parameters and the data. For the parameters whose conditional
distributions are known and are easy to sample from, we used Gibbs
sampling. To sample from full conditionals which are otherwise difficult to
sample from, the Metropolis-Hastings algorithm was used.

\begin{enumerate}
\item Full Conditional for $v_k$

% Classical IBP (Truncated)
Recall that the prior distribution for $v_k$ is
$v_k \mid \alpha \ind \Be(\alpha / K, 1)$, for $k = 1,...,K$, that is,
$p(v_k \mid \alpha) = \frac{\alpha}{K} v_k^{\alpha/K-1}$.
\begin{align*}
p(v_k \mid \y, \rest) &\propto p(v_k) \prod_{j=1}^J p(z_{j,k} \mid v_k) \\
&\propto \frac{\alpha}{K} v_k^{\alpha/K - 1} \prod_{j=1}^J 
v_k^{z_{j,k}} (1 - v_k)^{1 - z_{j,k}}\\
&\propto v_k^{\alpha/K + \sum_{j=1}^J z_{j,k}- 1}
(1 - v_k)^{J - \sum_{j=1}^J z_{j,k}}
\end{align*}
$$
\Rightarrow v_k \mid \y, \rest \sim \text{Be}\p{
  \alpha / K + \sum_{j=1}^J z_{j,k}, ~
  J + 1 - \sum_{j=1}^J z_{j,k}
}.
$$ 
We use ``$\rest$'' to denote all parameters except the parameter(s) that we
sample. For example, ``$\rest$'' implies $\widetilde{\btheta} \backslash \{v_k\}$ for updating $v_k$.

% fc-Z
\item Full Conditional for $z_{j,k}$

Let $S_k = \bc{(i, n): \lambda_{i,n} = k}$, the set of cells in samples taking
cell subpopulation $k$.
%%
% classic IBP 
\def\pzOne{
 v_k \prod_{(i,n) \in S_k} \sum_{\ell=1}^L \eta^1_{i,j,\ell} \cdot
  \phi(y_{i,n,j} \mid \mu^\star_{1,\ell}, \sigma^{2}_{i})
}
\def\pzZero{
(1-v_k) \prod_{(i,n) \in S_k} \sum_{\ell=1}^L \eta^0_{i,j,\ell} \cdot
  \phi(y_{i,n,j} \mid \mu^\star_{0,\ell}, \sigma^{2}_{i})
}
\begin{align*}
p(z_{j,k} = 1 \mid \y, \rest) &\propto p(z_{j,k} = 1 \mid v_k)
\prod_{(i,n) \in S_k} p(y_{i,n,j} \mid \bm\mus_1, \bet_{i,j}^1, \sigma^2_i) \\
&\propto \pzOne, \\
p(z_{j,k} = 0 \mid \y, \rest) &\propto p(z_{j,k} = 0 \mid v_k)
\prod_{(i,n) \in S_k} p(y_{i,n,j} \mid \bm\mus_0, \bet_{i,j}^0, \sigma^2_i) \\
&\propto \pzZero,
\end{align*}
where $\phi(y\mid m, s^2)$ denotes the probability density function of the
normal distribution with mean $m$ and variance $s^2$, evaluated at $y$. 
$$
\Rightarrow z_{j,k} \mid \y, \rest \sim \text{Ber}\p{
  \bk{1 + \frac{\pzZero}{\pzOne}}^{-1}
}.
$$ 

% fc-alpha
\item Full Conditional for $\alpha$
% classical IBP
\begin{align*}
p(\alpha \mid \y, \rest) &\propto p(\alpha) \times \prod_{k=1}^K p(v_k \mid
  \alpha) \\
&\propto \alpha^{a_\alpha - 1} \exp\bc{-b_\alpha \alpha} \times \prod_{k=1}^K 
\alpha~v_k^{\alpha/K} \\
&\propto \alpha^{a_\alpha + K -1} \exp\bc{-\alpha\p{b_\alpha - 
\sum_{k=1}^K \log v_k / K}} 
\end{align*}
$$
\Rightarrow \alpha \mid \y, \rest \sim 
\G\p{a_\alpha + K,~ b_\alpha - \sum_{k=1}^K \log v_k /K}.
$$

\item Full Conditional for $\lin$

\def\Ainjk{
  \sum_{\ell=1}^{L} \eta^{z_{j,k}}_{i,j,\ell} \cdot
  \phi(y_{i,n,j} \mid 
  \mus_{z_{j,k},\ell}, \sigma^2_{i})
}

The prior for $\lin$ is 
$$
p(\lin = k \mid \bw_i, \epsilon_i) =
\begin{cases}
\epsilon_i, &\text{if } k = 0\\
(1 - \epsilon_i) \cdot w_{i,k}, &\text{if } k \in \bc{1, \dots, K}. 
\end{cases}
$$
We thus have
%%%%
\begin{align*}
p(\lin=0\mid \y,\rest) &\propto p(\lin=0) ~ p(\y \mid \lin=0, \rest) \\
&\propto \epsilon_i \prod_{j=1}^J \phi(y_{i,n,j} \mid 0, s_\epsilon^2), \\
p(\lin=k\mid \y,\rest) &\propto p(\lin=k) ~ p(\y \mid \lin=k, \rest) \\
&
\propto (1 - \epsilon_i) w_{ik} \prod_{j=1}^J \p{ \Ainjk }, \mbox{ for } k =1, \ldots, K.
\end{align*}
We sample $\lin$ with probabilities proportional to 
$p(\lin=k \mid \y,\rest)$ for $k \in \bc{0, \dots, K}$.

\item Full Conditional for $\bm w_{i}$

The prior for $\bm{w}_i=(w_{i,1}, \ldots, w_{i,K})$ is $\bm w_i \sim \Dir(d/K, \cdots, d/K)$. The full conditional for $\bm{w}_i$ is:
\begin{align*}
p(\bm w_i \mid \rest) \propto&~~ p(\bm{w}_i) \times \prod_{n=1}^{N_i} p(\lin \mid \bm{w}_i)\\
%\propto&~~ p(\bm{w}_i) \times \prod_{n=1}^{N_i}\prod_{k=1}^K  w_{ik}^{1(\lin=k)}\\
%\propto&~~ \prod_{k=1}^K w_{ik}^{d/K-1} \times \prod_{n=1}^{N_i}\prod_{k=1}^K  w_{ik}^{1(\lin=k)}\\
\propto&~~ \prod_{k=1}^K w_{ik}^{\p{d/K + \sum_{n=1}^{N_i}1(\lin=k)}-1}.
\end{align*}
Therefore, 
$$
\bm{w}_i \mid \y,\rest ~\sim~
\Dir\p{d/K+\sum_{n=1}^{N_i}1(\lambda_{i,n}=1),...,d/K+\sum_{n=1}^{N_i}1(\lambda_{i,n}=K)}. 
$$
%Consequently, the full conditional for $\bm{W}_i$ can be sampled from
%directly from a Dirichlet distribution of the form above. \\

% fc-gam
\item Full Conditional for $\gamma_{i,n,j}$
%%%%%%%%%%55

For the cells with $\lin > 0$,
\begin{align*}
p(\gamma_{i,n,j}=\ell \mid \y, z_{j\lin}=z, \rest) &\propto p(\gamma_{i,n,j}=\ell)
  \times p(y_{i,n,j} \mid \gamma_{i,n,j}=\ell, \rest) \\
%&\propto p(\gamma_{i,n,j}=\ell) \times p(y_{i,n,j} \mid \mus_{z,\ell},
%  \sigma^2_{i}, \rest) \\
%
%&\propto \eta^z_{ij\ell} \times \N(y_{i,n,j} \mid \mus_{z,\ell}, \sigma^2_i) \\
&= \eta^z_{ij\ell} \times \phi(y_{i,n,j} \mid \mus_{z \ell}, \sigma^2_i).
%(\sigma^2_i)^{-1/2}
%\exp\bc{-\frac{(y_{i,n,j} - \mus_{z,\ell})^2}{2\sigma^2_i}}.
\end{align*}
Therefore, sample $\gamma_{i,n,j}$ with probabilities proportional to
$p(\gamma_{i,n,j}=\ell \mid \y,\rest)$ for $\ell = 1,...,L^{z_{j,\lin}}$.

% fc-delta
\item Full Conditional for $\delta_{z,\ell}$

For $\delta_{1,\ell}$, let $S_{1,i,\ell} = \bc{(i,n,j) : \p{z_{j,\lin} = 1
~\cap~ \gamma_{i,n,j} \ge \ell}}$ and $|S_{1,i,\ell}|$ the cardinality of
$S_{1,i,\ell}$.
%%%
\newcommand\dOnePostvarDenom{
  \frac{1}{\tau^2_1} +
  \sum_{i=1}^I\frac{|S_{1,i,\ell}|}{{\sigma^2_{i}}}
}
\newcommand\dOnePostMeanNum{
  \frac{\psi_1}{\tau^2_1} + 
  \sum_{i=1}^I \sum_{S_{1,i,\ell}}  
  \frac{g_{i,n,j}}{{\sigma^2_{i}}}
}
%%%%
\begin{align*}
p(\delta_{1,\ell} \mid \y, \rest) &\propto 
p(\delta_{1,\ell} \mid \psi_1, \tau^2_1) \times p(\y \mid \delta_{1,\ell},\rest) \\
&\propto
1(\delta_{1,\ell} \ge 0) \times
\exp\bc{\frac{-(\delta_{1,\ell} - \psi_1)^2}{2\tau^2_{1}}}
\\
& \hspace{5em} \times \prod_{i=1}^I\prod_{(i,n,j)\in S_{1i\ell}}
\exp\bc{-\p{y_{i,n,j} -
\sum_{r=1}^{\gamma_{i,n,j}} \delta_{1r}}^2 \bigg/ 2\sigma^2_{i}} \\
&\propto
\exp\bc{
  -\frac{(\delta_{1,\ell})^2}{2}\p{\dOnePostvarDenom} + 
  \delta_{1,\ell}\p{\dOnePostMeanNum}
} \\ 
& \hspace{5em} \times 1(\delta_{1,i,\ell} \ge 0),
\end{align*}
where $\ds g_{i,n,j} = y_{i,n,j} -
\sum_{r=1}^{\gamma_{i,n,j}} (\delta_{1,r})^{1(r \neq \ell)}$.

$$
\renewcommand\dOnePostvarDenom{
  1 + \tau^2_1\sum_{i=1}^I(|S_{1,i,\ell}|/{\sigma^2_{i}})
}
\renewcommand\dOnePostMeanNum{
  \psi_1 + \tau^2_1 \sum_{i=1}^I\sum_{S_{1,i,\ell}} (g_{i,n,j} /
  {\sigma^2_{i}})
}
\Rightarrow \delta_{1,\ell} \mid \y, \rest \ind \text{TN}^+\p{
  \frac{\dOnePostMeanNum}{\dOnePostvarDenom},
  \frac{\tau^2_1}{\dOnePostvarDenom}
}.
$$

Similarly, for $\delta_{0,\ell}$, let
$S_{0,i,\ell} = \bc{(i,n,j) : \p{Z_{j,\lin} = 
0 ~\cap~ \gamma_{i,n,j} \ge \ell}}$
and $|S_{0,i,\ell}|$ be the cardinality of $S_{0,i,\ell}$.
$$
\newcommand\dZeroPostvarDenom{
  1 + \tau^2_0 \sum_{i=1}^I (|S_{0,i,\ell}|/{\sigma^2_{i}})
}
\newcommand\dZeroPostMeanNum{
  \psi_0 + \tau^2_0 \sum_{i=1}^I \sum_{S_{1,i,\ell}} (g_{i,n,j} /
  {\sigma^2_{i}})
}
\Rightarrow \delta_{0,l} \mid \y, \rest \ind \text{TN}^+\p{
  \frac{\dZeroPostMeanNum}{\dZeroPostvarDenom},
  \frac{\tau^2_0}{\dZeroPostvarDenom}
},
$$
where $\ds g_{i,n,j} = -y_{i,n,j} -
\sum_{r=1}^{\gamma_{i,n,j}} (\delta_{0,r})^{1(r \neq \ell)}$.

% fc-sig2
\item Full Conditional for $\sigma^2_i$

Let $r_{i,n,j} = 1(\lin > 0)$, and let
$R_i = \sum_{n=1}^{N_i}\sum_{j=1}^J r_{i,n,j}$. We then have
\begin{align*}
p(\sigma^2_i \mid \y, \rest) &\propto p(\sigma^2_i) \times p(\y
  \mid \sigma^2_i, \rest) \\
&\propto (\sigma^2_i)^{-a_\sigma-1}
         \exp\bc{-\frac{b_\sigma}{\sigma^2_i}} 
\prod_{j=1}^J \prod_{n=1}^{N_i} \bc{
  \frac{1}{\sqrt{2\sigma^2_i}}
  \exp\bc{\frac{-(y_{i,n,j}-\mu_{i,n,j})^2}{2\sigma^2_i}}
} \\
&\propto (\sigma^2_i)^{-\p{a_\sigma + \frac{R_i}{2}}-1}
\exp\bc{-\p{\frac{1}{\sigma^2_i}}\p{b_\sigma + \sum_{j=1}^J \sum_{n=1}^{N_i} 
r_{i,n,j}\cdot\frac{(y_{i,n,j}-\mu_{i,n,j})^2}{2}
}}.
\end{align*}
$$
\Rightarrow \sigma^2_i \mid \y, \rest \ind
\IG\p{a_\sigma + \frac{R_i}{2}, ~~ b_\sigma + \sum_{j=1}^J \sum_{n=1}^{N_i} 
r_{i,n,j}\cdot\frac{(y_{i,n,j}-\mu_{i,n,j})^2}{2}
}.
$$

% fc-eta
\item Full Conditional for $\eta^z_{i,j}$

The prior for $\bm\eta^z_{i,j}$ is
$\bm \eta^z_{i,j} \sim \Dir_{L_z}(a_{\eta^z})$, for $z\in\bc{0,1}$.
So the full conditional for $\bm\eta^z_{i,j}$ is:
%%%
\def\etaind{
1\{(\gamma_{i,n,j}=\ell) ~\&~ (z_{j,\lin}=z) ~\&~ (\lin>0)\}
}
%%%
\begin{align*}
p(\bm \eta^z_{i,j} \mid \rest) \propto&~~ p(\bm{\eta}^z_{i,j}) \times
  \prod_{n=1}^{N_i} p(\gamma_{i,n,j} \mid \bm \eta^z_{i,j})\\
  \propto&~~ \prod_{\ell=1}^{L_z} \p{\eta^z_{i,j,\ell}}^{a_{\eta^z}-1} \times 
  \prod_{\ell=1}^{L_z} \prod_{n=1}^{N_i} \p{\eta^z_{i,j,\ell}}^{\etaind}\\
  \propto&~~ \prod_{\ell=1}^{L_z} \p{\eta^z_{i,j,\ell}}^{\p{a_{\eta^z}+
  \sum_{n=1}^{N_i} \etaind} - 1}.
\end{align*}
$$
\Rightarrow \bm{\eta}^z_{i,j} \mid \y,\rest ~\sim~ \Dir_{L_z}\p{a^*_1,...,a^*_{L_z}},
$$ where $a^*_\ell = a_{\eta^z}+\sum_{n=1}^{N_i}\etaind$.

% fc-epsilon
\item Full Conditional for $\epsilon_i$
\begin{align*}
  p(\epsilon_i \mid y, \rest) &\propto p(\epsilon_i) \prod_{n=1}^{N_i}
  \epsilon_i^{1(\lin=0)} (1-\epsilon_i)^{1(\lin>0)} \\
  &\propto \epsilon_i^{a_\epsilon - 1} (1-\epsilon_i)^{b_\epsilon-1}
  \epsilon_i^{\sum_{n=1}^{N_i}1(\lin=0)}
  (1-\epsilon_i)^{\sum_{n=1}^{N_i}1(\lin>0)} \\
  &\propto
  \epsilon_i^{a_\epsilon + \sum_{n=1}^{N_i}1(\lin=0) - 1}
  (1-\epsilon_i)^{b_\epsilon + \sum_{n=1}^{N_i}1(\lin>0) - 1}.
\end{align*}
$$
\Rightarrow 
\epsilon_i \mid y, \rest \sim \Be\p{
  a_\epsilon + \sum_{n=1}^{N_i}1(\lin=0),
  b_\epsilon + \sum_{n=1}^{N_i}1(\lin>0)}.
$$

% fc-y
\item Full Conditional for Missing $y_{i,n,j}$
%%%%
\begin{align*}
p(y_{i,n,j} \mid m_{i,n,j}=1, \rest) &\propto
p(m_{i,n,j} =1\mid y_{i,n,j}, \rest) ~
p(y_{i,n,j} \mid \rest) \\
&\propto
\rho_{i,n,j} 
\sum_{\ell=1}^{L} \eta^{z_{j,\lin}}_{i,j,\ell} \cdot \phi(y_{i,n,j} \mid
\mus_{z_{j,k}, \ell}, \sigma^2_{i}).
\end{align*}
Direct sampling from the full conditional of $y_{i,n,j}$ is difficult,
so we use a Metropolis step with a normal proposal distribution to
sample from the full conditional instead.

\end{enumerate}

\subsection{Variational Inference Implementation Details}\label{sec:vi}
Variational inference (VI) is a
popular alternative for fitting Bayesian models \citep{jordan1999introduction,
beal2003variational, wainwright2008graphical, blei2017variational}. VI tends to be faster and
more scalable with data size than the traditional MCMC method. In particular,
we utilize automatic differentiation variational inference (ADVI),
\citep{advi}, a derivation-free method. It is a gradient-based
stochastic optimization method and is amenable to common machine learning
techniques, such as stochastic gradient descent, which makes inference for
large datasets more tractable. For a comprehensive review of recent advances in
VI, see \cite{blei2017variational} and \cite{zhang2018advances}.

In VI, parameters of a tractable approximating \dquote{variational}
distribution are iteratively optimized until it \dquote{sufficiently}
resembles the target (posterior) distribution. The most common metric for
measuring the \dquote{closeness} of the target distribution to the
variational distribution is the Kullback-Leibler (KL) divergence
\citep{kullback1951information}. For our Bayesian feature allocation model
(FAM), minimizing the KL divergence between the variational distribution and
the posterior distribution is equivalent to maximizing the following evidence lower
bound (ELBO)
\begin{align}
    \elbo &= \E_Q\bk{\log p(\m, \y \mid \btheta) + \log p(\btheta) 
              - \log q(\btheta) - \log q(\y^\miss)} \nonumber \\
    &= \E_Q\bk{\log p(\m \mid \y, \btheta) + \log p(\y \mid \btheta) + \log p(\btheta)
               - \log q(\btheta) - \log q(\y^\miss)} \nonumber \\
    &= \E_Q\bk{\log p(\m \mid \y) + \log p(\y \mid \btheta) + \log p(\btheta)
               - \log q(\btheta) - \log q(\y^\miss)} \label{eq:elbo}.
\end{align}
$p(\bm m \mid \y)$ and $p(\y\mid\btheta)$ are the sampling distributions of
$m_{i,n,j}$ and $y_{i,n,j}$, and $p(\btheta)$ is the prior distribution for
all model parameters. $q(\btheta)$ is the mean-field variational distribution
for model parameters. For $q(\btheta)$, each model parameter is
transformed to the unconstrained space \citep{advi} and is assumed to have
a normal distribution \citep{advi}. $q(\y^{\miss}) = \prod_{i,n,j}
q(y_{i,n,j})^{1(m_{i,n,j}=0)}$ is an amortized  variational
distribution for the missing values\citep{vae}. Specifically, $q(y_{i,n,j}^\miss)$ is a
normal probability density function with mean $r_{i,j}$ and standard deviation
$s_{i,j}$. This simplification for the missing $y_{i,n,j}$ will produce imputed
values different from those under our Bayesian FAM, but yields acceptable
performance in our simulation studies at greatly reduced computational cost.
Computing the gradient (in gradient descent) requires the computation of the
ELBO using the entire dataset. This can be computationally prohibitive for large datasets. Instead, 
stochastic gradient descent (SGD) is used.  A mini-batch of size $B$ (much less than
the size of the full data set $N$) can be sampled at each iteration of the
SGD to compute the ELBO. The ELBO should be appropriately scaled by $N / B$.
This works well in practice provided that the size of the mini-batch is  
sufficiently large.

In our model, parameters of primary interest $\bZ$ and $\bm \lambda$ are discrete. Since ADVI is only valid for continuous parameters in
differentiable models, we let $z_{j,k} = 1(v_k > h_{j,k})$, where $v_k \mid
\alpha \sim \Be(\alpha / K, 1)$, and $h_{j,k} \sim \mbox{Unif}(0, 1)$, similar
to the construction of the dependent IBP in \cite{williamson2010dependent}.
We approximate the gradient of the indicator function with the gradient of
$\text{sigmoid}\p{(\logit(v_k) - \logit(h_{j,k})) \cdot 1000}$, which is smooth.
We marginalize over $\bm \lambda$ for VI, and then sample from
their full conditionals using the parameters estimated from the variational
distributions.

For completeness, we have included key terms in the computation of 
the ELBO using SGD.
%
% \hh what is $c$?? why sigmoid??? not logit link???  Is it $p$ or $\rho$???\ech
$p(\m \mid \y)$ is defined as
% P(m|y)
\begin{align*}
    p(\m \mid \y) &= \prod_{i=1}^I \prod_{n=1}^{N_i} p(\bm m_{i,n} \mid \bm y_{i,n}) \\
    &=  \prod_{i=1}^I \prod_{n=1}^{N_i}
        \prod_{j=1}^J \rho_{i,n,j}^{1 - m_{i,n,j}}
        \p{1 - \rho_{i,n,j}}^{m_{i,n,j}}\\
    &=  \prod_{i=1}^I \prod_{n=1}^{N_i}
        \prod_{j=1}^J \rho_{i,n,j}^{1 - m_{i,n,j}} c_{i,n,j} \\
    &=  \prod_{i=1}^I \prod_{n=1}^{N_i} \prod_{j=1}^J \rho_{i,n,j}^{1 - m_{i,n,j}}
        \prod_{i=1}^I \prod_{n=1}^{N_i} \prod_{j=1}^J c_{i,n,j} \\
    &=  C \prod_{i=1}^I \prod_{n=1}^{N_i} \prod_{j=1}^J \rho_{i,n,j}^{1-m_{i,n,j}},
\end{align*}
where $\rho_{i,n,j} = \sigmoid(\beta_{0,i} + \beta_{1,i} y_{i,n,j} +
\beta_{2,i} y_{i,n,j}^2)$, and
$C=\ds\prod_{i=1}^I \prod_{n=1}^{N_i} \prod_{j=1}^J c_{i,n,j}$ is a constant.
Computing $p(\m \mid \y)$ is computationally expensive when $N_i$ is large.
Hence, we can approximate it by only iterating through a subset of the data,
and scaling the relavant terms. The log of the resulting expression is:
\begin{align*}
    \log p(\m \mid \y) &=
      \log C + \sum_{i=1}^I \sum_{n=1}^{N_i} \sum_{j=1}^J
      (1-m_{i,n,j}) \log \rho_{i,n,j} \nonumber \\
    &\approx
      \log C + \sum_{i=1}^I \frac{N_i}{\abs{S_i}}
      \sum_{n\in S_i} \sum_{j=1}^J (1-m_{i,n,j}) \log \rho_{i,n,j}
      %\label{eq:pmgy}
\end{align*}
where $S_i$ is a subset of $\bc{1, \dots, N_i}$.
The likelihood term $p(\y\mid\btheta)$ is defined as
% P(y|theta)
\begin{align*}
    p(\y \mid \btheta) &= \prod_{i=1}^I \prod_{n=1}^{N_i} 
    \underbrace{\left\{
      \epsilon_i \prod_{j=1}^J \N(0, s^2_\epsilon) +
      (1-\epsilon_i) \sum_{k=1}^K w_{i,k} \prod_{j=1}^J \sum_{\ell=1}^{L_{z_{j,k}}}
      \eta_{i,j,\ell}^{z_{j,k}} \N(y_{i,n,j}\mid \mus_{z_{j,k}, \ell}, \sigma^2_i)
      \right\}
    }_\text{$=A_{i,n}$}.
\end{align*}
We thus have
\begin{align*}
    \log p(\y \mid \btheta) &= \sum_{i=1}^I \sum_{n=1}^{N_i} \log A_{i,n}
    \nonumber \\
    &\approx
    \sum_{i=1}^I \frac{N_i}{\abs{S_i}} \sum_{n\in S_i} 
    \log A_{i,n} ~~\text{(if using mini-batches)}
    %\label{eq:pygt}
\end{align*}
Finally, the variational distribution for the missing values in $\y$ is defined 
as 
% q(y)
\begin{align*}
  q(\y) &=
  \prod_{i=1}^I \prod_{n=1}^{N_i} \prod_{j=1}^J q(y_{i,n,j} \mid
  r_{i,j}, s_{i,j}) ^{m_{i,n,j}}
    \nonumber \\
    \Rightarrow
    \log q(\y) &= \sum_{i=1}^I \sum_{n=1}^{N_i} \sum_{j=1}^J
    m_{i,n,j}  \log q(y_{i,n,j} \mid r_{i,j}, s_{i,j}) \nonumber \\
    &\approx \sum_{i=1}^I \frac{N_i}{\abs{S_i}} \sum_{n\in S_i} \sum_{j=1}^J
    m_{i,n,j} \log q(y_{i,n,j} \mid r_{i,j}, s_{i,j})
    ~~\text{(if using mini-batches)}
\end{align*}
As previously noted, independent Gaussian variational distributions were placed
on all other model parameters $\btheta$ after they were transformed to have
support on $\mathbb{R}^{\text{dim}(\btheta)}$. Notably, the parameters with
support on simplexes (i.e. $\bm\eta$ and $\bw$) were transformed using the
stick breaking transformation \citep{stan2016stan}.

\begin{figure}[t]
  \centering
  \includegraphics[scale=.5]{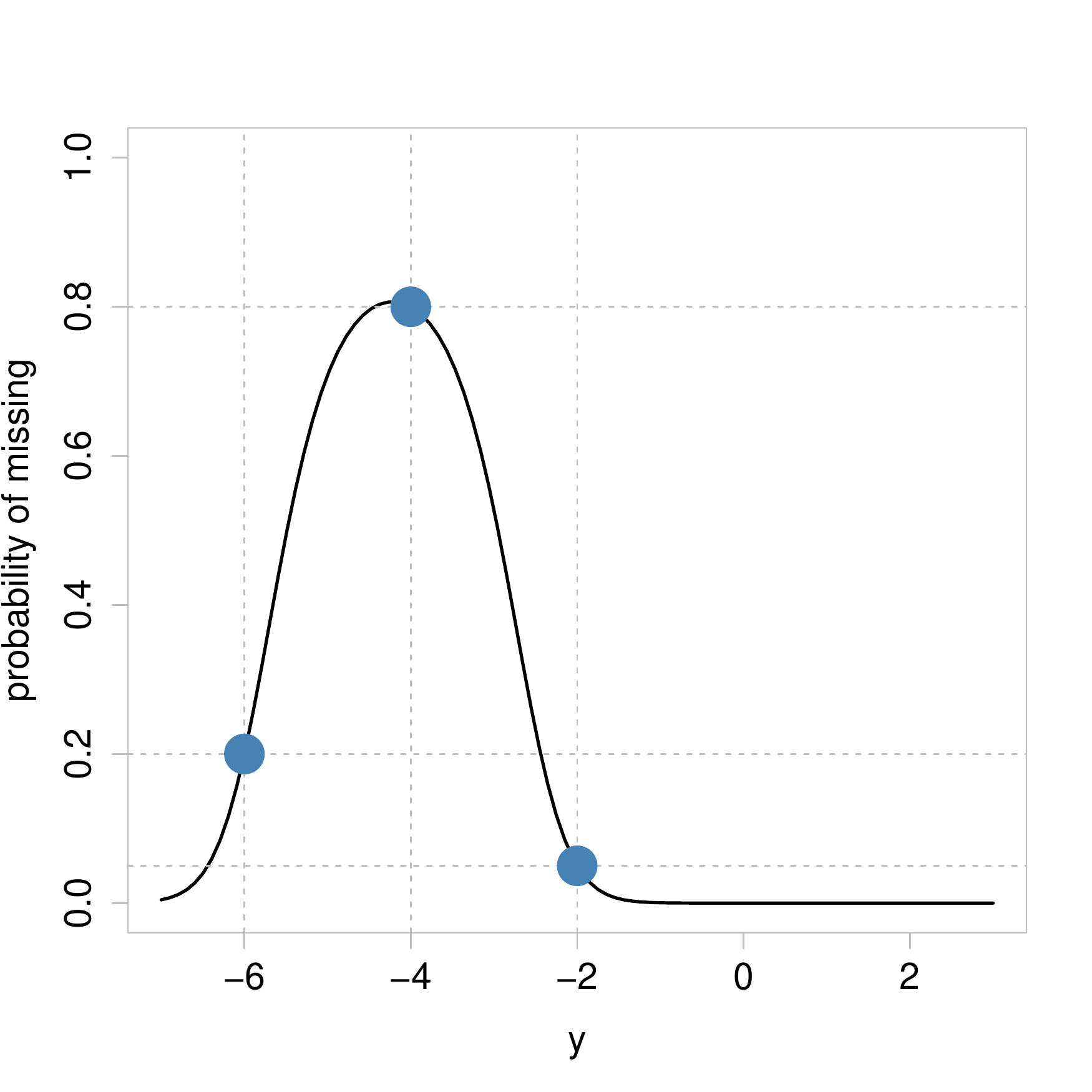}
  \caption{A quadratic data missingship mechanism for imputing missing data
  that passes through the points $(y_1=-6.0, p_1=0.2)$, $(y_2=-4.0,
  p_2=0.8)$, and $(y_3=-2.0, p_3=0.05)$.}
  \label{fig:prob-miss-eg}
\end{figure}

\section{Specification of Data Missingship Mechanism}\label{sec:missing-spec}
We discuss the approach used to specify the data missingship mechanism. Recall
that we assume a logit regression model for the probability $\rho_{i,n,j}$ for
the missing $y_{i,n,j}$ in \eqref{eq:link} of the main text,
$\logit(\rho_{i,n,j}) = \beta_{0,i} + \beta_{1,i} y_{i,n,j} + \beta_{2,i}
y_{i,n,j}^2$, with $\beta_{p,i} \in \mathbb{R}$, $p \in
\bc{0, 1, 2}$. To specify values of $\beta_{p,i}$,  we first select three points of $(\tilde{y}, \tilde{\rho})$ for each sample,  $(\tilde{y}_1, \tilde{\rho}_1)$, $(\tilde{y}_2, \tilde{\rho}_2)$, and $(\tilde{y}_3, \tilde{\rho}_3)$.  We let $\logit(\tilde{\rho}) =  \beta_{0,i} + \beta_{1,i} \tilde{y} + \beta_{2,i}\tilde{y}^2$ and solve for $\beta_{i,p}$.  We accommodate the subject knowldge that missing $y_{i,n,j}$ strongly indicates that the marker is not expressed in the selection of three points of $(\tilde{y}, \tilde{\rho})$, and the mechanism encourages imputed values to
take on negative values.
For instance,
Figure~\ref{fig:prob-miss-eg} shows an example of data missingship mechanism specified by selecting
$(-6.0, 0.2)$, $(-4.0, 0.8)$, and $(-2.0, 0.05)$ of $(\tilde{y}, \tilde{\rho})$.
This specification imputes values between -2 and -6 with large probability.  The mechanism thus strongly implies that the marker is not expressed.   We used empirical quantiles of negative values of observed $y$ to specify $\tilde{y}$.

%We first define $f_{i,n,j}$ to be
%\begin{align*}
%f_{i,n,j} &= P(m_{i,n,j} \mid \rho_{i,n,j}, y_{i,n,j}) \\
%&= \rho_{i,n,j}^{1-m_{i,n,j}} (1-\rho_{i,n,j})^{m_{i,n,j}} \\
%&= \p{\frac{1}{1+e^{-x_{i,n,j}}}}^{1-m_{i,n,j}}
%    \p{\frac{1}{1+e^{x_{i,n,j}}}}^{m_{i,n,j}},
%\end{align*}
%for $i = 1, \cdots, I$, where $x_{i,n,j} = \beta_{0,i} + \beta_{1,i}
%y_{i,n,j} + \beta_{2,i} y_{i,n,j}^2$, ~$\beta_{p,i} \in \mathbb{R}$, for $p \in
%\bc{0, 1, 2}$. We select three points to constrain the data missingship
% mechanism $(y_1, p_1)$, $(y_2, p_2)$, and $(y_3, p_3)$, and solve for
%$\bm\beta_i=(\beta_{0,1}, \beta_{1,i}, \beta_{2,i})$.
%
%As mentioned in the main text, the mechanism encourages imputed values to
%take on negative values where data are observed with high probability.
%For instance,
%
%Figure~\ref{fig:prob-miss-eg} shows an example data missingship mechanism where
%$(y_1=-6.0, p_1=0.2)$, $(y_2=-4.0, p_2=0.8)$, and $(y_3=-2.0, p_3=0.05)$.
%This specification limits imputed values to take on values less than 
%-2 and greater than -6, while allowing some probability of values below 
%-6. 
% \hh
% explain here how we selected those three points and why in few sentences.
% \ech
% and their distributions are very similar to those of observed $y_{i,n,j}<0$.

\def\cpo{\text{CPO}}
\def\lpml{\text{LPML}}
\def\data{\text{data}}

\section{Computation of LPML and DIC}\label{sec:lpml-dic}
% \hh I revised the equations for LPLM and DIC computation. check if the
% revised equations are correct and if you used the correct equations for your
% computation. \ech
% RESPONSE: They are correct.

We use the log pseudo marginal likelihood (LPML) and deviance criterion
information (DIC) to select the number of cell subpopulations ($K$) as discussed
in \S \ref{sec:prob-model} of the main text. LPML \citep{gelfand1994bayesian,
gelfand1992bayesian}) is defined as $\lpml = \sum_{i=1}^n \log\cpo_i$, where
$\cpo_i = \int f(\data_i \mid \data_{-i}, \theta)p(\theta \mid
\data_{-i})d\theta \approx \bk{\frac{1}{B}\sum_{b=1}^B \frac{1}{f(\data_i
\mid \theta^{(b)})}}^{-1}$, where $f(\data_i\mid\theta^{(b)})$ is the
likelihood evaluated at Monte Carlo sample $b$ of $B$
samples for observation
$i$, and $\cpo_i$ is the conditional predictive ordinates. The likelihood of cell $n$ in sample $i$ is
\begin{align}
f(\bm m_{i,n}, \bm y_{i,n} \mid \btheta) &=
\prod_{j=1}^J
\rho_{i,n,j}^{1-m_{i,n,j}} (1-\rho_{i,n,j})^{m_{i,n,j}}\cdot
\phi(y_{i,n,j} \mid \mu_{i,n,j}, \sigma^2_i) \nonumber\\
&\propto 
\prod_{j=1}^J
\rho_{i,n,j}^{1 - m_{i,n,j}}\cdot
\phi(y_{i,n,j} \mid \mu_{i,n,j}, \sigma^2_i), \label{eq:lpml-like-prop}
\end{align}
where $\phi(y\mid m, s^2)$ denotes the probability density function of the
normal distribution with mean $m$ and variance $s^2$, evaluated at $y$. Note
that $(1-\rho_{i,n,j})^{m_{i,n,j}}$ in \eqref{eq:lpml-like-prop} is dropped
since it remains constant for observed $y_{i,n,j}$. We then compute $\lpml$ as
\begin{eqnarray*}\label{eq:cpo}
% Normalized LPML
% \lpml &=& \frac{1}{\sum_{i=1}^I N_i}\sum_{i=1}^I\sum_{n=1}^{N_i} \log\cpo_{in} \\
%
\lpml &=& \sum_{i=1}^I\sum_{n=1}^{N_i} \log\cpo_{i,n}\\
&\approx&
\sum_{i=1}^I\sum_{n=1}^{N_i}
\log \left\{\frac{1}{B} \sum_{b=1}^B
\frac{1}{f(\bm m_{i,n}, \bm y_{i,n} \mid \btheta^{(b)})}\right\}^{-1} \\
&\propto&
\sum_{i=1}^I\sum_{n=1}^{N_i} \log \left\{\frac{1}{B} \sum_{b=1}^B 
\frac{1}{
\prod_{j=1}^J (\rho^{(b)}_{i,n,j})^{m_{inj}}\cdot
\phi(y_{i,n,j} \mid \mu^{(b)}_{i,n,j}, \sigma^{2, (b)}_{i})
}\right\}^{-1}.
\end{eqnarray*} 
Deviance is defined as as $D = -2\log f(\bm m, \bm y \mid \btheta)$, where
$f(\bm m, \bm y \mid \btheta)$
is the likelihood. The deviance criterion
information (DIC) \citep{dic} is computed as DIC = $\bar D - D(\bar\theta)$,
where $\bar D = \E\bk{D}$ is the posterior mean of the deviance, and
$\bar\btheta$ is the posterior mean of the parameters $\btheta$.
We compute the likelihood as
\begin{equation}\label{eq:dic-like}
f(\bm m, \bm y \mid \btheta) = \prod_{i=1}^I\prod_{n=1}^{N_i} \prod_{j=1}^J
\rho_{i,n,j}^{1-m_{i,n,j}} \cdot \phi(y_{i,n,j} \mid \mu_{i,n,j}, \sigma^2_i). 
\end{equation}
% \hh
% how did you find the posterior mean of $\bZ$, $\bw$ etc?
% \ech
The parameters that appear in the likelihood include $\mu_{i,n,j},\sigma^2_i$,
and the missing values of $y_{i,n,j}$. So $\bar\btheta$ can be obtained
by computing the posterior means of $\mu_{i,n,j}$, $\sigma_i^2$, 
and the missing $y_{i,n,j}$.

%\bch
%I temporarily moved all figures and tables to the end since they are
%floating around. We may move them back into the text once we finish editing.
%\ech

\section{Simulation Study}\label{sec:sim}

\subsection{Additional Results for Simulation 1}\label{sec:sim-1}
\begin{figure}[t]
\begin{center}
  \begin{tabular}{cc}
  \includegraphics[width=0.5\columnwidth]{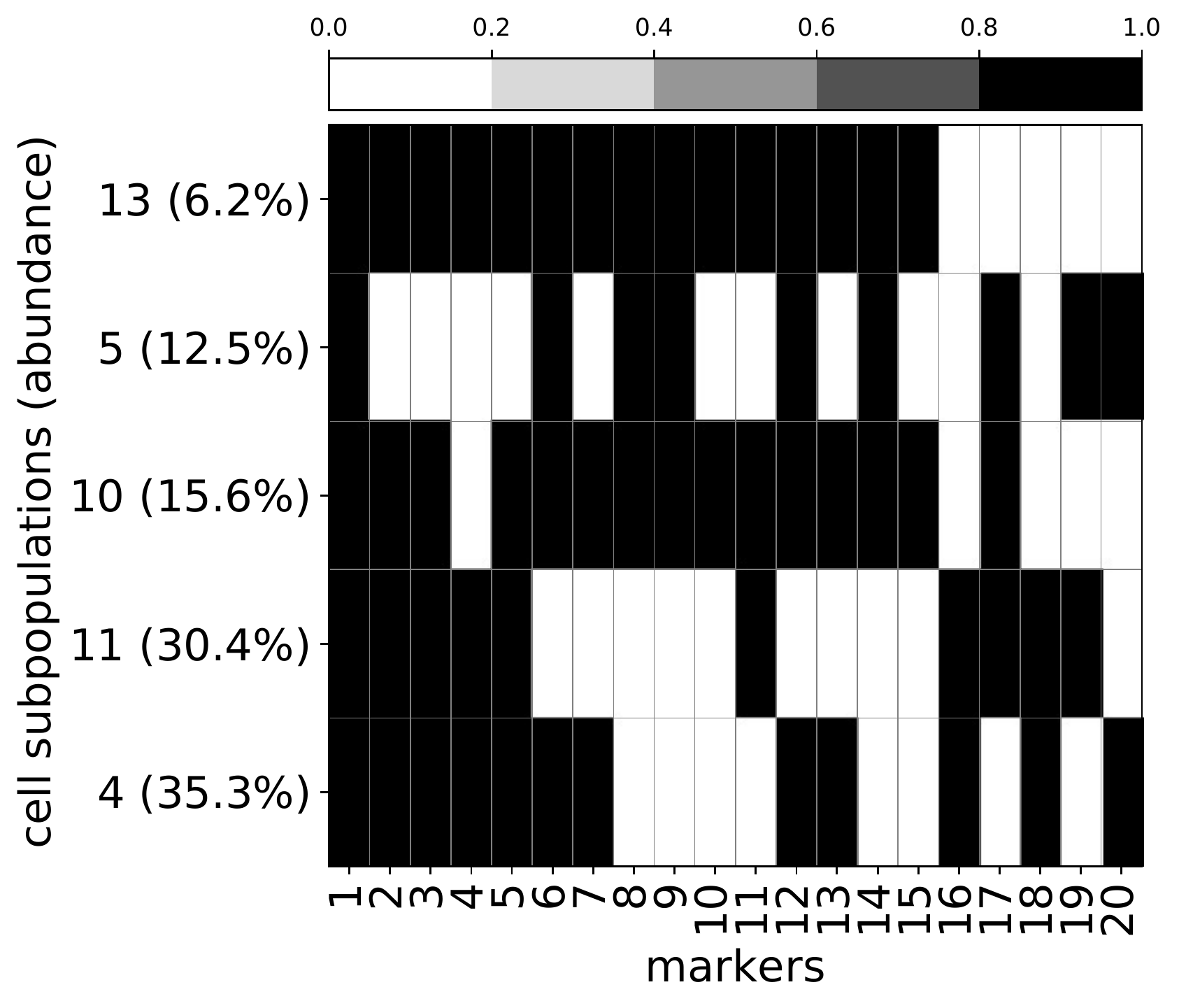} &
  \includegraphics[width=0.5\columnwidth]{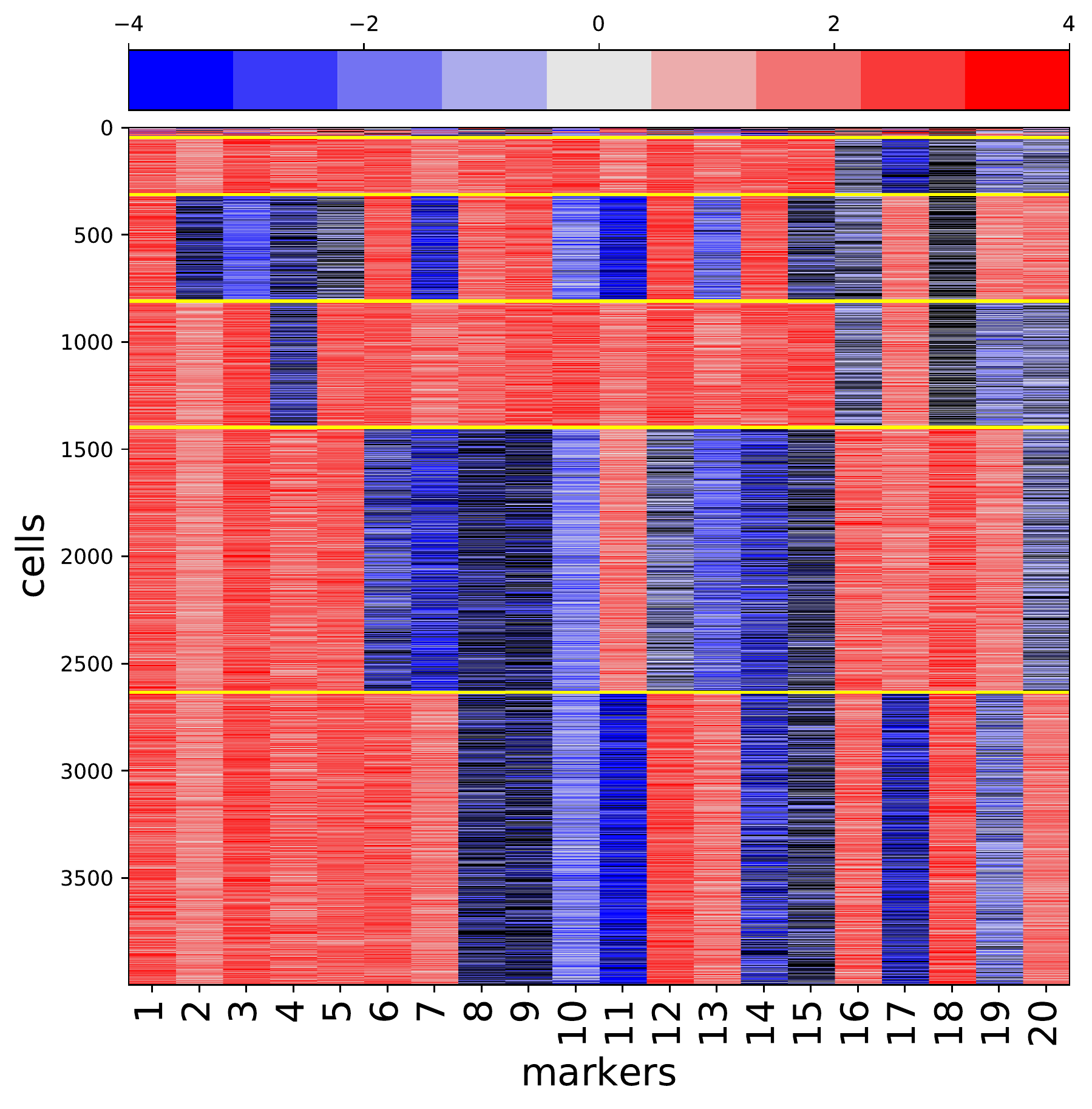} \\
  (a) $\hat{\Z}^\prime_1$ and $\hat{\bw}_1$ & (b) $y_{1nj}$\\
  \includegraphics[width=0.5\columnwidth]{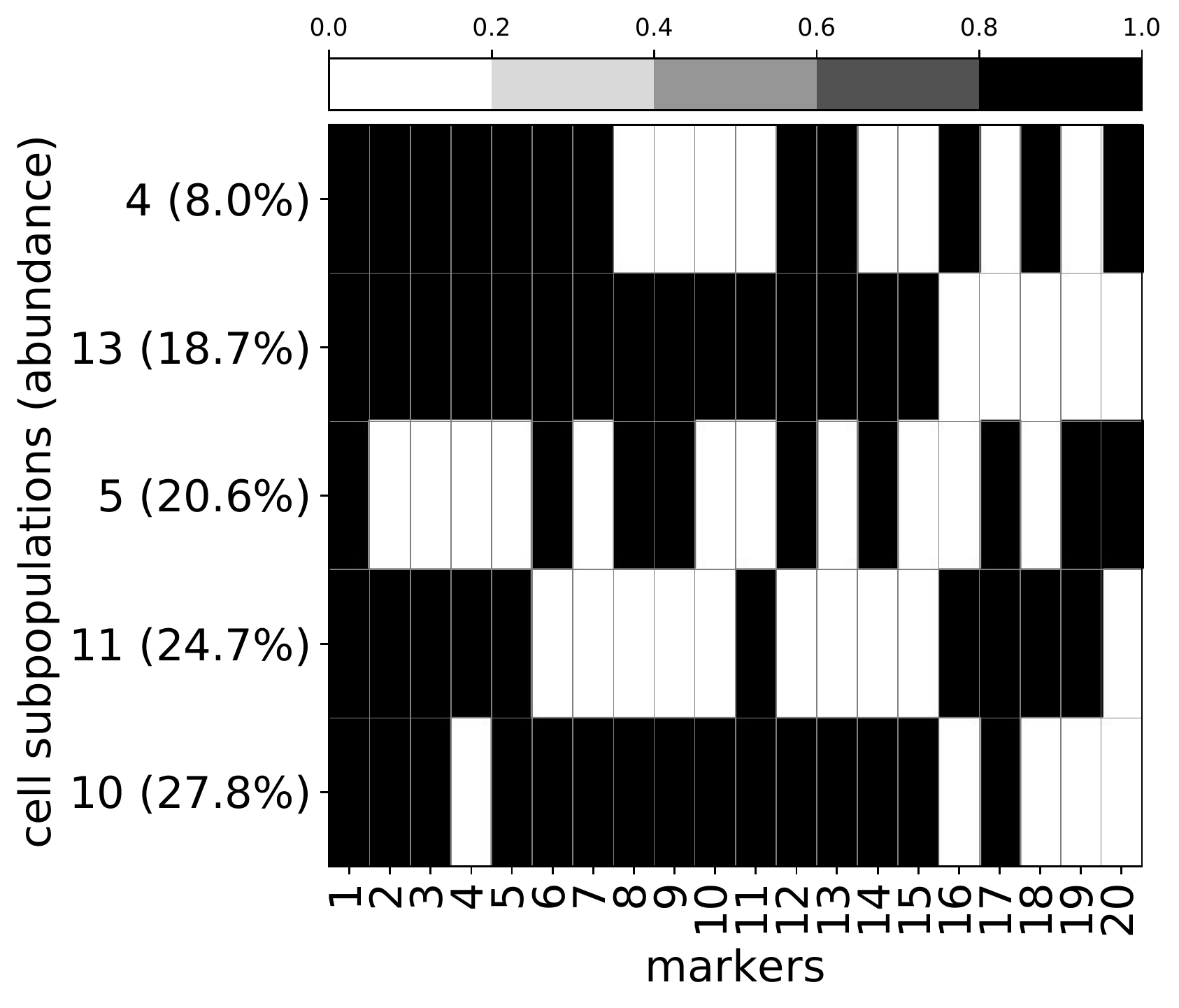} &
  \includegraphics[width=0.5\columnwidth]{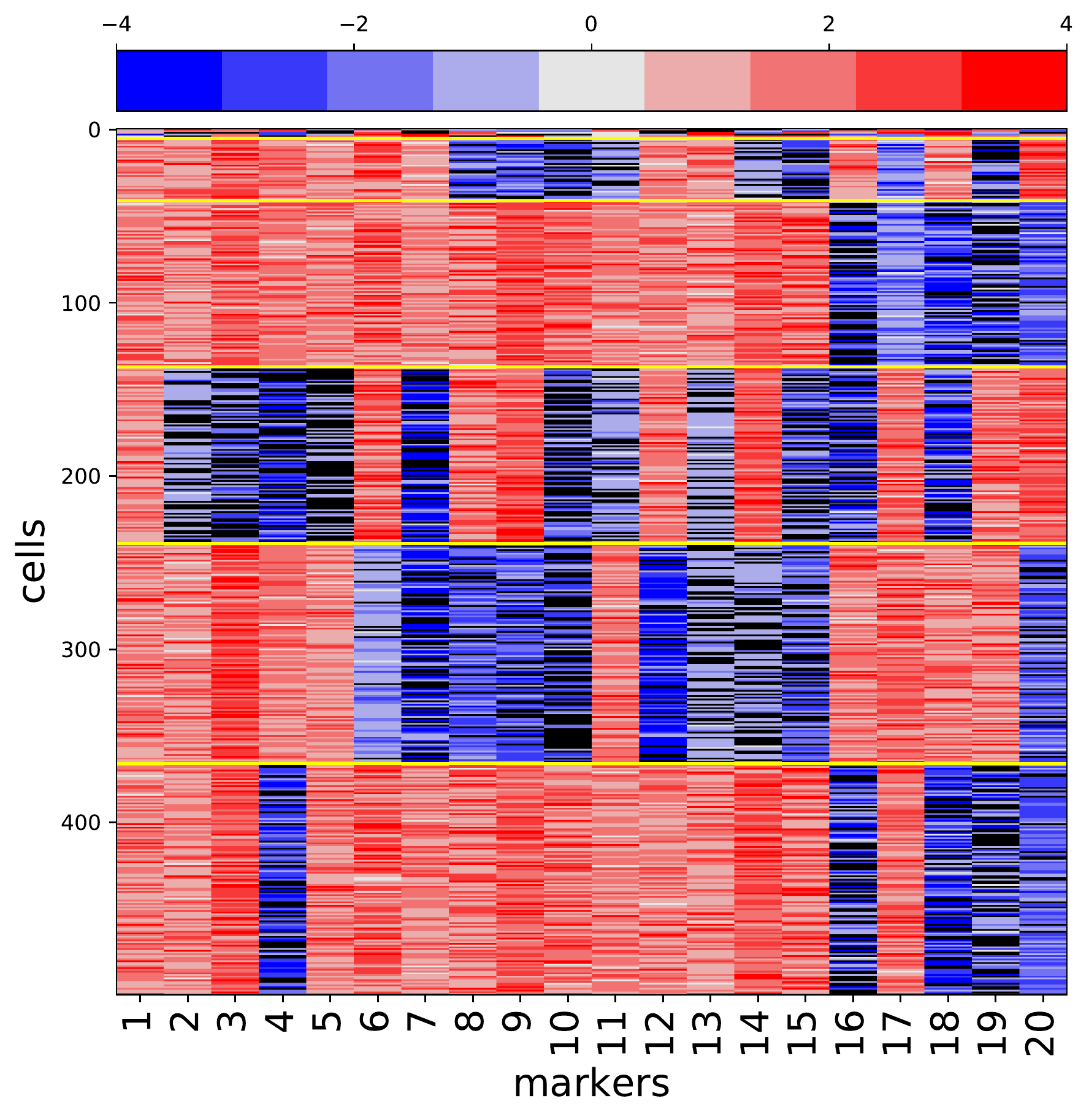} \\
  (c) $\hat{\Z}^\prime_2$ and $\hat{\bw}_2$ & (d) $y_{2nj}$\\
  \end{tabular}
  \vspace{-0.05in}
  \caption{[ADVI for Simulation 1]
  In (a) and (c), the transpose
  $\hat{\Z}^\prime_i$ of $\hat \bZ_i$ and $\hat{\bw}_i$ are shown for samples
  1 and 2, respectively, with markers that are expressed dented by black and
  not expressed by white. Only subpopulations with $\hat{w}_{i,k} > 1\%$ are
  included. Heatmaps of $\bm y_i$ are shown for sample 1 in (b) and sample 2
  in (d). Cells are ordered by posterior point estimates of their subpopulations,
  $\hat{\lambda}_{i,n}$. Cells are given in rows and markers are given in
  columns. High and low expression levels are represented by red and blue,
  respectively, and black represents missing values. Yellow horizontal lines
  separate cells into five subpopulations. Posterior estimates are obtained via ADVI.  }
  \label{fig:sim-vb-1}
\end{center}
\end{figure}

Here we present additional figures and tables for Simulation 1. 
Figure~\ref{fig:sim-vb-1} summarizes the results from the analysis
of Simulation 1 via ADVI. It contains the elementwise posterior means of $\bm Z$ 
and the posterior means of $\bm w_i$ (panels (a), (c), and (e)), and heatmaps 
of the simulated data $y_{i,n,j}$ sorted according to the posterior
mode of the cell subpopulation indicators $\hat\lambda_{i,n}$ (panels (b), (d), and (f)). 
% MOVE fig:sim-vb-1 HERE
Table~\ref{tab:missmechsen-sim} contains the three data missingship
mechanisms (MM) used in Simulation 1. MM0 is the default mechanism.
Recall that we used empirical $\tilde{\bm q}$-quantiles to specify $\tilde{\bm y}$. Different $\tilde{\bm q}$ yields different values of $\bm \beta$.  
 Three different sets of $\tilde{\bm q}$ are used for the sensitivity analysis, while fixing $\tilde{\bm \rho}$. 
For each mechanism, the LPML and DIC are shown in the last two columns of the table. 
% MOVE tab:missmechsen-sim HERE
Figures \ref{fig:Z-w-sim1-missmechsen-1} and \ref{fig:Z-w-sim1-missmechsen-2}
respectively summarize the results for the analysis of Simulation 1 under
data missingship mechanism I and II, done via MCMC. The figures contain the
posterior estimate of $\bm Z$ and $\bm w$ in panels (a), (c), and (e), and heatmaps of the
simulated data $y_{i,n,j}$ sorted according to the posterior estimate of the
cell subpopulation indicator $\hat\lambda_{i,n}$ in panels (b), (d), and (f). 
% MOVE fig:Z-w-sim1-missmechsen-1 and fig:Z-w-sim1-missmechsen-2 HERE.

\subsection{Simulation 2}\label{sec:sim-2}
An additional simulation study, Simulation 2, that assumes a larger simulated
dataset and a more complex cell subpopulation structure, was performed. The
dataset was simulated in a manner similar to Simulation 1 in
\S~\ref{sec:sim-study} of the main text, but the data size is larger with $N=(40000,
5000, 10000)$, and has more cell subpopulations with $K^\true=10$. We first specify $\bZ^\true$ and simulated $\bw^\true_i$ from a Dirichlet
distribution with parameters being some random permutation of $(1, \ldots,
K)$. Table~\ref{tab:sim2-tr} illustrates $\bZ^\true$ and $\bw^\true$.
Parameters $\mu^{\star, \true}_{0}$, $\mu^{\star, \true}_{1}$, and
$\sigma^{2, \true}_i$ are set in the same way as Simulation 1.
We fit the model over a grid for $K$, for $K$ from 2 to 20 in increments
of 2. For all models, we fixed $L_0=5$ and $L_1=5$. Recall that
$L_0^\true=L_1^\true=3$. All other parameter specifications, MCMC
initialization, and MCMC specifications were done in the same way as
Simulation 1.

% big sim: results
The LPML, DIC, and calibration metric for $K$ are presented in Figure
\ref{fig:metrics-sim2}. The metrics indicate that the model with $\hat{K}=10$
fits the data best and achieves a balance between good model fit and low
model complexity. Figure \ref{fig:sim2-post} shows posterior estimates of the
clusterings for each sample for the large simulated dataset, along with
posterior estimates of the subpopulations present ($\hat \Z_i$) and their
abundances ($\hat \bw_i$) in each sample. The red, blue, and black cells
represent high, low, and non-observed expression levels, respectively.
Hotizontal yellow lines separate cells into clusters.  The simulation truth for the cell subpopulations in $\bZ^\true$
 is recovered by $\hat{\Z}$, and $\hat \bw_i$ is close to $\bw^\true$.

% FlowSOM
Figure~\ref{fig:sim2-FlowSOM-Z} shows estimated clusterings for each sample
$\y_i$ using FlowSOM. The largest cluster in
sample 1 shown in panel (a) contains a mixture of high and low expression levels for marker
9, resulting in poor performance of clustering cells.  This undesired behavior is not observed in the FAM.
%
% ARI
%We again used the adjusted rand index (ARI) to assess the accuracy of
%cluster assignments produced by the FAM and FlowSOM, based on the simulation
%truth. Table \ref{tab:ari-sim2} shows the ARI by sample for each method.
%Our method produced higher ARI for every sample. The ARI in sample 1
%is especially low for FlowSOM, as the two similar subpopulations that were grouped
%together make up a large portion of the cells in that sample. 

% ADVI Sim 2
Figure~\ref{fig:sim-vb-2} summarizes the posterior inference obtained via ADVI. The posterior mean of $\bm Z$ 
and the posterior mean of $\bm w_i$ are in panels (a), (c), and (e), and heatmaps 
of the simulated data $y_{i,n,j}$ sorted according to the posterior
mode of the cell subpopulations $\hat\lambda_{i,n}$ in panels (b), (d), and (f). The posterior inference covers the simulation truth well.  
% MOVE fig:sim-vb-2 HERE

% big sim: miss mech sen 
We performed the sensitivity analysis to the specification of the
data missingship mechanism after selecting $K=10$ via DIC and LPML. Table
\ref{tab:missmechsen-sim2} summarizes the missingship mechanisms used in the
sensitivity analysis. Again, we note that inference on $\Z$ and $\bm w$ do not
change significantly across the various missing mechanisms. However, the fit
(in terms of LPML and DIC) on the observed data was highest for missingship
mechanism II, which encourages imputing values that are more negative, as it
best matched the simulation truth.
Figures \ref{fig:Z-w-sim2-missmechsen-1} and \ref{fig:Z-w-sim2-missmechsen-2}
respectively summarize the results for the analysis of Simulation 1 under
data missingship mechanism I and II, done via MCMC. The figures contain the
posterior estimate of $\bm Z$ and $\bm w$ in panels (a), (c), and (e), and heatmaps of the
simulated data $y_{i,n,j}$ sorted according to the posterior estimate of the
cell subpopulation indicators $\hat\lambda_{i,n}$ in panels (b), (d), and  (f). \\
% MOVE fig:Z-w-sim2-missmechsen-1 and fig:Z-w-sim2-missmechsen-2 HERE.

\section{Additional Results for Analysis of Cord Blood Derived NK Cell Data}
This section contains additional figures and tables for the CB NK cell data 
analysis presented in \S~\ref{sec:cb-analysis} of the main text.
%
% Tables and figures
% 
% Table 5: marker names
% Table 6: proportion of cells in each cluster estimated by FLowSOM for CB
% Table 7: miss mech used in UCB study
% Table 8: Values for beta in UCB NK cell data analysis.
%
% Fig 11: MM1
% Fig 12: MM2
% Fig 13: ADVI CB
%
Table~\ref{tab:marker-codes} lists the marker names and numbers for
each marker included in the CB derived NK data analysis.  Figure~\ref{fig:CB-tsne} visualizes the CB NK cell data in a two-dimensional space using a data visualization technique “t-SNE (t-Distributed Stochastic Neighbor Embedding)” \citep{maaten2008visualizing, van2014accelerating}.  The two dimensional embeddings are learned separately for each sample. Cells are represented with different symbols and colors by their posterior estimate $\hat{\lambda}_{in}$ of the cell clustering.  All cells in the samples are used to obtain the embeddings, but cells in the subpopulations with $\hat{w}_{ik}\geq 0.05$ are included in the plots for better illustration.   

%Two-dimensional t-SNE’s (Maaten and Hinton, 2008) were learned separately on each sample of the CB data. The learned embeddings were then colored by estimated clusterings from via the MCMC analysis of the CB data. Due to the size of the data (38636, 9555, and 4827 cells respectively in samples 1, 2, and 3), the Barnes–Hut algorithm was used to approximate the exact t-SNE. Moreover, since our analyses (via MCMC) revealed that 21 subpopulations were required to adequately explain the data, and plotting 21 clusters was visually overwhelming, only clusters comprising more than 5% of cells in each sample are included in the graphs, though the t-SNE’s were fit on every cell of each sample.

%able~\ref{tab:cb-flowsom} summarizes the proportion of cells that belong in
%each cluster and sample, obtained from analyzing the CB derived NK data using FlowSOM.
%Note that 13 clusters are learned.

Table~\ref{tab:missmechsen-cb} contains the three data missingship
mechanisms (MM) used in analyzing the CB derived NK data. MM0 is the default mechanism.
Each mechanism is defines the parameters $\bm\beta$ through the
quantiles of the negative observed values in each sample $\tilde{\bm q}$, 
and probability that a record is missing at those quantiles $\tilde{\bm\rho}$.
For each mechanism, the LPML and DIC are shown. 
Table~\ref{tab:missmechsen-cb-beta} list the implied $\bm\beta$ for
each data missingship mechanism. 
% MOVE tab:missmechsen-cb HERE
% MOVE tab:missmechsen-cb-beta HERE

Figures \ref{fig:Z-w-CB-missmechsen-1} and \ref{fig:Z-w-CB-missmechsen-2}
respectively summarize the results for the analysis of the CB NK cell data
under data missingship mechanism I and II, done via MCMC. The posterior
estimate of $\bm Z$ and $\bm w$ are shown in panels (a), (c), and (e), and
heatmaps of the simulated data $y_{i,n,j}$ sorted according to the posterior
estimate of the cell subpopulations $\hat\lambda_{i,n}$ in panels (b), (d), and
(f)). 
% MOVE fig:Z-w-CB-missmechsen-1 and fig:Z-w-CB-missmechsen-2 HERE.

Figure~\ref{fig:cb-vb-Z} summarizes the results from the analysis of the UCB NK
cell data via ADVI. The posterior mean of $\bm Z$ and the posterior mean of
$\bm w_i$ are in panels (a), (c), and  (e)), and heatmaps of the simulated data
$y_{i,n,j}$ sorted according to the posterior mode of the cell subpopulations
$\hat\lambda_{i,n}$ in panels (b), (d), and (f). 
% MOVE fig:cb-vb-Z HERE

\clearpage
% \bibliographystyle{natbib}
% \bibliography{main}

\clearpage
%%%%%%%%%%%%%%%%%%%%%%%%
%% Simulation 1 - ADVI 1

%% Simulation 1 - ADVI 2
\begin{figure}[thb]
\begin{center}
  \begin{tabular}{cc}
  \includegraphics[width=0.5\columnwidth]{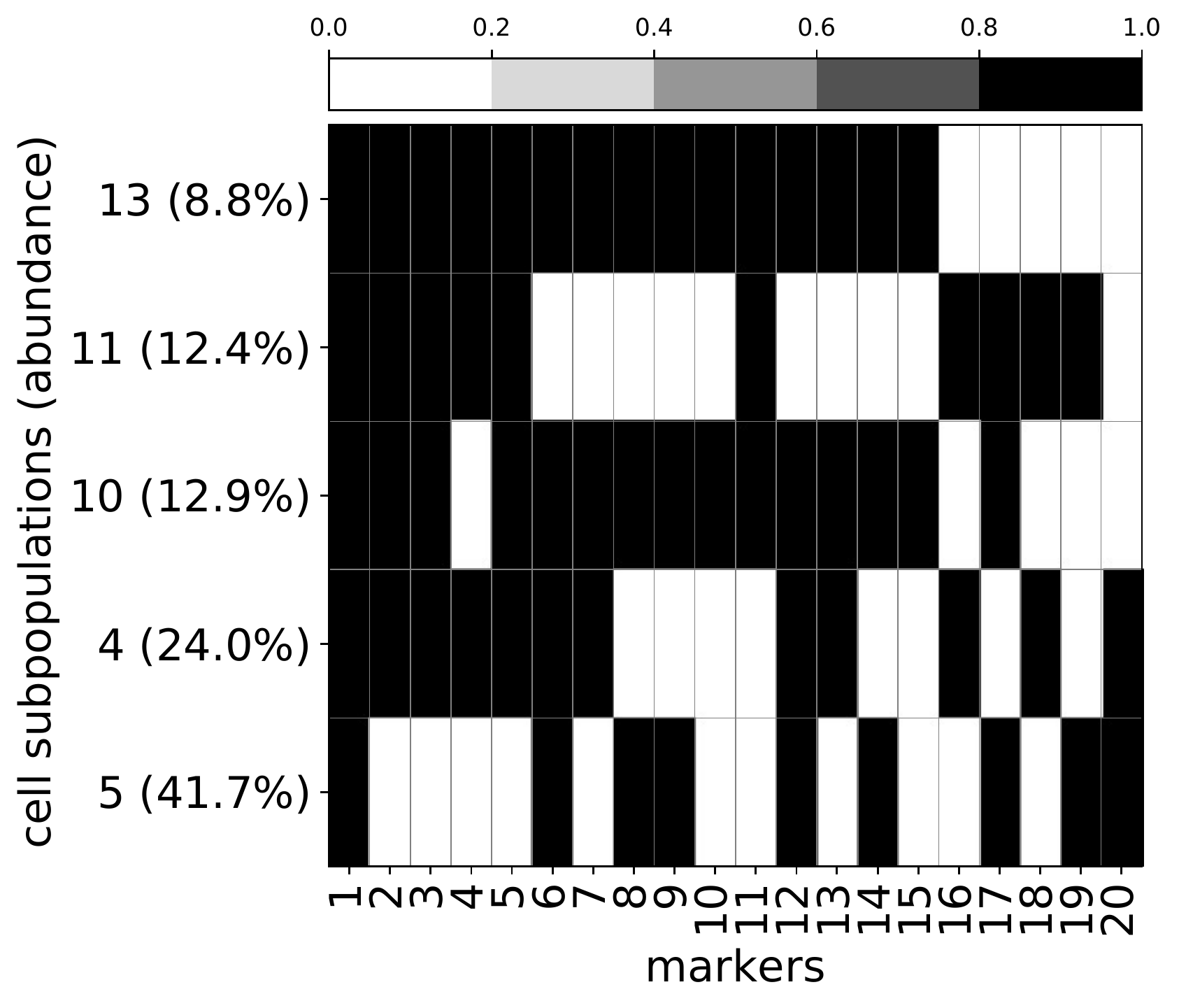} &
  \includegraphics[width=0.5\columnwidth]{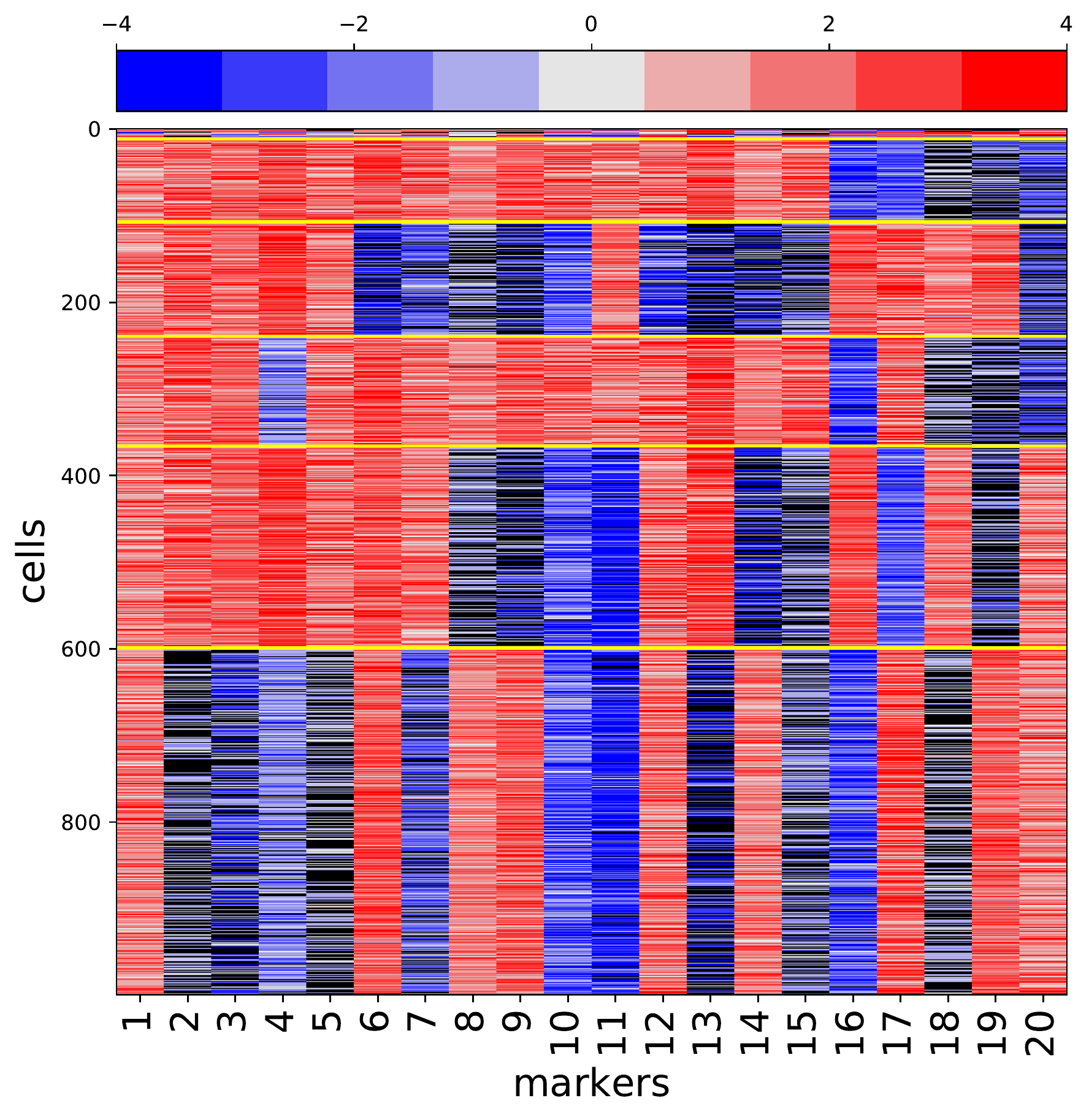} \\
  (e) $\hat{\Z}^\prime_3$ and $\hat{\bw}_3$ & (f) $y_{3nj}$\\
  \end{tabular}
  \vspace{-0.05in}
  \caption*{Figure~\ref{fig:sim-vb-1} continued: [ADVI for Simulation 1]
	In
  (e), the transpose $\hat{\Z}^\prime_i$ of $\hat \bZ_i$ and $\hat{\bw}_i$
  are shown for sample 3, with markers that are expressed dented by black and
  not expressed by white. Only subpopulations with $\hat{w}_{i,k} > 1\%$ are
  included. Heatmaps of $\bm y_i$ for sample 3 is shown in (f). Cells are
  ordered by posterior point estimates of their subpopulations,
  $\hat{\lambda}_{i,n}$. Cells are given in rows and markers are given in
  columns. High and low expression levels are represented by red and blue,
  respectively, and black represents missing values. Yellow horizontal lines 
  separate cells into five subpopulations. Posterior estimates are obtained via ADVI.  }
\end{center}
\end{figure}
%%%%%%%%%%%%%%%%%%%%%%%%%

%% Simulation 1 - missing speci
\clearpage
\begin{table}[t]
  \centering
  \begin{tabular}{c|cccc}
    \hline
    Data Missingship & $\tilde{\bm q}$ & Probability of Missing $(\tilde{\bm\rho})$ & LPML & DIC \\
    Mechanism & & & & \\
    \hline
    0  & (0\%, 25\%, 50\%) & (5\%, 80\%, 5\%) & -16.728 & 172989 \\
    I  & (0\%, 20\%, 40\%) & (5\%, 80\%, 5\%) & -16.681 & 172914 \\
    II & (0\%, 15\%, 30\%) & (5\%, 80\%, 5\%) & -16.462 & 170971 \\
    \hline
  \end{tabular}
  \caption[Data Missingship Mechanism Specifications]{
  Data missingship mechanisms used for Simulation 1.
  $\tilde{\bm q}$-quantiles of the negative observed values in each sample are
  used to specify $\tilde{\bm y}$, and $\tilde{\bm\rho}$ are the probability of
  missing at those $\tilde{\bm y}$.  Three different sets of $\tilde{\bm q}$
  and $\tilde{\bm \rho}$ are used to examine the sensitivity to the missingship
  mechanism specification. LPML and DIC are shown in the last two columns under
  each of the specification.}
  \label{tab:missmechsen-sim}
\end{table}

%% Simulation 1 
\clearpage
\begin{figure}[t]
  \centering
  \begin{tabular}{ccc}
    \includegraphics[width=.3\columnwidth]{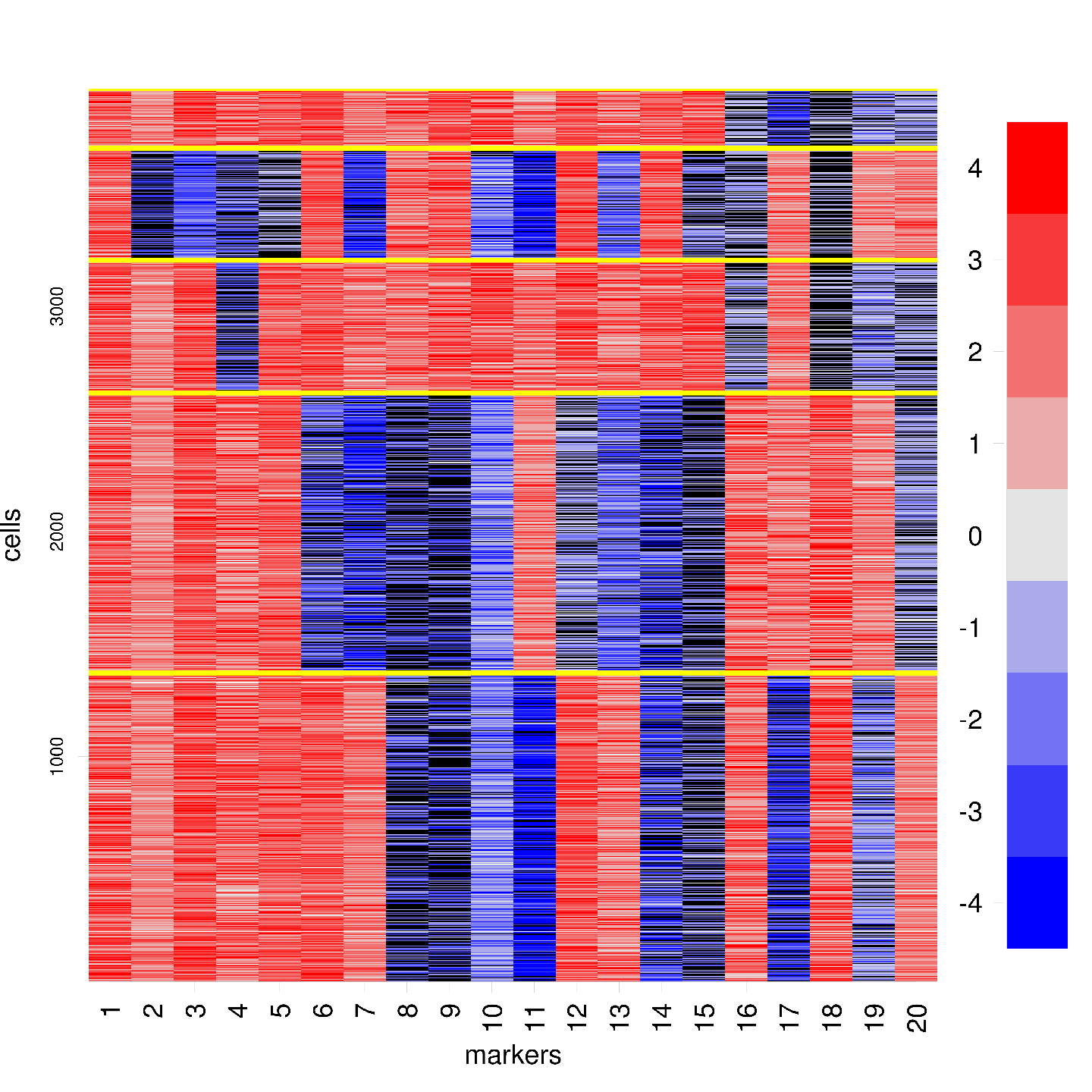} &
    \includegraphics[width=.3\columnwidth]{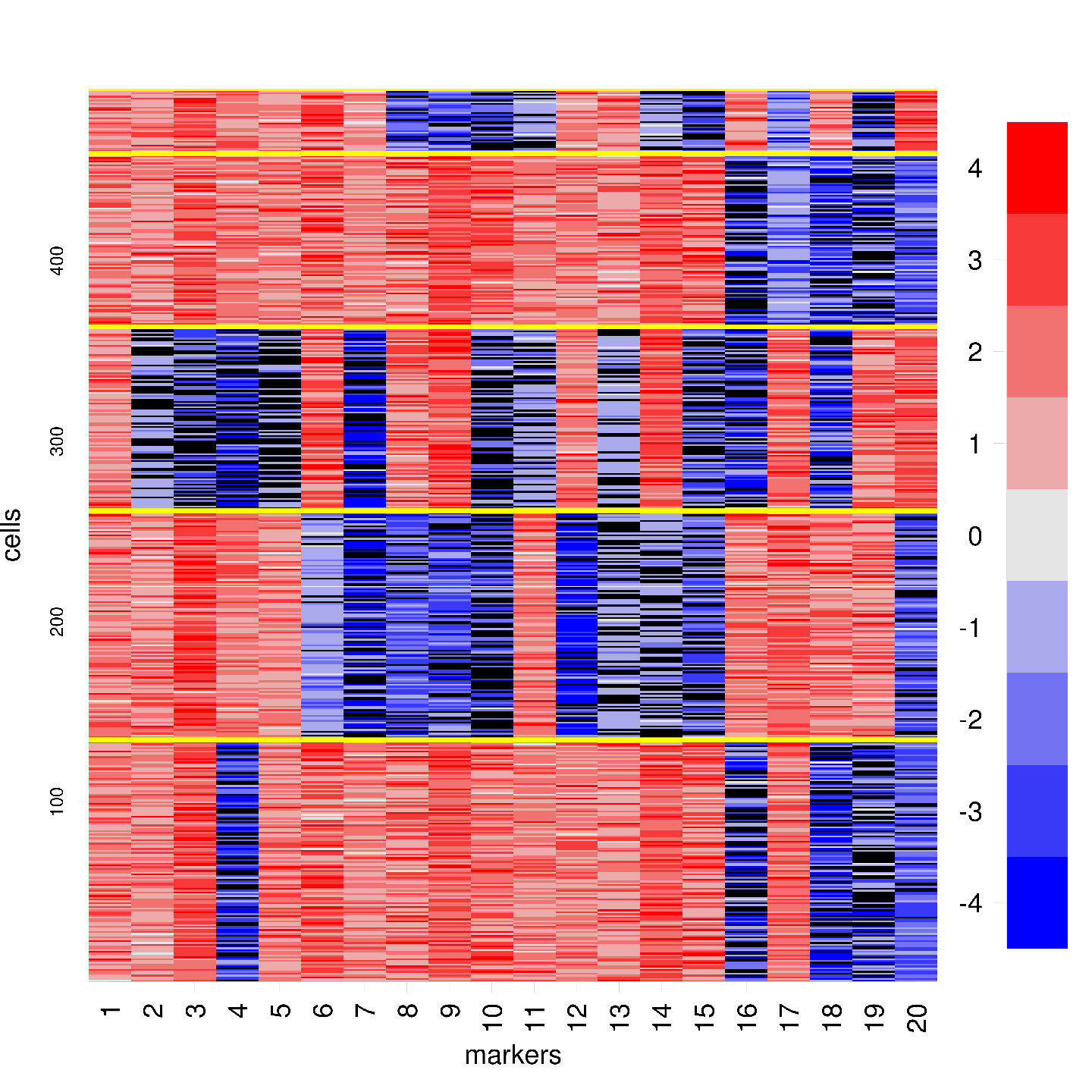} &
    \includegraphics[width=.3\columnwidth]{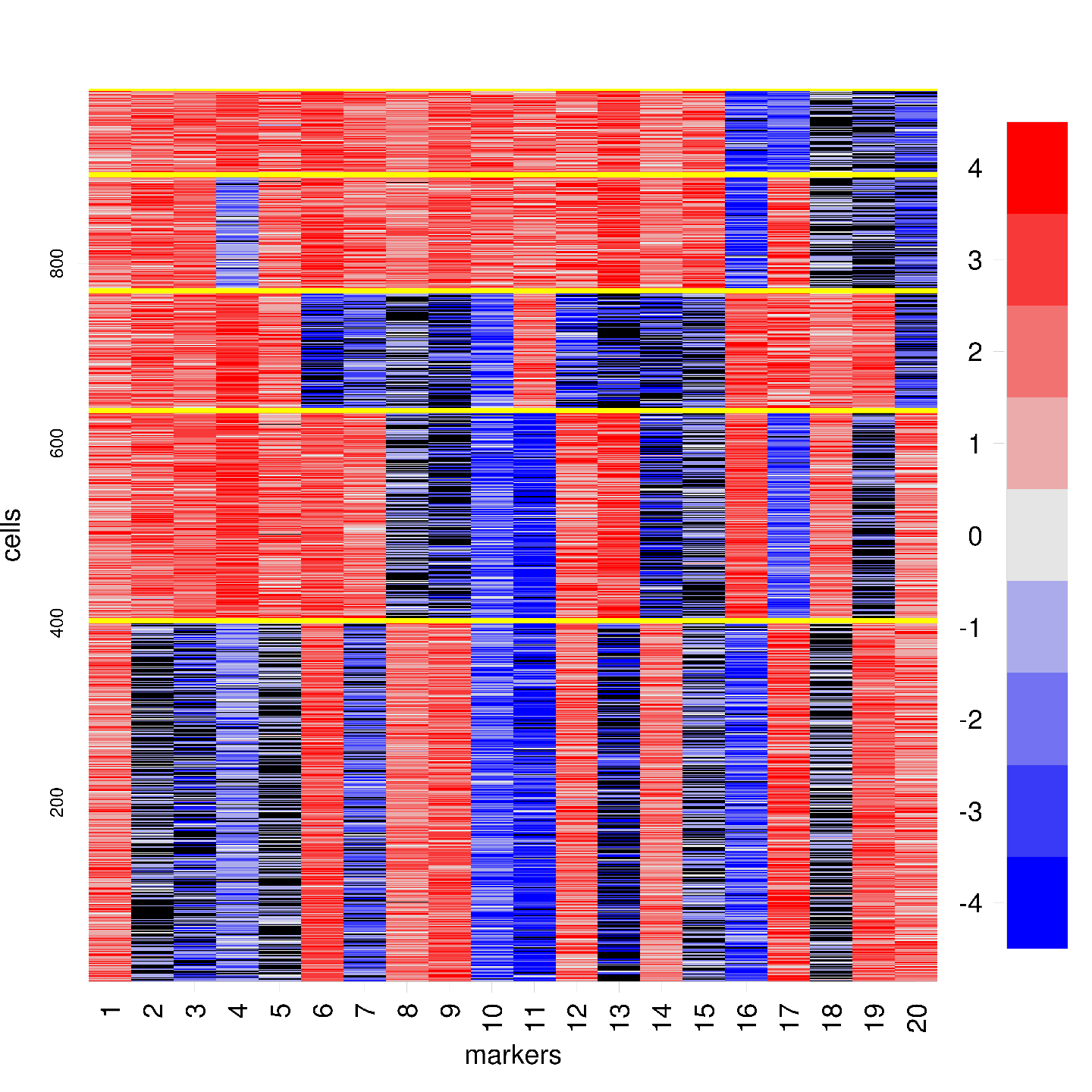} \\
  	(a) heatmap of $y_{1nj}$ & (b) heatmap of $y_{2nj}$ & (c) heatmap of $y_{3nj}$\\    
    \includegraphics[width=.3\columnwidth]{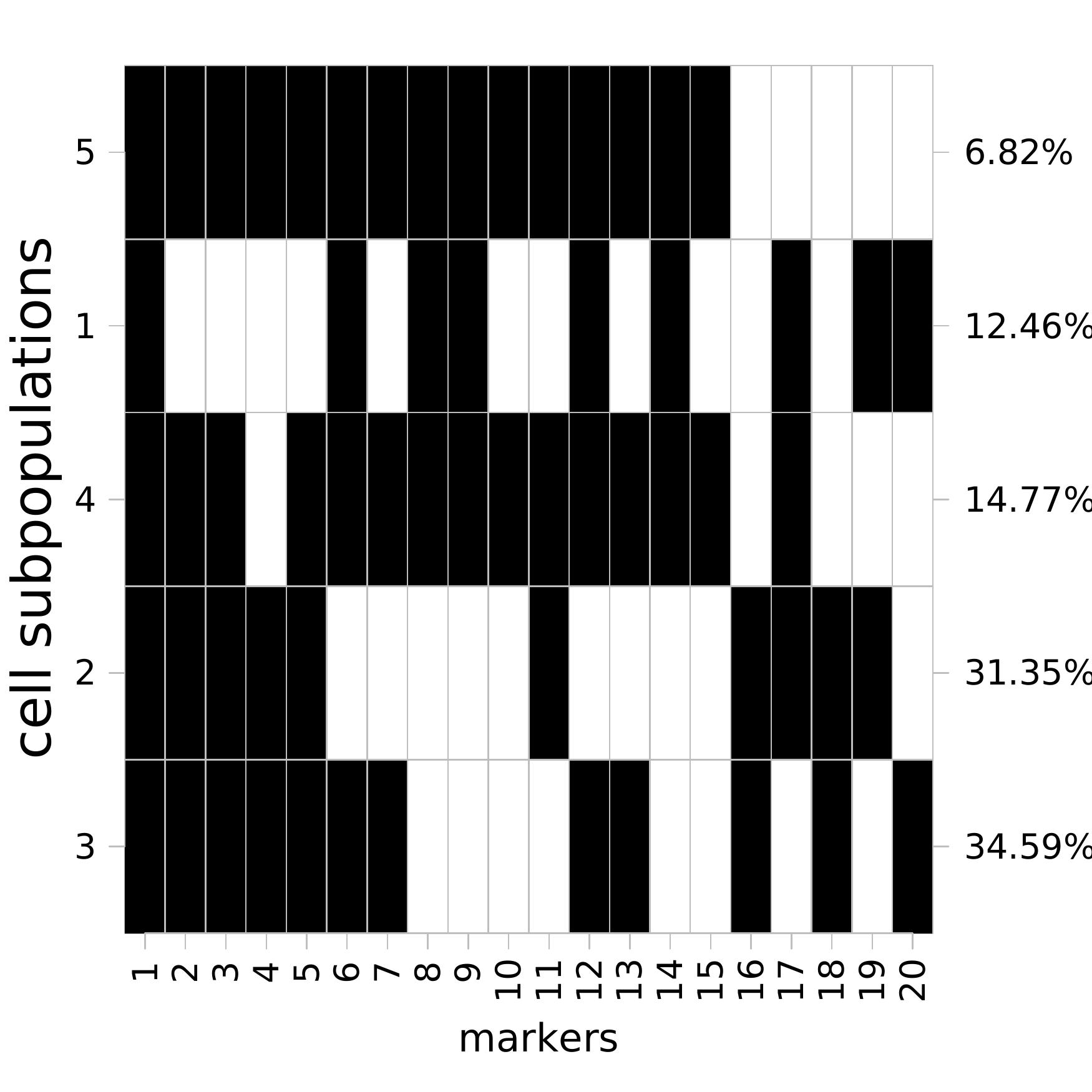} &
    \includegraphics[width=.3\columnwidth]{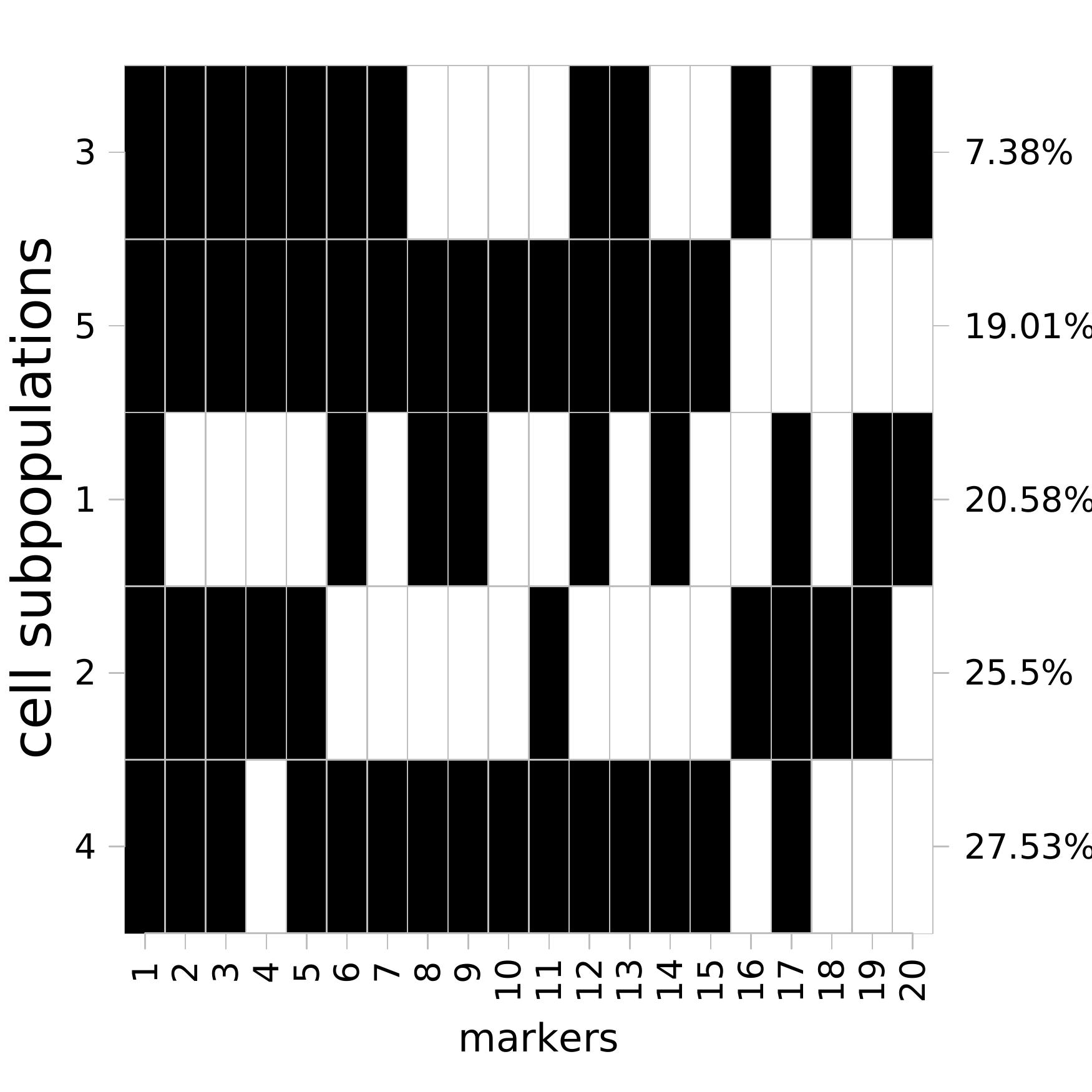} &
    \includegraphics[width=.3\columnwidth]{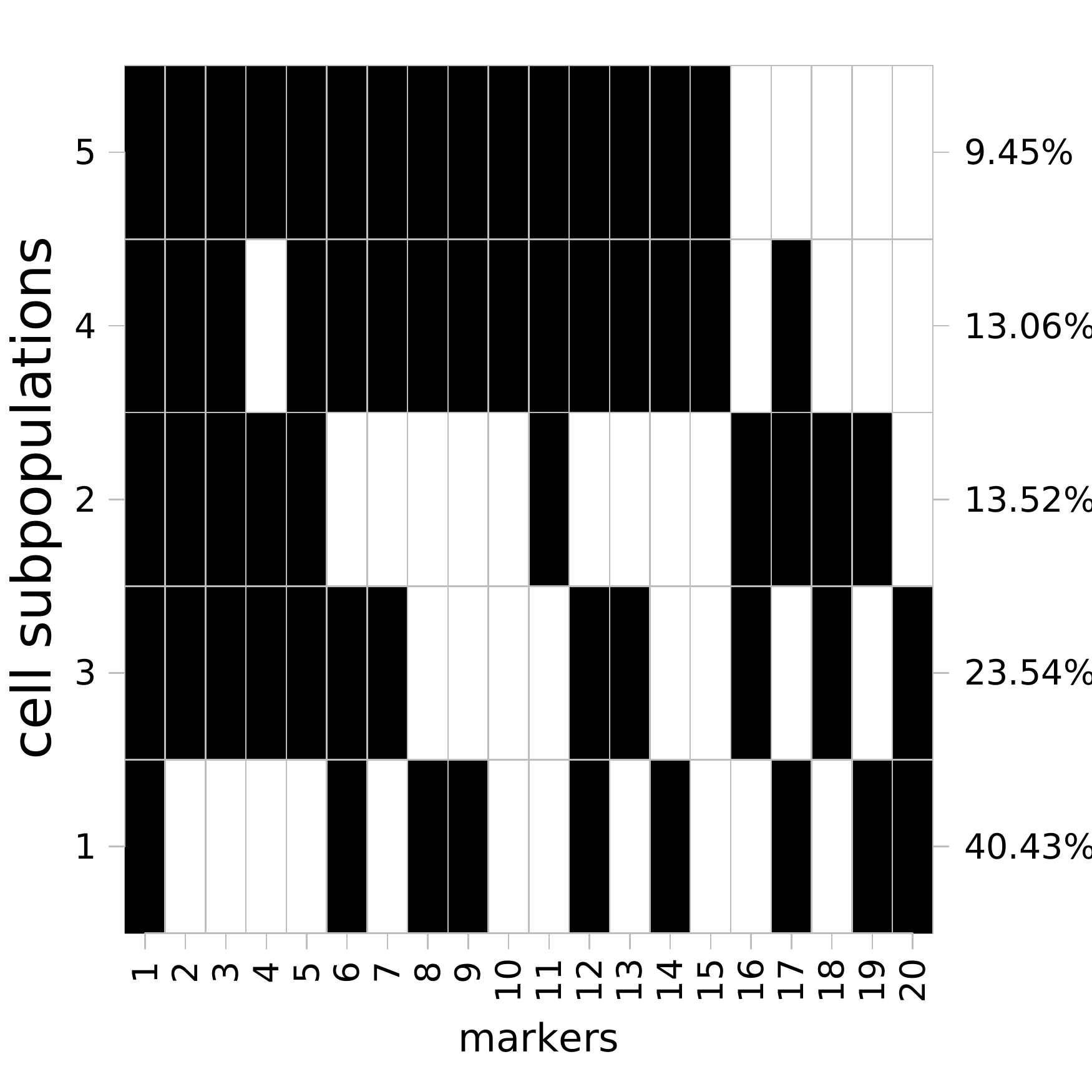} \\
    (d) $\hat{\Z}^\prime_1$ \& $\hat{\bw}_1$ & (e) $\hat{\Z}^\prime_2$ \& $\hat{\bw}_2$ & (c) $\hat{\Z}^\prime_3$ \& $\hat{\bw}_3$ \\
  \end{tabular}
  \caption{Data missingship mechanism sensitivity analysis for Simulation 1.
  Specification I is used for $\bm \beta$. Heatmaps of $\y_i$ are shown in
  (a)-(c) for samples 1-3, respectively. Cells are rearranged by the posterior
  point estimate of cell clustering, $\hat{\lambda}_{i,n}$. Cells and markers
  are in rows and columns, respectively. High and low expression levels are in
  red and blue, respectively, and black is used for missing values. Yellow
  horizontal lines separate cells by different subpopulations.
  $\hat{\Z}^\prime_i$ and $\hat{\bw}_i$ are shown for each of the samples in
  (d)-(f). We include only subpopulations with $\hat{w}_{i,k} > 1\%$.}
  \label{fig:Z-w-sim1-missmechsen-1}
\end{figure}

%% Simulation 1
\clearpage 
\begin{figure}[t]
  \centering
  \begin{tabular}{ccc}
    \includegraphics[width=.3\columnwidth]{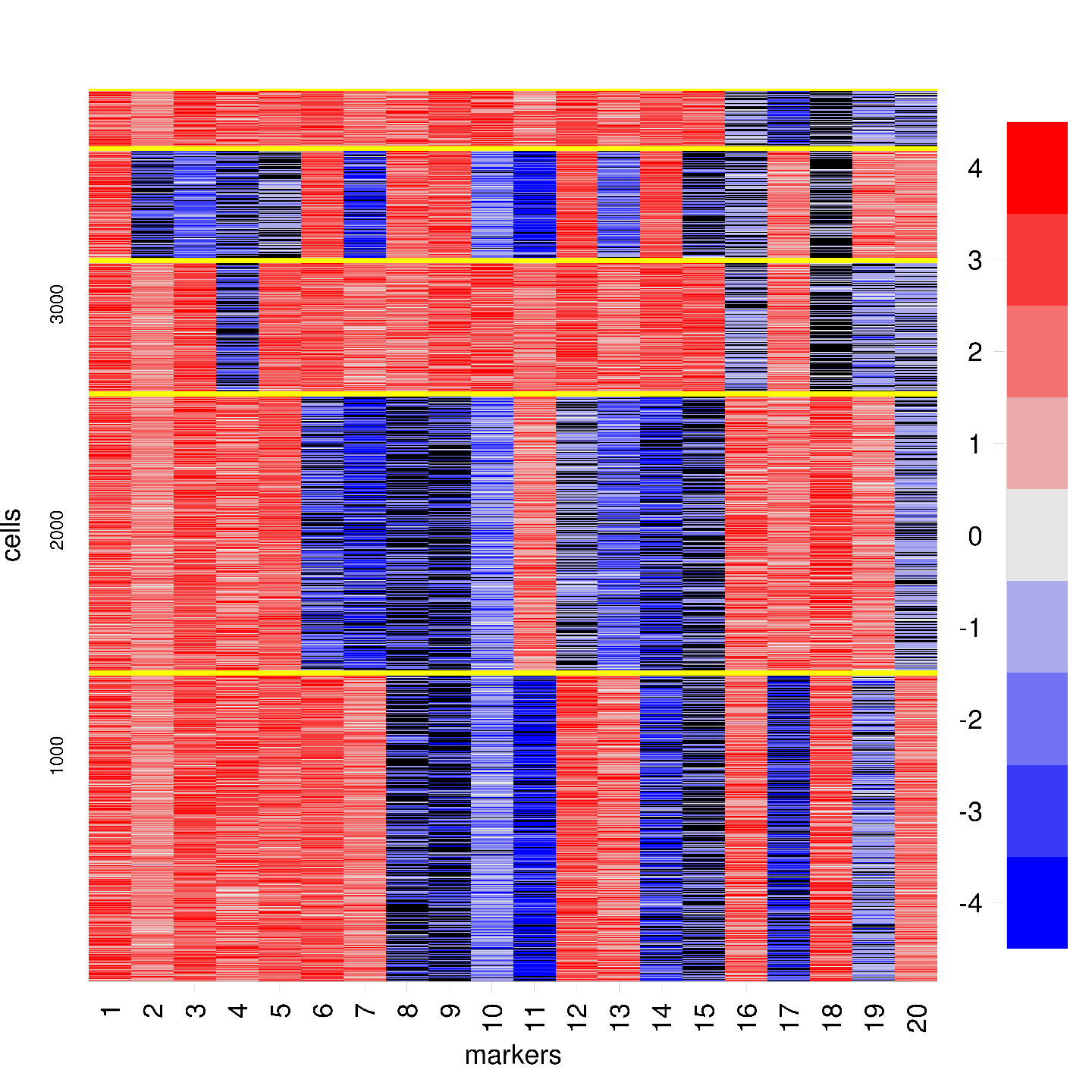} &
    \includegraphics[width=.3\columnwidth]{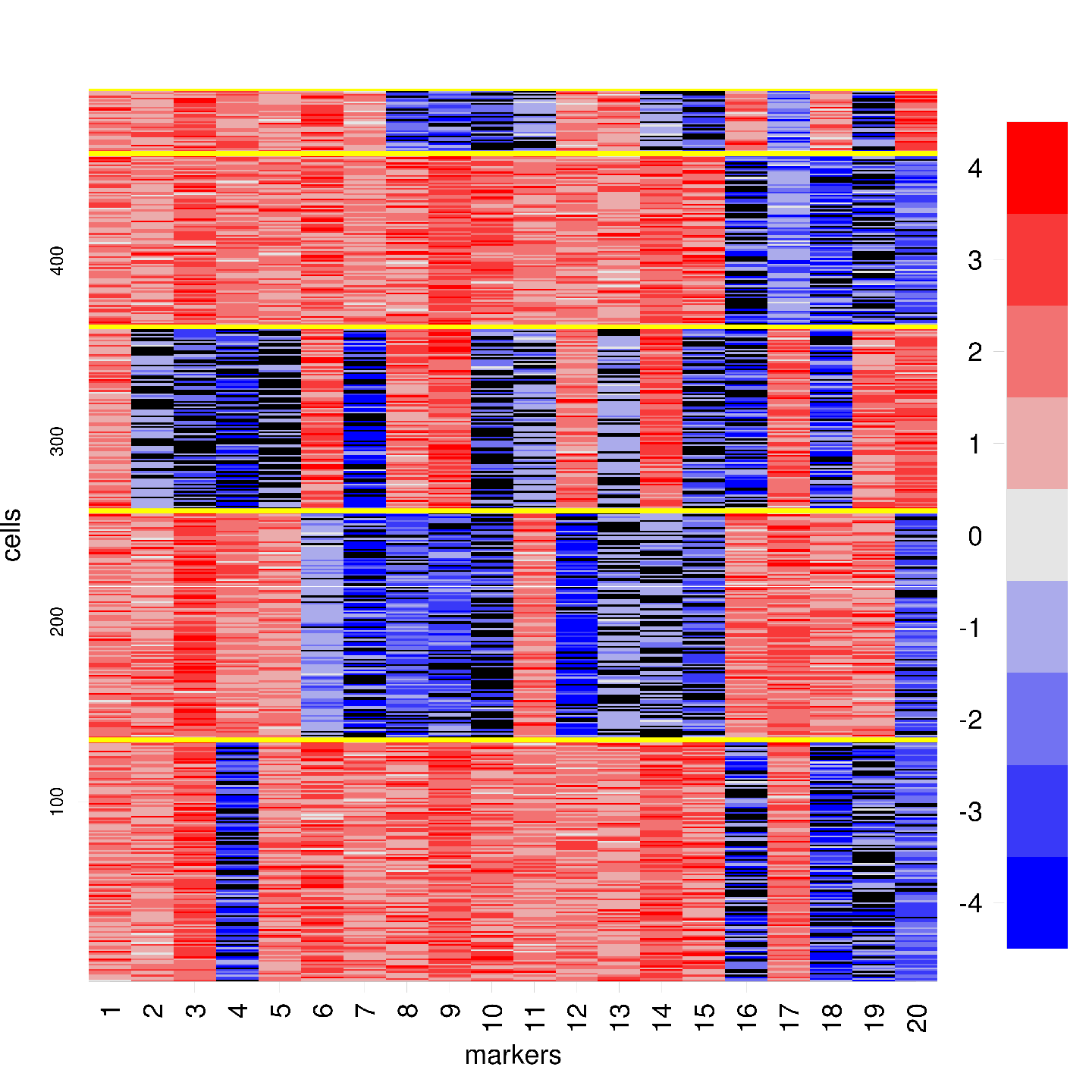} &
    \includegraphics[width=.3\columnwidth]{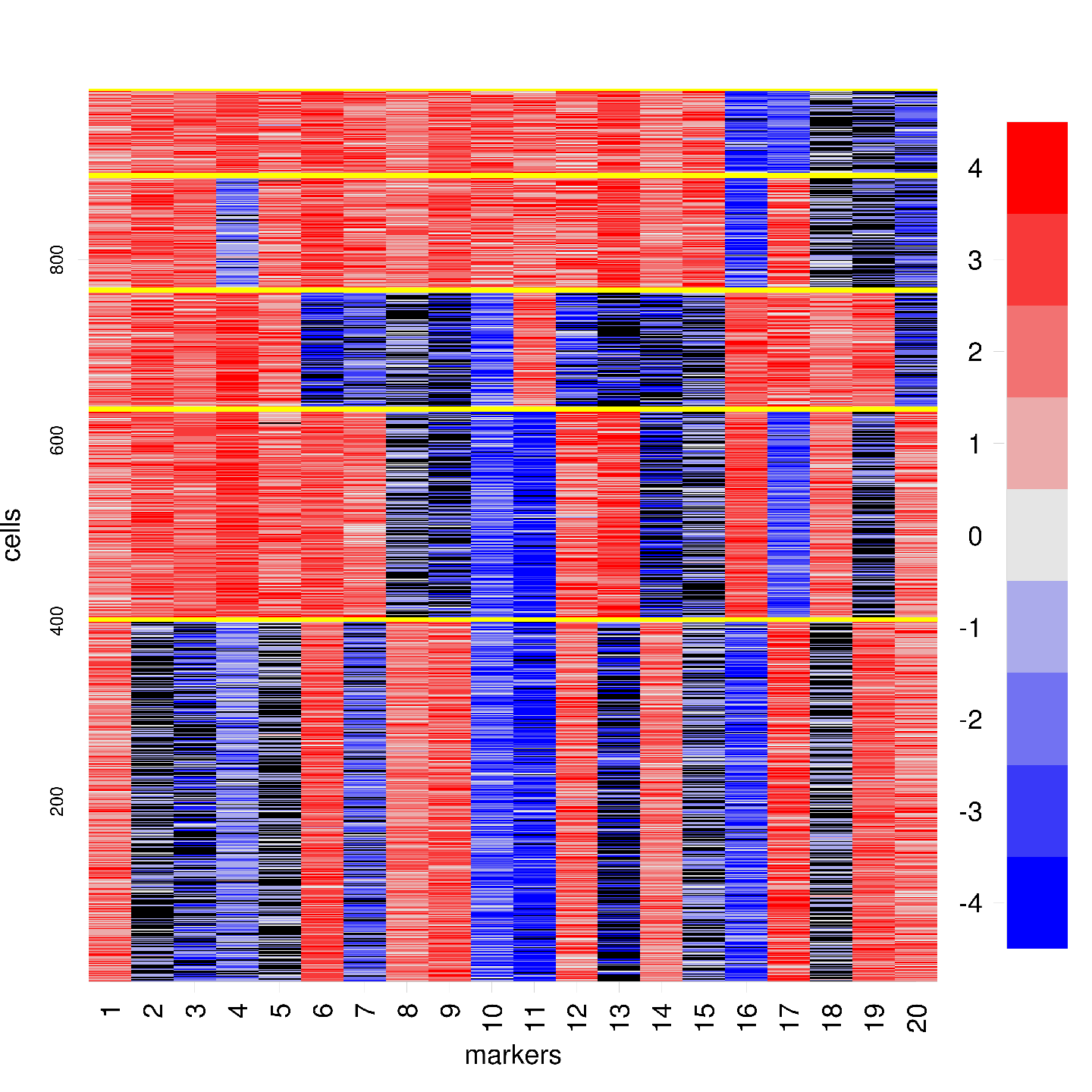} \\
  	(a) heatmap of $y_{1nj}$ & (b) heatmap of $y_{2nj}$ & (c) heatmap of $y_{3nj}$\\    
    \includegraphics[width=.3\columnwidth]{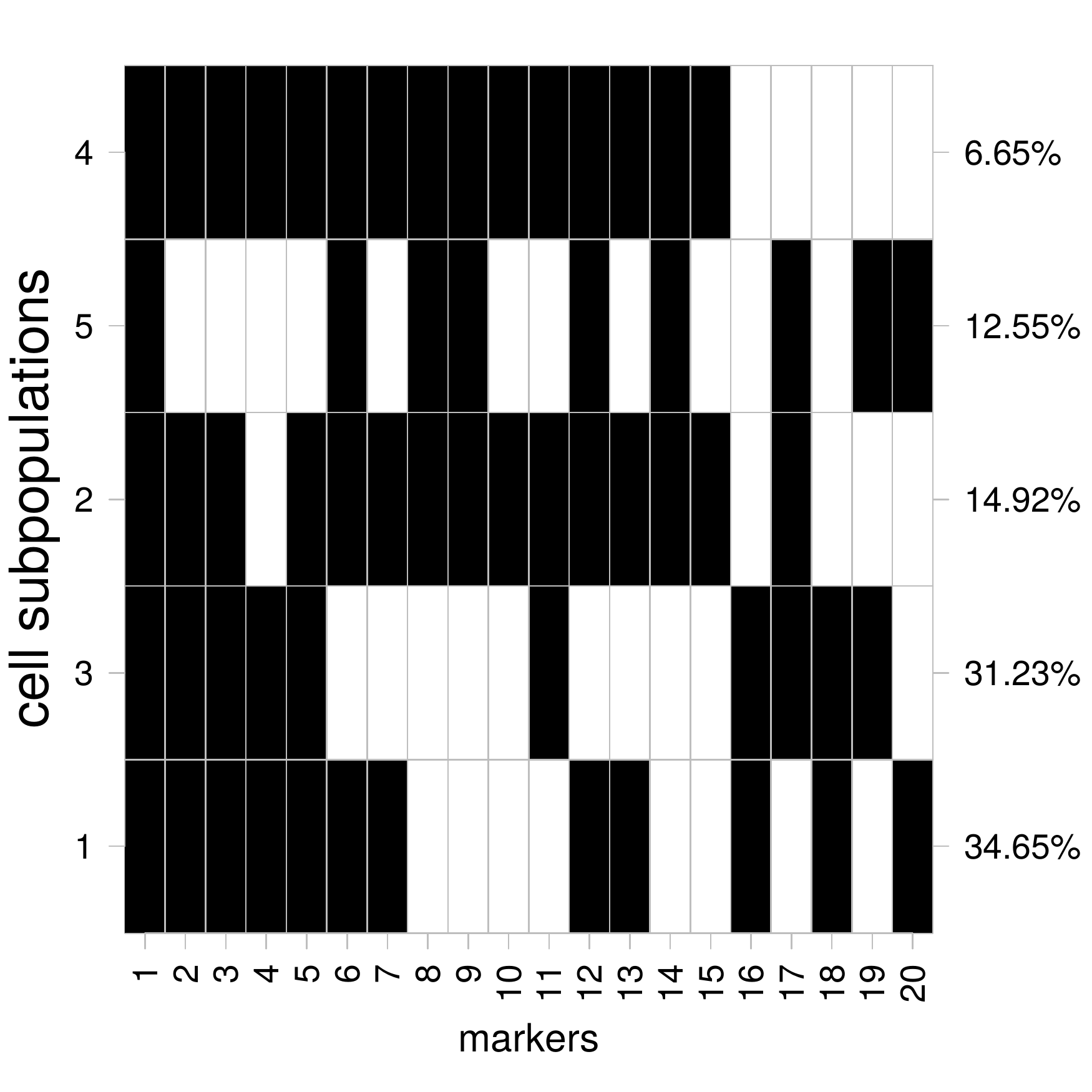} &
    \includegraphics[width=.3\columnwidth]{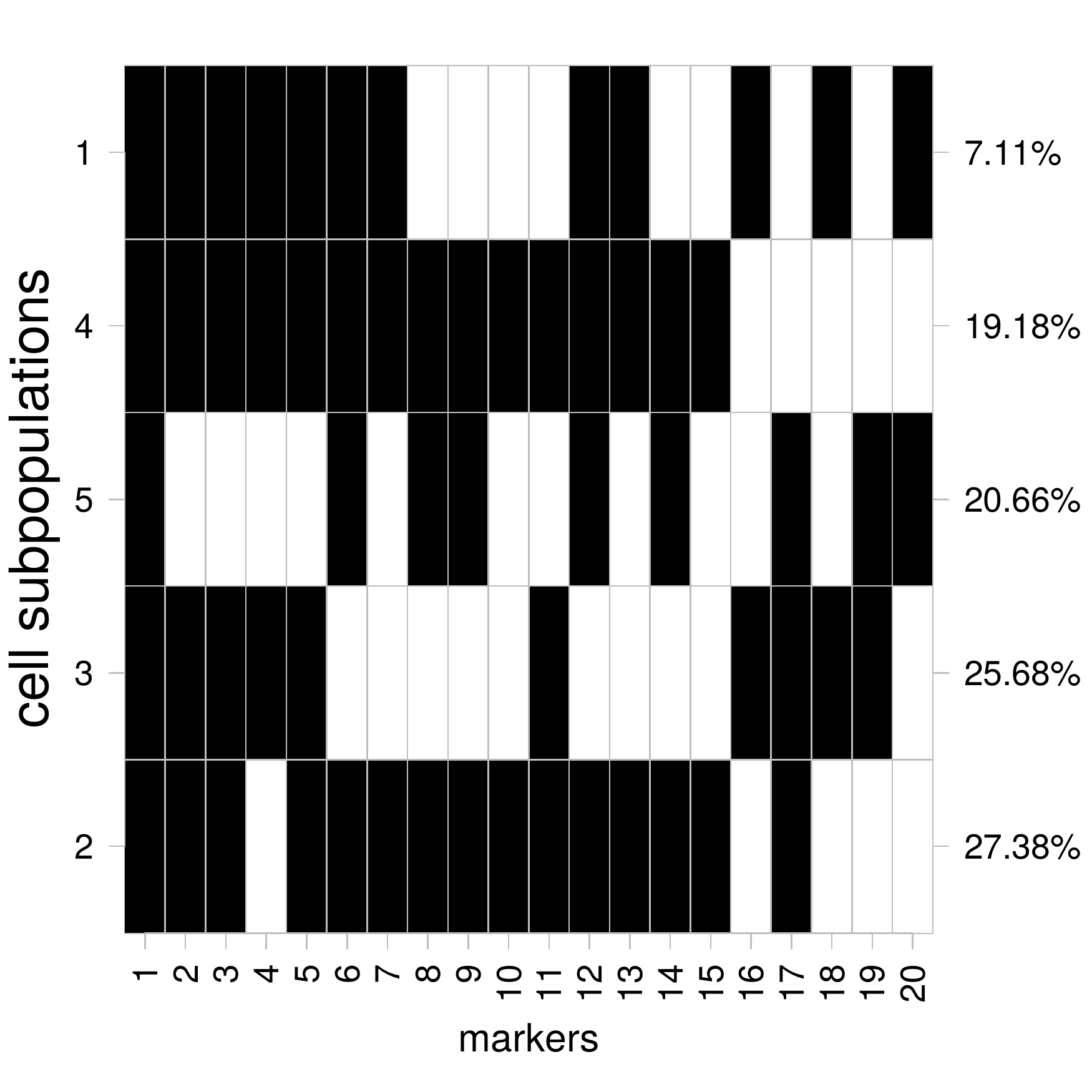} &
    \includegraphics[width=.3\columnwidth]{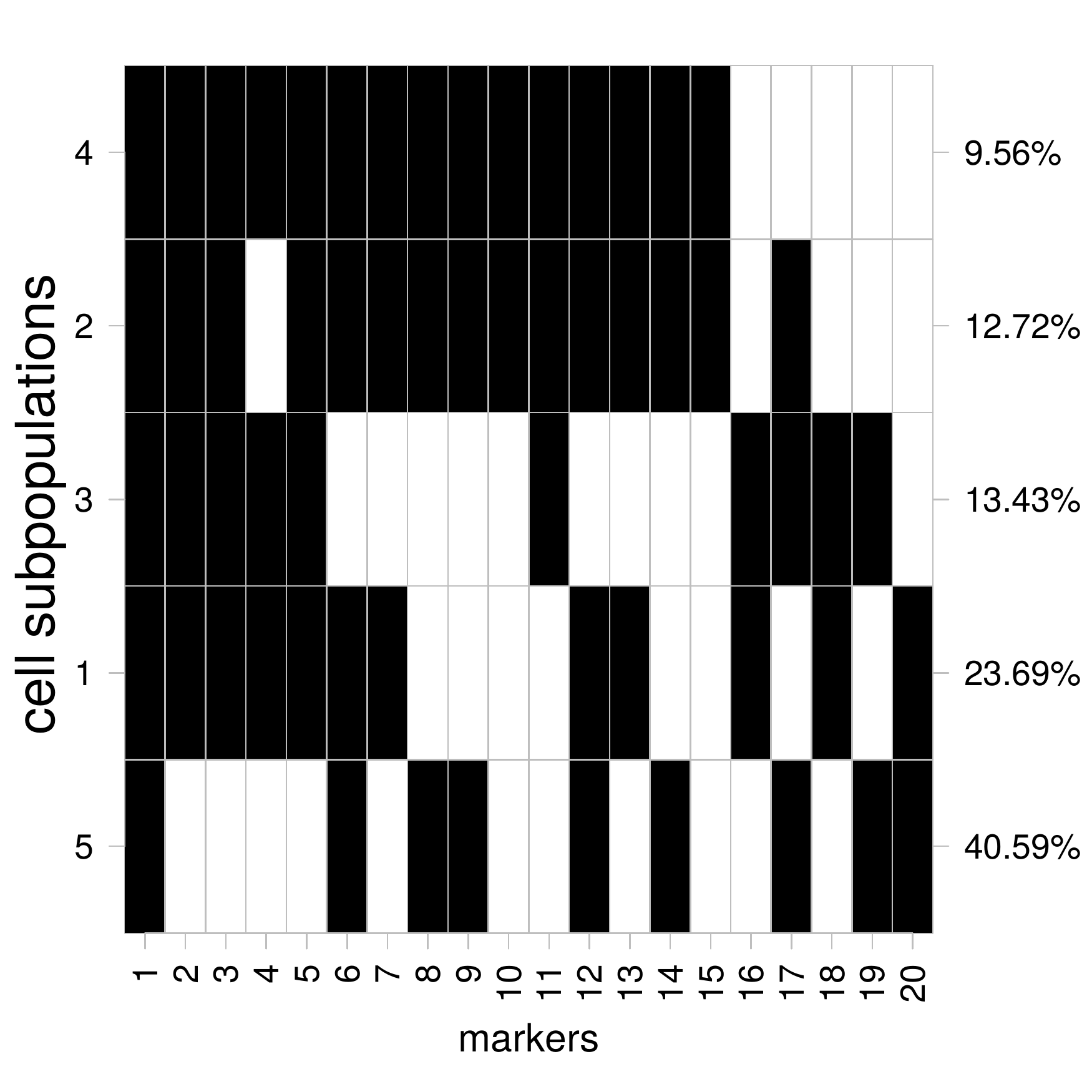} \\
    (d) $\hat{\Z}^\prime_1$ \& $\hat{\bw}_1$ & (e) $\hat{\Z}^\prime_2$ \& $\hat{\bw}_2$ & (c) $\hat{\Z}^\prime_3$ \& $\hat{\bw}_3$ \\
  \end{tabular}
  \caption{Data missingship mechanism sensitivity analysis for Simulation 1.
  Specification II is used for $\bm \beta$. Heatmaps of $\y_i$ are shown in
  (a)-(c) for samples 1-3, respectively. Cells are rearranged by the
  posterior point estimate of cell clustering, $\hat{\lambda}_{i,n}$. Cells
  and markers are in rows and columns, respectively. High and low expression
  levels are in red and blue, respectively, and black is used for missing
  values. Yellow horizontal lines separate cells by different subpopulations.
  $\hat{\Z}^\prime_i$ and $\hat{\bw}_i$ are shown for each of the samples in
  (d)-(f).We include only subpopulations with $\hat{w}_{i,k} >1\%$.}
  \label{fig:Z-w-sim1-missmechsen-2}
\end{figure}

\clearpage
%% Simulation 2 - truth
\begin{table}[t!]
    \begin{subtable}{.5\linewidth}
      \centering
        \begin{tabular}{c}
        \includegraphics[width=0.95\columnwidth, height=0.35\textheight]{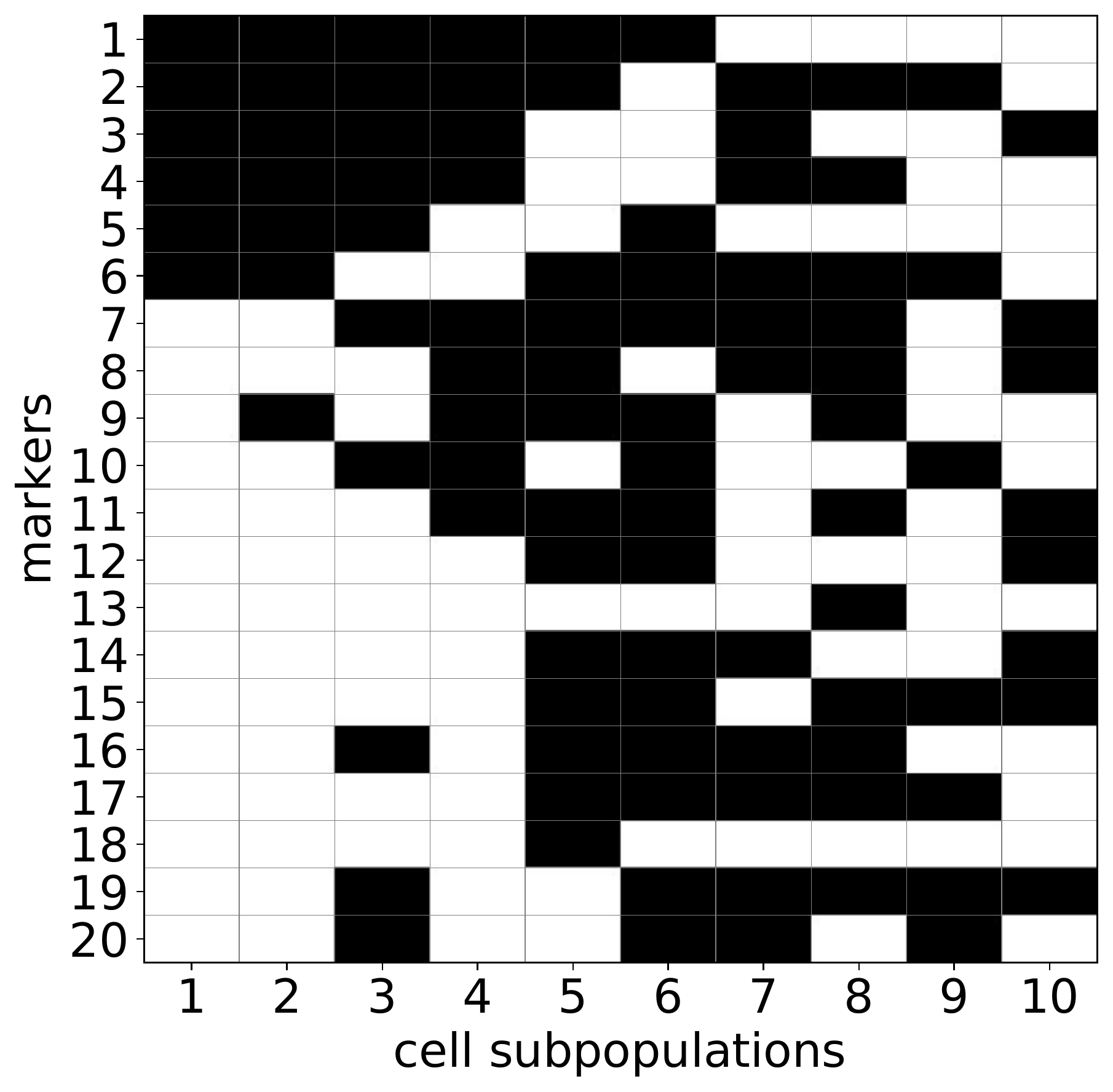}\\
        \end{tabular}
        \caption{ $\Z^\true$}
    \end{subtable}%
    \begin{subtable}{.5\linewidth}
      \centering
	\begin{tabular}{|c|rrr|}
	\hline
   subpopulations	&	sample 1	&	sample 2	&	sample 3	\\
   \hline
   $k=1$	&	0.136	&	0.160	&	0.033	\\
   $k=2$	&	0.132	&	0.021	&	0.128	\\
   $k=3$	&	0.111	&	0.037	&	0.257	\\
   $k=4$	&	0.157	&	0.084	&	0.110	\\
   $k=5$	&	0.044	&	0.183	&	0.049	\\
   $k=6$	&	0.046	&	0.111	&	0.142	\\
   $k=7$	&	0.215	&	0.045	&	0.142	\\
   $k=8$	&	0.072	&	0.109	&	0.001	\\
   $k=9$	&	0.018	&	0.109	&	0.099	\\
   $k=10$	&	0.065	&	0.135	&	0.035	\\	
   \hline
   \end{tabular}
   \caption{$\bw^\true$}
   \end{subtable} 
   \caption{[Simulation 2] $\Z^\true$ and $\bw^\true$ are illustrated in (a)
   and (b), respectively. $K^\true=10$, $J=20$, $I=3$ and $N=(40000, 5000,
   10000)$ are assumed. Black and white in (a) represents $z^\true_{j,k}=1$
   and 0, respectively.}
\label{tab:sim2-tr}
\end{table}

\clearpage
%% sim2 - LPML & DIC
%%%%%%%%%%%%%%%%%%%%%%%%
\begin{figure}
  \begin{center}
    \begin{tabular}{ccc}
    \includegraphics[width=.3\columnwidth]{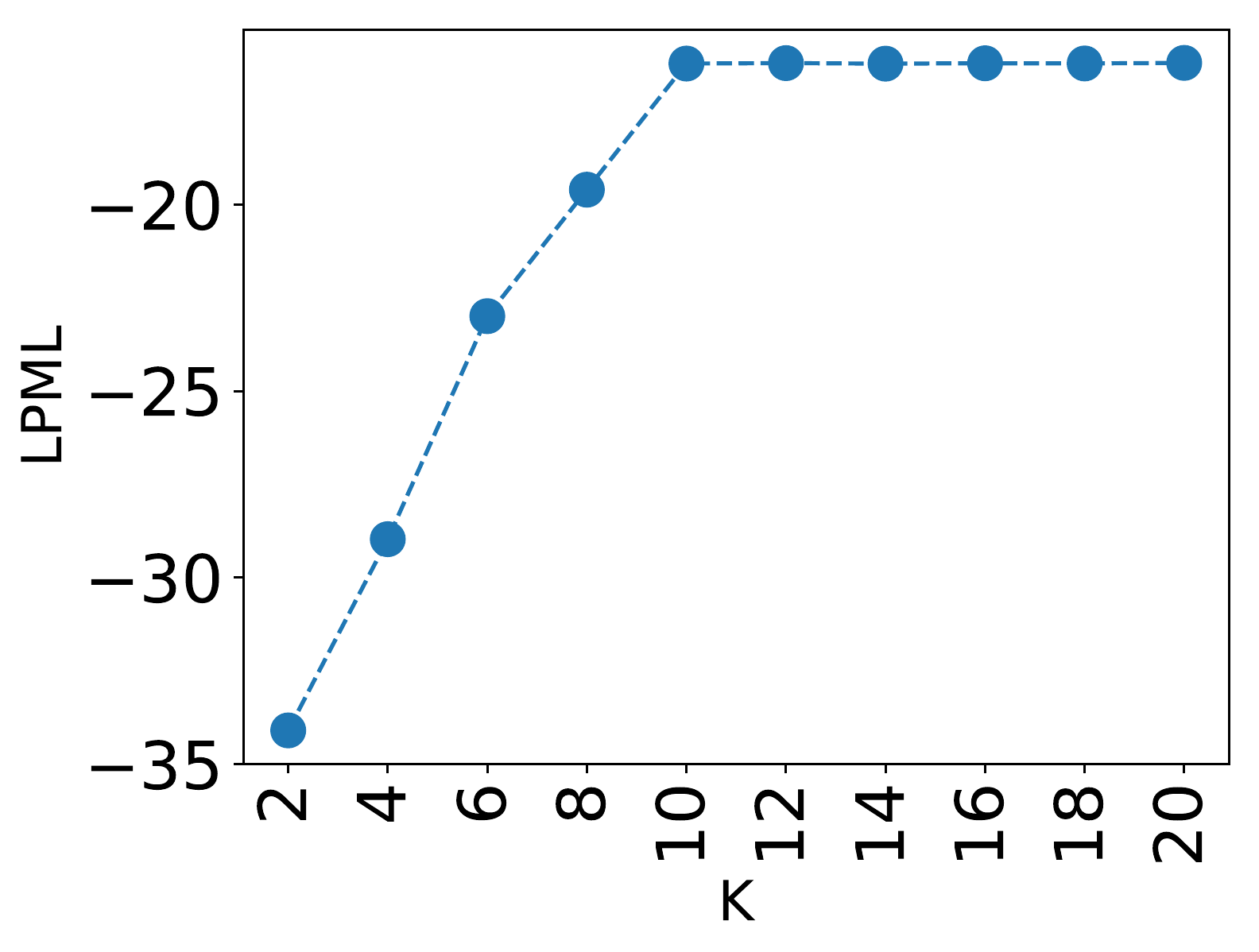} &
    \includegraphics[width=.3\columnwidth]{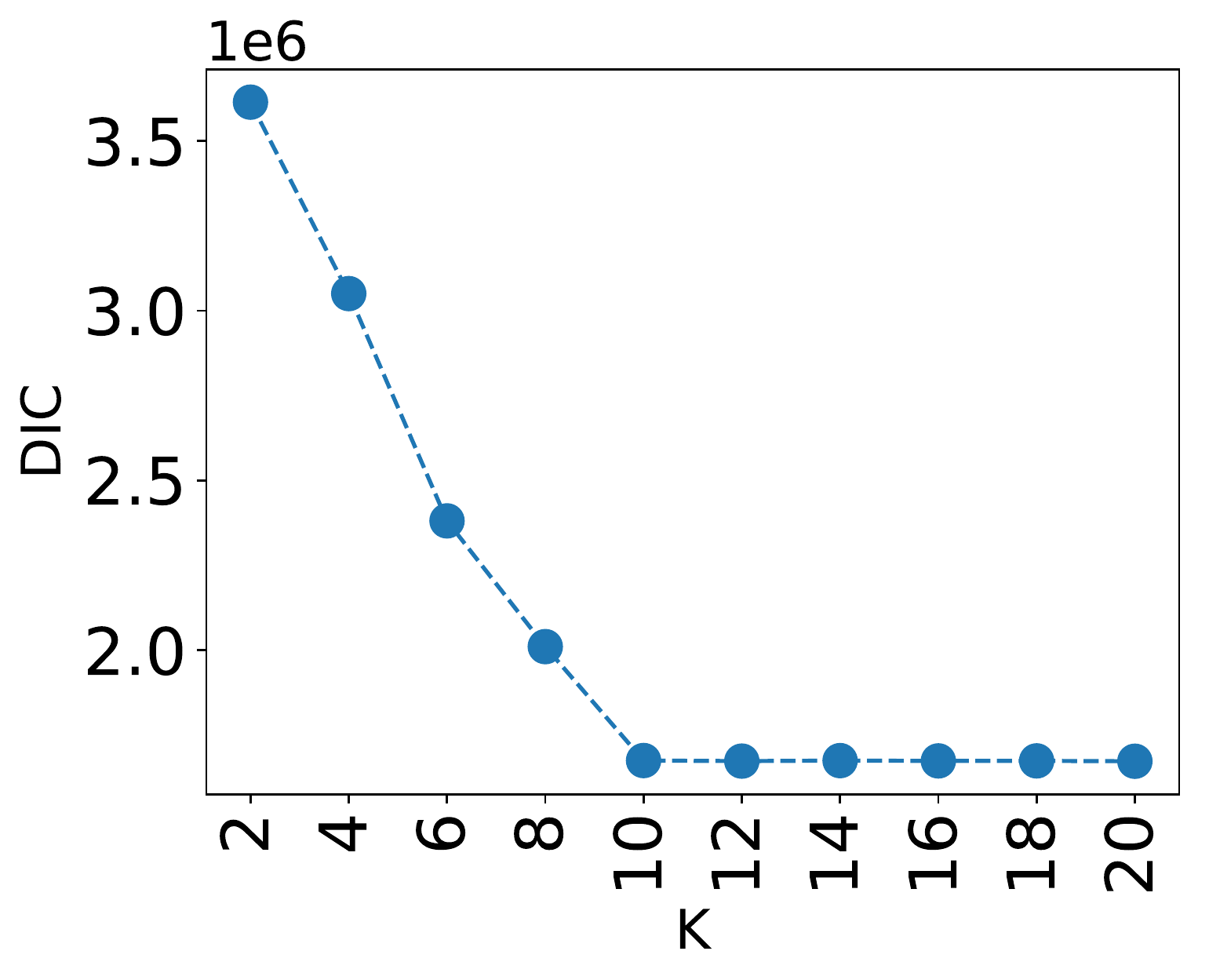} &
    \includegraphics[width=.3\columnwidth]{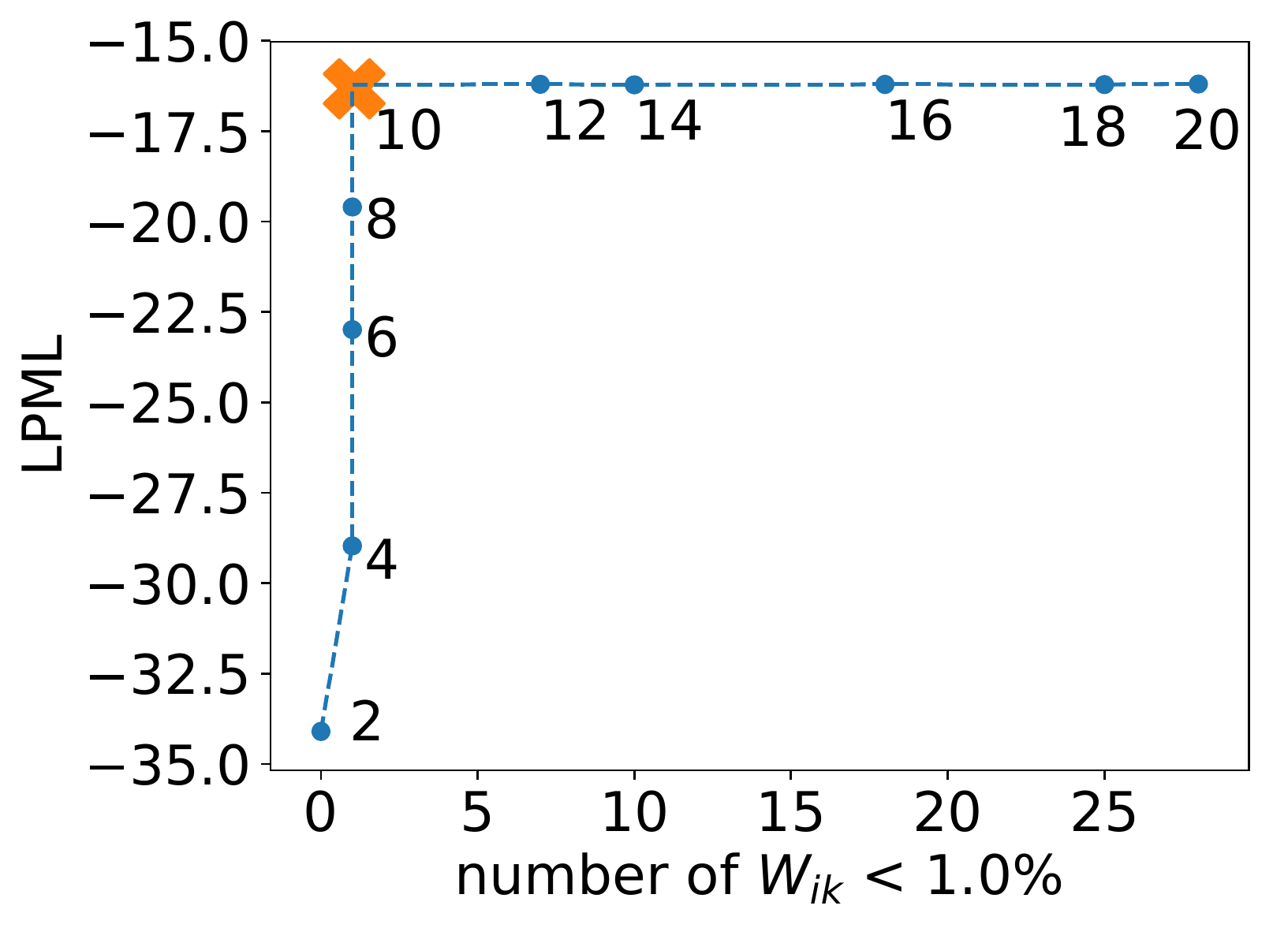} \\
    {(a) LPML} &  {(b) DIC} &  {(c) Calibration of $K$} \\
    \end{tabular}
  \end{center}
  \caption{[Simulation 2] Plots of (a) LPML, (b) DIC, and (c) calibration
  metric, for $K=2, 4, \dots, 20$, for large simulated data suggest that
  $\hat{K}=10$ is sufficient to explain the latent cell subpopulations.}
  \label{fig:metrics-sim2}
\end{figure}

%%%%%%%%%%%%%%%%%%%%%%%%
\clearpage
%% Simulation 2
\begin{figure}[t]
%\begin{figure}[th!]
  \begin{center}
  \begin{tabular}{cc}
  \includegraphics[width=0.5\columnwidth]{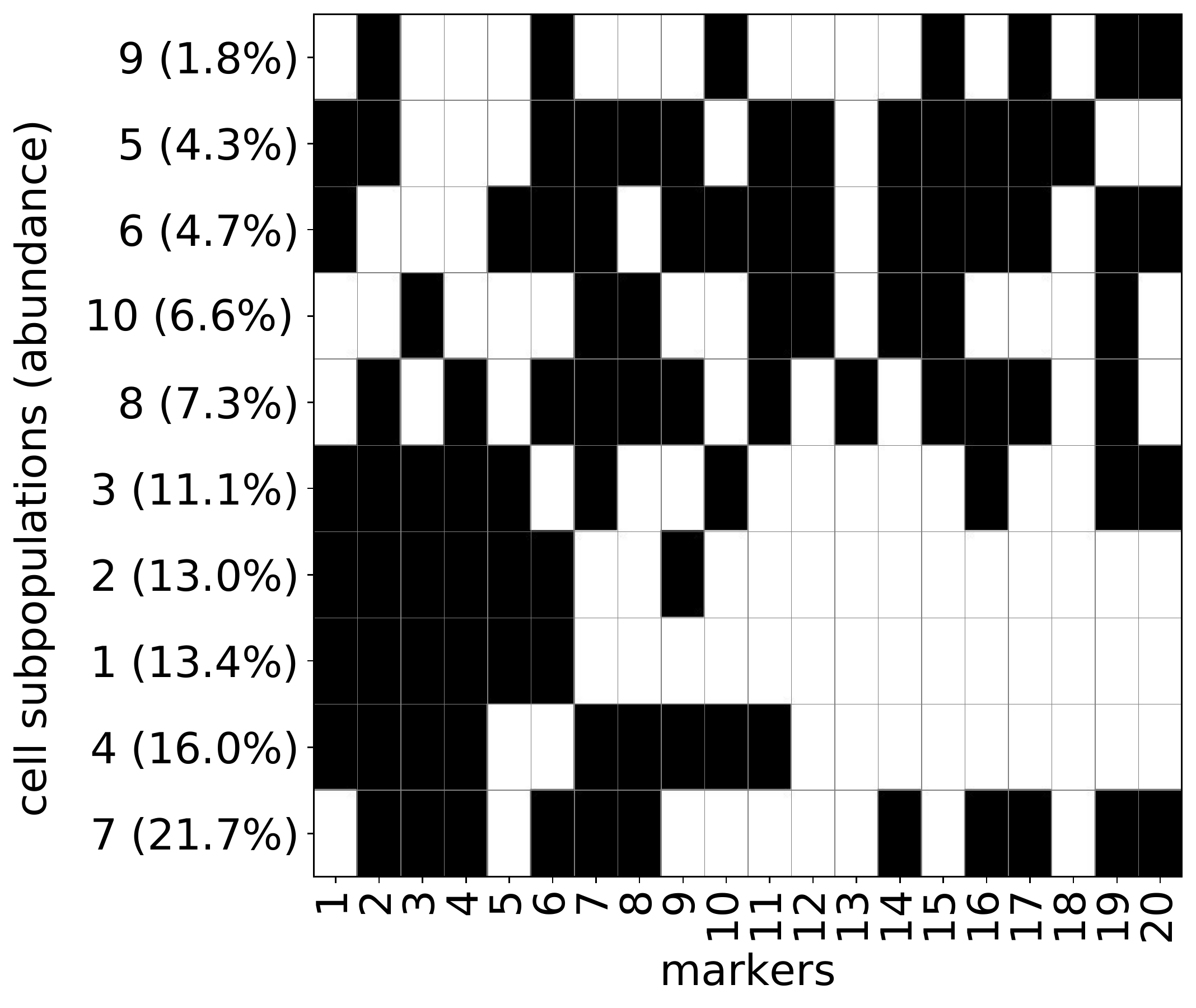} &
  \includegraphics[width=0.5\columnwidth]{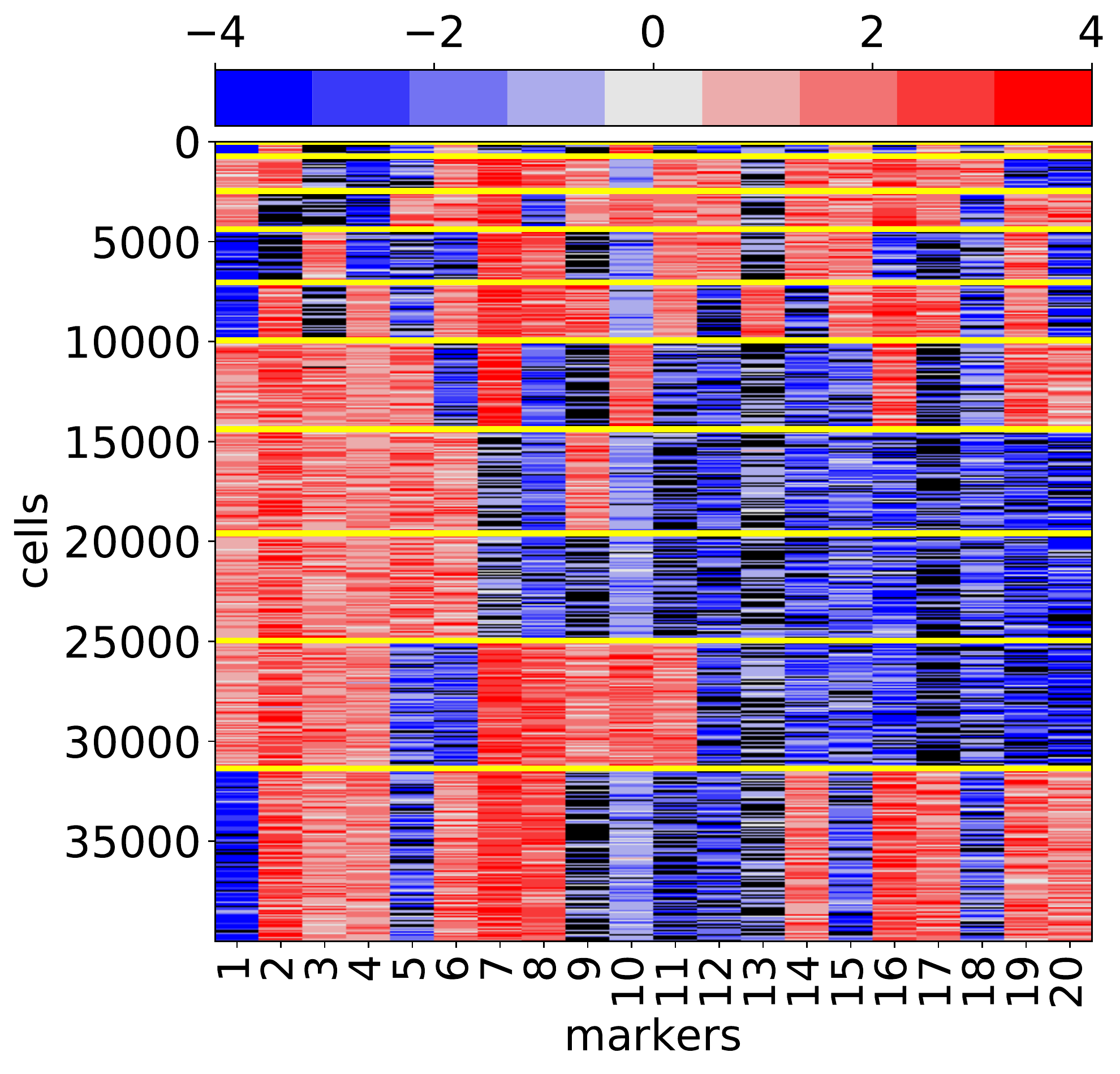} \\
  {(a) $\hat{\Z}_1$ \& $\hat{\bw}_1$} & {(b) $y_{1nj}$} \\
  \includegraphics[width=0.5\columnwidth]{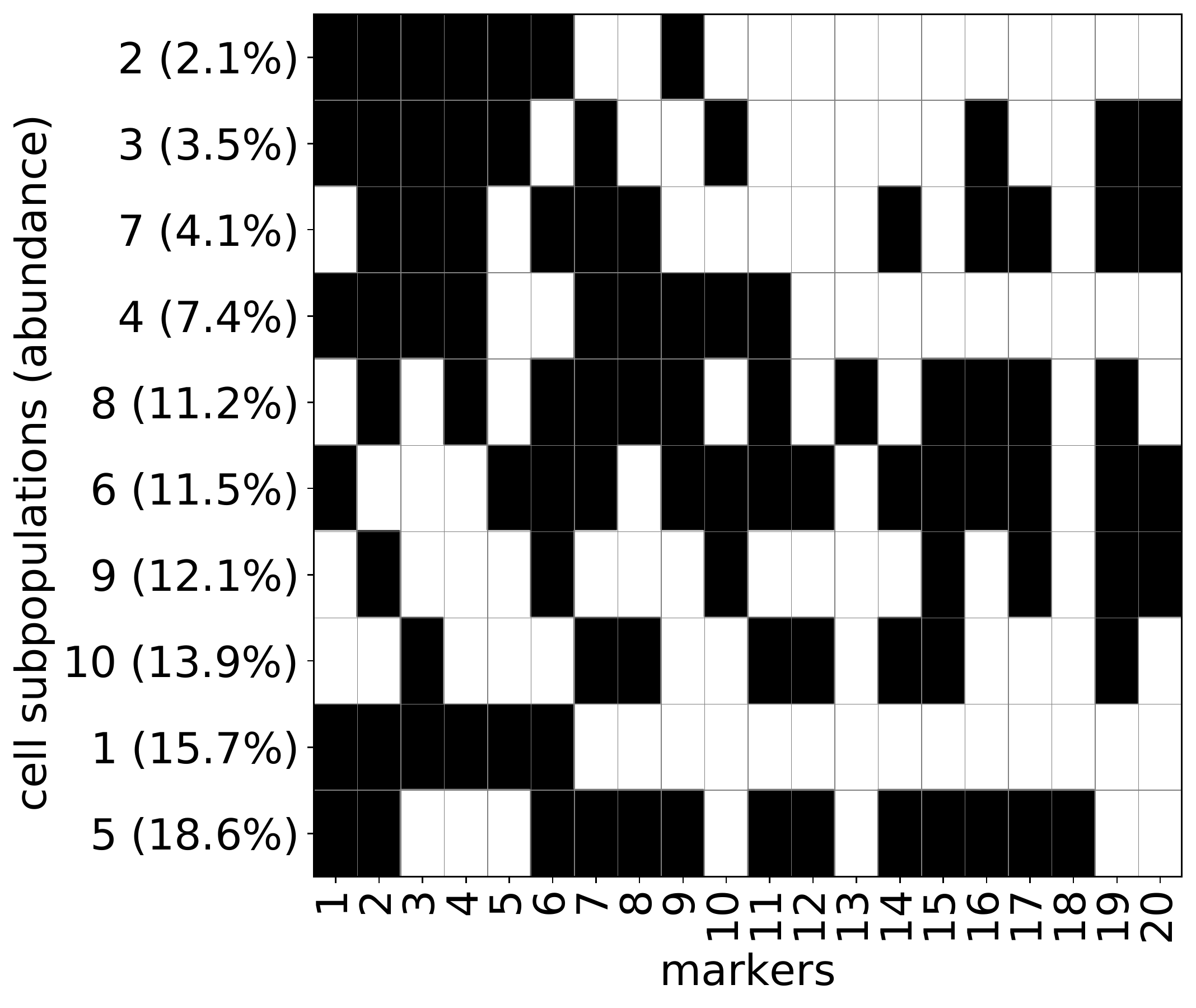} &
  \includegraphics[width=0.5\columnwidth]{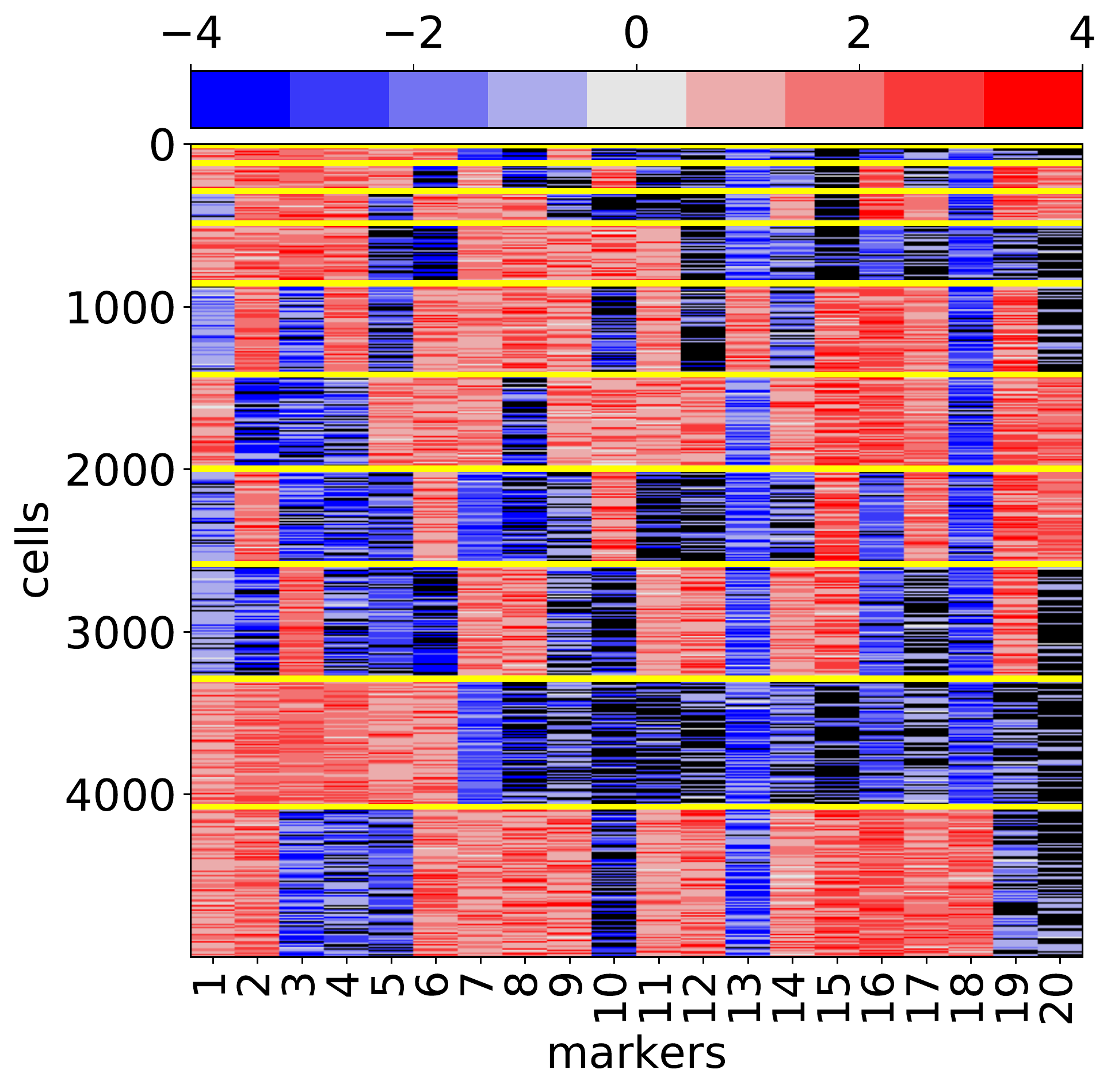} \\
  {(c) $\hat{\Z}_2$ \& $\hat{\bw}_2$} &  {(d) $y_{2nj}$}\\
  \end{tabular}
  \end{center}
  \vspace{-0.05in}
  \caption{Results of Simulation 2. In (a) and (c), 
  $\hat{\Z}^\prime_i$ and $\hat{\bw}_i$ are shown for samples
  1 and 2, respectively, with markers that are expressed dented by black and
  not expressed by white. Only subpopulations with $\hat{w}_{i,k} > 1\%$ are
  included. Heatmaps of $\bm y_i$ are shown for sample 1 in (b) and sample 2
  in (d). Cells are ordered by posterior point estimates of their subpopulations,
  $\hat{\lambda}_{i,n}$. Cells are given in rows and markers are given in
  columns. High and low expression levels are represented by red and blue,
  respectively, and black represents missing values. Yellow horizontal lines
  separate cells into five subpopulations.}
\label{fig:sim2-post}
\end{figure}
%%%%%%%%%%%%%%%%%%%%%%%%%

%%%%%%%%%%%%%%%%%%%%%%%%
\clearpage
% sim2 
\begin{figure}[h!]
%\begin{figure}[th!]
  \begin{center}
  \begin{tabular}{cc}
  \includegraphics[width=0.5\columnwidth]{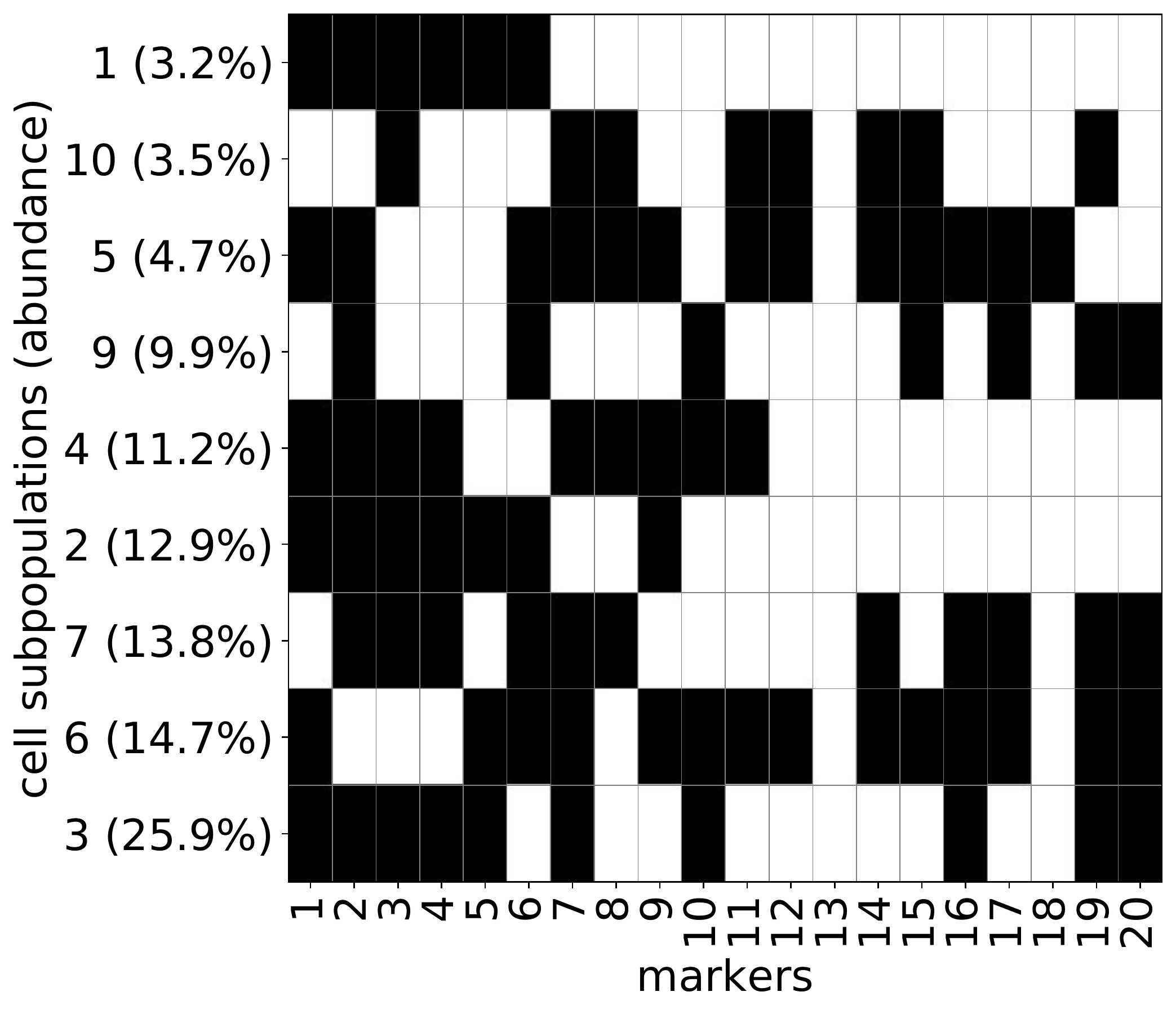} &
  \includegraphics[width=0.5\columnwidth]{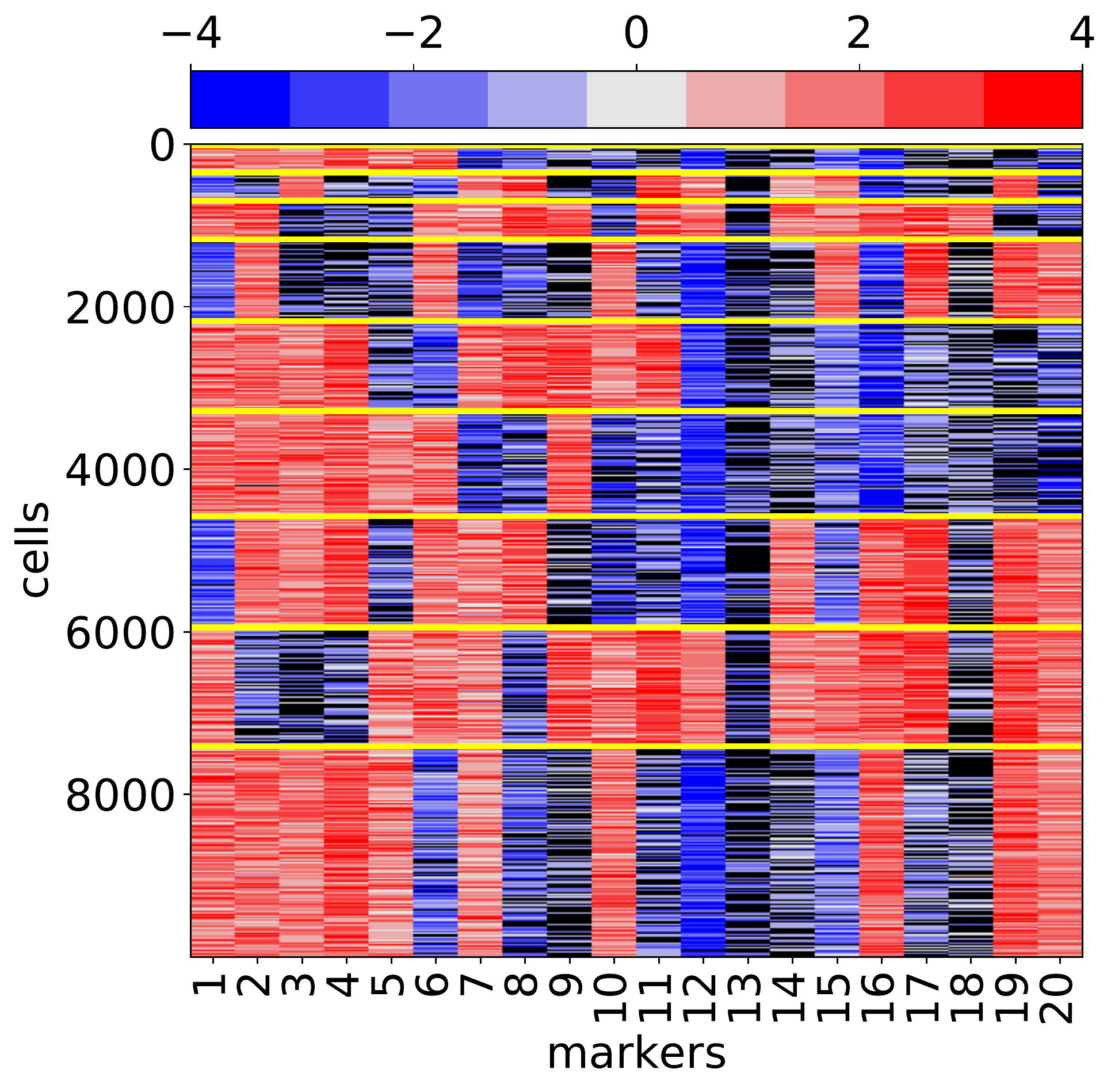} \\
  {(e) $\hat{\Z}_3$ \& $\hat{\bw}_3$} & {(f) $y_{3nj}$}\\
  \end{tabular}
  \end{center}
  \vspace{-0.05in}
  \caption*{Figure~\ref{fig:sim2-post}. Results of Simulation 2 (continued) In
  (e), $\hat{\Z}^\prime_i$ and $\hat{\bw}_i$ are shown for sample 3, with
  markers that are expressed dented by black and not expressed by white. Only
  subpopulations with $\hat{w}_{i,k} > 1\%$ are included. Heatmaps of $\bm y_i$
  for sample 3 is shown in (f). Cells are ordered by posterior point
  estimates of their subpopulations, $\hat{\lambda}_{i,n}$. Cells are given in
  rows and markers are given in columns. High and low expression levels are
  represented by red and blue, respectively, and black represents missing
  values. Yellow horizontal lines separate cells into five subpopulations.}
\end{figure}
%%%%%%%%%%%%%%%%%%%%%%%%%

%%%%%%%%%%%%%%%%%%%%%%%%
%% Simulation 2 - ADVI 1
\begin{figure}[t]
\begin{center}
  \begin{tabular}{cc}
  \includegraphics[width=0.5\columnwidth]{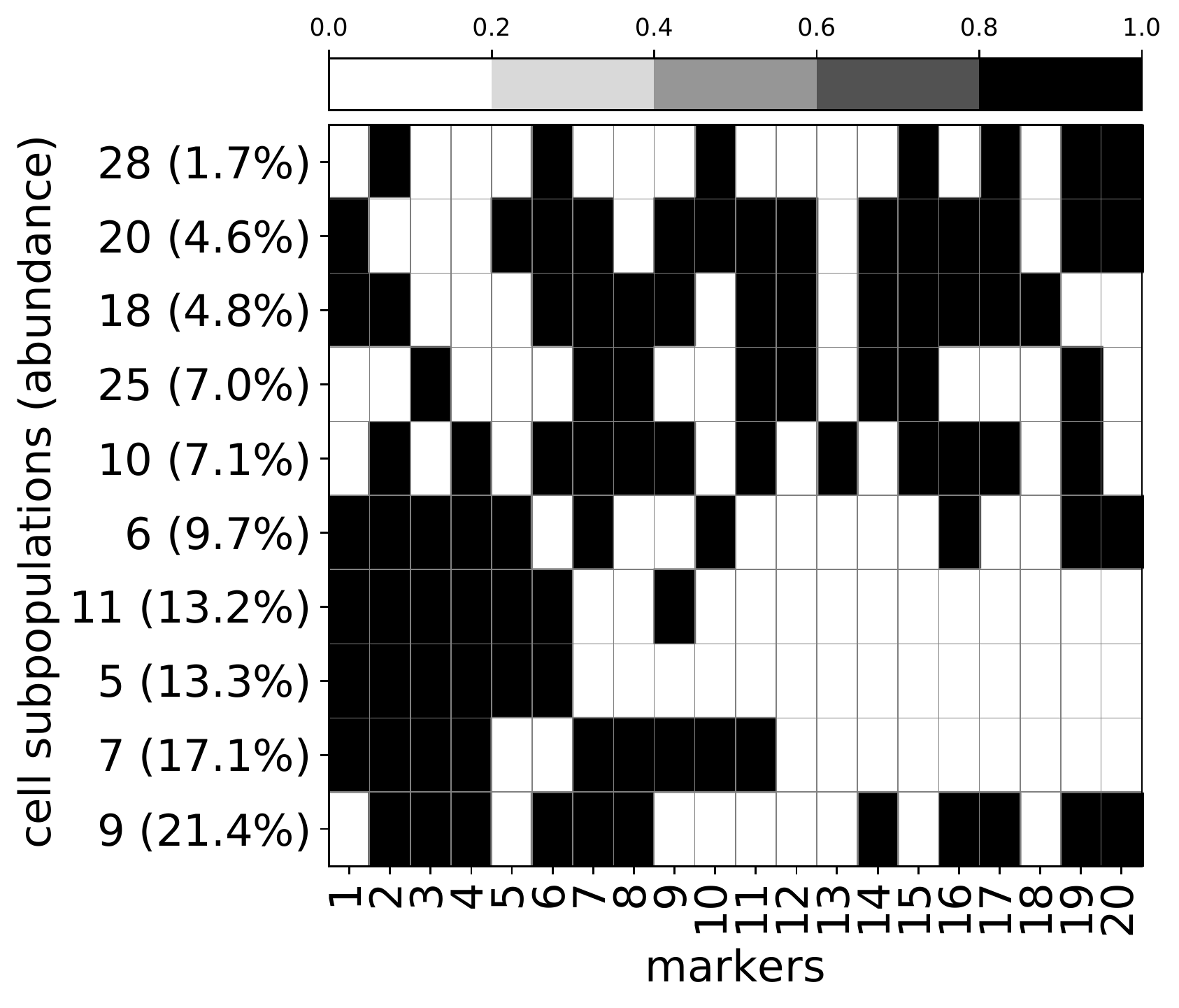} &
  \includegraphics[width=0.5\columnwidth]{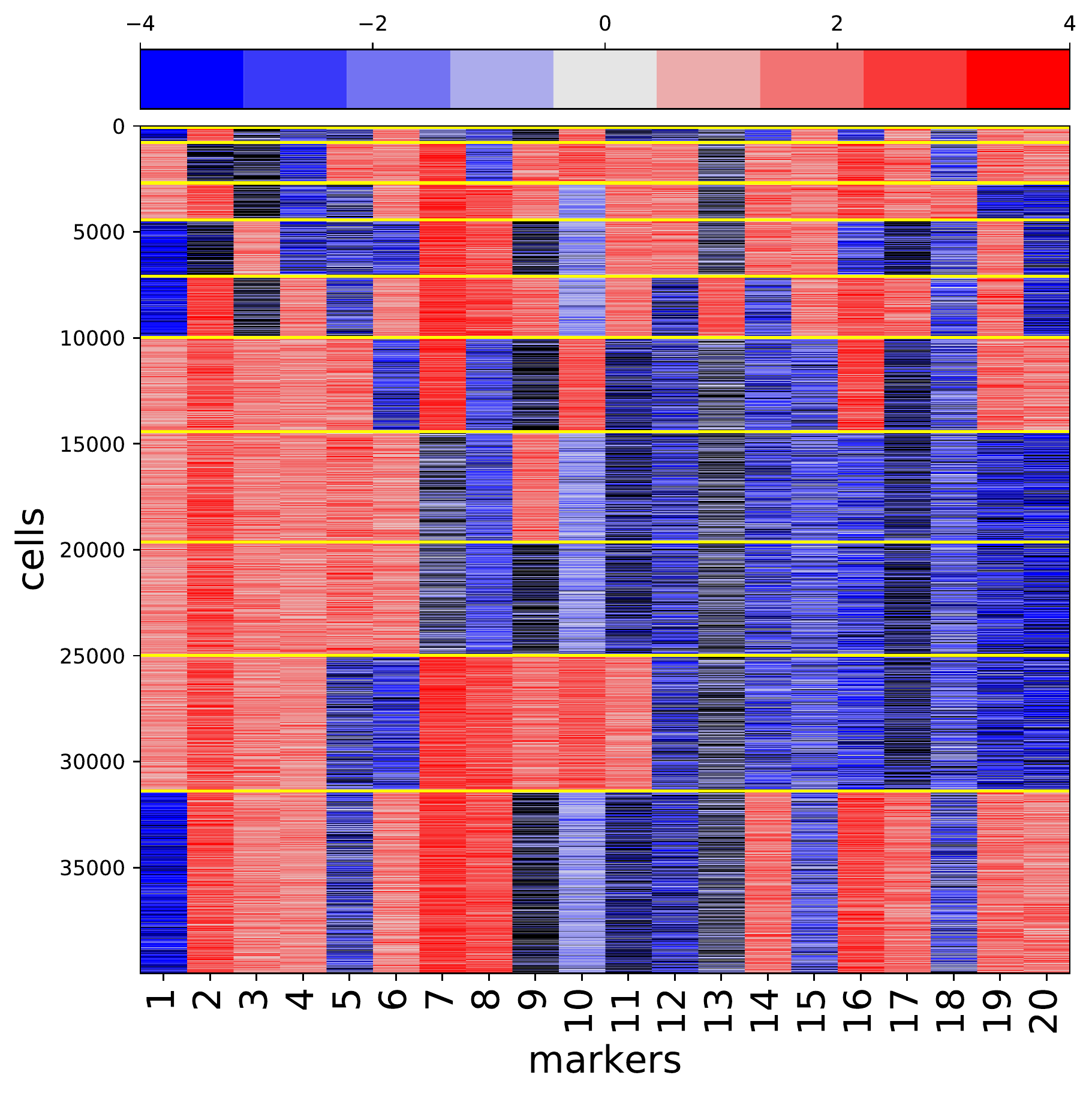} \\
  (a) $\hat{\Z}^\prime_1$ and $\hat{\bw}_1$ & (b) $y_{1nj}$\\
  \includegraphics[width=0.5\columnwidth]{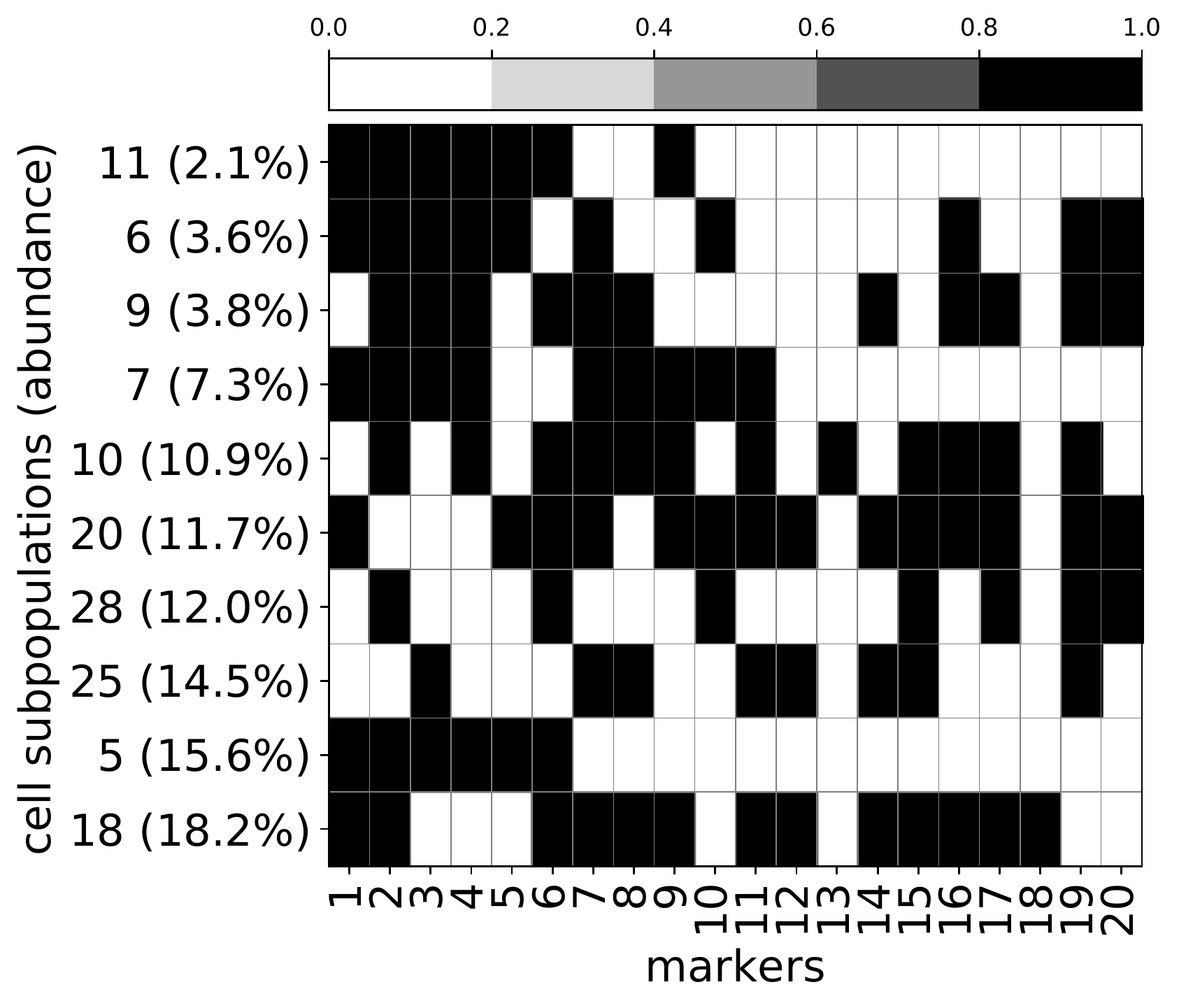} &
  \includegraphics[width=0.5\columnwidth]{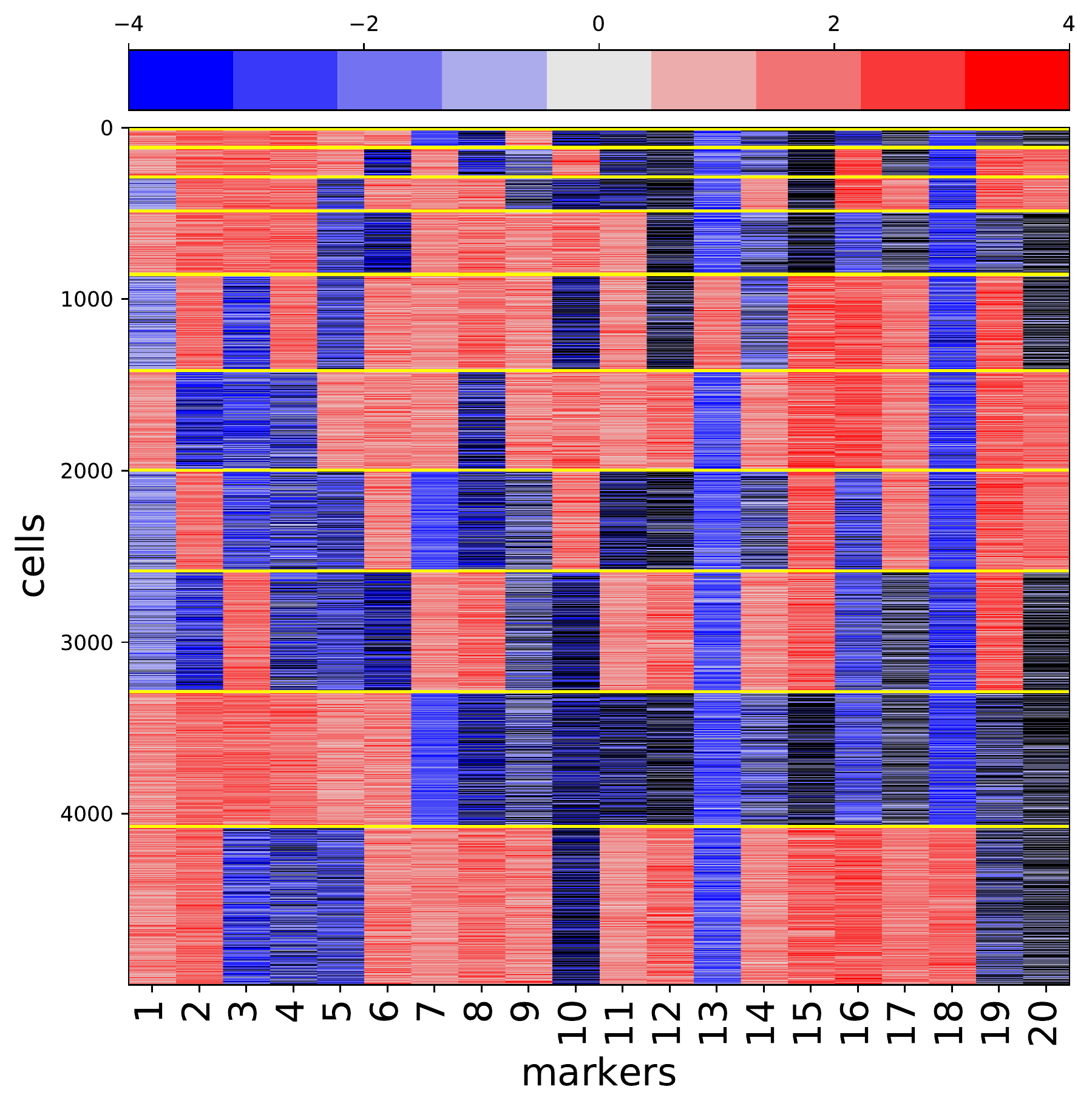} \\
  (c) $\hat{\Z}^\prime_2$ and $\hat{\bw}_2$ & (d) $y_{2nj}$\\
  \end{tabular}
  \vspace{-0.05in}
  \caption{[ADVI for Simulation 2]
  In (a) and (c), the transpose
  $\hat{\Z}^\prime_i$ of $\hat \bZ_i$ and $\hat{\bw}_i$ are shown for samples
  1 and 2, respectively, with markers that are expressed dented by black and
  not expressed by white. Only subpopulations with $\hat{w}_{i,k} > 1\%$ are
  included. Heatmaps of $\bm y_i$ are shown for sample 1 in (b) and sample 2
  in (d). Cells are ordered by posterior point estimates of their subpopulations,
  $\hat{\lambda}_{i,n}$. Cells are given in rows and markers are given in
  columns. High and low expression levels are represented by red and blue,
  respectively, and black represents missing values. Yellow horizontal lines
  separate cells into five subpopulations. Posterior estimates are obtained via ADVI. }
  \label{fig:sim-vb-2}
\end{center}
\end{figure}

%% Simulation 2 - ADVI 2
\begin{figure}[thb]
\begin{center}
  \begin{tabular}{cc}
  \includegraphics[width=0.5\columnwidth]{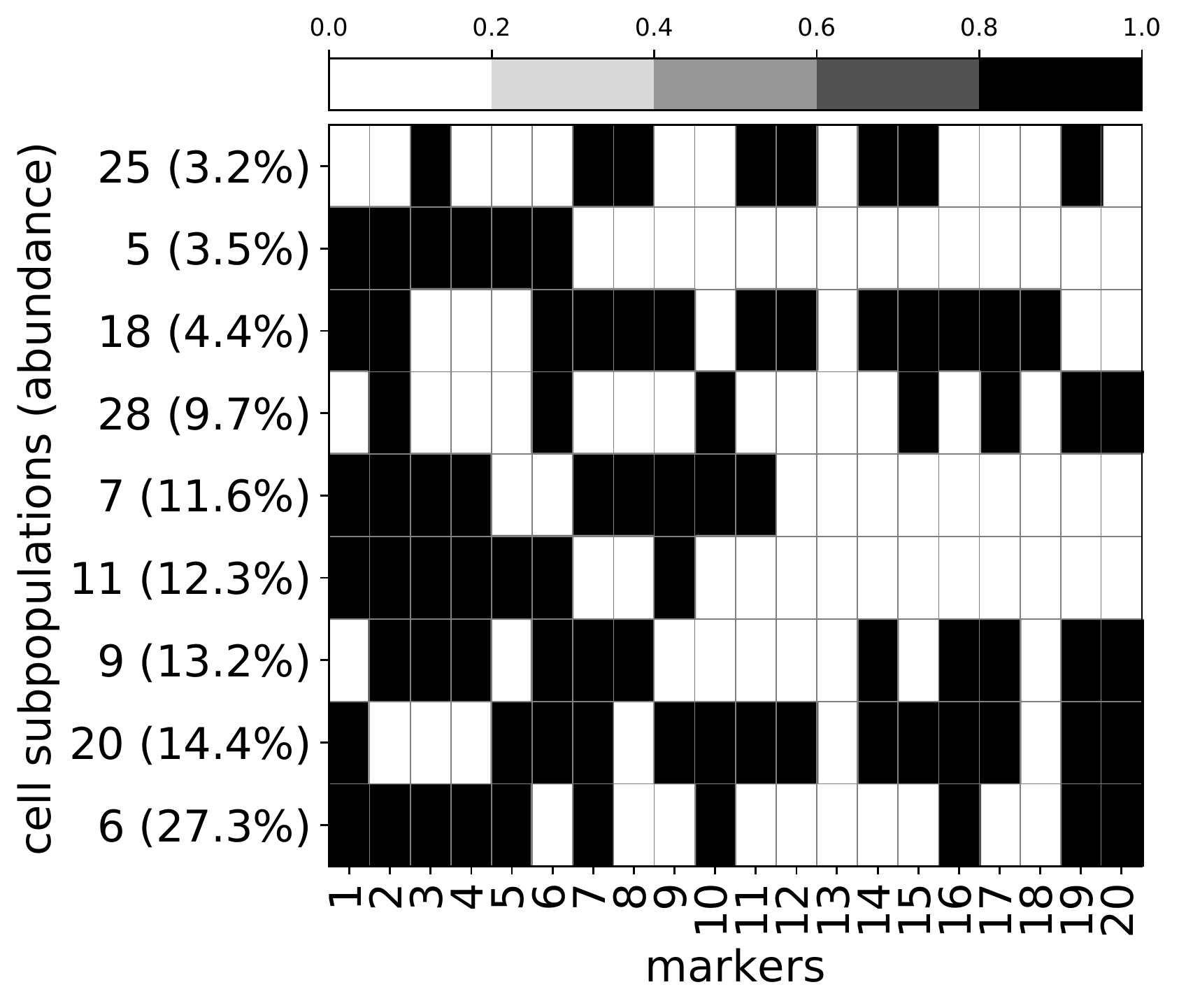} &
  \includegraphics[width=0.5\columnwidth]{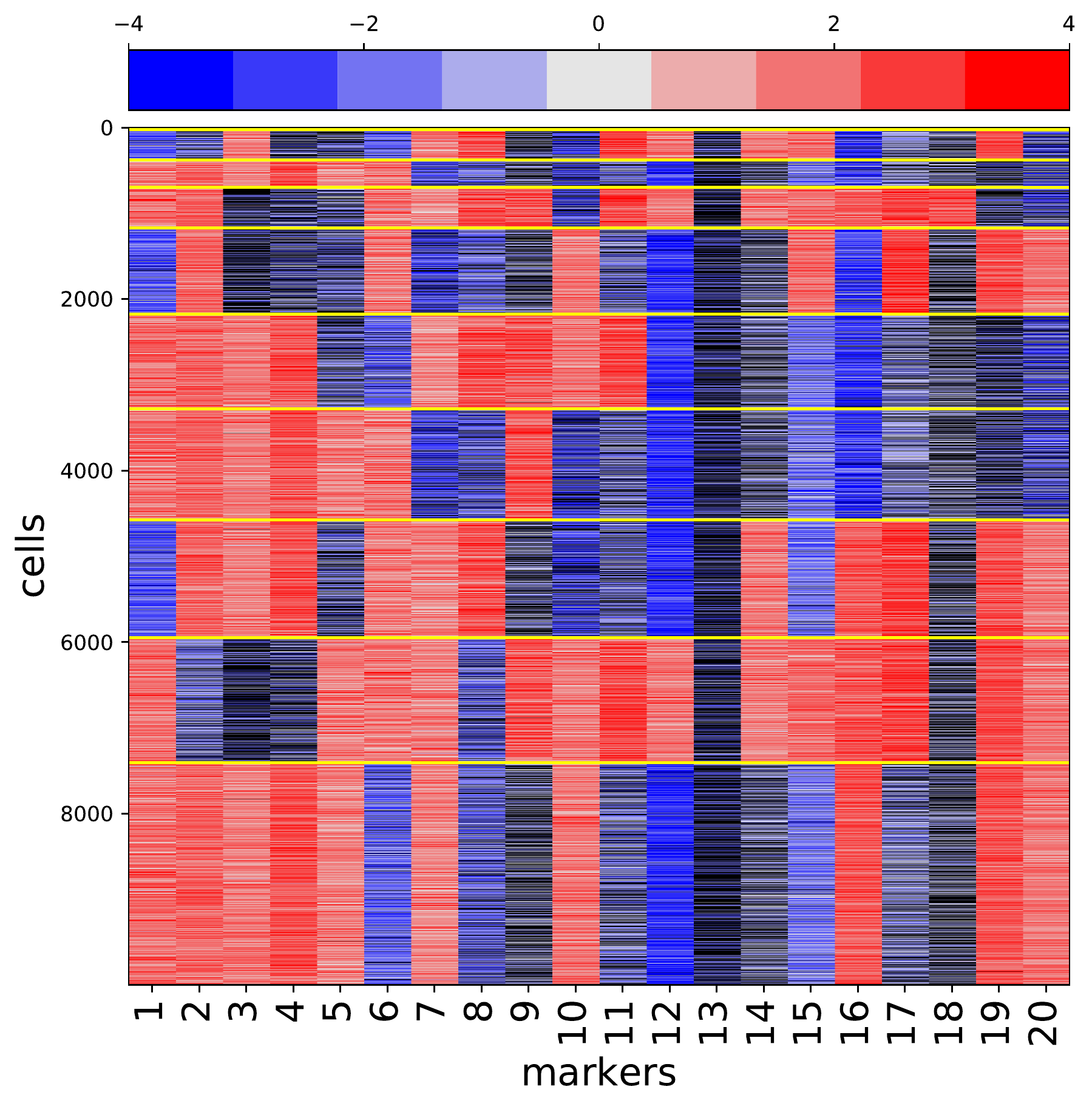} \\
  (e) $\hat{\Z}^\prime_3$ and $\hat{\bw}_3$ & (f) $y_{3nj}$\\
  \end{tabular}
  \vspace{-0.05in}
  \caption*{Figure~\ref{fig:sim-vb-2} continued: [ADVI for Simulation 2]
	In
  (e), the transpose $\hat{\Z}^\prime_i$ of $\hat \bZ_i$ and $\hat{\bw}_i$
  are shown for sample 3, with markers that are expressed dented by black and
  not expressed by white. Only subpopulations with $\hat{w}_{i,k} > 1\%$ are
  included. Heatmaps of $\bm y_i$ for sample 3 is shown in (f). Cells are
  ordered by posterior point estimates of their subpopulations,
  $\hat{\lambda}_{i,n}$. Cells are given in rows and markers are given in
  columns. High and low expression levels are represented by red and blue,
  respectively, and black represents missing values. Yellow horizontal lines 
  separate cells into five subpopulations. Posterior estimates are obtained via ADVI.}
\end{center}
\end{figure}
%%%%%%%%%%%%%%%%%%%%%%%%%

\clearpage
%% Simulation 2 - missing
\begin{table}[t]
  \centering
  \begin{tabular}{c|cccc}
    \hline
    Missing Mechanism & $\tilde{\bm q}$ &
    Probability of Missing $(\bm\rho)$ & LPML & DIC \\
    \hline
    0   & (0\%, 25\%, 50\%) & (5\%, 80\%, 5\%) & -16.215 & 1675117 \\
    I   & (0\%, 20\%, 40\%) & (5\%, 80\%, 5\%) & -16.052 & 1662834 \\
    II  & (0\%, 15\%, 30\%) & (5\%, 80\%, 5\%) & -15.771 & 1640255 \\
    \hline
  \end{tabular}
    \caption[Missingness Mechanism Specifications for Simulation 2]{Missingness mechanisms used for
  Simulation 2. $\tilde{\bm q}$-quantiles of the negative observed
  values in each sample are used to specify $\tilde{\bm y}$, and $\bm\rho$ are the probability of missing at $\tilde{\bm y}$.  Three different sets of $\tilde{\bm q}$ and $\tilde{\bm \rho}$ are used to examine the sensitivity to the missingship mechanism specification. LPML and DIC are shown in the last two columns under each of the specification. }
  \label{tab:missmechsen-sim2}
\end{table}

%% Simulation 2 - missing I
\begin{figure}[t]
  \centering
  \begin{tabular}{ccc}
    \includegraphics[width=.3\columnwidth]{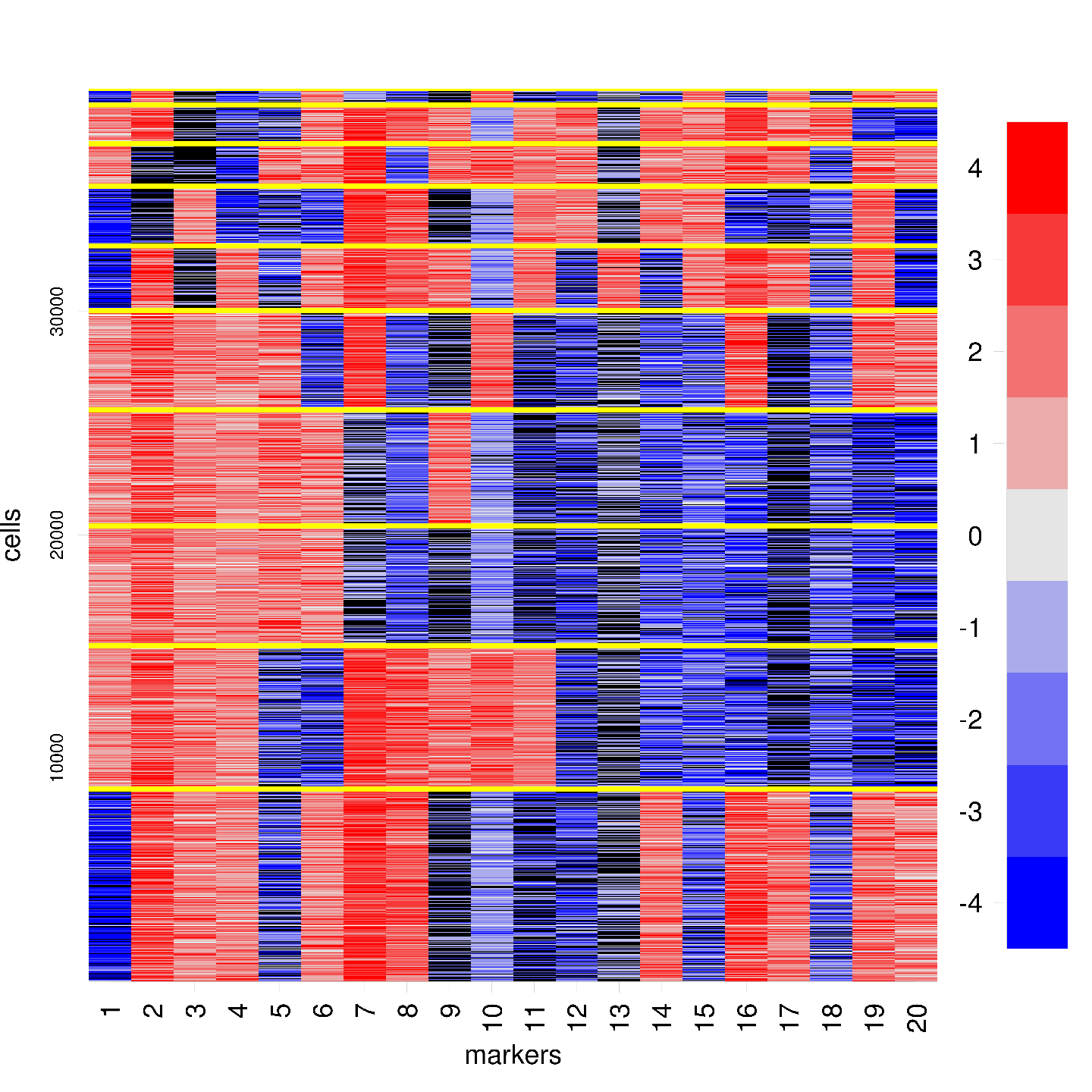} &
    \includegraphics[width=.3\columnwidth]{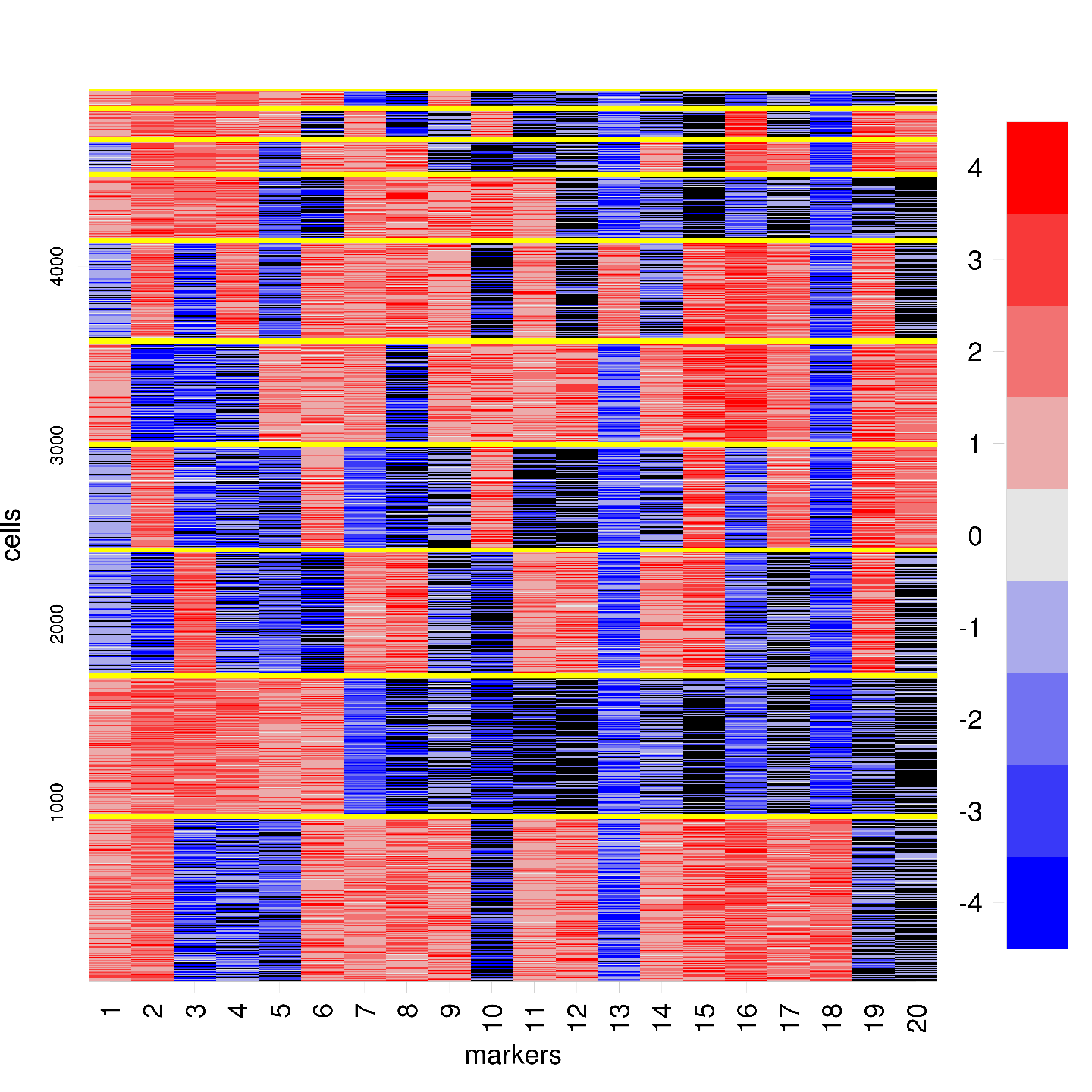} &
    \includegraphics[width=.3\columnwidth]{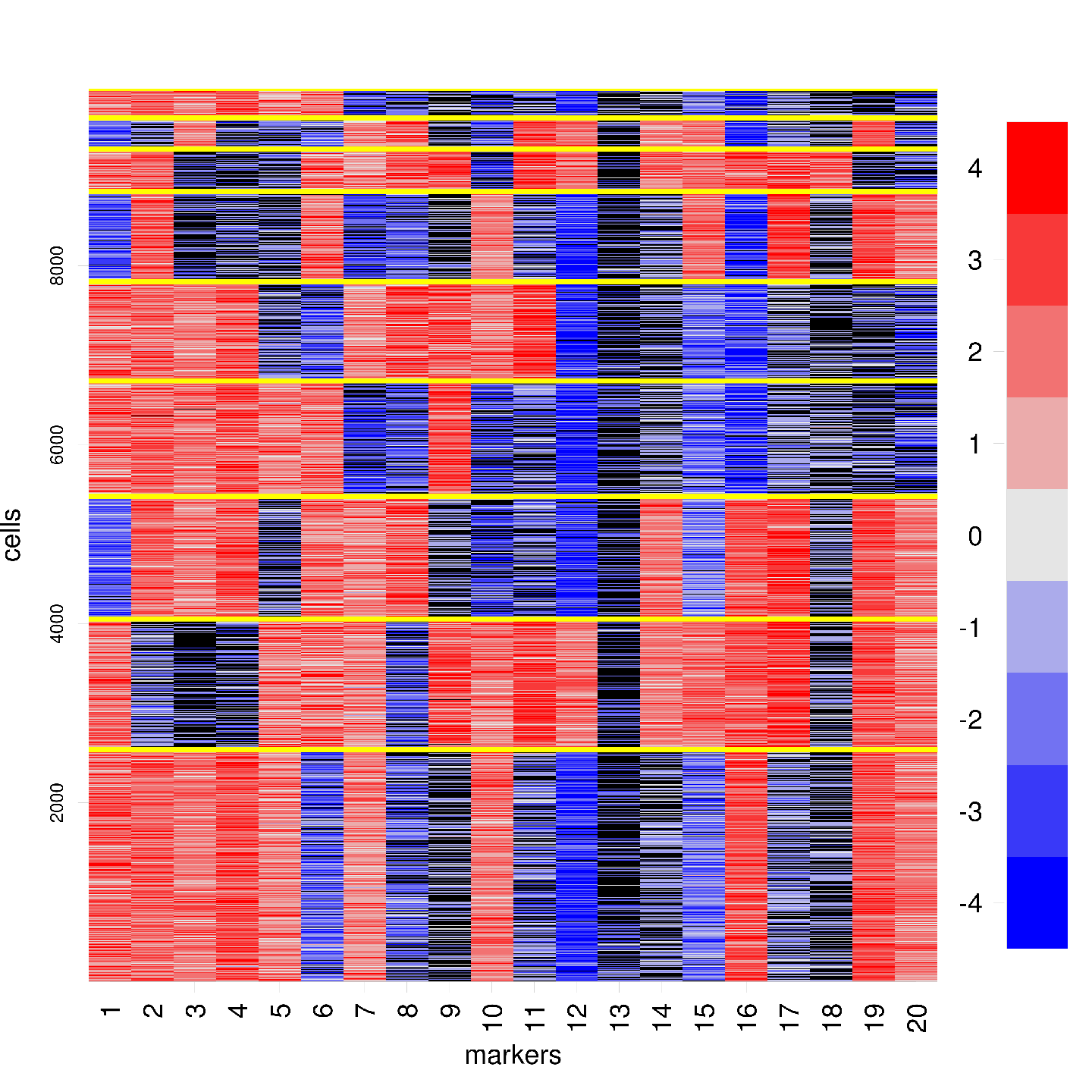} \\
	(a) heatmap of $y_{1nj}$ & (b) heatmap of $y_{2nj}$ & (c) heatmap of $y_{3nj}$\\    
    \includegraphics[width=.3\columnwidth]{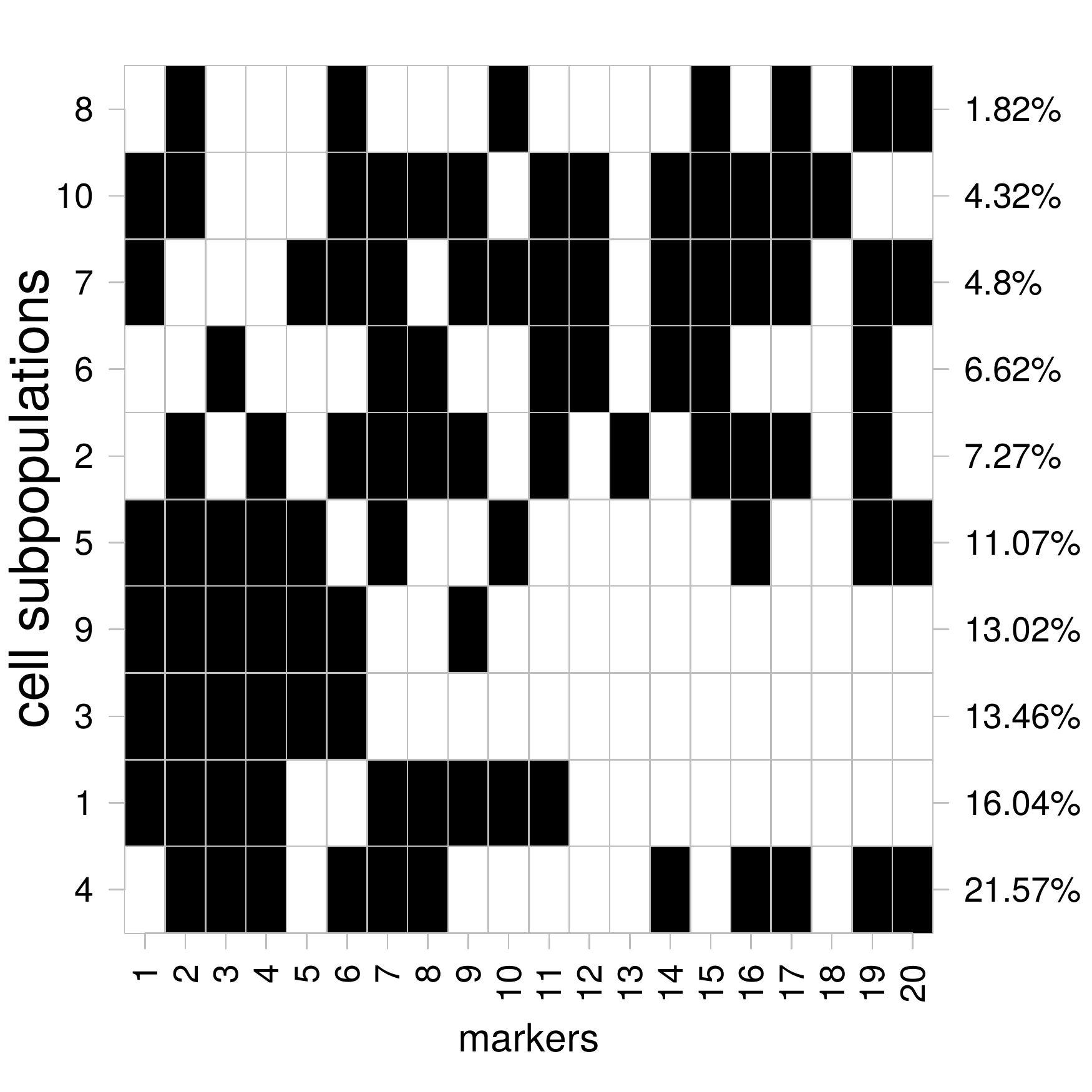} &
    \includegraphics[width=.3\columnwidth]{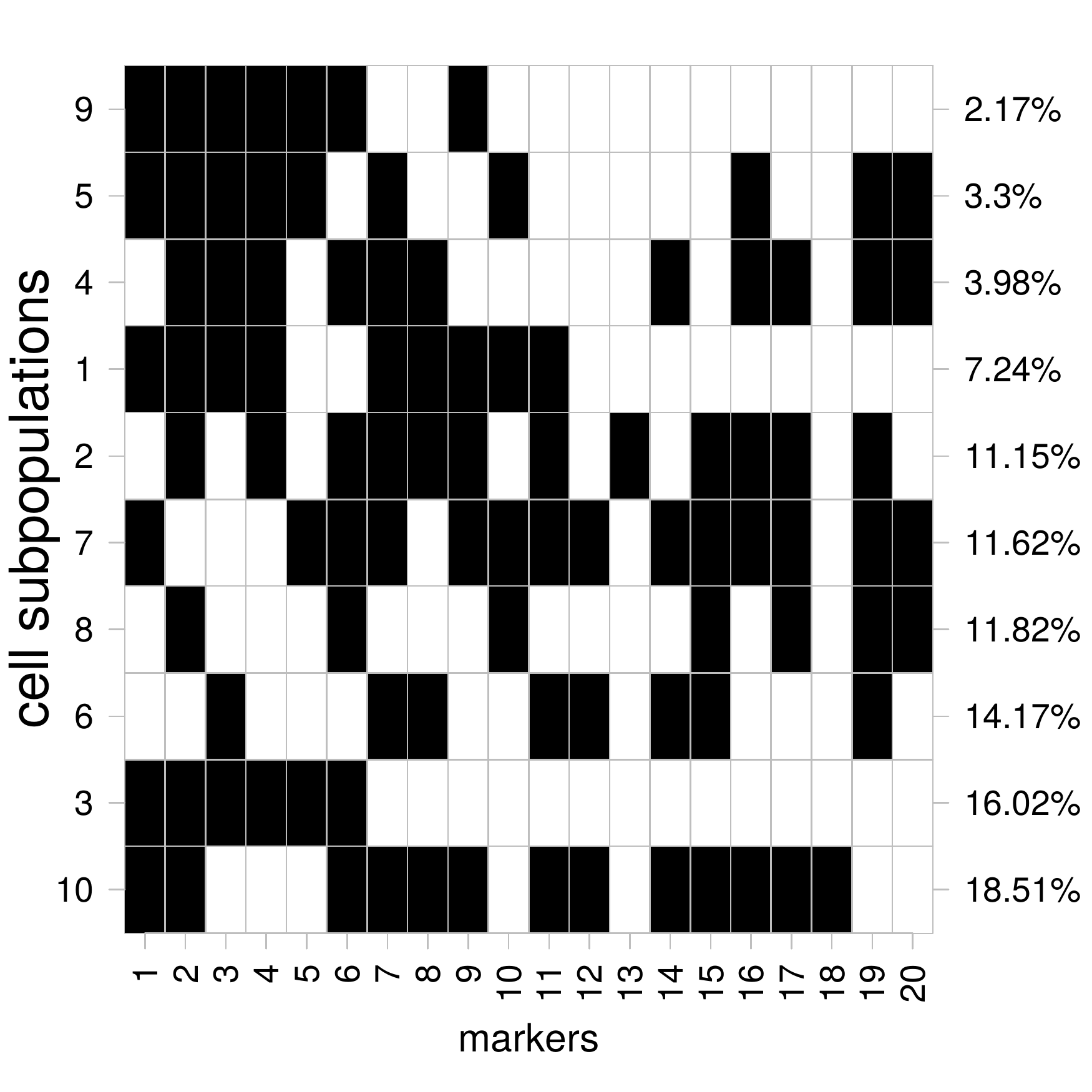} &
    \includegraphics[width=.3\columnwidth]{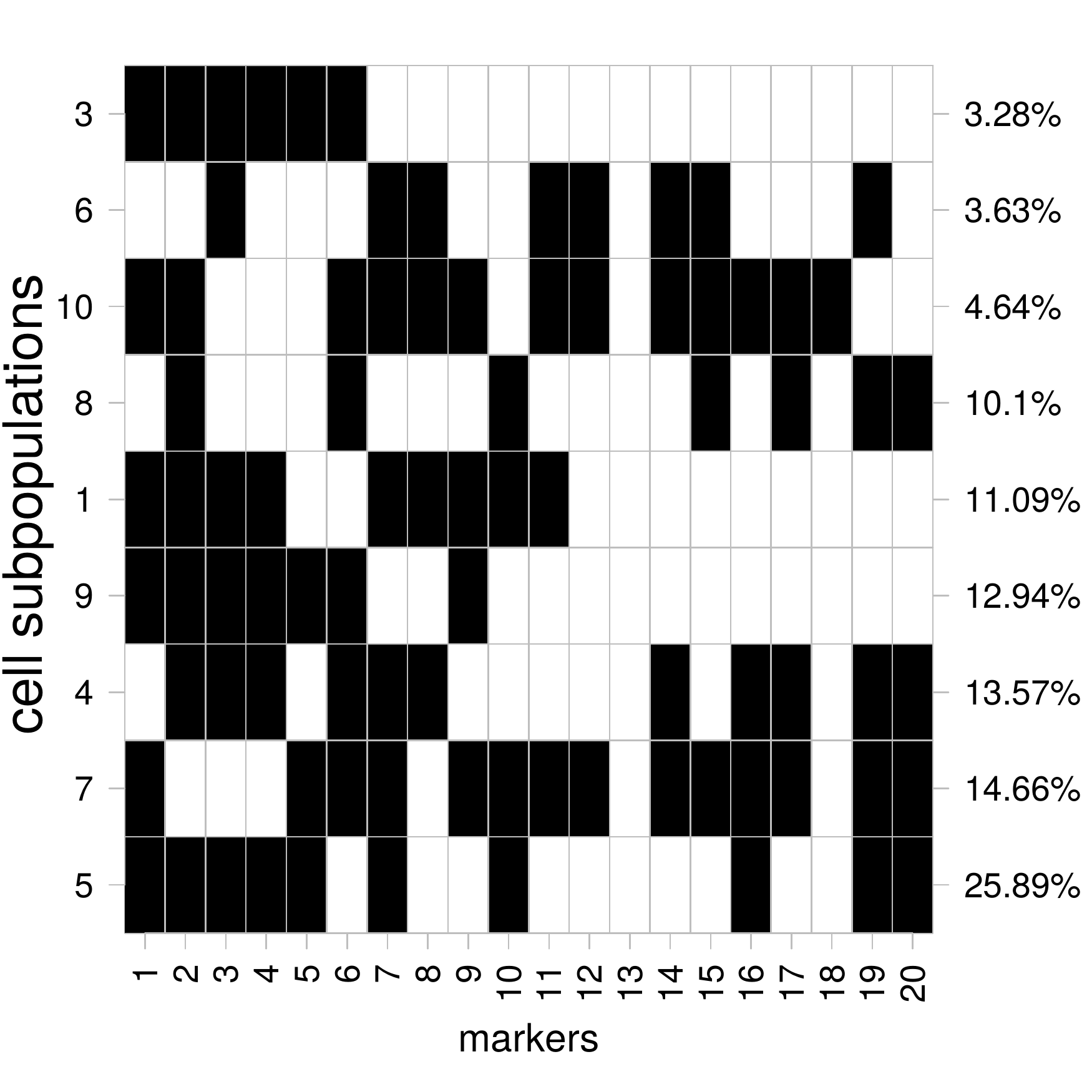} \\
    (d) $\hat{\Z}^\prime_1$ \& $\hat{\bw}_1$ & (e) $\hat{\Z}^\prime_2$ \& $\hat{\bw}_2$ & (c) $\hat{\Z}^\prime_3$ \& $\hat{\bw}_3$ \\
  \end{tabular}
  \caption{Data missingship mechanism sensitivity analysis for Simulation 2.
  Specification I is used for $\bm \beta$. Heatmaps of $\y_i$ are
  shown in (a)-(c) for samples 1-3, respectively. Cells are rearranged by the
  posterior point estimate of cell clustering, $\hat{\lambda}_{i,n}$. Cells
  and markers are in rows and columns, respectively. High and low expression
  levels are in red and blue, respectively, and black is used for missing
  values. Yellow horizontal lines separate cells by different subpopulations. 
  $\hat{\Z}^\prime_i$ and $\hat{\bw}_i$ are shown
  for each of the samples in (d)-(f). We include only subpopulations with
  $\hat{w}_{i,k} > 1\%$.}
  \label{fig:Z-w-sim2-missmechsen-1}
\end{figure}

%% Simulation 2 - missing II
\begin{figure}[t]
  \centering
  \begin{tabular}{ccc}
    \includegraphics[width=.3\columnwidth]{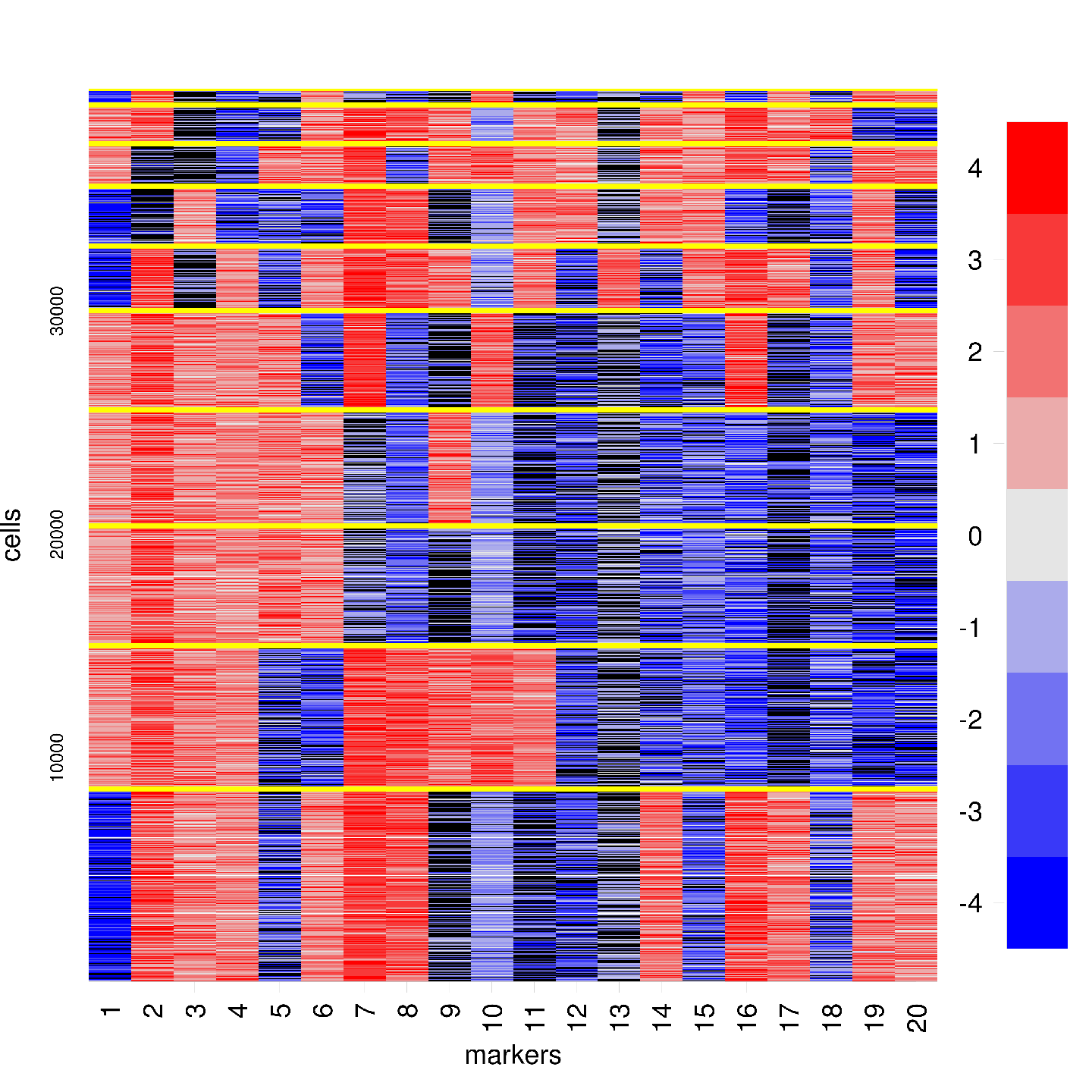} &
    \includegraphics[width=.3\columnwidth]{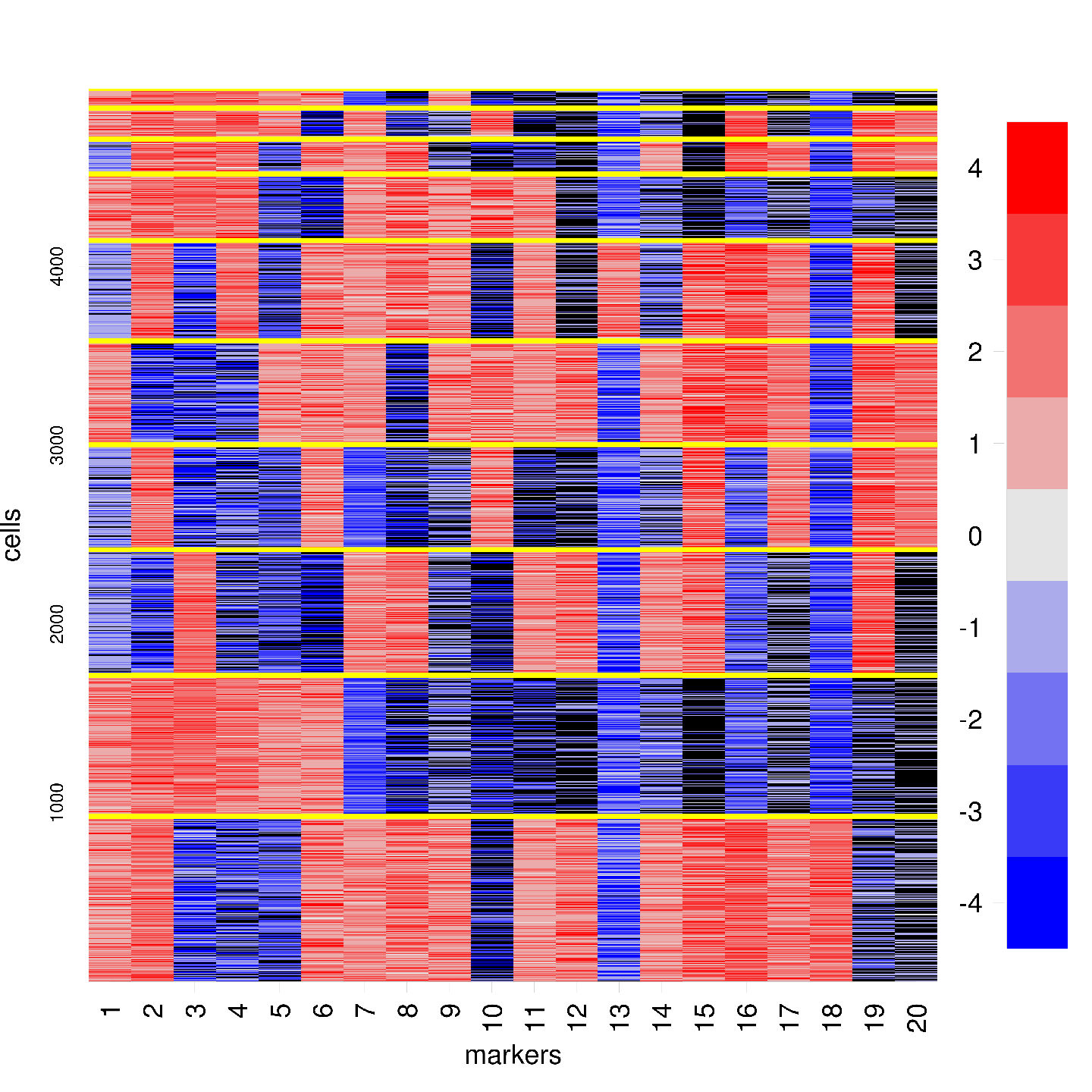} &
    \includegraphics[width=.3\columnwidth]{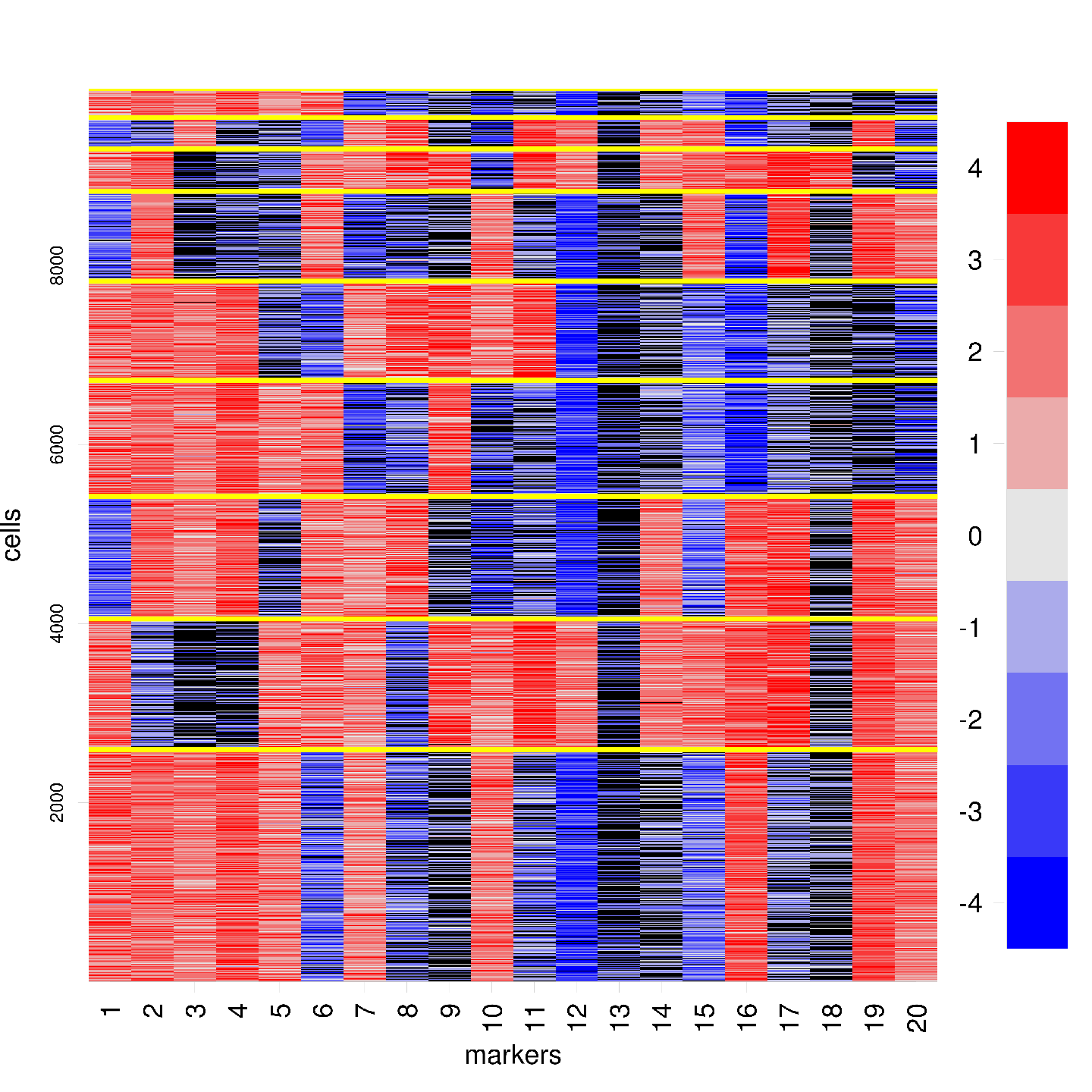} \\
  	(a) heatmap of $y_{1nj}$ & (b) heatmap of $y_{2nj}$ & (c) heatmap of $y_{3nj}$\\    
    \includegraphics[width=.3\columnwidth]{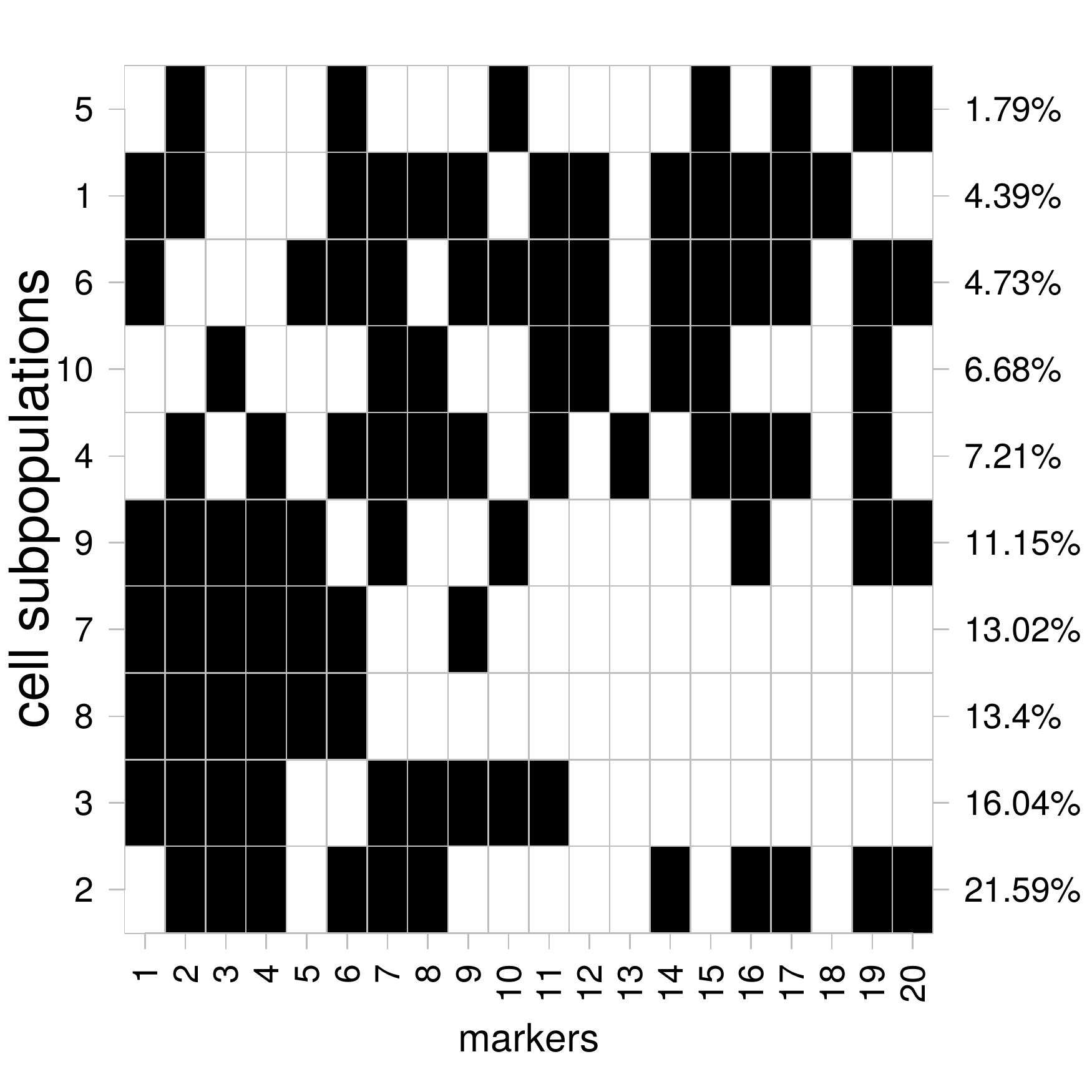} &
    \includegraphics[width=.3\columnwidth]{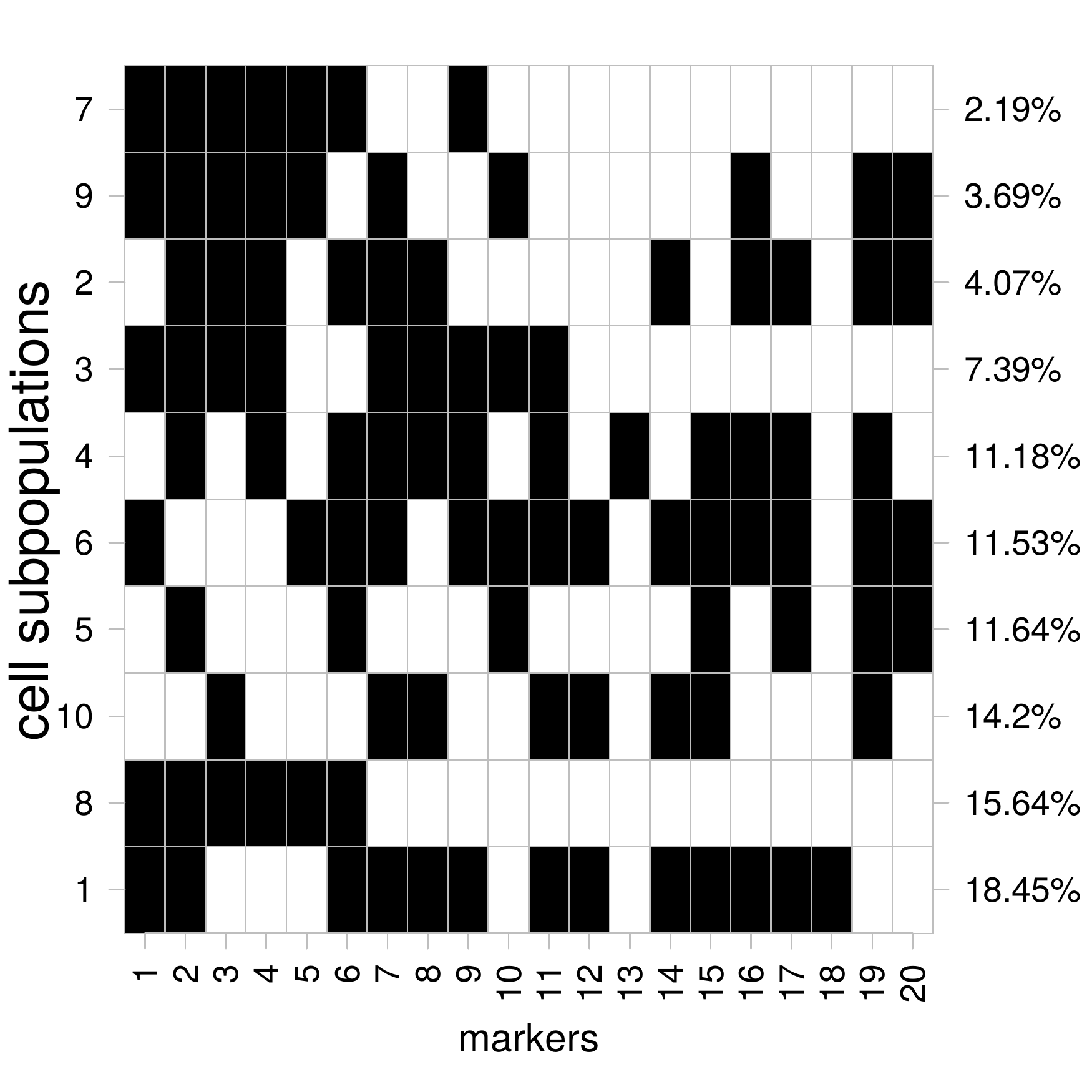} &
    \includegraphics[width=.3\columnwidth]{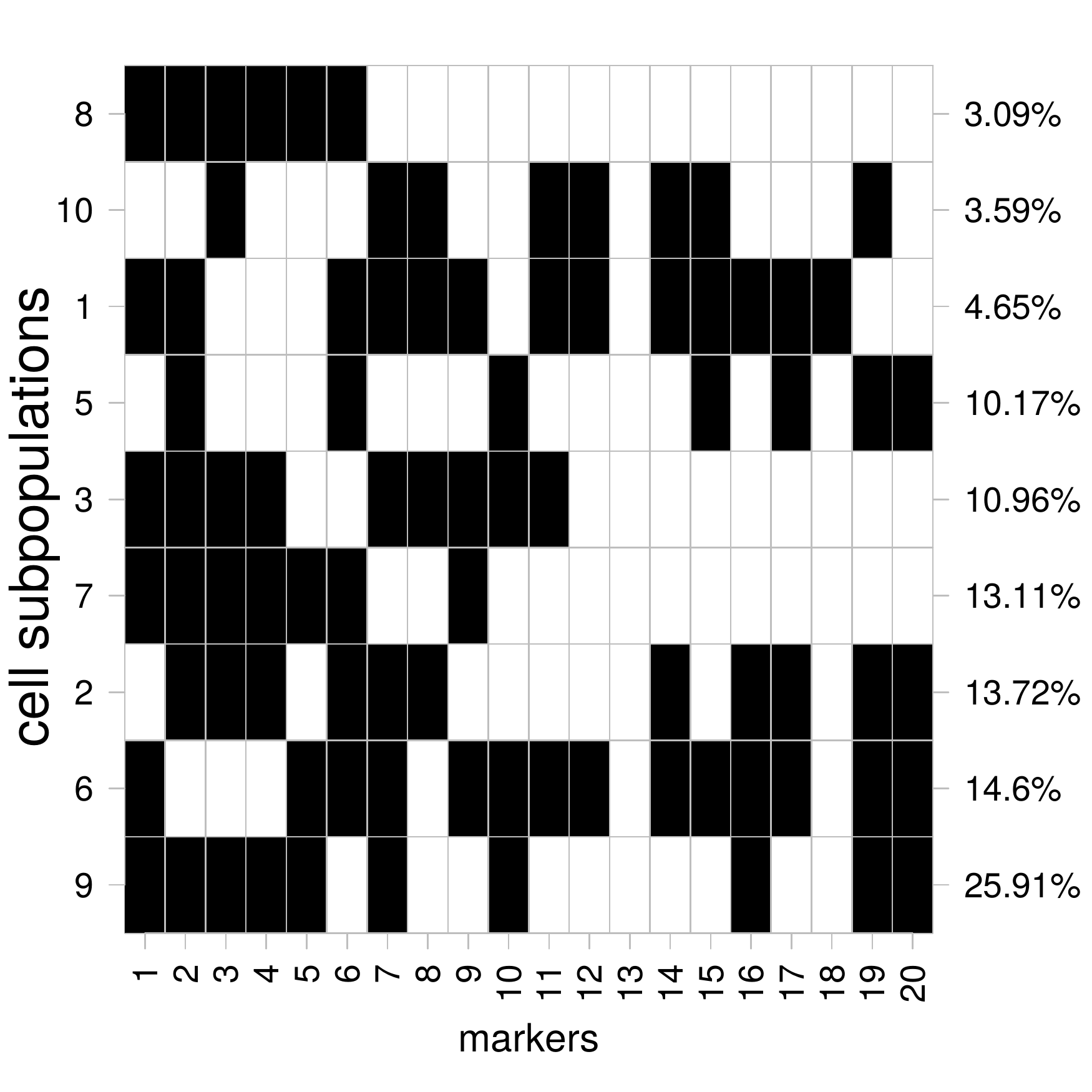} \\
    (d) $\hat{\Z}^\prime_1$ \& $\hat{\bw}_1$ & (e) $\hat{\Z}^\prime_2$ \& $\hat{\bw}_2$ & (c) $\hat{\Z}^\prime_3$ \& $\hat{\bw}_3$ \\
  \end{tabular}
  \caption{Data missingship mechanism sensitivity analysis for Simulation 2.
  Specification II is used for $\bm \beta$. Heatmaps of $\y_i$ are shown in
  (a)-(c) for samples 1-3, respectively. Cells are rearranged by the
  posterior point estimate of cell clustering, $\hat{\lambda}_{i,n}$. Cells
  and markers are in rows and columns, respectively. High and low expression
  levels are in red and blue, respectively, and black is used for missing
  values. Yellow horizontal lines separate cells by different subpopulations.
  $\hat{\Z}^\prime_i$ and $\hat{\bw}_i$ are shown for each of the samples in
  (d)-(f). We include only subpopulations with $\hat{w}_{i,k} > 1\%$.}
  \label{fig:Z-w-sim2-missmechsen-2}
\end{figure}

%%%%%%%%%%%%%%%%%%%%%%%%
\clearpage
% sim2 -flowsom
\begin{figure}[h]
\begin{center}
  \begin{tabular}{ccc}
  \includegraphics[width=0.3\columnwidth]{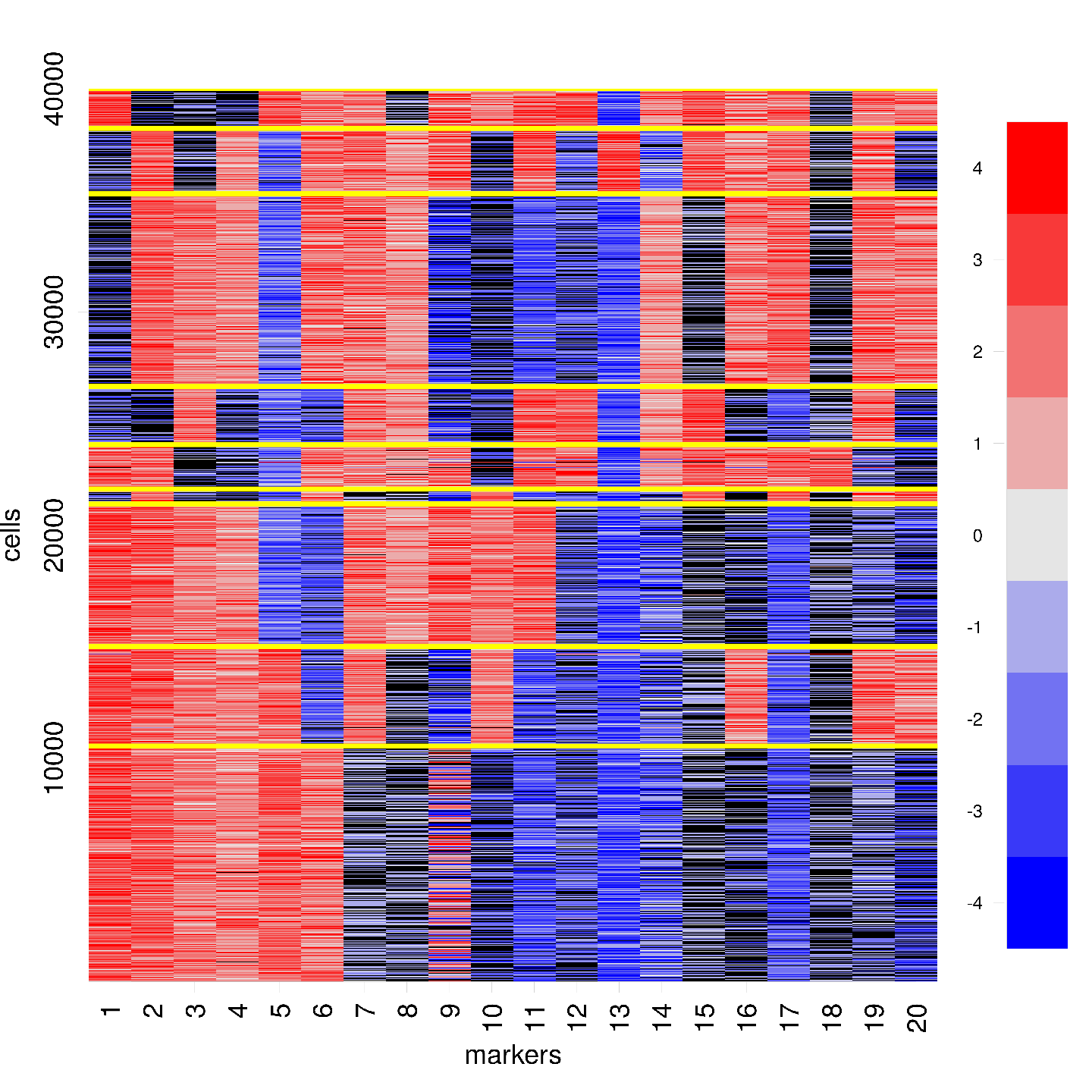}&
  \includegraphics[width=0.3\columnwidth]{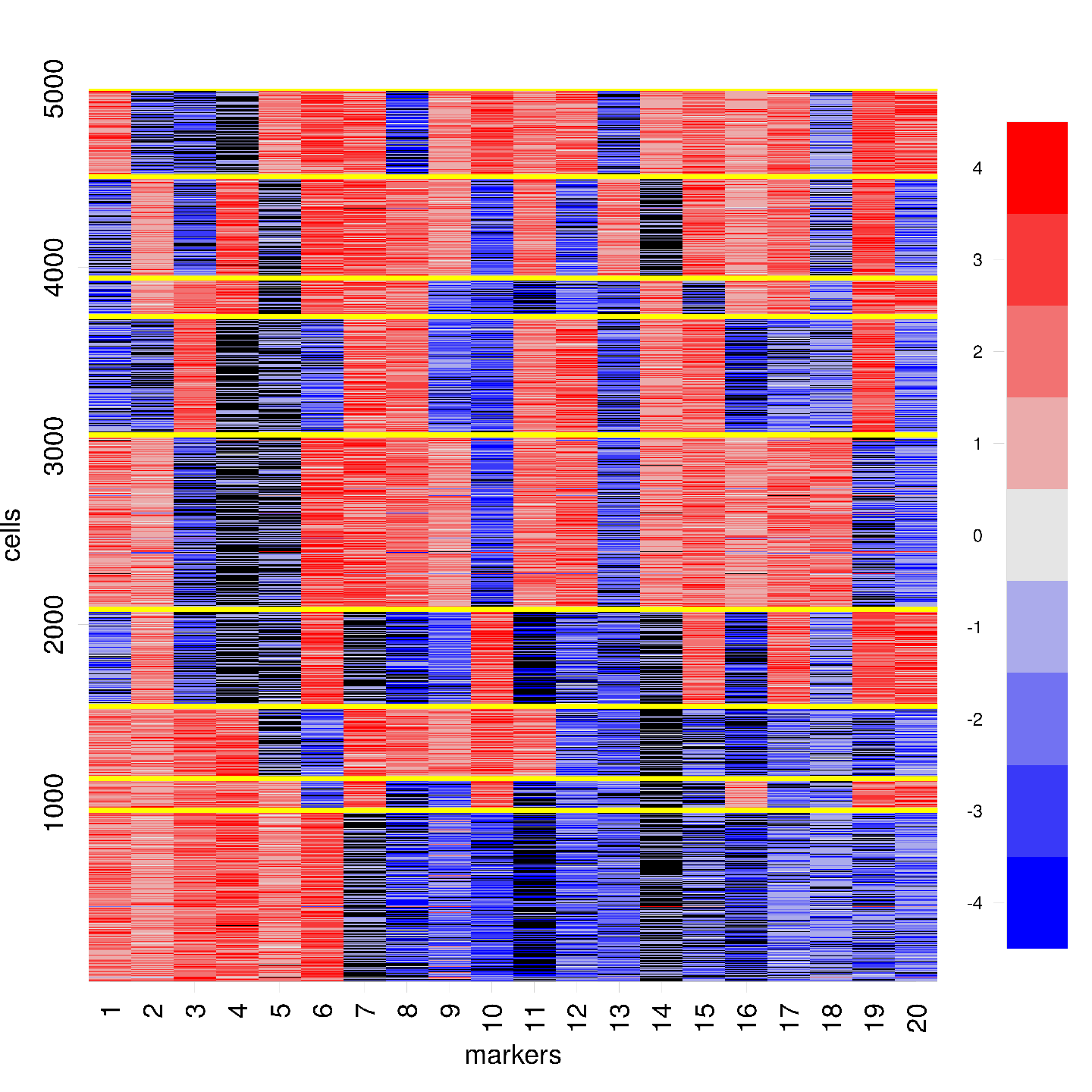}&
  \includegraphics[width=0.3\columnwidth]{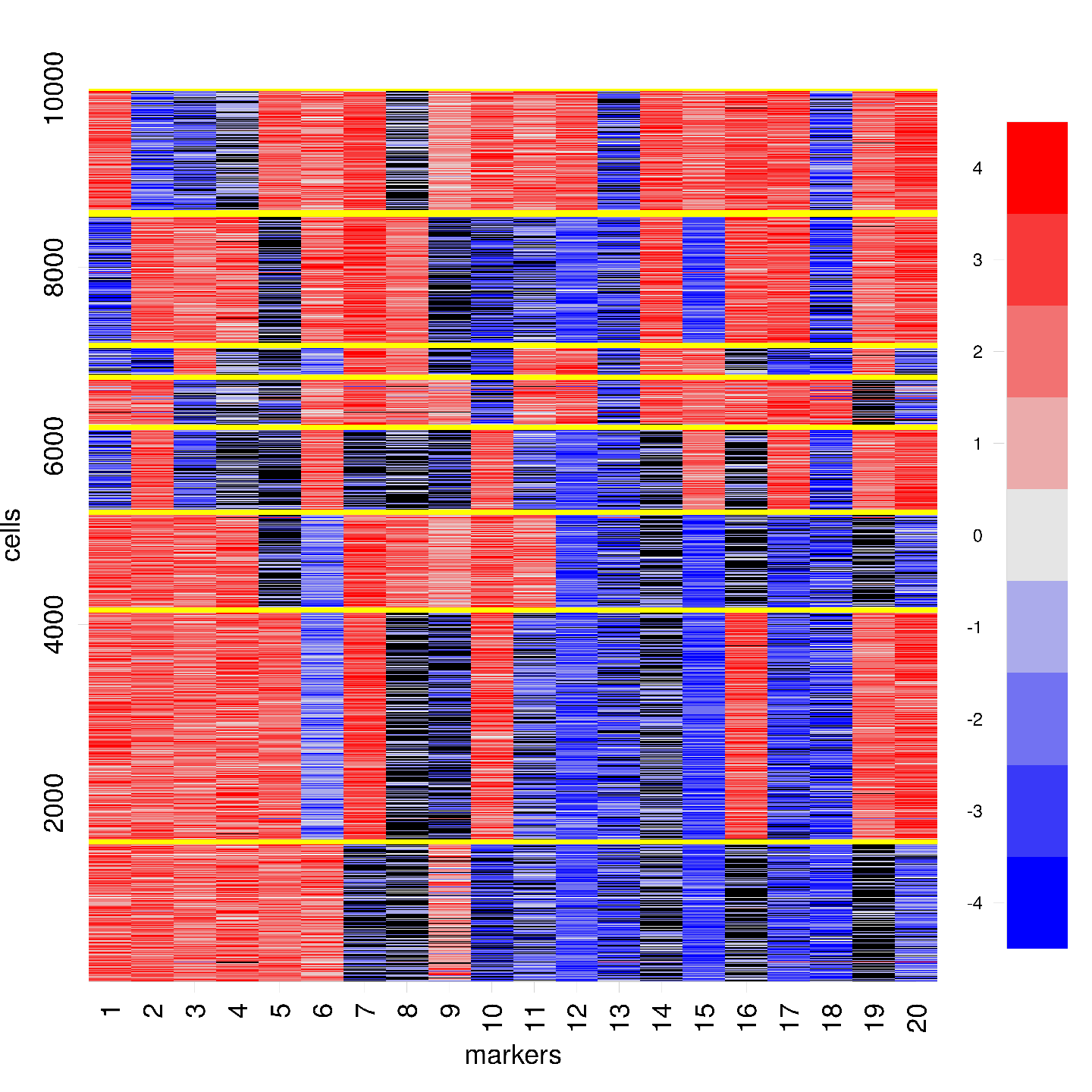}\\
  (a) Sample 1 & (b) Sample 2 & (c) Sample 3\\
  \end{tabular}
  \vspace{-0.05in}
  \caption{\small[FlowSOM for Simulation 2] Heatmaps of $\y_{i}$ for
  Simulation 2. Samples 1-3 are in (a)-(c), respectively. The cells are
  sorted by the cluster labels $\lambda_{i,n}$ for each sample, estimated by
  FlowSOM.} 
  \label{fig:sim2-FlowSOM-Z}
\end{center}
\end{figure}
%%%%%%%%%%%%%%%%%%%%%%%%%

%
%\begin{table}[H]
%  \begin{center}
%  \begin{tabular}{|c|ccc|c|}
%    \hline 
%    Method & Sample 1 ARI & Sample 2 ARI & Sample 3 ARI & Elapsed Time \\
%    \hline 
%      FAM ($K=21$)    & 0.999 & 0.996 & 0.999 & 8 days \\
%      FlowSOM & 0.858 & 0.940 & 0.959 & 20 seconds\\
%    \hline
%  \end{tabular} 
%  \end{center}
%  \caption{ARI for FAM and FlowSOM by sample for Simulation 2. ARI ranges
%  between 0 and 1, with values closer to 1 indicating estimated clusters are closer
%  to the truth.}
%  \label{tab:ari-sim2}
%\end{table}

\clearpage

\begin{table}[h]
      \centering
      \begin{tabular}{|c|c|}
      \hline
      % \textbf{Marker Number} & \textbf{Marker Name} \\
      \textbf{Marker} & \textbf{Marker} \\
      \textbf{Number} & \textbf{Name} \\
      \hline
      1  & 2B4      \\
      2  & KIR2DL3     \\
      3  & KIR3DL1     \\
      4  & CD158B   \\
      \hline
      5  & CD16     \\
      6  & CD27     \\
      7  & CD62L    \\
      8  & CD8      \\
      \hline
      9  & CD94     \\
      10 & DNAM1    \\
      11 & EOMES    \\
      12 & KLRG1    \\
      \hline
      13 & NKG2A    \\
      14 & NKG2C    \\
      15 & NKG2D    \\
      16 & NKP30    \\
      \hline
      17 & SIGLEC7  \\
      18 & TBET     \\
      19 & TIGIT    \\
      20 & ZAP70    \\
      \hline
      \end{tabular}
      \caption{Marker names and numbers for each marker referenced in the CB NK cell data.}
      \label{tab:marker-codes}
\end{table}

%%%%%%%%%%%%%%%%%%%%%%%%
\clearpage
% sim2 -flowsom
\begin{figure}[h]
\begin{center}
  \begin{tabular}{cc}
  \includegraphics[width=0.47\columnwidth]{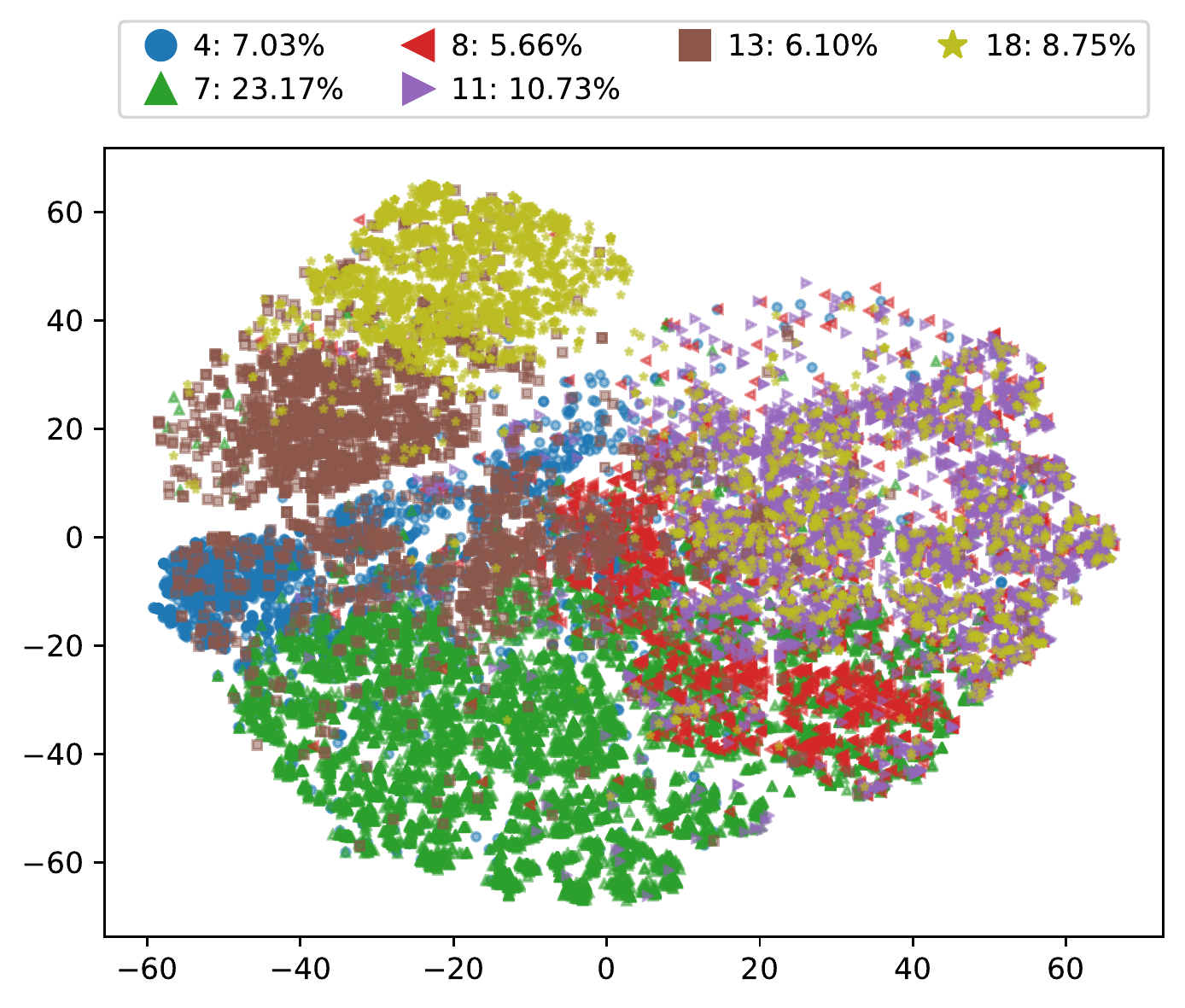}&
  \includegraphics[width=0.47\columnwidth]{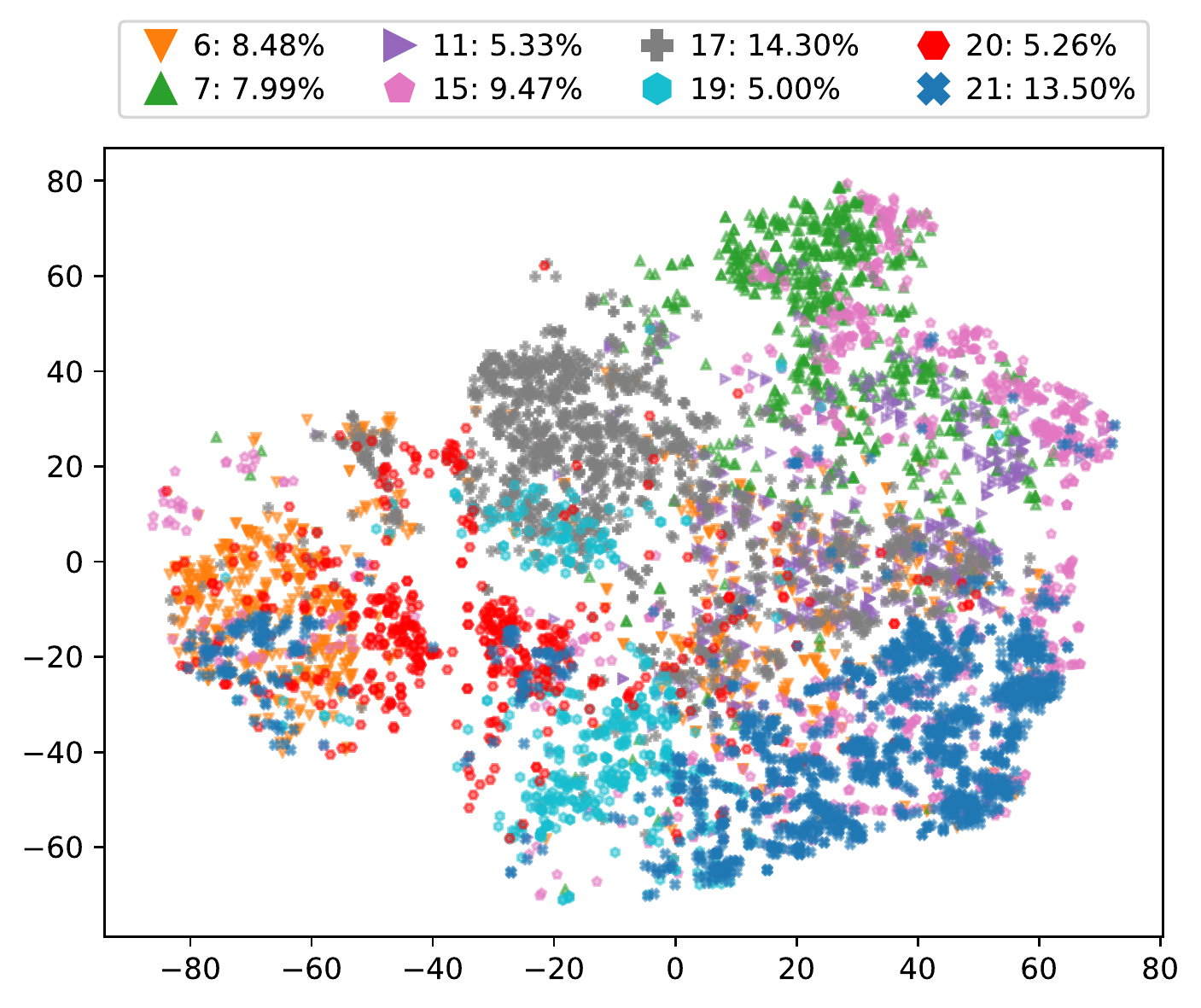}\\
  (a) Sample 1 & (b) Sample 2 \\
  \includegraphics[width=0.47\columnwidth]{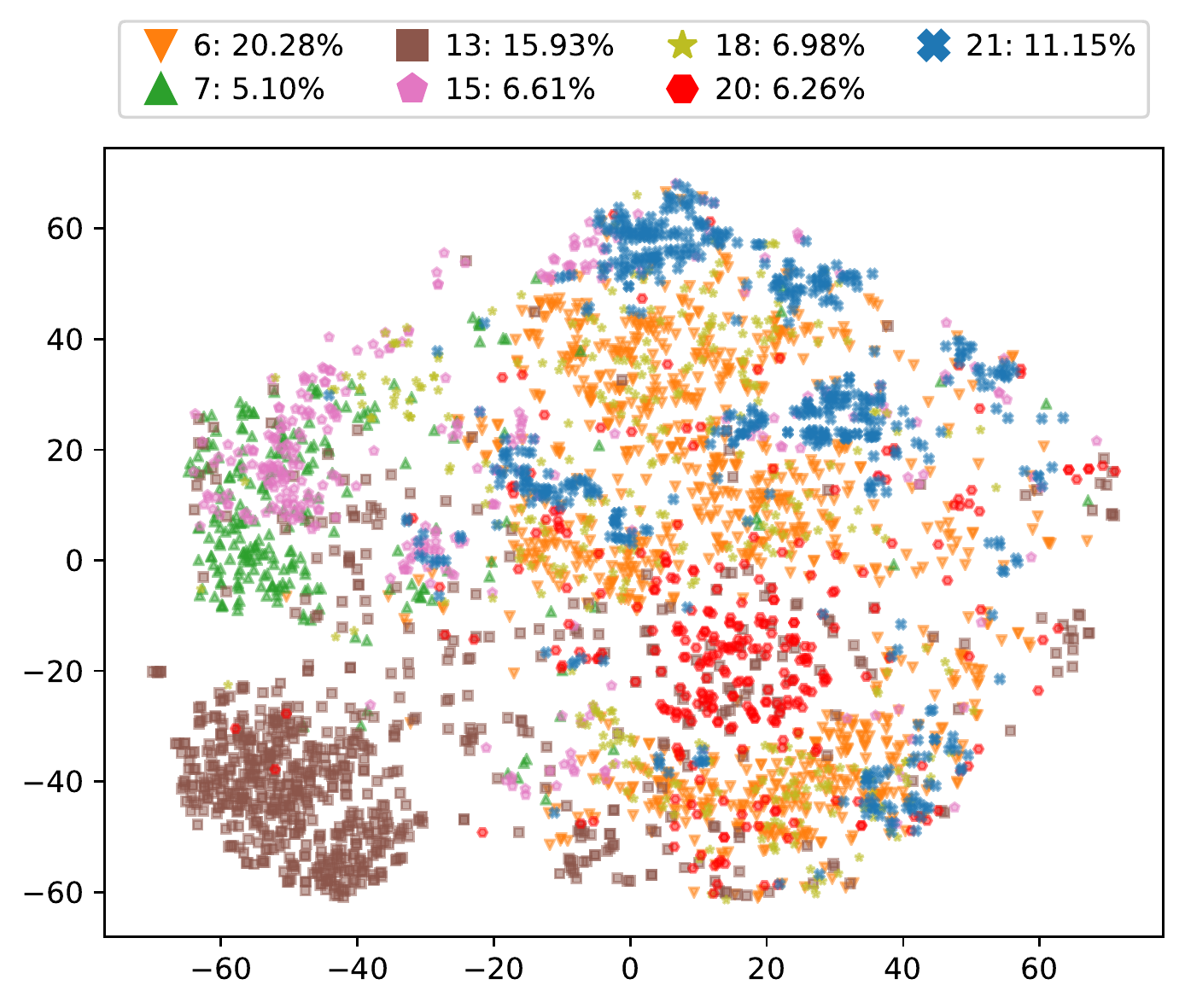}&\\
  (c) Sample 3 & \\
  \end{tabular}
  \vspace{-0.05in}
  \caption{\small[Plots of t-SNE's for the CB data] The CB data is visualized using two-dimensional t-SNE’s that are learned separately on each sample, where each point represents a cell.  Cells in different subpopluations estimated by the FAM are marked by different symbols and colors.  On the top of the scatterplots, the subpopulation numbers are listed with their corresponding symbols and colors.  All cells are used to obtain t-SNE embeddings, but only cell subpopulations belonging to subpopluations with $\hat{w}_{ik} \geq 0.05$ are included in the plots for better illustration.  } 
  \label{fig:CB-tsne}
\end{center}
\end{figure}
%%%%%%%%%%%%%%%%%%%%%%%%%

% latex table generated in R 3.4.4 by xtable 1.8-3 package
% Tue May 28 11:15:26 2019
%\begin{table}[ht]
%  \centering
%  \begin{tabular}{c|rrr}
%    \hline
%            & & Proportions &\\ 
%    Cluster & Sample 1  & Sample 2 & Sample 3 \\ 
%    \hline
    % sorted by abundance in each sample
%    1 & 0.367 & 0.538 & 0.541 \\ 
%    2 & 0.320 & 0.154 & 0.112 \\ 
%    3 & 0.102 & 0.152 & 0.089 \\ 
%    4 & 0.068 & 0.055 & 0.058 \\ 
%    5 & 0.035 & 0.030 & 0.044 \\ 
%    6 & 0.031 & 0.027 & 0.043 \\ 
%    7 & 0.030 & 0.011 & 0.027 \\ 
%    8 & 0.029 & 0.008 & 0.026 \\ 
%    9 & 0.011 & 0.008 & 0.023 \\ 
%    10 & 0.004 & 0.006 & 0.012 \\ 
%    11 & 0.002 & 0.004 & 0.011 \\ 
%    12 & 0.001 & 0.003 & 0.010 \\ 
%    13 & 0.000 & 0.003 & 0.005 \\ 
    % sorted by cluster. Cluster ordering is the same in each sample.
    % 1 & 0.367 & 0.538 & 0.541 \\ 
    % 2 & 0.035 & 0.027 & 0.012 \\ 
    % 3 & 0.029 & 0.055 & 0.044 \\ 
    % 4 & 0.031 & 0.004 & 0.005 \\ 
    % 5 & 0.004 & 0.003 & 0.010 \\ 
    % 6 & 0.001 & 0.008 & 0.089 \\ 
    % 7 & 0.068 & 0.006 & 0.026 \\ 
    % 8 & 0.030 & 0.008 & 0.023 \\ 
    % 9 & 0.102 & 0.030 & 0.027 \\ 
    % 10 & 0.320 & 0.152 & 0.112 \\ 
    % 11 & 0.002 & 0.154 & 0.043 \\ 
    % 12 & 0.011 & 0.003 & 0.011 \\ 
    % 13 & 0.000 & 0.011 & 0.058 \\ 
%    \hline
%  \end{tabular}
%  \caption[CB FlowSOM W]{Proportions of cells in each cluster (rows) estimated by FlowSOM for CB NK cell data.}
%  \label{tab:cb-flowsom}
%\end{table}

\clearpage
\begin{table}[h]
  \centering
  \begin{tabular}{|c|cccc|}
    \hline
    Data Missingship & $\tilde{\bm q}$ & Probability of Missing $(\bm\rho)$ & LPML & DIC \\
    Mechanism &  &  & & \\
    \hline
    0   & (0\%, 25\%, 50\%) & (5\%, 80\%, 5\%) & -24.90 & 2569097 \\
    I   & (0\%, 20\%, 40\%) & (5\%, 80\%, 5\%) & -24.93 & 2569098 \\
    II  & (0\%, 15\%, 30\%) & (5\%, 80\%, 5\%) & -24.98 & 2569098 \\
    \hline
  \end{tabular}
  \caption[Different data missingship mechanisms in UCB NK cell data
  analysis]{
$\tilde{\bm q}$-quantiles of the negative observed
  values in each sample are used to specify $\tilde{\bm y}$, and $\bm\rho$ are the probability of missing at $\tilde{\bm y}$.  Three different sets of $\tilde{\bm q}$ and $\tilde{\bm \rho}$ are used to examine the sensitivity to the missingship mechanism specification. LPML and DIC are shown in the last two columns under each of the specification.
  }
  \label{tab:missmechsen-cb}
\end{table}

%%% CB NK cell data: beta values
\begin{table}[h]
  \centering
  \begin{tabular}{|c|c|rrr|}
    \hline
    Data Missingship Mechanism & $\beta$ & Sample 1 & Sample 2 & Sample 3 \\
    \hline
    \hline
    0 & $\beta_0$ & -15.35 & -15.73 & -13.66 \\
     & $\beta_1$ & -10.39 & -10.20 &  -9.60 \\
     & $\beta_2$ &  -1.38 &  -1.34 &  -1.30 \\
    \hline
    \hline
    I & $\beta_0$ & -20.40 & -21.50 & -18.21 \\
     & $\beta_1$ & -12.60 & -12.76 & -11.62 \\
     & $\beta_2$ &  -1.61 &  -1.61 &  -1.51 \\
    \hline
    \hline
    II & $\beta_0$ & -27.43 & -29.21 & -25.26 \\
     & $\beta_1$ & -15.52 & -15.86 & -14.62 \\
     & $\beta_2$ &  -1.90 &  -1.91 &  -1.81 \\
    \hline
  \end{tabular}
  \caption{Values for $\beta$ used for the sensitivity analysis to the missinghsip mechanism in CB NK cell data analysis.}
  \label{tab:missmechsen-cb-beta}
\end{table}

%% CB-sens I
\begin{figure}[t]
  \centering
  \begin{tabular}{ccc}
    \includegraphics[width=.3\columnwidth]{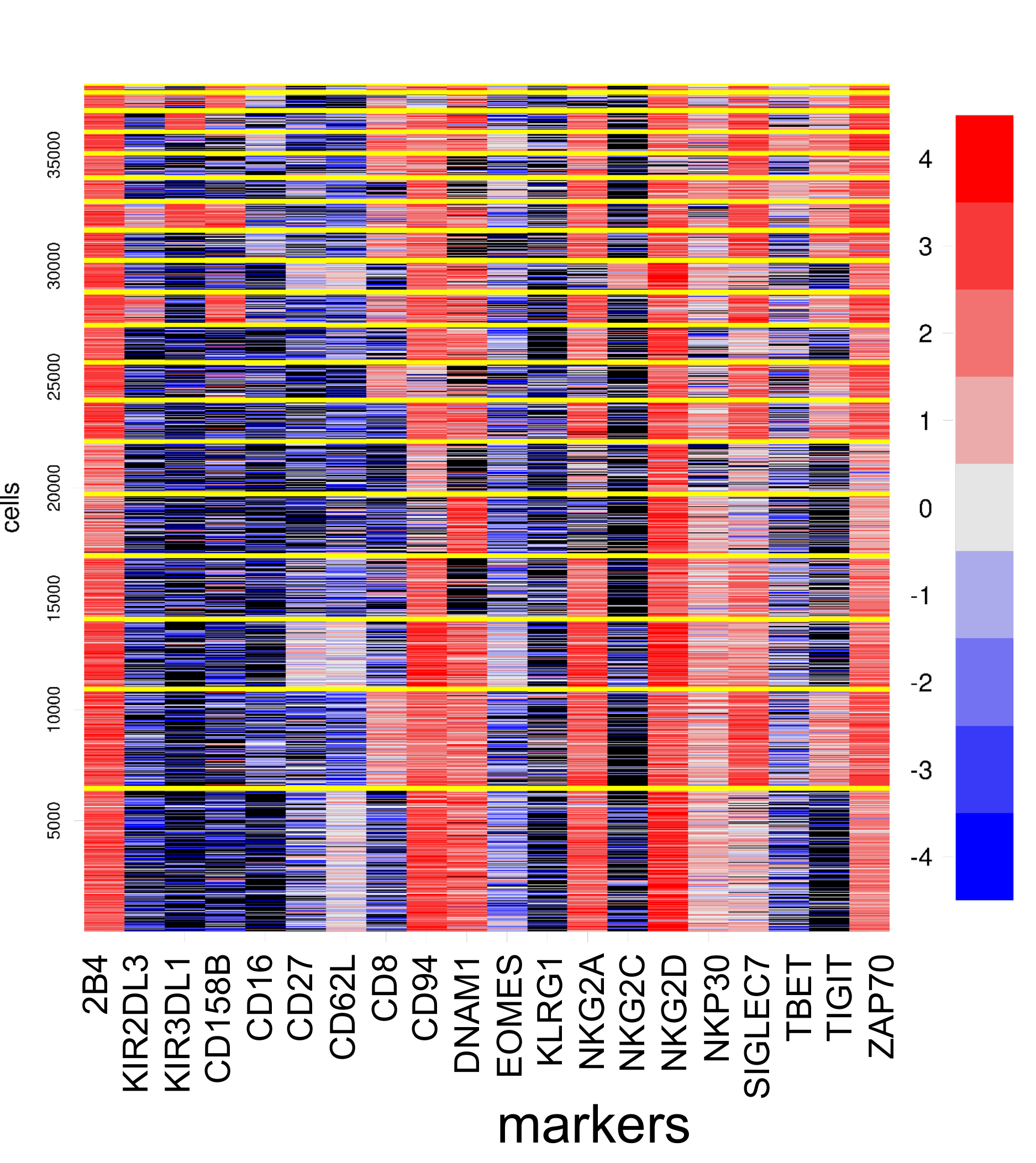} &
    \includegraphics[width=.3\columnwidth]{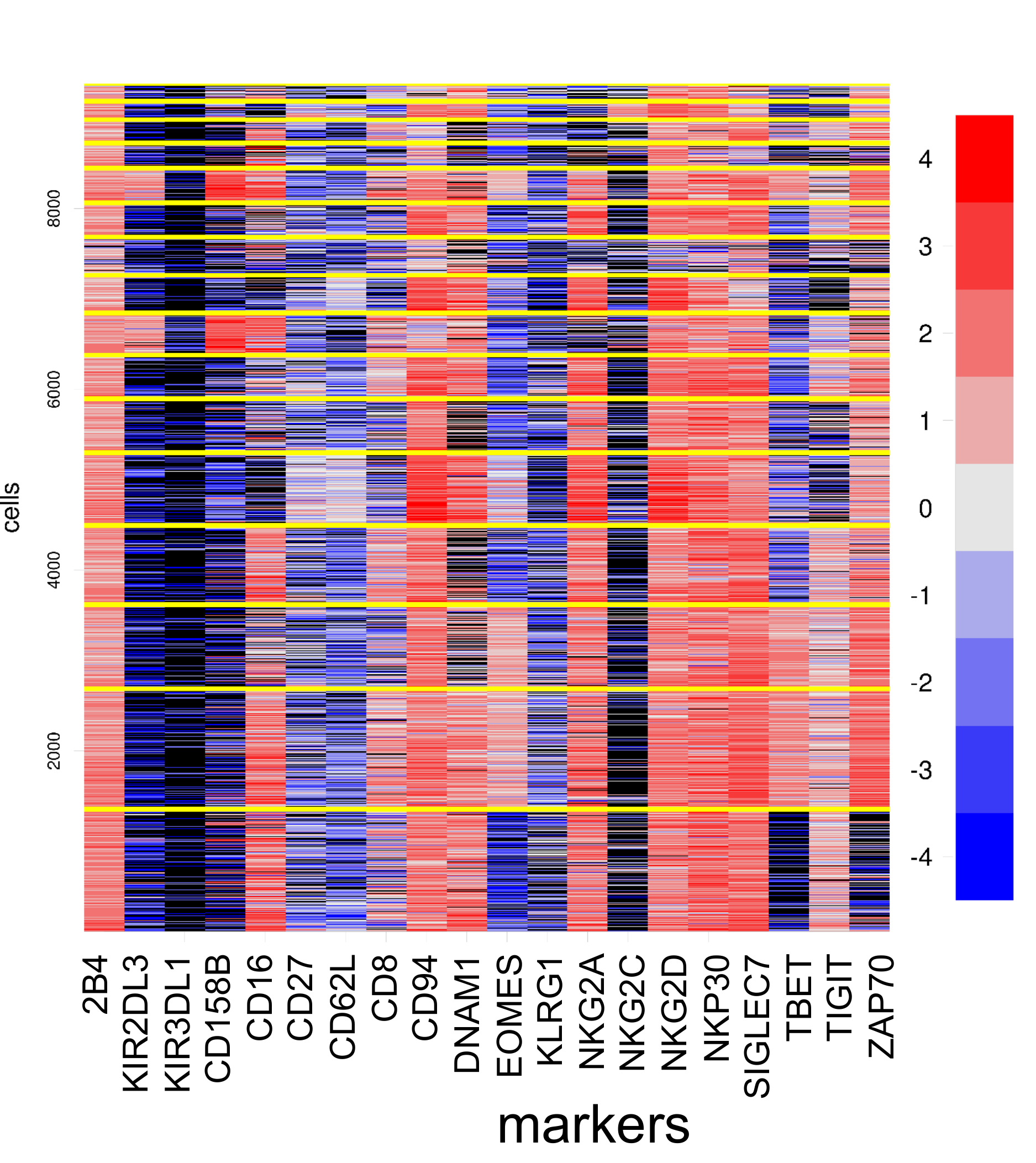} &
    \includegraphics[width=.3\columnwidth]{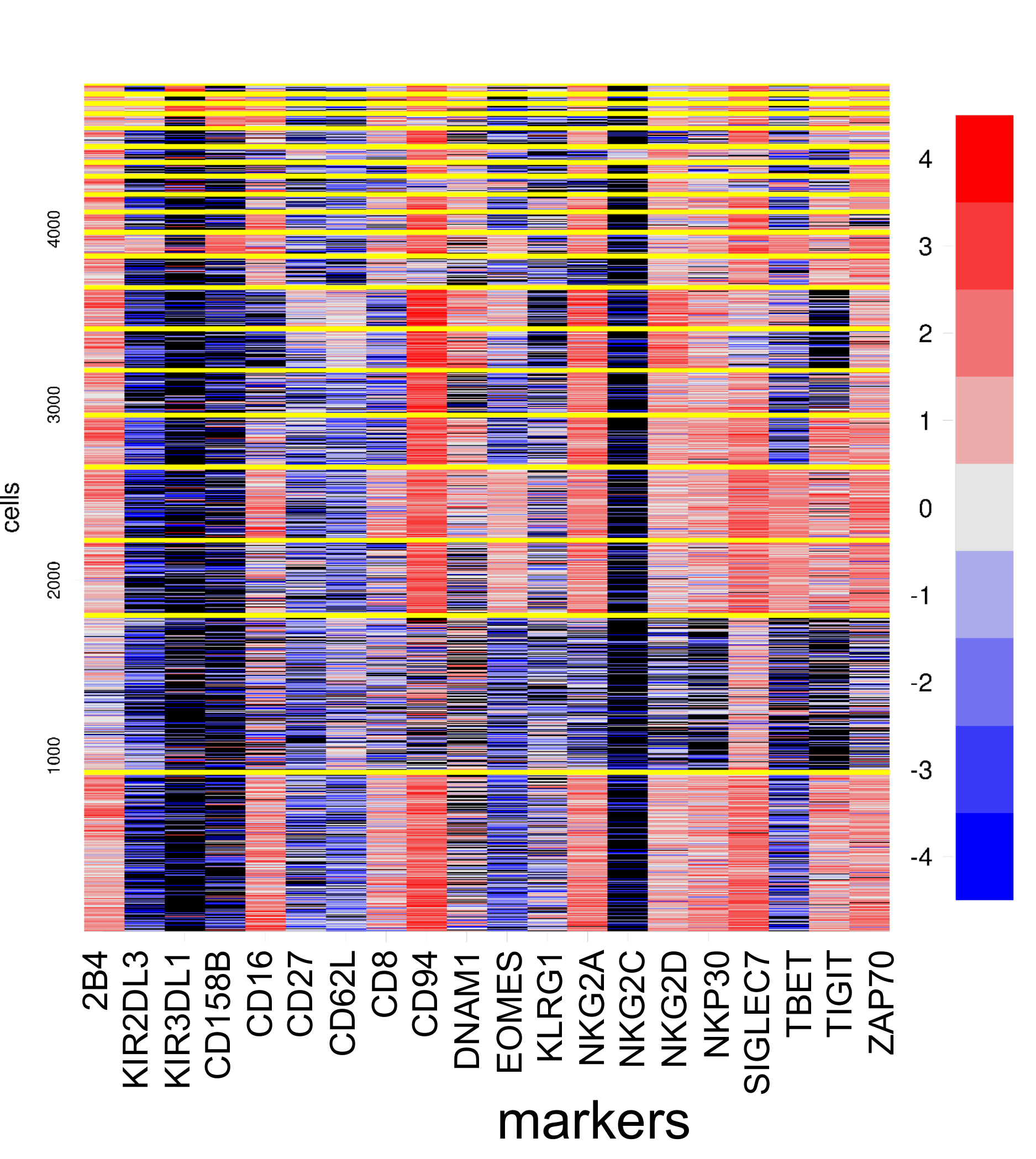} \\
  	(a) heatmap of $y_{1nj}$ & (b) heatmap of $y_{2nj}$ & (c) heatmap of $y_{3nj}$\\    
    \includegraphics[width=.3\columnwidth]{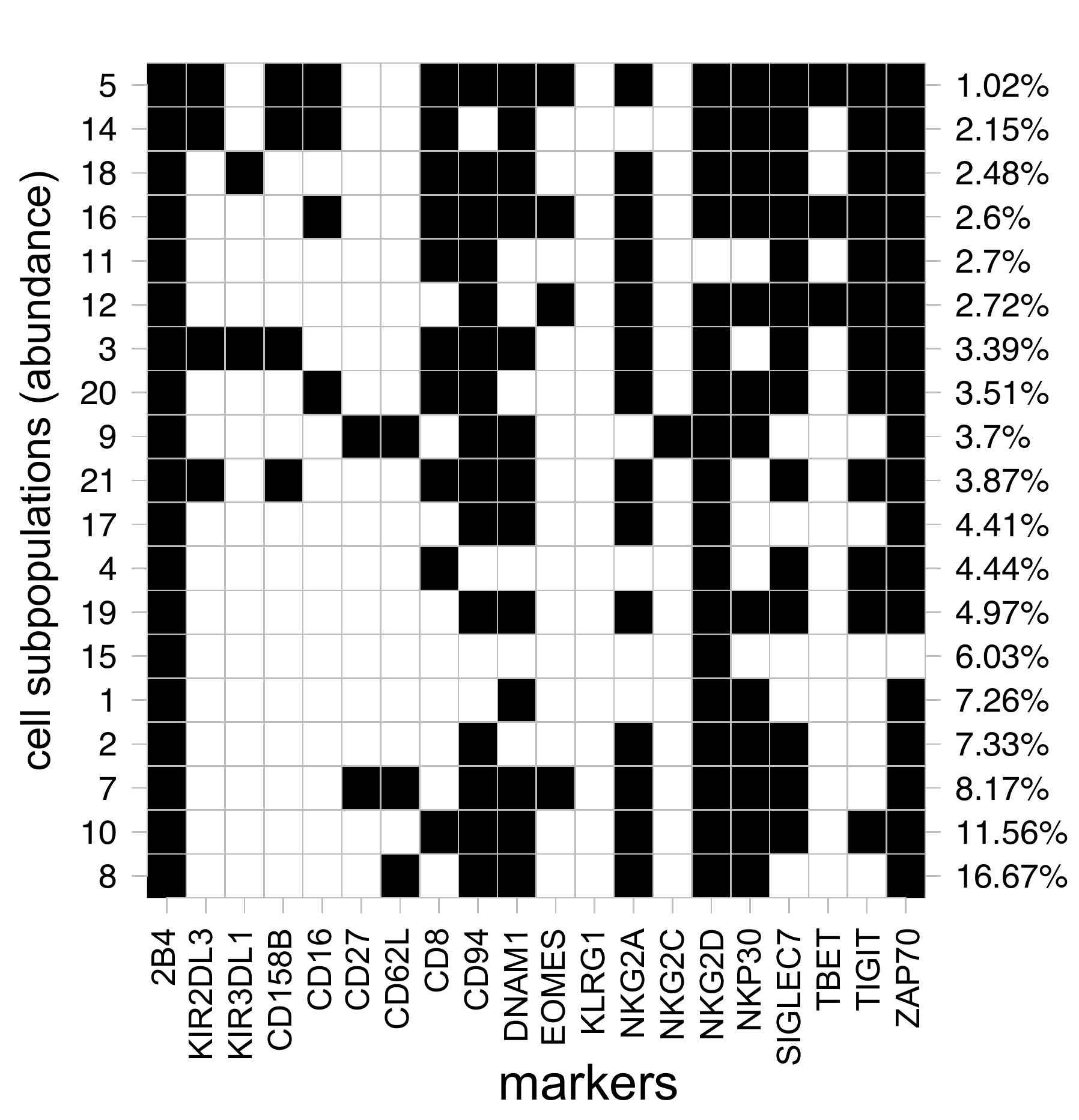} &
    \includegraphics[width=.3\columnwidth]{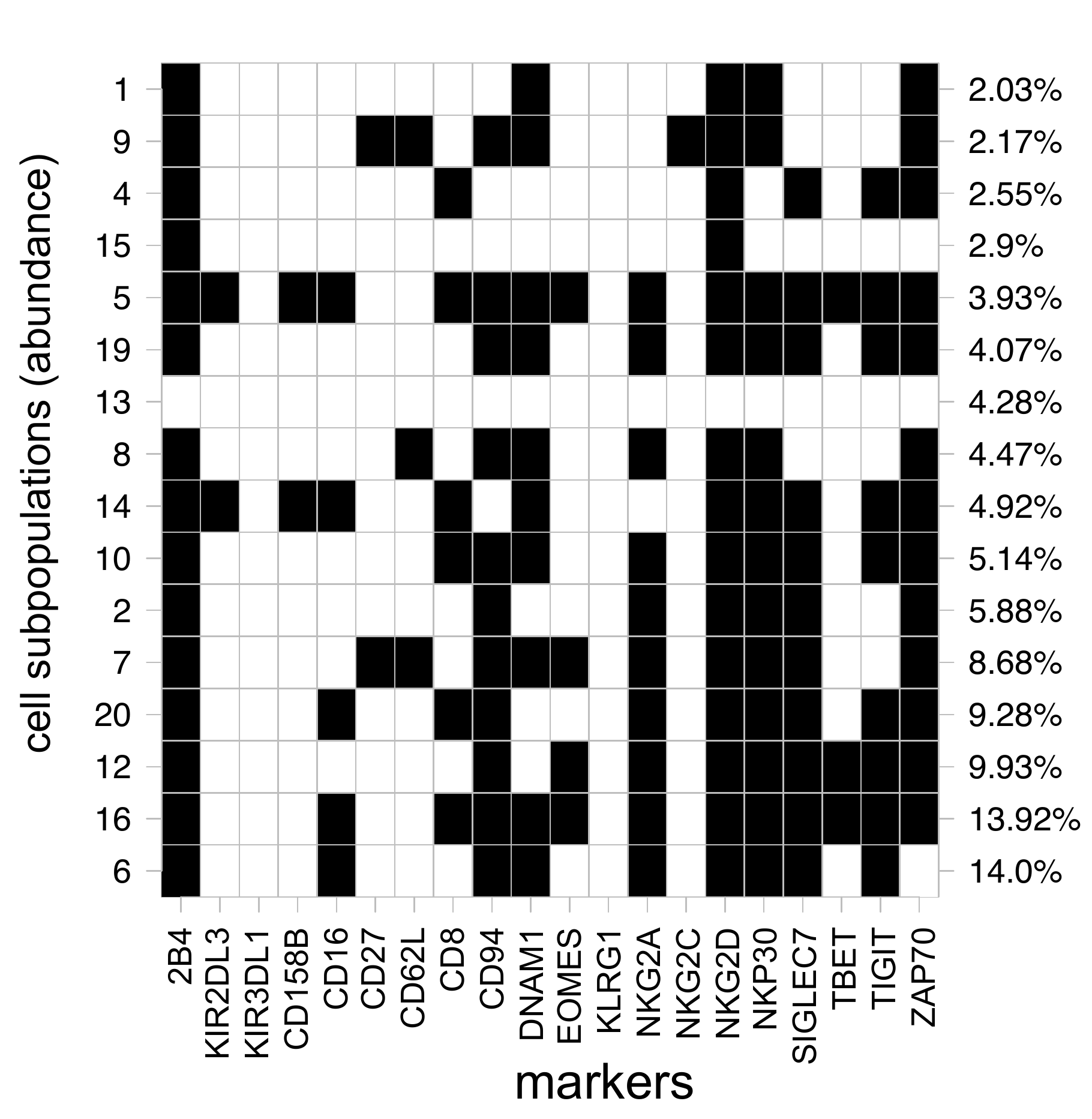} &
    \includegraphics[width=.3\columnwidth]{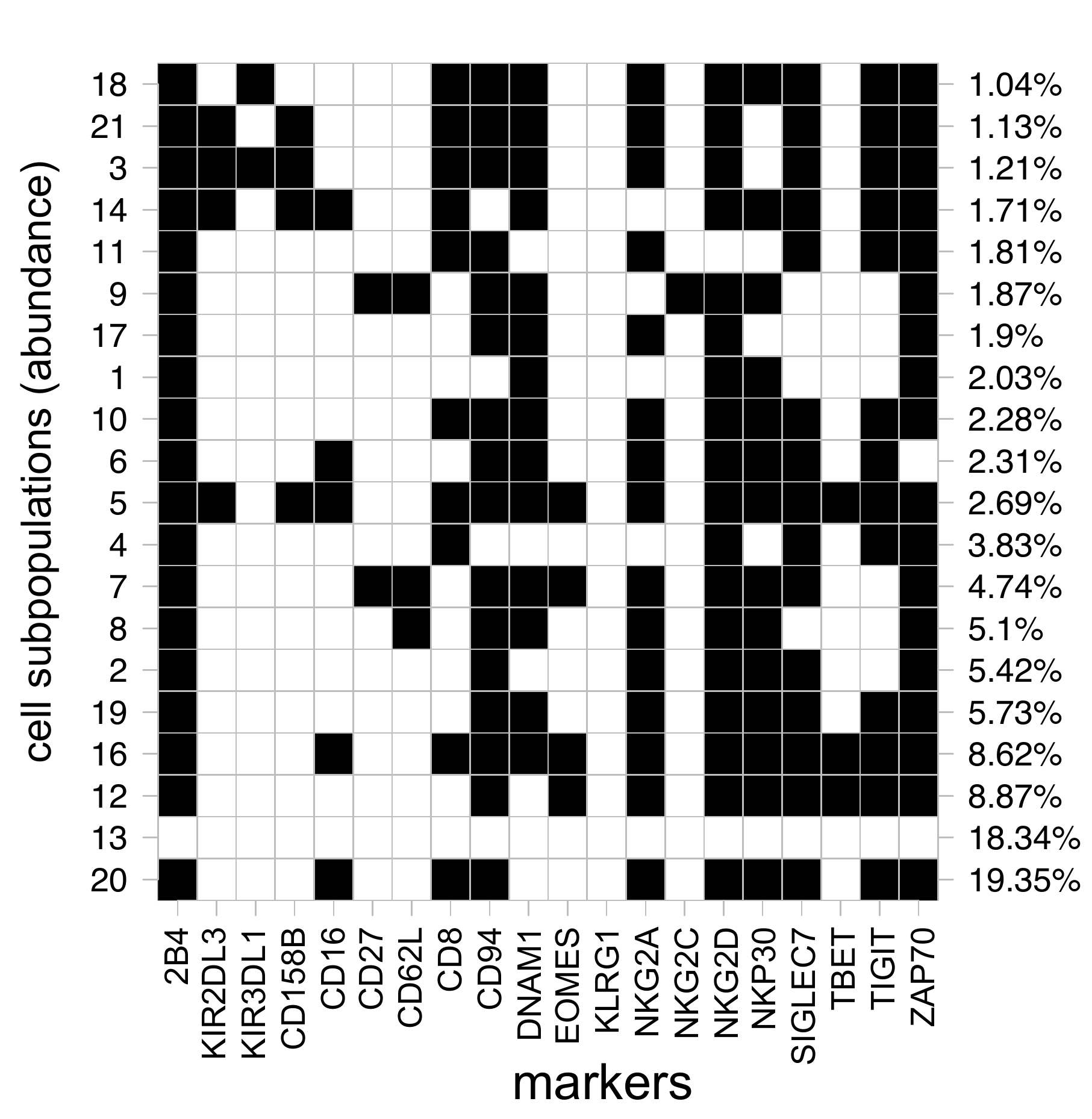} \\
    (d) $\hat{\Z}^\prime_1$ \& $\hat{\bw}_1$ & (e) $\hat{\Z}^\prime_2$ \& $\hat{\bw}_2$ & (c) $\hat{\Z}^\prime_3$ \& $\hat{\bw}_3$ \\
  \end{tabular}
  \caption{Data missingship mechanism sensitivity analysis for CB NK cell data
  analysis. Specification I is used for $\bm \beta$. Heatmaps
  of $\y_u$ are shown in (a)-(c) for samples 1-3, respectively. Cells are
  rearranged by the posterior point estimate of the cell clusterings
  $\hat{\lambda}_{i,n}$. Cells and markers are in rows and columns,
  respectively. High and low expression levels are in red and blue,
  respectively, and black is used for missing values. Yellow horizontal lines
  separate cells by different subpopulations. $\hat{\Z}^\prime_i$
  and $\hat{\bw}_i$ are shown for each of the samples in
  (d)-(f). We include only subpopulations with $\hat{w}_{i,k} > 1\%$.}
  \label{fig:Z-w-CB-missmechsen-1}
\end{figure}

%% CB-sens II
\begin{figure}[t]
  \centering
  \begin{tabular}{ccc}
    \includegraphics[width=.3\columnwidth]{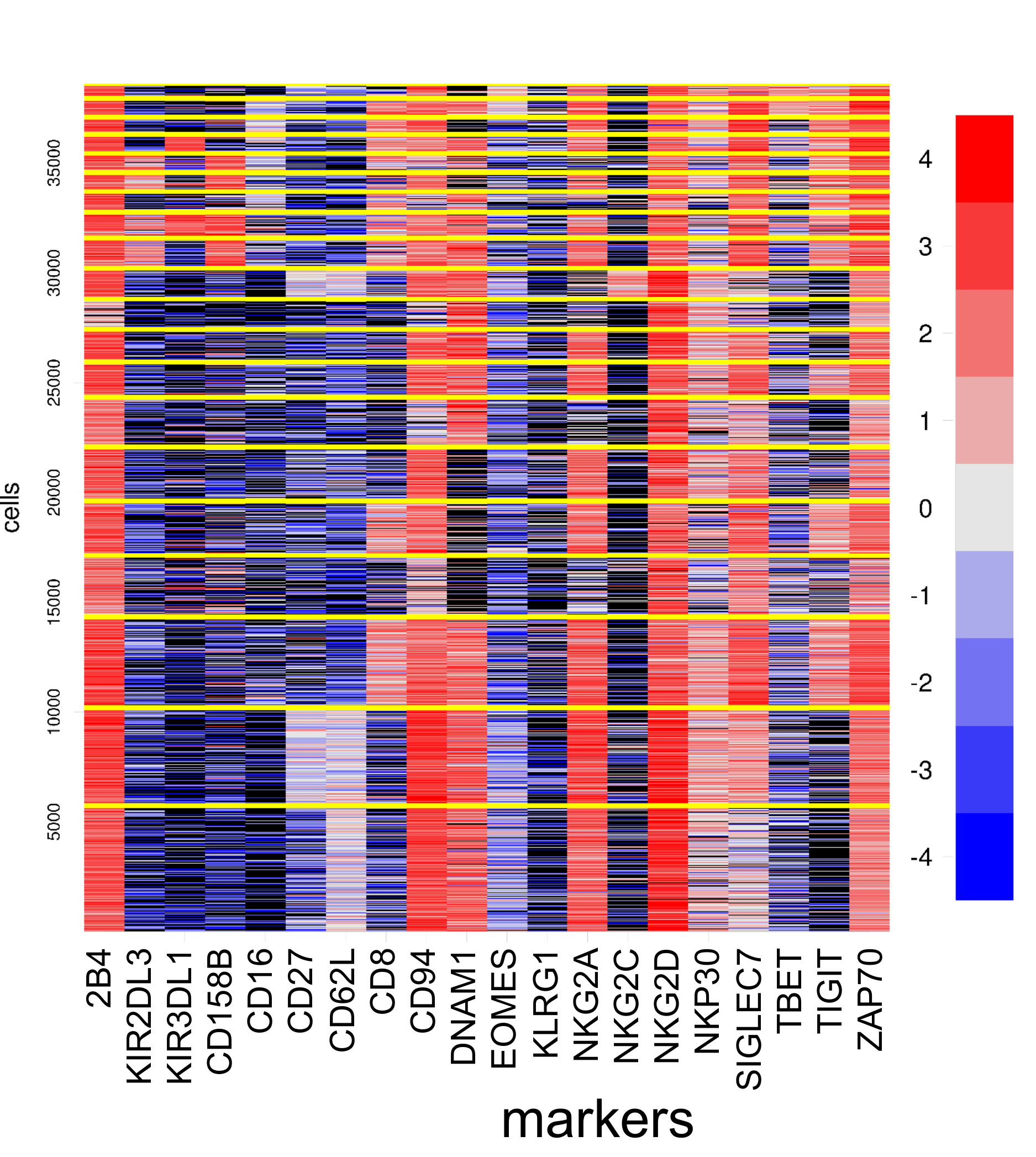} &
    \includegraphics[width=.3\columnwidth]{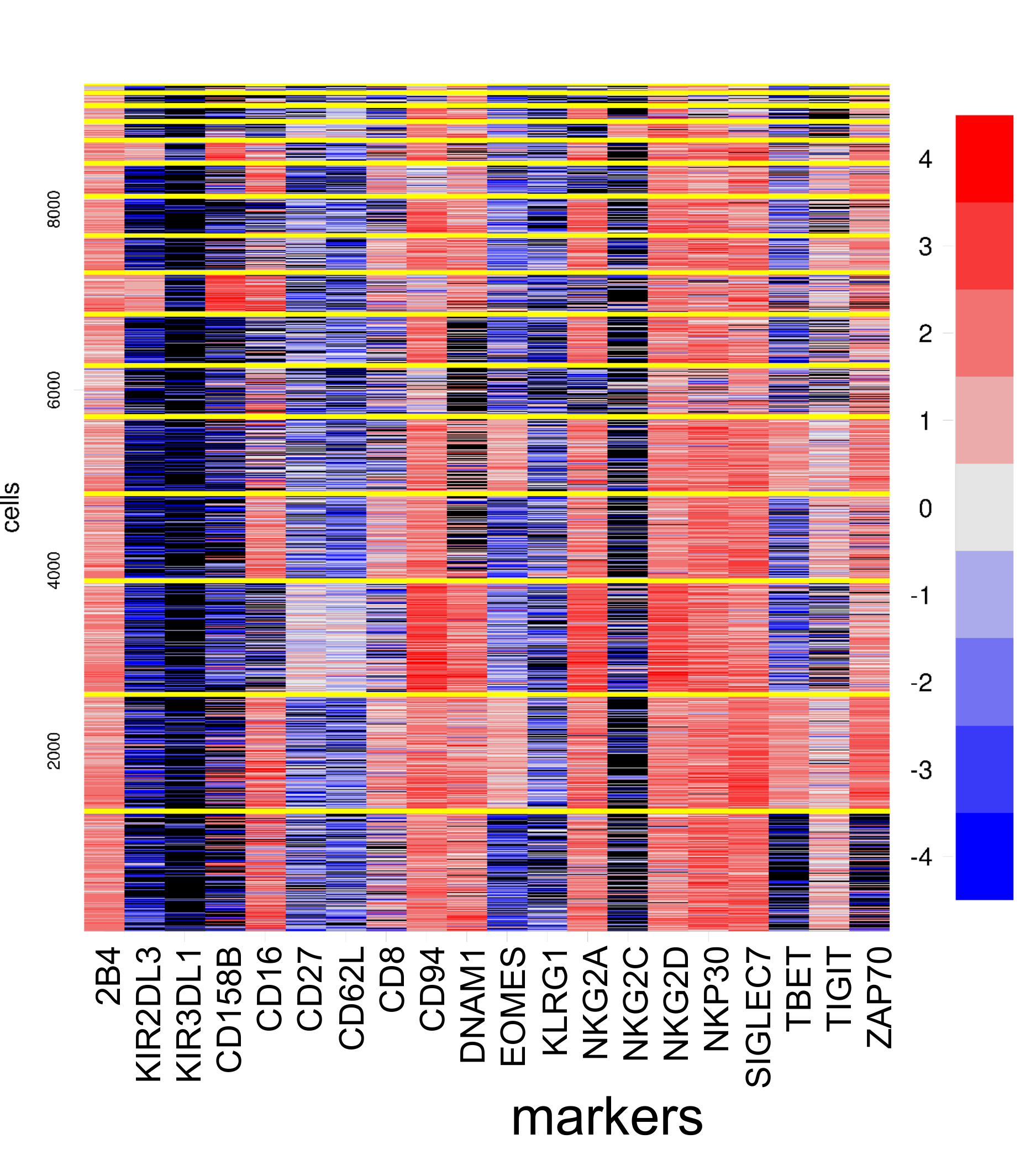} &
    \includegraphics[width=.3\columnwidth]{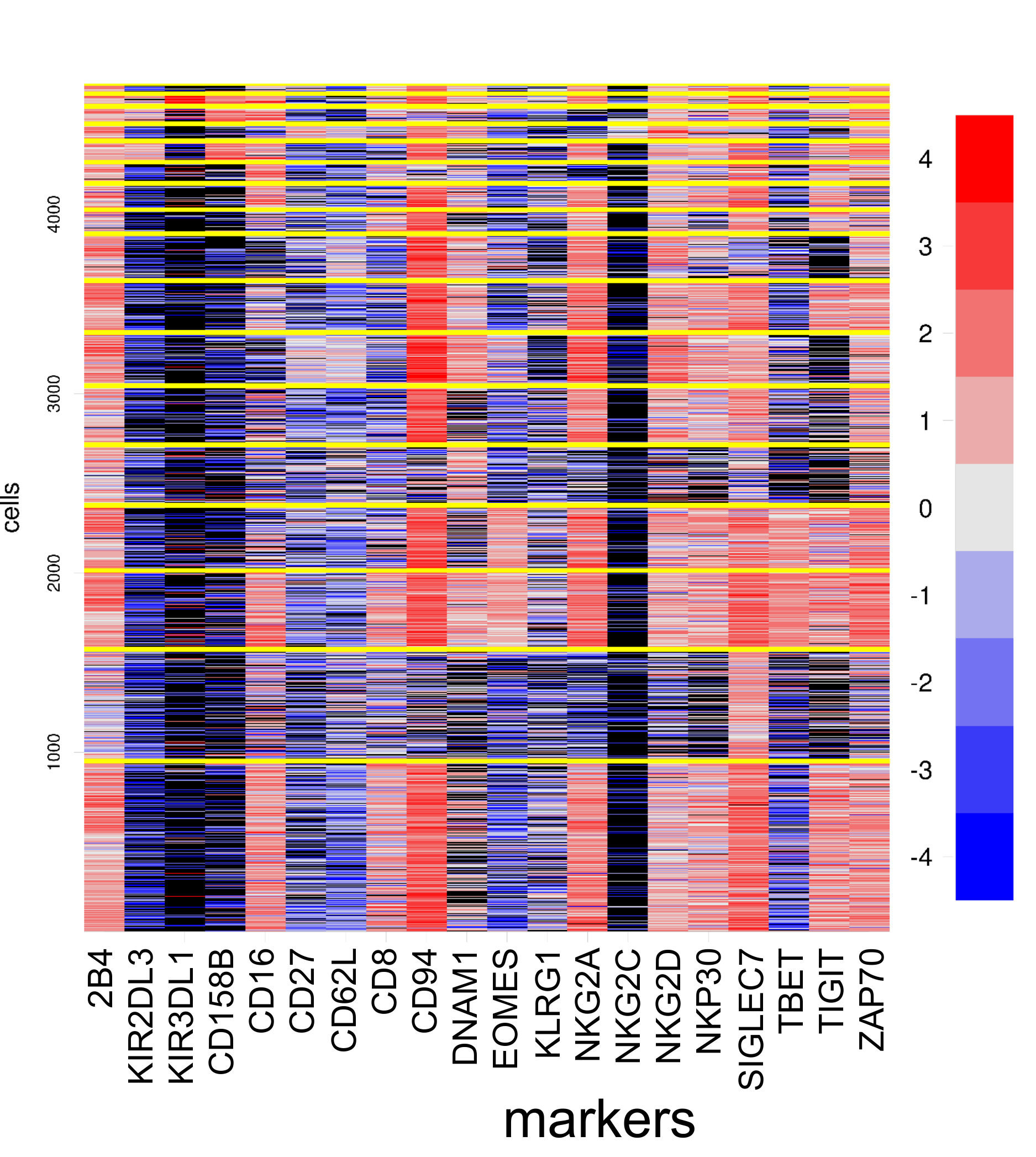} \\
	  (a) heatmap of $y_{1nj}$ & (b) heatmap of $y_{2nj}$ & (c) heatmap of $y_{3nj}$\\    
    \includegraphics[width=.3\columnwidth]{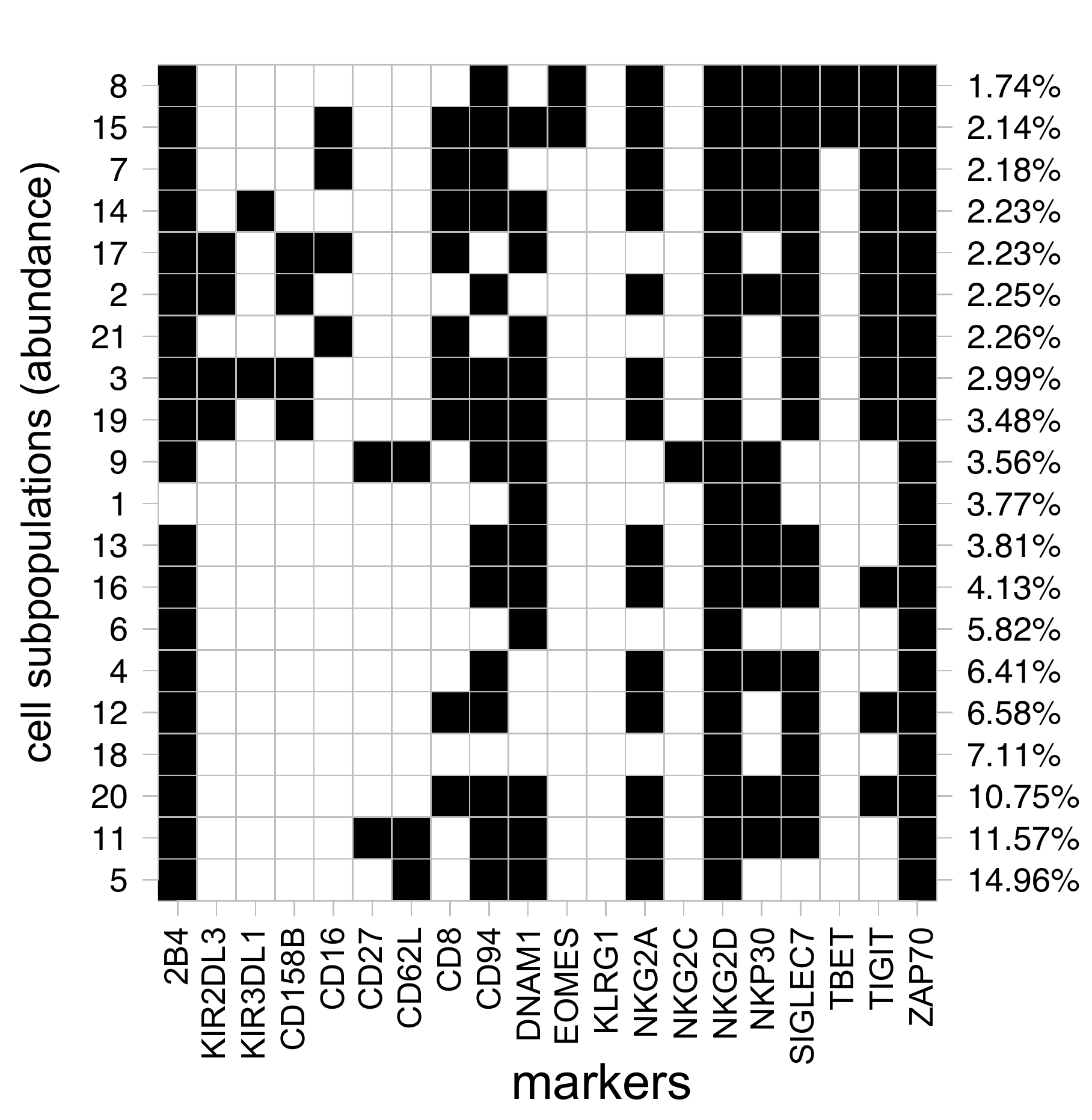} &
    \includegraphics[width=.3\columnwidth]{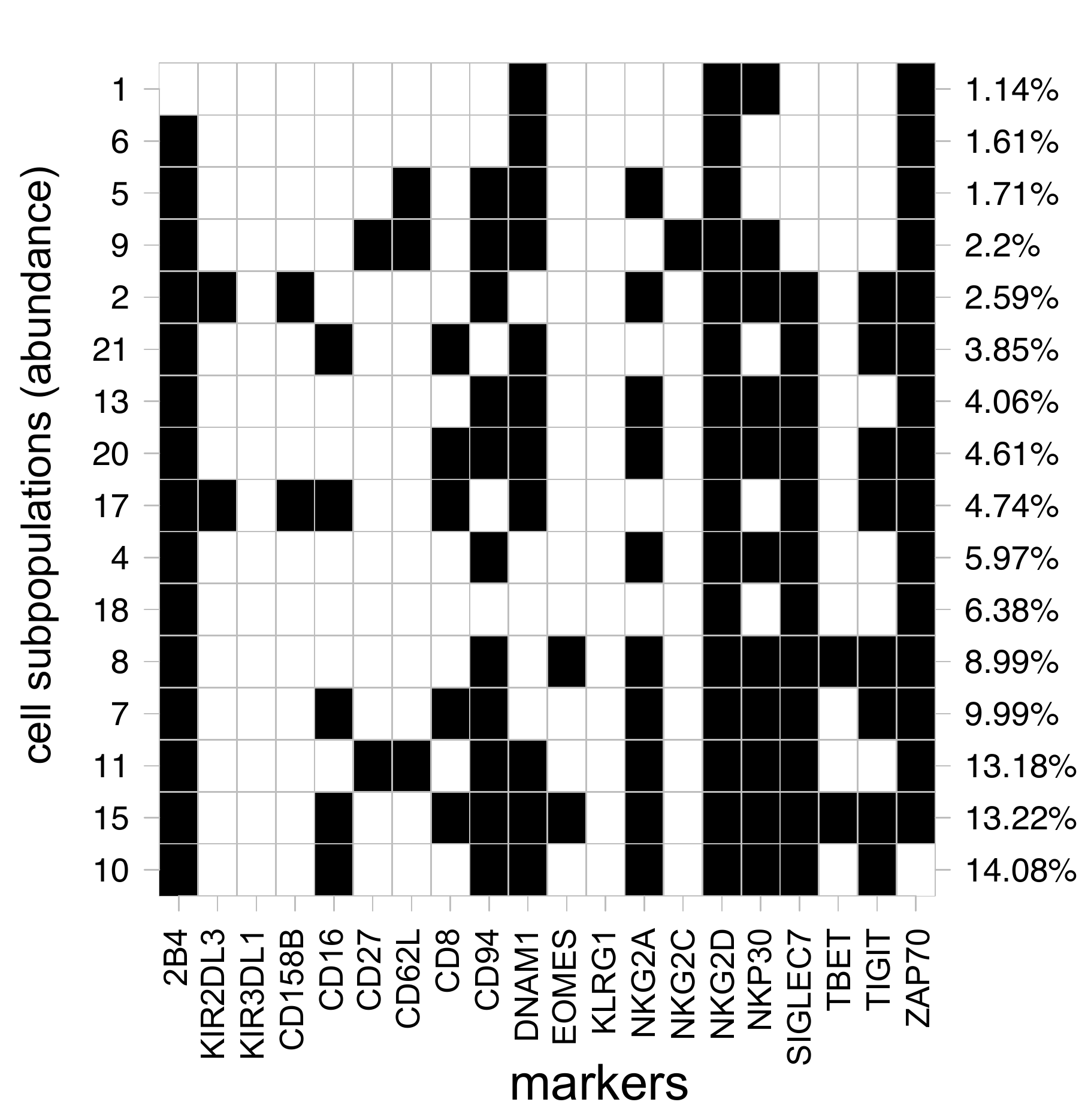} &
    \includegraphics[width=.3\columnwidth]{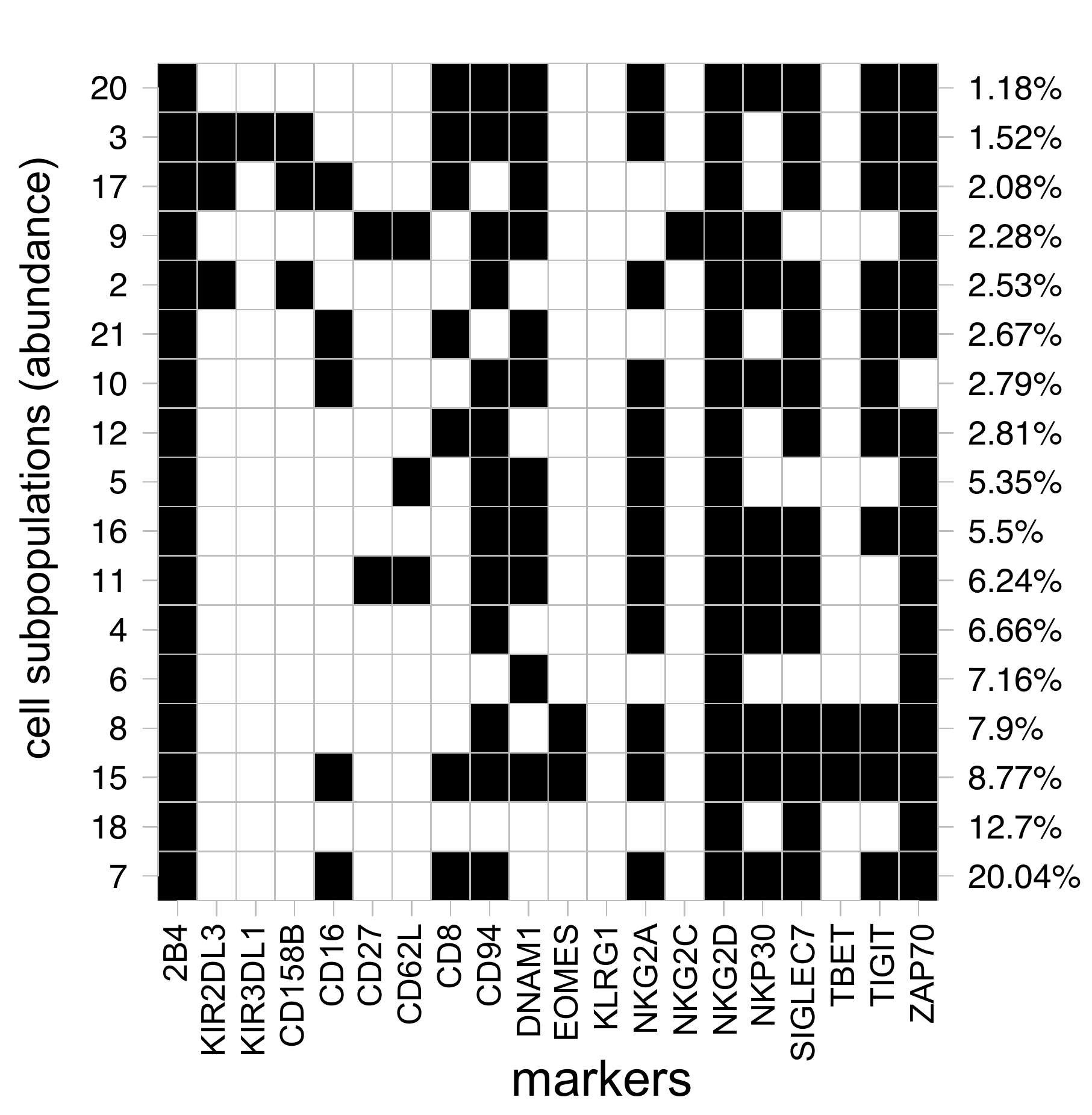} \\
    (d) $\hat{\Z}^\prime_1$ \& $\hat{\bw}_1$ & (e) $\hat{\Z}^\prime_2$ \& $\hat{\bw}_2$ & (c) $\hat{\Z}^\prime_3$ \& $\hat{\bw}_3$ \\
  \end{tabular}
  \caption{Data missingship mechanism sensitivity analysis for CB NK cell data
  analysis. Specification II is used for $\bm \beta$. Heatmaps of $\y_i$ are
  shown in (a)-(c) for samples 1-3, respectively. Cells are rearranged by the
  posterior point estimate of the cell clusterings $\hat{\lambda}_{i,n}$.
  Cells and markers are in rows and columns, respectively. High and low
  expression levels are in red and blue, respectively, and black is used for
  missing values. Yellow horizontal lines separate cells by different
  subpopulations. $\hat{\Z}^\prime_i$ and $\hat{\bw}_i$ are shown for each of the
  samples in (d)-(f). We include only subpopulations with $\hat{w}_{i,k} > 1\%$.}
  \label{fig:Z-w-CB-missmechsen-2}
\end{figure}

%%%%%%%%%%%%%%%%%%%%%%%%
\begin{figure}[t!]
  \begin{center}
  \begin{tabular}{cc}
  \includegraphics[width=0.5\columnwidth]{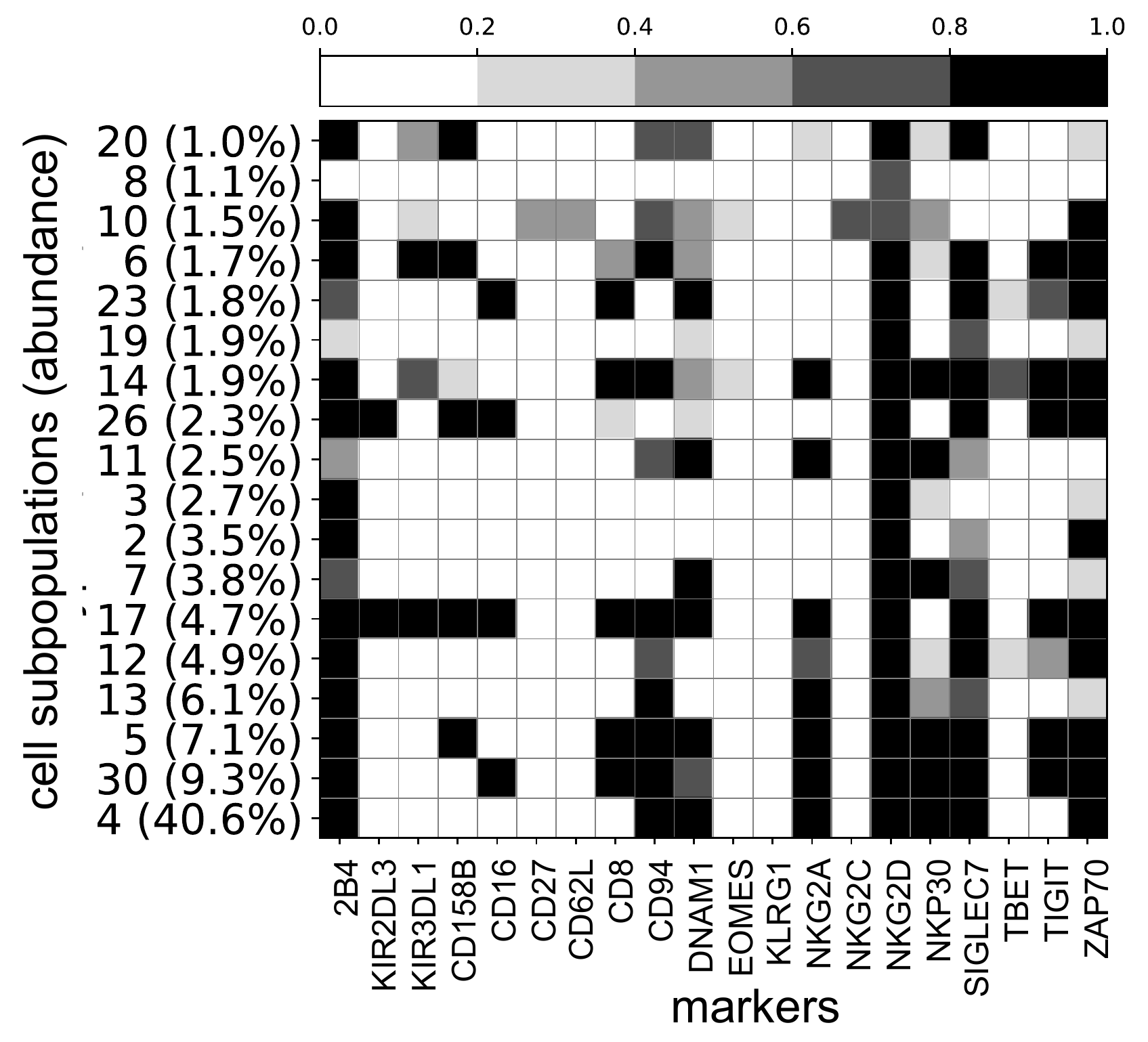} &
  \includegraphics[width=0.5\columnwidth]{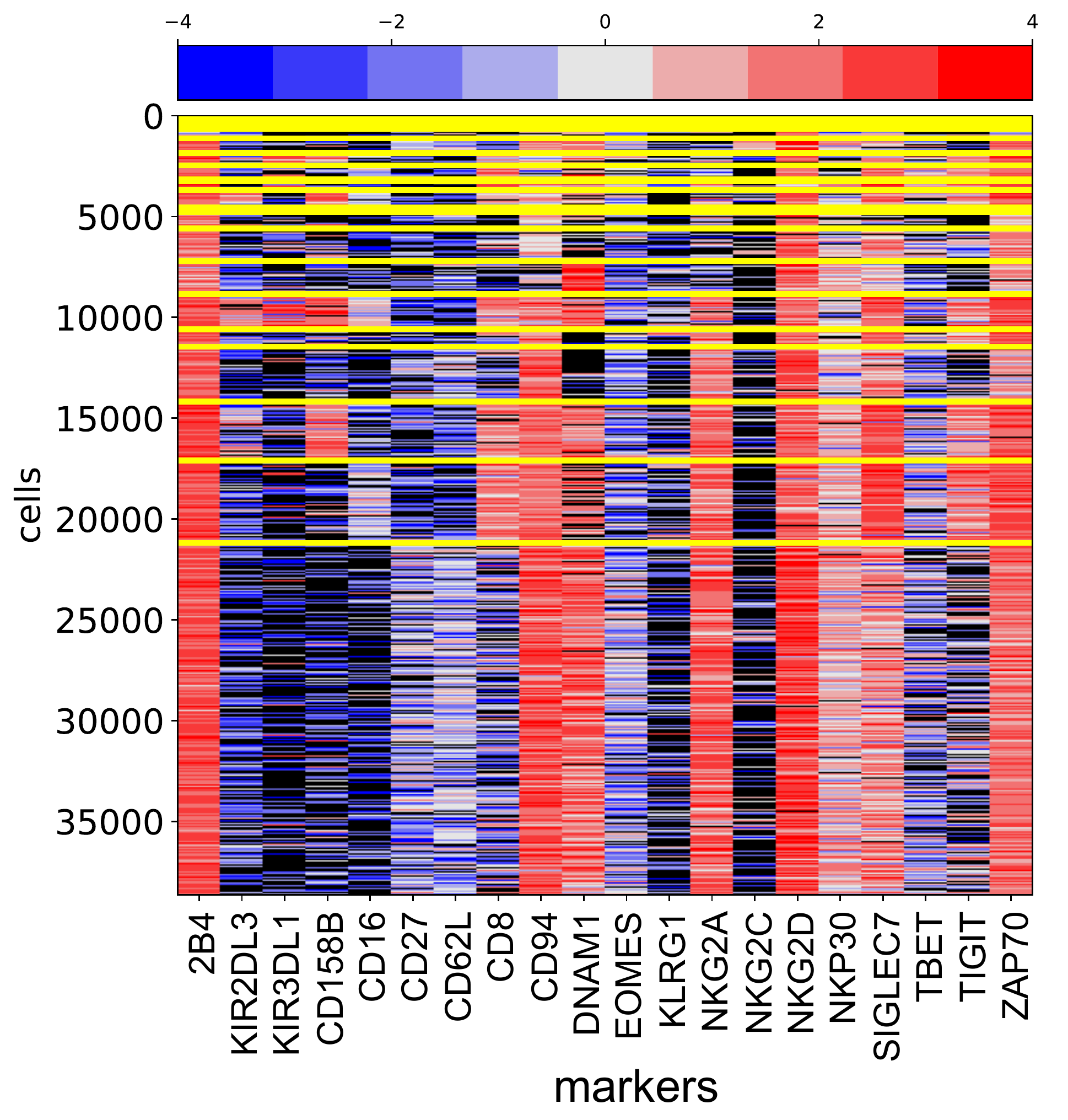}\\
   (a) $\hat{\Z}^\prime_1$ and $\hat{\bw}_1$ & (b) $y_{1nj}$\\
  \includegraphics[width=0.5\columnwidth]{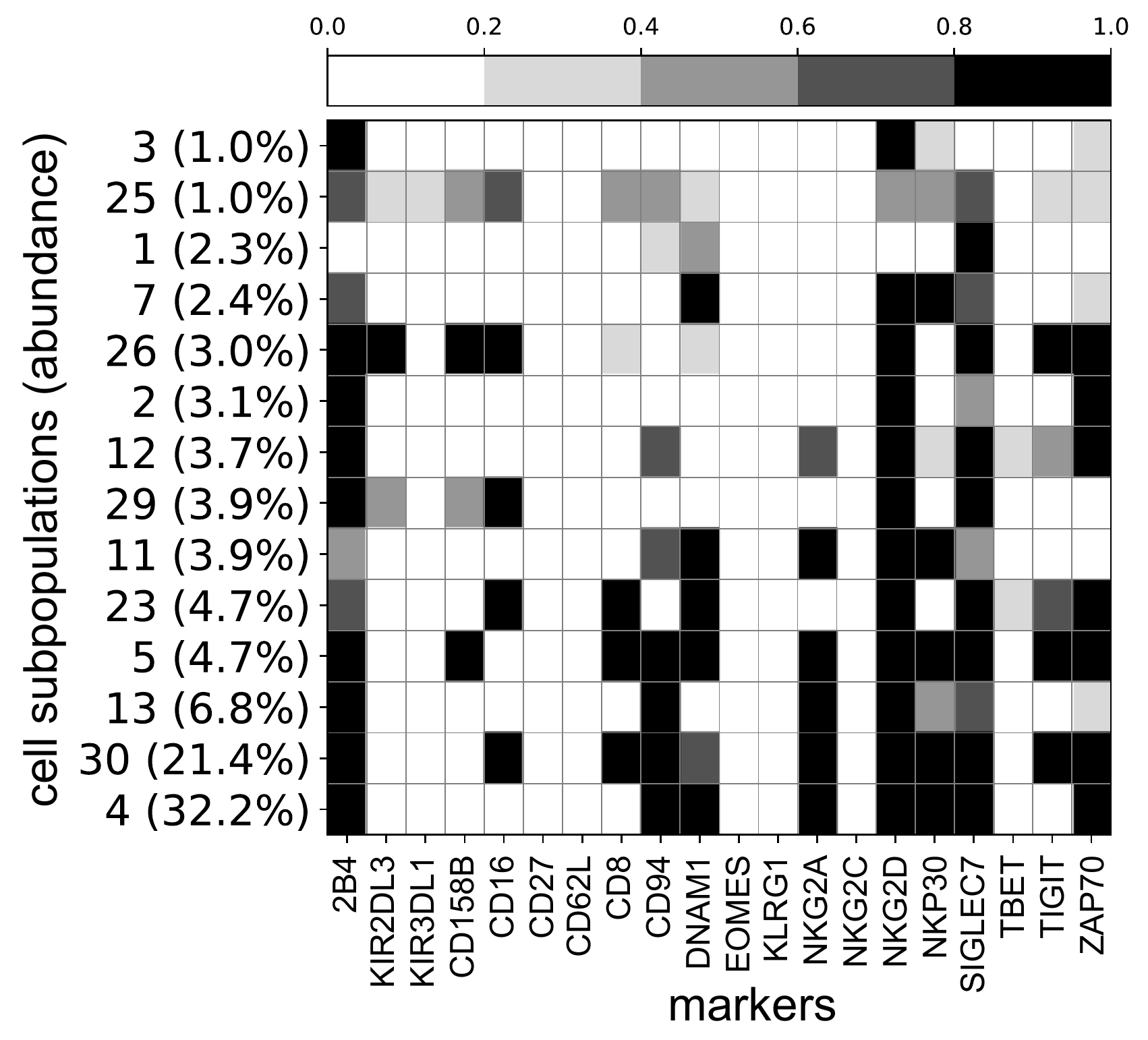} &
  \includegraphics[width=0.5\columnwidth]{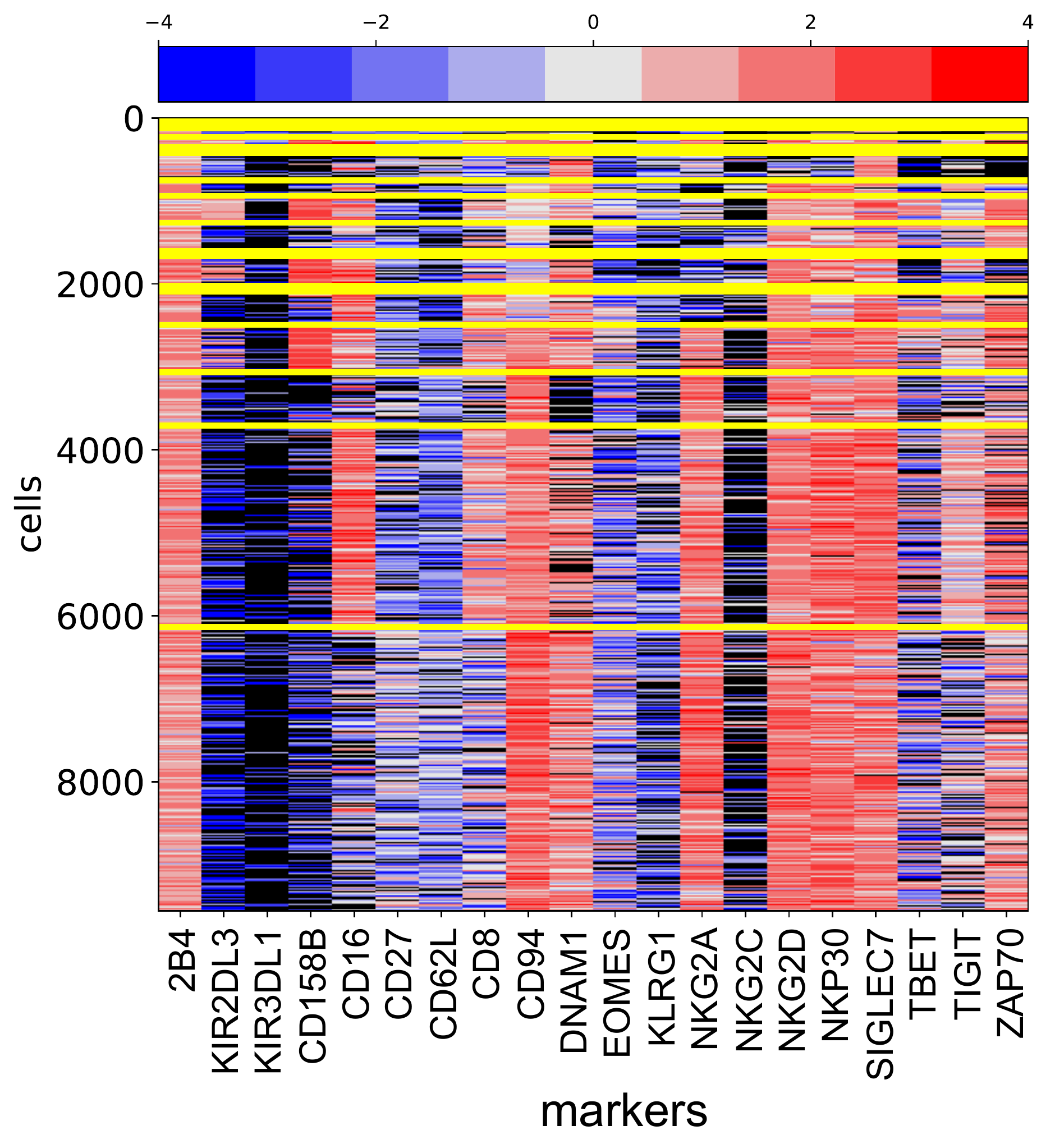}\\
   (c) $\hat{\Z}^\prime_2$ and $\hat{\bw}_2$ & (d) $y_{2nj}$\\
  \end{tabular}
  \end{center}
  \vspace{-0.05in}
  \caption{\small[CB NK cell data]
  Inference obtained by VI is illustrated. 
  $\hat{\Z^\prime}_i$ and $\hat{\bw}_i$ of samples 1 and 2 are
  illustrated in panels (a) and (c), respectively, with markers that are
  expressed dented by black and not expressed by white. Only subpopulations with
  $\hat{w}_{i,k} > 1\%$ are included. Heatmaps of $\y_i$ are shown in panels (b)
  and (d) for samples 1 and 2, respectively. Cells and markers are in rows
  and columns, respectively. Each column contains the expression levels of a
  marker for all cells in the sample. High and low expression levels are red
  are blue, respectively. Missing values are black. Cells are rearranged
  by the corresponding posterior estimate of their subpopulation indicator,
  $\hat{\lambda}_{i,n}$. Yellow horizontal lines separate cells by different
  subpopulations.}
  \label{fig:cb-vb-Z}
% \label{fig:cb-post}
\end{figure}
%%%%%%%%%%%%%%%%%%%%%%%%%

%%%%%%%%%%%%%%%%%%%%%%%%
\begin{figure}[t!]
  \begin{center}
  \begin{tabular}{cc}
  \includegraphics[width=0.5\columnwidth]{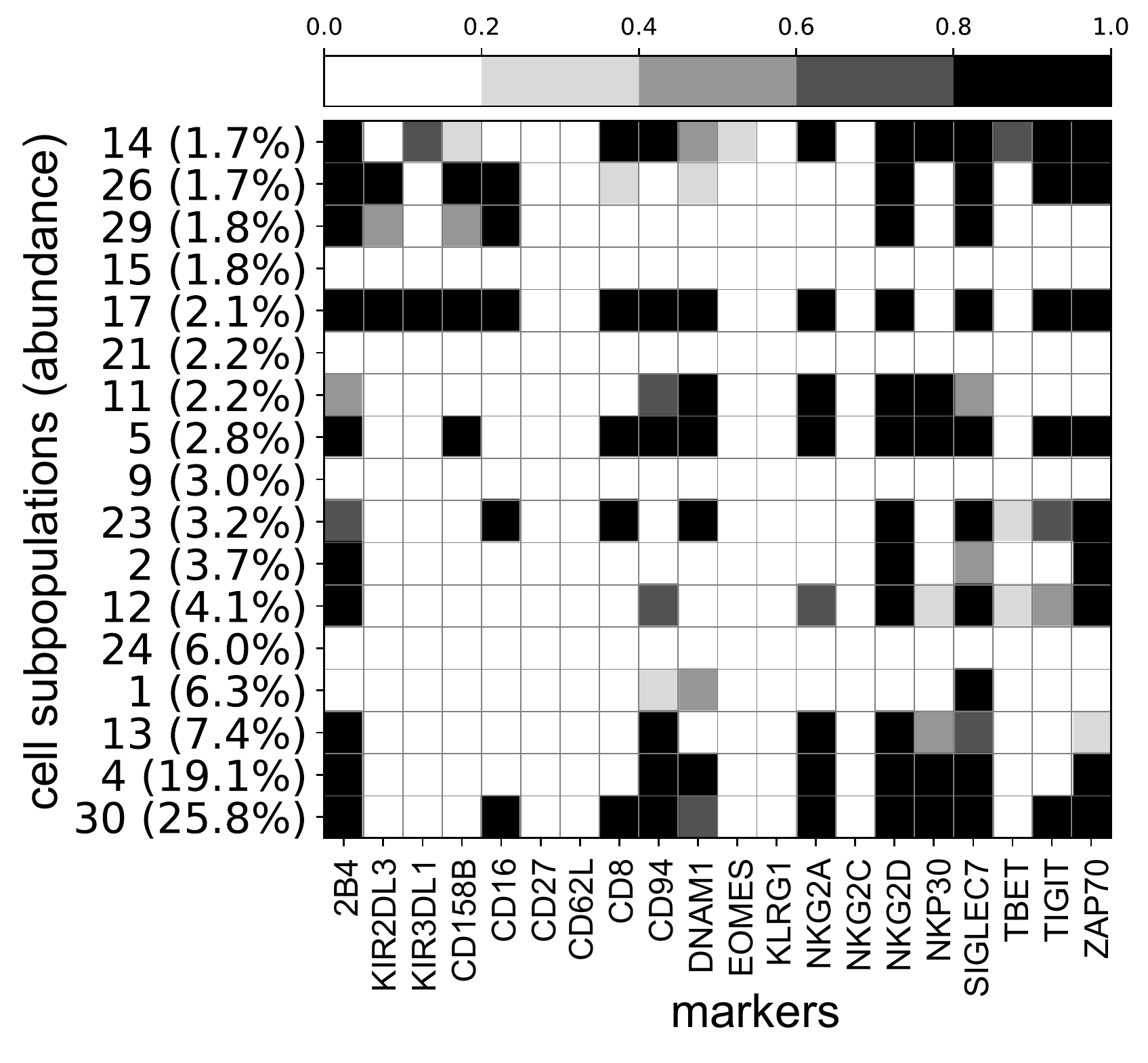} &
  \includegraphics[width=0.5\columnwidth]{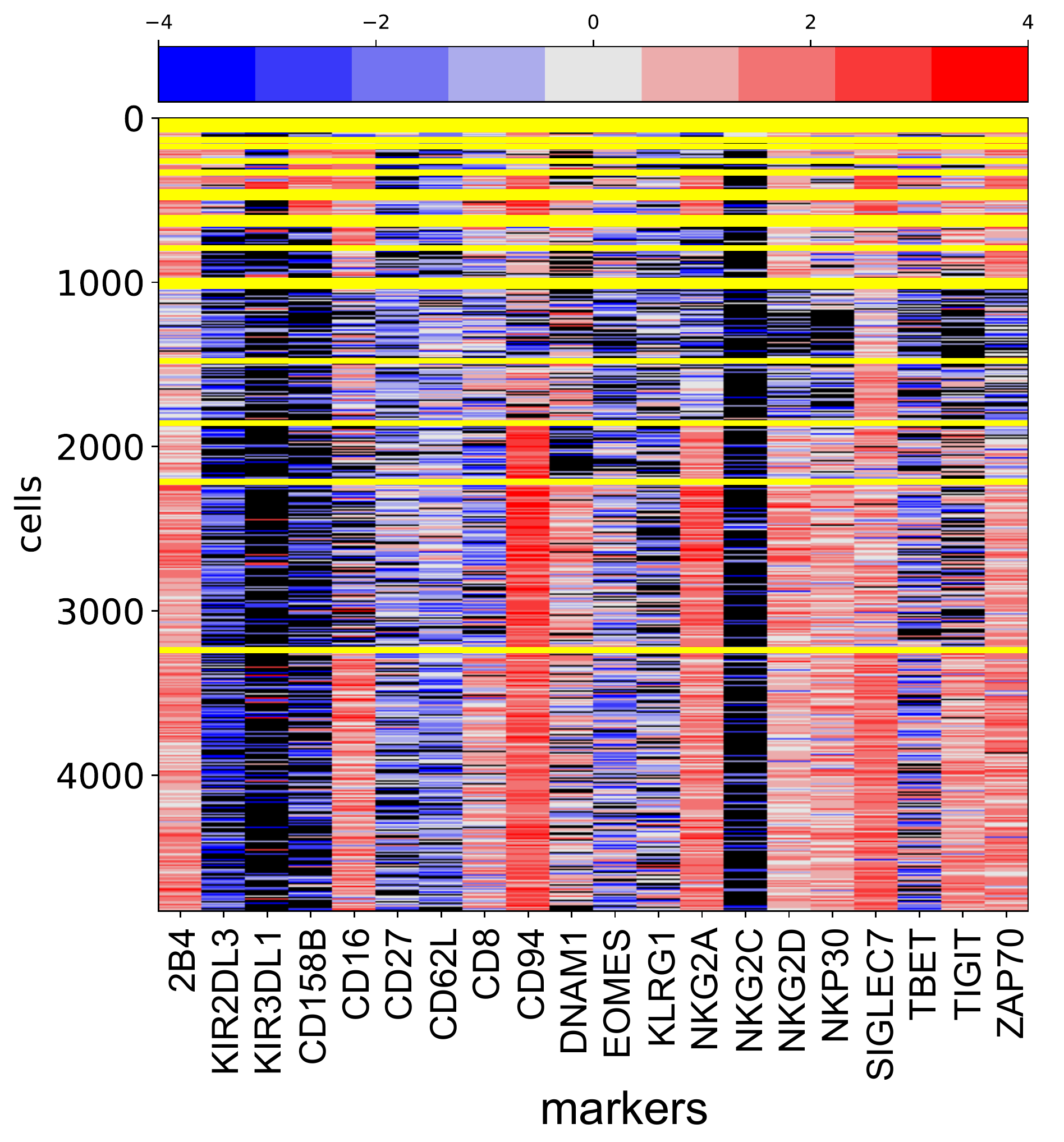}\\
   (e) $\hat{\Z}^\prime_3$ and $\hat{\bw}_3$ & (f) $y_{3nj}$\\
  \end{tabular}
  \end{center}
  \vspace{-0.05in}
  \caption*{Figure~\ref{fig:cb-vb-Z} continued: [CB NK cell data]
  Inference obtained by VI is illustrated. 
  $\hat{\Z^\prime}_i$ and $\hat{\bw}_i$ of sample 3 illustrated in panel
  (e), with markers that are expressed dented by black and not expressed by
  white. Only subpopulations with $\hat{w}_{i,k} > 1\%$ are included. Heatmaps
  of $\y_i$ are shown in panels (b) and (d) for samples 1 and 2,
  respectively. Cells and markers are in rows and columns, respectively.
  Each column contains the expression levels of a marker for all cells in
  the sample. High and low expression levels are red are blue,
  respectively. Missing values are black. Cells are rearranged by the
  corresponding posterior estimate of their subpopulation indicator,
  $\hat{\lambda}_{i,n}$. Yellow horizontal lines separate cells by
  different subpopulations.}
\end{figure}
%%%%%%%%%%%%%%%%%%%%%%%%%
%%% END_OF_SECTIONS -- DO NOT REMOVE!!! %%%

%xxx%\makeatletter\@input{supp-aux.tex}\makeatother

\clearpage
\bibliographystyle{natbib}
\bibliography{main.bib}

%xxx%\makeatletter\@input{main-aux.tex}\makeatother
\end{document}